\pdfoutput=1
\documentclass{aa}  

\usepackage{natbib}
\usepackage{graphicx}
\usepackage{amsmath}
\usepackage{float}
\usepackage{longtable}
\usepackage[toc,page]{appendix}
\usepackage{subfig}
\usepackage{multicol}
\usepackage{multirow}
\usepackage{xcolor}
\usepackage{lscape}

\usepackage{txfonts}
\begin{document}

\title{Exocomets: A spectroscopic survey}

	\author{I. Rebollido
          \inst{1}
\and       C. Eiroa\inst{1,2}
\and       B. Montesinos\inst{3}
\and       J. Maldonado \inst{4}
\and       E. Villaver\inst{1}
\and       O. Absil\inst{5}
\and       A. Bayo\inst{6,7}
\and       H. Canovas\inst{8}
\and       A. Carmona\inst{9}
\and       Ch. Chen\inst{10}
\and       S. Ertel\inst{11}
\and       Th. Henning\inst{12}
\and       D. P. Iglesias\inst{6,7}
\and       R. Launhardt\inst{12}
\and       R. Liseau\inst{13}
\and       G. Meeus\inst{1}
\and       A. Mo\'or\inst{14,15}
\and       A. Mora\inst{16}
\and       J. Olofsson\inst{6,7}
\and       G. Rauw\inst{5}
\and       P. Riviere-Marichalar\inst{17}
       }
   \institute{Dpto. F\'\i sica Te\'orica, Universidad Aut\'onoma de Madrid, 
Cantoblanco,28049 Madrid, Spain,
              \email{isabel.rebollido@uam.es}
\and Observatorio Astr\'onomico de Calar Alto, CAHA, 04550 G\'ergal, Almer\'ia, Spain
\and Centro de Astrobiolog\'ia (CAB, CSIC-INTA), ESAC Campus Camino Bajo del Castillo, s/n, Villanueva de la 
Ca\~nada, 28692, Madrid, Spain
\and INAF, Osservatorio Astronomico di Palermo, Piazza del Parlamento 1, 90134 Palermo, Italy
\and STAR Institute, Universit\'e de Li\`ege, F.R.S.-FNRS, 19c All\'ee du Six Ao\^ut, B-4000 Li\`ege, Belgium
\and Instituto de Física y Astronomía, Facultad de Ciencias, Universidad de 
Valparaíso,  5030 Casilla, Valparaíso, Chile
\and N\'ucleo Milenio de Formaci\'on Planetaria - NPF, Universidad de Valpara\'iso, Av. Gran Breta\~na 1111, Valpara\'iso, Chile
\and European Space Astronomy Centre (ESA), PO Box, 78, 28691 Villanueva de la Ca\~nada, Madrid, Spain
\and Universit\'e de Toulouse, UPS-OMP, IRAP, Toulouse F-31400, France
\and Space Telescope Science Institute, 3700 San Martin Drive, Baltimore, MD 21212, USA
\and Steward Observatory, Department of Astronomy, University of Arizona,  Tucson, AZ 85721, USA
\and Max-Planck-Institut für Astronomie (MPIA), Königstuhl 17, D-69117 Heidelberg, Germany
\and Department of Space, Earth and Environment, Chalmers University of Technology, Onsala Space Observatory, 439 92 Onsala, Sweden
\and Konkoly Observatory, Research Centre for Astronomy and Earth Sciences, Konkoly-Thege Mikl\'os \'ut 15-17, 1121 Budapest, Hungary
\and ELTE E\"otv\"os Lor\'and University, Institute of Physics, P\'azm\'any P\'eter s\'et\'any 1/A, 1117 Budapest, Hungary
\and Aurora Technology B.V. for ESA, ESA-ESAC, Villanueva de la Ca\~nada, 28691, Madrid, Spain
\and Observatorio Astronómico Nacional (OAN-IGN)-Observatorio de Madrid, Alfonso XII, 3, 28014 Madrid, Spain}

   \date{}

 \abstract
 {While exoplanets are now routinely detected, the detection of small bodies in extrasolar systems remains challenging. Since the discovery of sporadic events interpreted as exocomets (Falling Evaporating Bodies) around $\beta$ Pic in the early 80s, only $\sim$20 stars have been reported to host exocomet-like events.}
 {We aim to expand the sample of known exocomet-host stars, as well as to monitor the hot-gas environment around stars with previously known exocometary activity.}
 {We have obtained high-resolution optical spectra of a heterogeneous sample of 117 main-sequence stars in the spectral type range from B8 to G8. The data have been collected in 14 observing campaigns expanding over 2 years from both hemispheres. We have analysed  the Ca {\sc ii} K\&H and Na {\sc i} D lines in order to search for non-photospheric absorptions originated in the circumstellar environment, and for variable events that could be caused by outgassing of exocomet-like bodies. }
 {We have detected non-photospheric absorptions towards 50\% of the sample, attributing a circumstellar origin to half of the detections (i.e. 26\% of the sample). Hot circumstellar gas is detected in the metallic lines inspected via narrow stable absorptions, and/or variable blue-/red-shifted absorption events. Such variable events were found in 18 stars in the Ca {\sc ii} and/or Na {\sc i} lines;  6 of them are reported in the context of this work for the first time. In some cases the variations we report in the Ca {\sc ii} K line are similar to those observed in $\beta$ Pic.  While we do not find a significant trend with the age or location of the stars, we do find that the probability of finding CS gas in stars with larger $v\sin i$ is higher. We also find a weak trend with the presence of near-infrared excess, and with anomalous ($\lambda$ Boo-like) abundances, but this would require confirmation by expanding the sample. }
 {} 

   \keywords{stars: general - planetary systems - comets:general - ISM: clouds - circumstellar matter
                            }

   \maketitle
%
\section{Introduction}
Main sequence (MS) stars are known to host a complex circumstellar (CS) environment populated by a plethora of planets, debris discs, and minor bodies, inherited from the physics that regulates the formation of stars. Later, the mutual dynamical interaction among those bodies and their host stars determines the evolution of the planetary systems.

Since the discovery of a giant planet orbiting the solar-type star 51 Peg \citep{mayor95}, thousands of planets have been detected, which make up planetary systems with diversified architectures (see e.g. The Extrasolar Planets Encyclopaedia\footnote{exoplanet.eu}). 
Minor bodies, such as asteroids and comets, are expected in these planetary systems. Their study is relevant as they provide clues for understanding the formation and dynamical evolution of planetary systems \citep[e.g.][]{armitage10}. However, evidences of their presence are practically limited to indirect observations, such as the detection of circumstellar dust and gas in debris discs and, to a considerable less extent, to more direct evidences such as observations of some  metallic line absorptions.

Hundreds of debris discs are currently known to surround MS stars across practically all spectral types, and ages from around 10 Myr to several Gyr.  Debris discs are detected as thermal
emission at \mbox{mid-/far-IR/(sub-)mm} wavelengths, as well as scattered light in the optical and near-IR, from small dust particles, which are  mainly originated in collisions among km-sized planetesimals and other destructive processes 
\citep[e.g][and references therein]{matthews14}.
It has been suggested that at least a fraction of debris discs posseses both warm (T $\sim$ 200 K) and cold (T $\lesssim$ 100 K) dust belts \citep[e.g.][]{ballering13,chen14,kennedy14,pawellek14,morales16}, reminiscent of the Solar asteroid and Kuiper belts respectively. The two-belt structure could be created by a chain of planets, while comets scattered by those planets could constitute a relevant source feeding the warm exozodiacal belt \citep{schuppler16,geiler17,marino18a}. In our own solar system cometary material from Jupiter family comets is responsible for replenishing the zodiacal cloud \citep{nesvorny10}.

Interferometric studies \citep{absil06,absil13,ertel14,nunhez17} have revealed near-IR excesses  also attributed in most cases to the combination of thermal emission and scattered light from small submicron-sized, hot (T $\sim$ 1500-2000 K) dust particles located within $\sim$ 0.01 - 1 AU from the stars \citep[depending on the luminosity, see][]{kirchschlager17}, very close to the dust sublimation zone. 
Again, cometary bodies scattered inwards from an outer reservoir is a likely scenario for the origin and persistence of the hot dust \citep{ beustmorbidelli00,thebaultbeust01,bonsor14,marboeuf16,faramaz017,marino18a,sezestre19}. 

Significant amounts of cold gas at several tens AU from the central stars (most young A-type stars) have been detected around $\sim$ 20 bright debris discs \citep[e.g.][]{moor15a,moor17,riviere12,riviere14,roberge13,greaves16,liemansifry16,marino16}. 

The cold gas most likely  has a secondary origin \citep[e.g.][]{marino16,matra17a,kral18}, but in some cases a remnant of the primordial protoplanetary disc has been suggested \citep{kospal13,moor15a,kral17b}.
A variety of physical processes, including outgassing of cometary objects, have been invoked to explain the secondary cold gas, all of them related to the presence of planetesimals \citep[e.g.][and references therein]{matthews14}. In parallel, UV/optical high resolution spectroscopy reveals non-photospheric stable absorption features of elements such as C, O, Ca, Na or Fe at or close to the radial velocity of the star, as well as weak sporadic red- or blue-shifted absorption events with respect to the stellar radial velocity  \citep{vidalmadjar94,vidalmadjar17,lagrange98,brandeker04,roberge06,Iglesias18,Rebollido18}. This gas would be hot with temperatures $\sim$ 1000 -- 2000 K \citep{hobbs88,beust98,vidalmadjar17}. The transient absorptions have been interpreted as hot gas released by the evaporation of exocomets grazing or falling onto the star, the "Falling Evaporating Bodies" scenario or FEBs  \citep[e.g.][]{ferlet87,beust90}, while additionally grain evaporation in the circumstellar disc has also been proposed to explain the stable absorption components \citep[e.g.][]{fernandez06}. A trend between the detection of this hot gas and the edge-on orientation of cold-gas-bearing debris discs has been found by \cite{Rebollido18}, who attributed it to a geometrical effect. It is remarkable that this hot gas would be located at distances $\lesssim$ 0.5 AU from the star, i.e., at similar distances as the hot exozodiacal dust.

The first direct  evidence for the presence of exocomets (FEBs) was found around $\beta$ Pic, which remains unique and the best studied 
theoretically and observationally 
\citep[e.g.][and references therein]{ferlet87,beust91,kiefer14b,vidalmadjar17}. 
Several hundreds of cometary transits in $\beta$ Pic have been detected, revealing two families of exocomets with distinct dynamical and compositional properties; one likely formed by old comets strongly depleted in volatiles, and a second one related to the recent fragmentation of one or few parent bodies \citep{kiefer14b}. It is worth to note that very recently $\beta$ Pic transiting exocomets have likely been detected by means of $TESS$ broadband photometry \citep{zieba19}. Those exocomets would have been driven into the vicinity of the star by a larger body, i.e., a planet \citep{beust91,beustmorbidelli00}. Thus, FEBs constitute an indicator of the plausible presence of planets; we recall that the Jupiter-like planet $\beta$ Pic b, was later  revealed by imaging \citep{lagrange10}.

In addition to $\beta$ Pic, absorption events, mainly in the Ca {\sc ii} K line accompanied in most cases by a stable component, have been detected towards more than 20 stars \cite[e.g.][]{redfield07a,kiefer14a,welsh15,welsh18}. Those stars are usually young (< 100 Myr) A-type stars, but transient features have also been found around older stars. Also,  \cite{welsh19} have recently reported the first detection of an exocomet-like event  with a 2.9$\sigma$ signal around the F2 V type star $\eta$ Crv. 
In addition, $Kepler$ photometric light curves have been explained as due to transiting exocomets in a few F- and later spectral type stars \citep{rappaport18,ansdell19} and maybe Boyajian's star \citep{boyajian16}. 
Exocomet-host stars have large projected rotational velocities, most with $v \sin i \gtrsim  $100 km/s, in principle suggesting that the systems are viewed close to edge-on, which is the favoured orientation to detect non-photospheric absorptions from comet-like bodies passing in front of the star, and consistent with the trend suggested by \cite{Rebollido18}. In some cases, the stars hosting exocomets are associated with a debris disc. We note, however, that scattering of exocomets by eccentric planets can take place in planetary systems with low luminous, non-detectable debris discs with flux levels comparable to the Kuiper belt \citep{faramaz017}.

This work presents the observational results of a high spectral resolution survey of a large sample of stars with the primary aim of detecting and monitoring non-photospheric absorption features due to the passing of exocomets in front of the stars, and as secondary goal to assess any potential trend between the presence of exocomets and the properties of the host stars. The paper is structured as follows. Section 2 describes the sample of stars, the criteria to select them, and some basic properties. Section 3 presents our spectroscopic observations and data analysis. Section 4 presents our spectroscopic results concerning non-photospheric stable and variable features, their plausible interstellar (ISM) or circumstellar (CS) origin, and comments to some individual stars. Section 5 discusses the detection of the non-photospheric absorptions regarding some stellar properties, as well as with respect to the selection criteria. Finally, Section 6 presents the conclusions of this work.

\section{Sample}
\label{sect:sample}

As mentioned above the primary goals of our study  are the detection and  monitoring of variable, non-photospheric absorption features that, as in the case of $\beta$ Pic, could be attributed  to the evaporation of solid bodies in the immediate surroundings  of main-sequence stars. The observed stellar sample is formed by a heterogeneous and biased set of 117 MS stars in the spectral type range from B8 to G8, aiming at optimising those goals. The targets have been selected according to the following criteria: i) stars with previously reported Ca {\sc ii} H\&K and/or Na {\sc i} D events attributed to exocomets; ii) debris disc stars, seen edge-on when known; iii) debris disc stars where the presence of cold gas has been detected at far-IR and \mbox{(sub-)mm} wavelengths; iv) stars with near-IR excesses that could be due to hot dust; v) stars belonging to young associations, namely Upper Scorpius (UpSco), Tucana-Horologium (Tuc-Hor), and the $\beta$ Pictoris moving group (BPMG); vi) shell stars with circumstellar Ti {\sc ii} absorptions; vii) $\lambda$ Bootis stars. The Ti {\sc ii} stars have been selected because those lines denote the presence of discs seen at nearly right angles to the rotational axes \citep{abt73,abt97}, i.e. hot discs seen near edge-on - we note that in general shell stars are a heterogenous group of late B- to early F-type with the distinct characteristic of enhanced lines of Fe {\sc ii} and Ti {\sc ii} denoting the presence of a gaseous circumstellar shell \citep[][and references therein]{gray09}. In addition, $\lambda$ Bootis stars are A and early F spectral type stars strongly depleted in heavier elements (such as Fe, Al, Mg, Ca...), and relatively normal abundances of volatile elements like C, N, O, and S \citep{paunzen04}, and some of them show clear evidences of accreting CS gas in their UV/optical spectra \citep{grady96a,holweger99}. The $\lambda$ Bootis abundance pattern is most likely due to selective accretion of the volatile elements onto the star, material that could be provided by exocomets, although other sources could also be an alternative \citep{jura15,draper16}. We note that some stars in the sample share several of the selection criteria. 
 
\begin{figure}[h]
\centering
\includegraphics[width=0.5\textwidth]{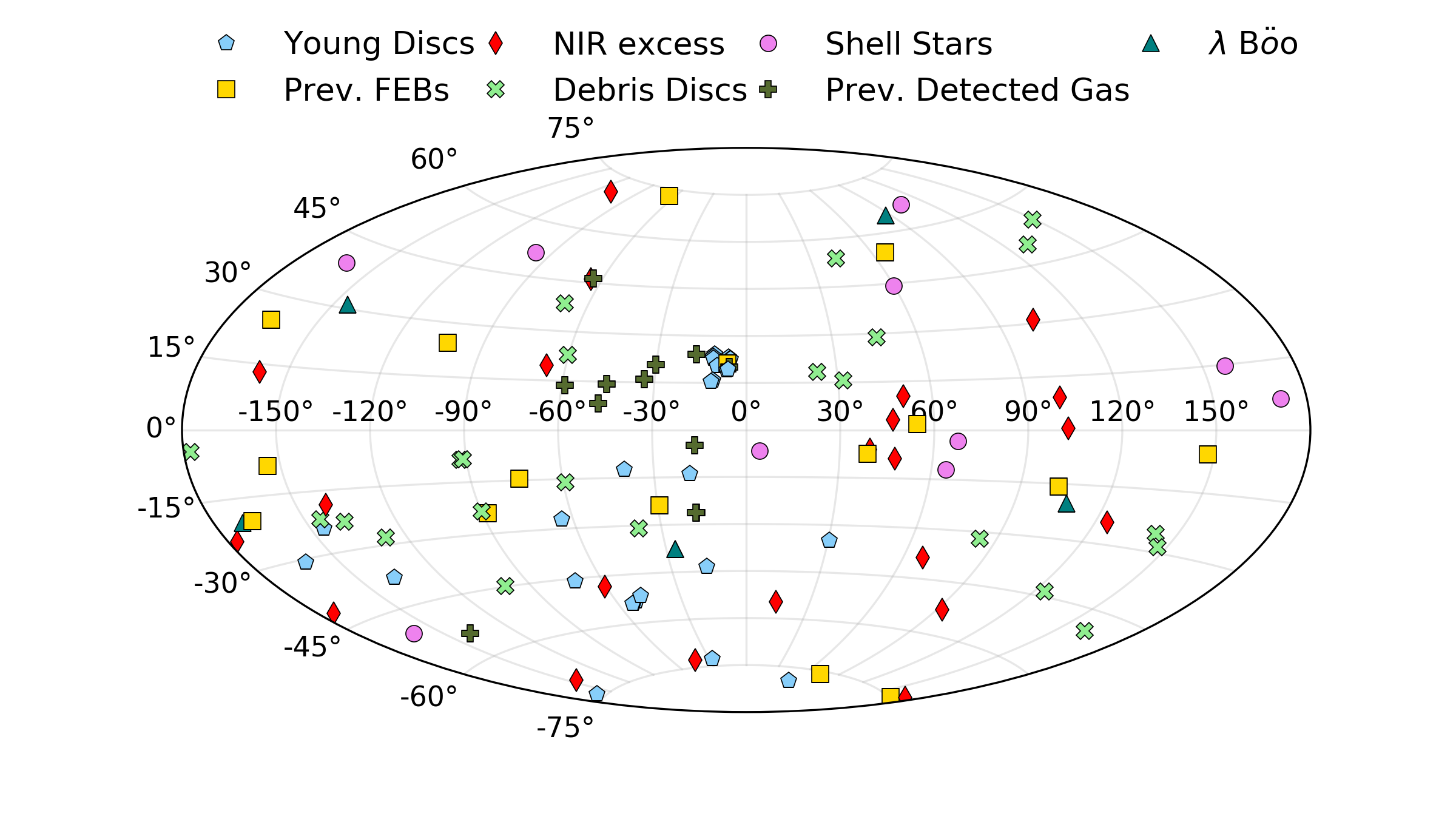}
\caption{All sky plot of the sample in galactic coordinates. Although most stars are in the Southern Hemisphere, there is no preferential spatial location, with the exception of stars in the UpSco and Tuc-Hor young associations.}
\label{fig:allsky}
\end{figure}

Table \ref{tab:sample} lists the observed stars. Columns 1 to 9 provide HD number, other names, 2000.0 equatorial coordinates, spectral types, distances, apparent $V$-magnitudes, $B-V$ colour indexes, and radial velocities v$_{\rm rad}$, all taken from SIMBAD \citep{simbad}. In addition, column 10 gives ages; column  11 gives the fractional luminosity of the dust, L$_{\rm  IR}$ /L$_{\rm star}$, for those stars where a debris disc has been  detected; column 12 gives the corresponding stellar association; and column 13 gives the  primary selection  criteria.  The corresponding references  for columns 10-13 are given within brackets. Fig. \ref{fig:allsky} shows the spatial distribution of the sample in galactic coordinates. Given the characteristics of the selection criteria, there is no preferencial location with the exception of stars in the UpSco and Tuc-Hor young associations, although most stars are in the Southern Hemisphere.
  
\section{Observations and data analysis}

\begin{table*}[t]
\centering
 \caption{\label{tab:obs_instruments}Instruments      and     observing
  campaings.} 
\begin{tabular}{lllll}
\hline
\hline
 &HERMES          & FIES                & FEROS              & HEROS$^!$ \\
 \hline 
Resolution & $\sim$ 85000 & $\sim$ 67000 & $\sim$ 48000 & $\sim$ 20000\\
Range (nm) & 377-900 & 364-736 & 352-920 & 374-884\\
\hline
Observations & 03-06/Sep/2015 & 26-27/Jan/2016 &21-24/Oct/2015 & Aug 2015\\
& 20-23/Dec/2015&16-19/Jul/2016&25-28/Mar/2016 & Sept 2015\\
& 27-30/Jan/2016&03/Mar/2016$^*$& 31/Mar-08/Apr/2017 & Oct 2015\\
& 03-06/Mar/2016&21/Mar/2016$^*$& 23/Sep-01/Oct/2017& Nov 2015\\
& 08-11/May/2016& & & Dec 2015\\
& 11-14/Jul/2016& & & Jan 2016\\
& 06-03/Apr/2017& & & \\
& 28/Mar-03/Apr/2017 & & & \\
\hline
\end{tabular}\\
(*) On March 3$^{rd}$ and 21$^{st}$ during  service mode  only one  spectrum was obtained each night. (!) HEROS spectra were taken in robotic mode during several months as complementary observations.
\end{table*}

High-resolution observations were taken in a series of campaigns from August 2015 to September  2017 from both the Northern and Southern hemispheres, using different  fiber-fed spectrographs and telescopes. The spectrographs HERMES \citep{Raskin2011} attached  at the 1.5 m Mercator Telescope, and FIES \citep{FIES} at the 2.5 m Nordic Optical Telescope (NOT) were used in El Roque de los  Muchachos Observatory (Canary Islands, Spain). In La Silla Observatory (Chile), the FEROS spectrograph \citep{FEROS} at the   MPG/ESO 2.2 telescope was used. Complementary observations were  obtained at La Luz Observatory (M\'exico) with the TIGRE  telescope and the HEROS spectrograph \citep{tigre}. Table \ref{tab:obs_instruments} summarizes the resolutions and wavelength ranges of the various instruments, and the observing campaigns. A total of 1575 spectra for the 117 stars were obtained; a detailed observing log with the specific dates (UT) of the spectra for each star and the corresponding spectrograph is given in the Appendix in Table \ref{tab:log}. Due to the nature of the irregular, sporadic exocometary-like events, we aimed at obtaining time-series spectra, with at least one spectrum per night per object when possible. Integration times never exceeded 30 minutes and were mainly selected depending on the telescope and brightness of the star with the general goal of obtaining a S/N ratio $\gtrsim$ 100 in the Ca {\sc ii} H\&K lines. That goal was not always achieved due to either poor weather conditions or the faintness of the star. Signal to noise (S/N) of the median spectra for each star measured in the continuum close to Ca {\sc ii} K line is given in Table \ref{tab:log}.

Data reduction was performed by the available pipelines of  the different spectrographs. The reduction includes the  usual steps for echelle spectra as bias subtraction, flat-field correction, cosmic ray removal, and  order extraction; wavelength calibration  is carried out by means of Th-Ar lamp spectra.  In addition, barycentric corrections for the HERMES and FIES spectra  were carried out as the corresponding pipelines do not include such correction.
All spectra were normalised and the continuum set to 1.0 in the regions between spectral lines.

\subsection{Telluric subtraction}
  
\begin{figure}[h]
\centering
\includegraphics[width=0.5\textwidth]{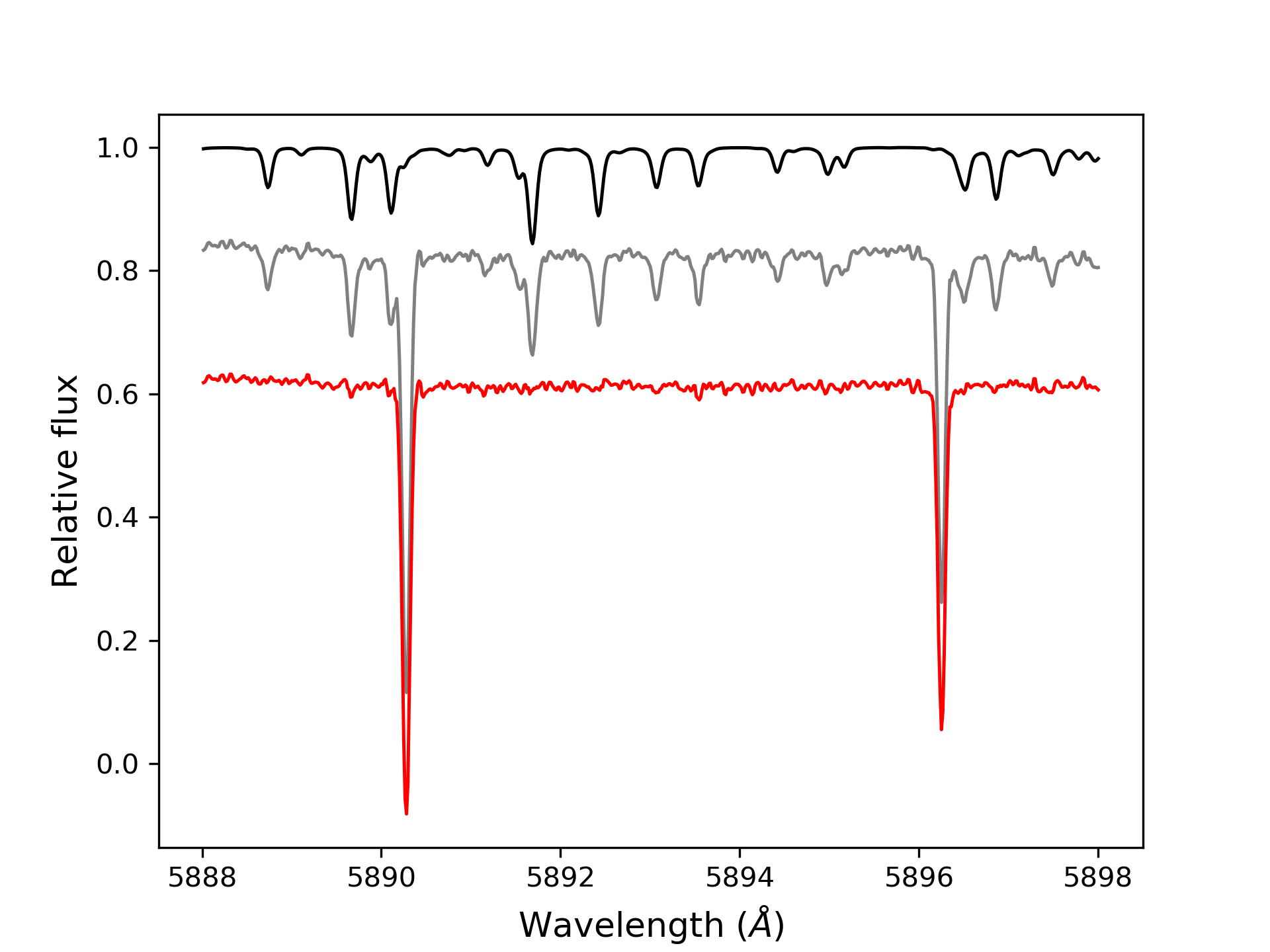}
\caption{Example of telluric subtraction in HD 21620 in the {\mbox{Na {\sc i} D}} spectral region. Black  line shows the MOLECFIT atmospheric model; grey line shows the observed spectrum; and red  line shows the final telluric-free HD 21620 spectrum.}
\label{fig:telluric}
\end{figure}

The observed wavelength range includes regions of the visible spectra heavily affected  by telluric lines;  in particular the  region around the 5890/5896 \AA ~Na {\sc i} D lines, which are highly relevant for both interstellar and  circumstellar absorptions. Removal of telluric lines was performed by means of MOLECFIT\footnote{http://www.eso.org/sci/software/pipelines/skytools/molect }(\citealt{Molecfit1}, \citealt{Molecfit2}), a tool that generates an atmosphere model accounting for the most  common absorbing molecules in the optical range (H$_2$O, O$_2$, O$_3$). Residuals after subtraction of the atmosphere model in the telluric line region are comparable to the noise level, and therefore negligible. An example of telluric subtraction is shown in Fig. \ref{fig:telluric}.

\subsection{Stellar parameters}
\label{sect:synthe}

Stellar parameters $T_{\rm eff}$, $\log g$ and $v \sin i$ for the early type stars (up to F2) in the sample were computed using the procedure described by \cite{Rebollido18}. Solar abundances were used initially to iterate the solutions for all objects, the solutions being consistent with that metallicity for most of the stars. For the 15 objects with [M/H]$\leq\!-0.5$ according to our estimates, eight of them, namely HD 31295, HD 74873, HD 110411, HD 125162 ($\lambda$ Boo), HD 183324, HD 198160 (HR 7959), HD 198161 (HR 7960) and HD 221756, were classified as $\lambda$ Boo stars by \cite{murphy2015}, consistently with their expected underabundace in heavier elements.

For cooler stars (later than F2), a different approach was used to calculate parameters. In those cases we have followed the procedure described in detail by \cite{maldonado17} and \cite{maldonado18}, which is based on the iron ionisation and excitation equilibrium, and match of the curve of growth conditions.

\begin{figure}[h]
\centering
\includegraphics[width=0.5\textwidth]{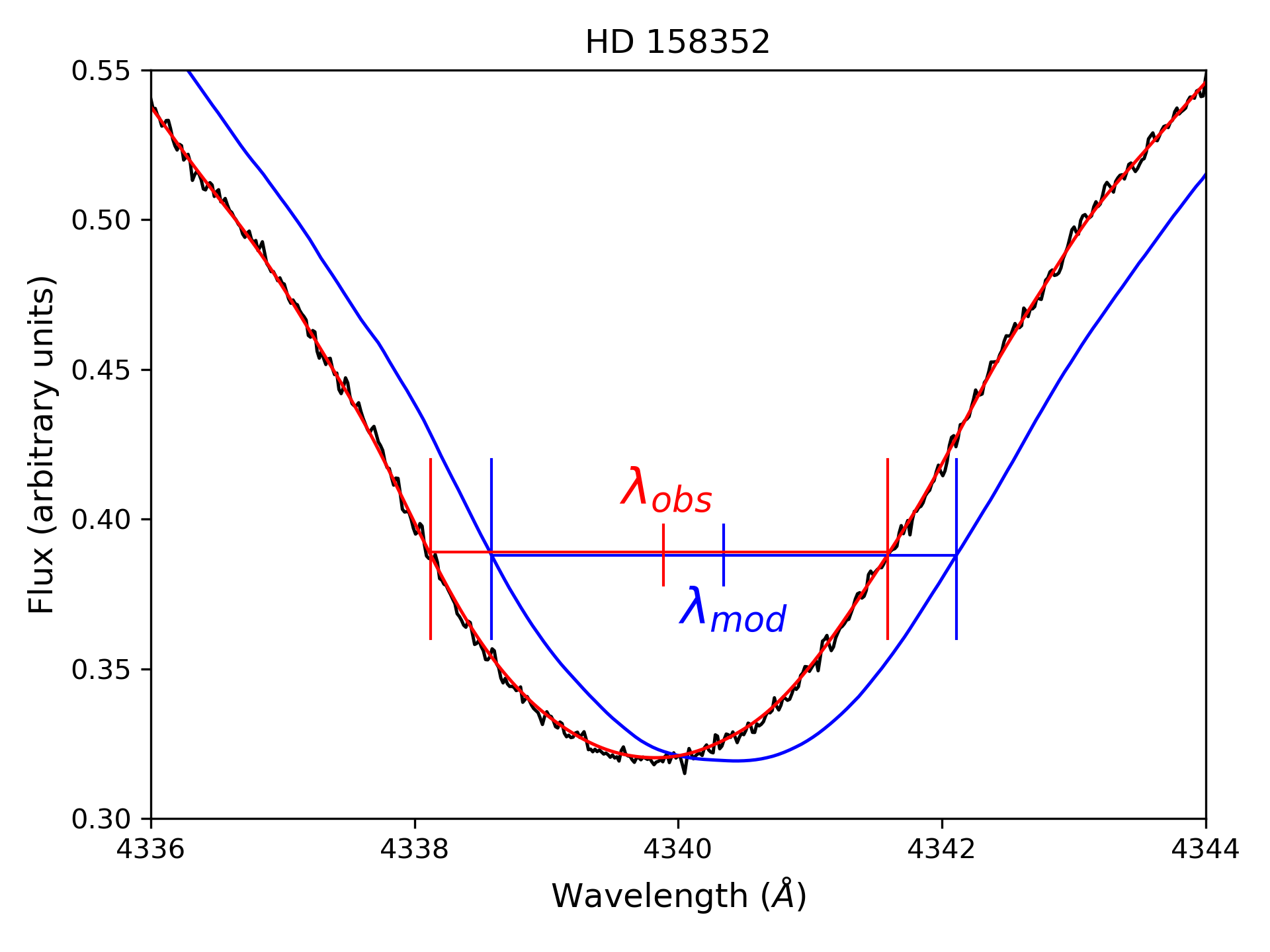}
\caption{Radial velocities are measured as the difference between the synthetic model at rest (blue line) and the smoothed spectra at the radial velocity of the star (red line). The method is illustrated in the plot by means of the  observed H$\gamma$ line in the star HD 158352 (black line). A detailed description is given in the text. }
\label{fig:vrad_calc}
\end{figure}

Radial velocities (v$_{\rm rad}$) were estimated by measuring the  shift between the synthetic spectrum, which is computed using a database of rest wavelengths, and the observed spectrum, corrected  for barycentric velocity. Individual shifts  were measured  for the  Balmer lines from H$\gamma$ to H9. H$\epsilon$, which is blended with Ca {\sc ii} H, was excluded. Lines bluer than H10 were  discarded since the lower part of the absorptions were usually  noisy. Most of the stars have large values of the projected  rotational velocity, $v\sin i$,  therefore the cores of the lines, and in particular the Balmer ones, are fairly rounded; thus, a direct evaluation of the wavelength at which the minimum intensity occurs introduces large uncertainties. The procedure we have followed is to slightly smooth the spectra, and then take as the reference wavelength that of the bisector corresponding to 10\% of the line intensity measured from the bottom of the  absorption. 
The same was done on the synthetic profile and the difference was converted into v$_{\rm rad}$; the results do not change significantly if instead of using the synthetic spectrum, the rest laboratory wavelength of the particular Balmer line was taken as reference. The v$_{\rm rad}$ uncertainties come from  the standard deviation of the set of displacements. Fig. \ref{fig:vrad_calc} shows the method explicitly.

\subsection{Non-photospheric absorptions}
\label{results}

The Ca {\sc ii} H\&K and Na {\sc i} D lines were visually inspected to search for the presence of non-photospheric, stable or variable, absorption features at the core of the photospheric lines, which would suggest the presence of circumstellar gas.  A median spectra (when data from different telescopes were available, all spectra were converted to the FEROS resolution in order to construct the median) was constructed for each star to improve the signal to noise ratio in order to analyse the potential stable absorption, This analysis has been carried out by estimating the photospheric contribution with splines fitted to the bottom of the lines, and then dividing the observed profiles by the estimated photospheric lines, and finally fitting Gaussians to the residuals. 

A total of 60 stars show stable, non-photospheric Ca {\sc ii} and/or Na {\sc i} absorptions (Table \ref{tab:narrow}); at the same time, irregular, variable features are seen in individual spectra of 18 stars (Table  \ref{tab:feblist}). For these latter stars, when no variations were detected in a range of hours/days, a median was constructed including all the consecutive non-variating spectra.
The stable absorptions, when arising in the CS environment (see Sect \ref{sect:origin}), appear at the radial velocity (v$_{\rm rad}$) of the stars or close to it, while the variable ones appear red- and/or blue-shifted.
  
In the following we present a separate analysis for  stars only showing stable non-photospheric absorptions, and stars with variable components regardless of whether they also have stable absorptions.
\section{Results}

\subsection{Narrow, stable absorption features}
\label{sect:stablegas}

As mentioned above, a total of 60 out of the 117 stars in the sample show narrow, stable absorption features superimposed on the photosphere, either in the Ca {\sc  ii} and/or the Na {\sc i} lines. 
Fig. \ref{fig:na_absorptions} shows the observed line profiles for both the Ca {\sc ii} H\&K and Na {\sc i} D lines of the 60 stars, along with their residuals, once the photosphere has been divided. Table \ref{tab:narrow} lists radial velocities (RVs), equivalent widths (EWs), column densities (N), and the velocity dispersion (FWHM of the Gaussian fits) of the non-photospheric median stable absorptions, as estimated for the {\mbox{Ca {\sc ii} H\&K}} and Na {\sc i} D lines. EWs were calculated with respect to the adjacent continuum, once divided the photospheric contribution. The EW ratio for the Ca {\sc ii} and Na {\sc i} doublets expected from the atomic properties of the transition (range 0.5 -- 1.0  in both cases encompassing the optically thin and optically thick regimes) is not always maintained, due to the EW uncertainties and the very complex geometry and composition of the CS gas. Column densities were estimated following \cite{somerville88} in order to deal with the saturated lines.
Uncertainties for the features' radial velocities are estimated as two pixels of the median spectra corresponding to $\sim$2.5 km/s in Ca {\sc ii} lines and {\mbox{$\sim$1.5}} km/s in Na {\sc i} lines, and being this value consistent with the FWHM of the lines of the calibration lamps. In the case of the EWs, we estimate an uncertainty of 10\% in our measurements. Uncertainties for column densities were calculated as propagation of the EW uncertainties according to the formulas for the non-saturation and saturation cases \citep[Eq. 1 and 7 in][respectively]{somerville88}. In some stars, most previously classified as shell stars, the Ca {\sc ii} K line has a sharp, very pronounced triangular-like profile consisting in a very narrow core and broader wings, which cannot be fitted by either a Gaussian or a Voigt profile; in those cases, we have measured the EWs of the excess absorption with respect to the photospheric line and the velocity dispersion as the FWHM of the mentioned excess absorption, i.e., the velocity dispersion at one half of the absorption depth.

\subsubsection{Origin of the non-photospheric stable absorptions}
\label{sect:origin}

Stable non-photospheric absorption features might originate in the close-in CS environment of the star, or in the warm and cold clouds of the local interstellar medium, ISM \citep[e.g.][]{redfield08}. In order to try to decipher the origin, we have compared the radial velocity of those features with the radial velocities (v$_{\rm rad}$) of the corresponding stars (Table \ref{tab:parameters} and Table \ref{tab:narrow}) and  with the velocity vectors of the local ISM (v$_{\rm ISM}$) towards the line of sight of  each star, as given for the ISM clouds in Redfield \& Linsky's Colorado model\footnote{http://sredfield.web.wesleyan.edu/}. Velocities and names of the clouds are also given in Table \ref{tab:narrow}. Nonetheless, a sound ascription to any of both CS or ISM origins is ambiguous in some cases; for instance, when the stellar v$_{\rm rad}$ and the ISM v$_{\rm ISM}$ along its line of sight are very close, or when there is no identified ISM Colorado cloud along the line of sight towards some stars with narrow Ca {\sc  ii} and Na {\sc i} absorptions. Further, the properties of the non-photospheric absorptions do not clearly discriminate the plausible origin \citep{redfield07b} since the observed EWs in the ISM of the Ca {\sc ii} H\&K and of the Na {\sc i} D lines and their ratios vary by several orders of magnitude \citep[e.g][]{welty96,redfield08,welsh10}. We note that Na {\sc i} is found in cold ISM gas ($\sim$ 50 K) typically at distances larger than $\sim$ 80 pc \citep{welsh10}, although is also occasionally found at shorter distances \citep[e.g.][]{bertin93,welty94}; whereas Ca {\sc ii} appears in much warmer medium ($\sim$ 5000 K) and is usually detected at much shorter distances \citep[e.g.][]{redfield2008}.   

In this context, and taking into account the mentioned caveats, we identify stars with CS gas those that satisfy any of the following criteria: i) stars with variable absorptions (see section  \ref{variable}); ii) stars where the non-photospheric absorption shares its velocity with the radial velocity of the star but not with any  ISM  Colorado  cloud; iii) stars with Ti {\sc ii} absorption lines and shell stars where the Ca {\sc ii} K line has a sharp, triangular-like profile. We point out that some stars have more than one non-photospheric  feature that can be independently identified with the velocity of the star or of the ISM. Also, for some stars the absorptions  neither coincides with v$_{\rm rad}$ nor with v$_{\rm ISM}$. 
When possible, an inspection for nearby stars was made in a field of 5$^{\rm o}$ in radius and distances $\pm$ 20 pc around each star.
A column in Table \ref{tab:narrow} shows our guess for the origin of the non-photospheric absorptions.

\subsubsection{Comments on individual stars}
\label{sect:individual_stars}

In this section we discuss the plausible origin of the stable non-photospheric absorptions for some individual stars or group of stars. Stars with variable absorption features detected in this work are discussed in section \ref{variable}. HD 9672 (49 Cet), HD 32297, HD 110058, HD 131488, HD 131835, HD 138813, HD 146897, HD 156623, HD 172555 and HD 181296 ($\eta$ Tel) were analysed in \cite{Rebollido18}; therefore, we do not repeat the discussion here. We note, however, that Na {\sc i} D features in the stars HD 9672, HD 156623, and HD 172555 were erroneously omitted by \cite{Rebollido18}. A corrigendum can be found in \cite{rebollido19a}. We point out that this mistake does not affect the main conclusions of \cite{Rebollido18}. The Na {\sc i} D features are nonetheless included in Table \ref{tab:narrow} and shown in Fig. \ref{fig:na_absorptions}. In HD 9672 the Na {\sc i} D absorption appears at the velocity of the Ca {\sc ii} feature, the radial velocity of the star and the velocity vector of the local interstellar cloud (LIC) cloud. In the case of HD 156623 the Na {\sc i} D feature appears in emission at the stellar v$_{\rm rad}$. For HD 172555 two Na {\sc i} D2 components appear, one close to the G cloud in agreement with \cite{kiefer14a}, and a second one very broad and variable (see section \ref{variable}). 

\begin{itemize}

\item 
HD  2884, HD 2885, and HD 3003 are members of the Tuc-Hor association. All three stars have similar proper motions, radial velocities and parallaxes. The projected angular separation between HD 2884 and HD 2885 is $\sim 27\arcsec$ while the one between HD 2884 and HD 3003 is $\sim 9\arcmin$. Thus, they likely form a physical multiple system \citep[see also][]{eggleton08,tokovinin08,howe09}. HD 2885 itself is a binary candidate \citep{lagrange09}. Our spectra show a weak narrow Ca {\sc ii} K absorption in HD 2884, a weak Na {\sc i} D absorptions towards HD 2885, and weak Ca {\sc ii} K and Na {\sc i} D absorptions towards HD 3003 \citep[see also][]{Iglesias18}. All these absorptions are close to the velocity vector of the Vel cloud and do not coincide  with the v$_{\rm  rad}$ of the stars (Table \ref{tab:narrow} and Fig. \ref{fig:na_absorptions}). HD 224392 is another Tuc-Hor star located at a similar distance from the Sun as the multiple system and at an angular separation of $\sim 4\deg$.
Its spectrum shows non-photospheric absorptions of both Ca {\sc ii} and Na {\sc  i}. The velocity of the Ca {\sc ii} feature is close to the Vel cloud and to the Ca {\sc ii} absorptions in HD 2884 and HD 3003; the velocity of the Na {\sc  i} is close to the Cet Colorado cloud that passes at $< 20^{\sc o}$ the line of sight of HD 224392. All these facts suggest an ISM origin for the Ca {\sc ii} and Na {\sc  i} features; however, this ascription is controversial.  First, we note that neither Ca {\sc ii} nor Na {\sc i} non-photospheric features appear simultaneously in HD 2884 and HD 2885. 
At the same time, Na {\sc i} is detected towards HD  3003 but not towards HD 2884. Thus, since all three stars are located very close, in particular HD 2884 and HD 2885, a CS origin for the observed features cannot be excluded. Given all these facts, we find ambiguous to ascribe the non-photospheric features in all 4 stars to either a CS or ISM origin and further observations are needed to elucidate their origin.

\item HD 5267 has a strong non-photospheric feature at {\mbox{$\sim$  --5.0  km/s}} in both  Ca {\sc ii} and Na {\sc i} lines. The velocity of this feature does not coincide with v$_{\rm rad}$ (9.5 km/s) or with the v$_{\rm  ISM}$  of any Colorado cloud. At the same time, there is a  weak Ca {\sc ii} K feature at $\sim$10.0 km/s, also tentatively detected in the Na {\sc i} D2 line, close to stellar radial velocity and the velocity vector, 11.44 km/s, of the LIC cloud (Table \ref{tab:narrow}). Unfortunately there are no spectra of field stars in the ESO archive that can be used to discriminate the origin of the component at --5 km/s. We  note, however, that this feature is $\sim$15 km/s blue-shifted relative to the radial velocity of the star, and that HD 5267 has  a similar effective temperature  as HD 181296, T$_{\rm eff}$ $ \sim$10500 K. HD 181296 has a stable component $\sim$20 km/s blue-shifted relative to the stellar velocity, most likely of CS origin as suggested in \cite{Rebollido18} \citep[see][for other examples]{chen03}. Thus, while the origin of the $\sim$10 km/s component is ambiguous, the origin of the --5 km/s is unknown.

\item HD 16978 is another Tuc-Hor star reported by  \cite{welsh18}  as having a variable non-photospheric Ca {\sc  ii} K profile within 3 observations. Our spectra show such absorption with a similar EW but does not show the mentioned variability. The radial velocity of the feature coincides with the stellar v$_{\rm  rad}$, and  it differs  $\sim$4 km/s from the  v$_{\rm  ISM}$ of the Vel cloud. Given the variability of the feature profile noted  by \cite{welsh18}, which might be indicative of a transient event, the more plausible origin is CS, although we cannot definitively exclude an ISM origin.

\item The stars HD 71043, HD 71722, and HD 105850  have {\mbox{Ca {\sc ii}}} and Na {\sc i} features close to the velocity vectors of Colorado clouds, and far away form the stellar v$_{\rm  rad}$, supporting an ISM origin, in agreement with \cite{Iglesias18}. This is also the case for the star HD 188228 where only a weak Na {\sc i} absorption is seen close to the velocity of the G and Vel clouds.

\item HD 118232 (24 CVn), HD 125162 ($\lambda$ Bootis), and HD 221756 have non-photospheric absorption features not coinciding either with the stellar radial velocities or any known ISM Colorado clouds. Due to their high declinations ($> +40^\circ$) we have not found any spectra of field stars in ESO archive which could help to elucidate the origin of the absorptions. Nonetheless, we find an ISM origin is the most plausible one given the remarkable shift in velocity between the features and the photospheric lines.

\item HD 145689 (HIP 79797) is a member of the Argus association and shows a weak Ca {\sc  ii} absorption at a velocity of --11.9 km/s, between the radial velocity of the star at --7.1  km/s and the velocity of the G Colorado cloud at --17.21 km/s. The F0 V field star HD 147787, located at a distance of 40 pc and a projected separation of $\sim$ 4$^\circ$ from HD 145689, does not show any non-photospheric feature in its spectrum. Thus, given this fact and since the difference between the Ca {\sc ii} feature and the stellar v$_{\rm rad}$ is $\sim 2 \sigma$ we cautiously assign a CS origin to the Ca {\sc ii} absorption.

\item 
HD 158352 (HR 6507) has two Ca {\sc ii} H\&K and Na {\sc i} D absorption components  relatively close to the stellar v$_{\rm rad}$. There are no Colorado clouds along the line of sight of  the star, but it  passes near ($<  20^\circ$) several clouds with similar velocity vectors. Our spectra reveal a faint narrow  Ca {\sc ii} 8542 \AA ~absorption, and several faint Fe {\sc ii} and Ti {\sc ii} lines characteristics of shell stars. \cite{Iglesias18} assign the Ca {\sc ii} H\&K absorptions  an ISM  origin based on two field stars, HD 156208 and V 2373 Oph, that have absorptions  similar to those of  HD 158352. We find, however, that the ISM origin is ambiguous. HD 156208 and V 2373 Oph are at distances $\sim$ 217 pc and $\sim$ 476 pc, respectively, i.e. considerably larger than the distance to HD 158352 of $\sim$ 63 pc. Therefore, while an ISM cloud could be located closer than 63 pc, it cannot be excluded that some ISM clouds responsible for the Ca {\sc ii} H\&K and Na {\sc i} D are located between HD 158352 and the other two stars. On the other hand, \cite{welsh15} observed two Ca {\sc ii} K absorption components and attributed one of them a CS origin as it coincides with the stellar v$_{\rm rad}$. Thus, while CS gas is certainly present around the star, the origin of the Ca {\sc ii} H\&K and Na {\sc i} D absorptions is not completely obvious. We note that \cite{lagrange09} identify this star as a binary candidate. 

\item HD 168646 (HR 6864) has a pronounced triangular profile in the Ca {\sc ii} at the bottom of the stellar lines, as well as many strong shell lines of Ti {\sc ii}, Fe {\sc ii}, and strong cores in the Balmer lines. The Na {\sc i} D lines present a strong feature close to the Ca {\sc ii} one, and a weaker one at a velocity of $\sim$ --22 km/s, away from the Aql cloud. A very weak third component might appear only in the D2 line at a velocity of $\sim$ 11 km/s. If real, this weak feature seems to have small variations in flux; we cannot make, however, any sound statement as we only have spectra taken in a single epoch. All features have velocities far away from any ISM Colorado cloud. In any case, the strong shell spectrum  clearly traces the presence of CS gas.  

\item HD  177724 ($\zeta$  Aql) and HD 210418 have weak Ca {\sc  ii} non-photospheric absorptions, each at their respective stellar radial velocities and close to some Colorado clouds. We have not found any useful field star in the ESO archive which could help to elucidate the origin of the features. Thus, the ascription to any of both CS or ISM origins is ambiguous .
   
\item HD 196724 is a candidate shell star \citep{hauck00} with low rotational velocity, which shows a weak Ca {\sc ii}  absorption at the stellar v$_{\rm rad}$ and also close to the v$_{\rm ISM}$ of the Mic and Aql clouds. It might have weak Na {\sc i} absorption but the spectra are too noisy to make a firm conclusion. The weak Ca {\sc ii}  feature does not reveal a sharp, pronounced profile as it is seen in the Ca {\sc ii} K\&H lines in other shell stars. Thus, we find ambiguous the origin of the Ca {\sc ii}  (and Na {\sc i} if real) absorptions.

\item HD 198160 (HR 7959) is a $\lambda$ Bootis type star with a very weak Ca {\sc ii} K absorption at the stellar v$_{\rm rad}$ and the Vel cloud v$_{\rm ISM}$. It forms a binary  system with another  $\lambda$ Bootis star, HD 198161 \citep{holweger95,faraggiana99}, both components at a projected angular separation of just  2.3$\farcs$. Weak Ca {\sc  ii} K and Na {\sc i} absorptions are detected towards HD 198161. We also note that a very weak Na {\sc i} absorption is present towards HD 198160.  
Thus, while a CS origin cannot be excluded, we find an ISM origin more realistic given the fact that we detect very similar absorptions towards both stars. However, we note that \cite{holweger95,holweger99} favour a CS origin around the binary. 

\item 
Stars belonging to the Upper Scorpius subgroup in the Scorpius-Centaurus association deserve particular attention. There are 13 stars in our sample belonging to this subgroup (Table \ref{tab:sample}). All of them are located in a region of 5 sq.  deg., and between 110 and  150 pc. The stars show non-photospheric Ca {\sc ii} and Na {\sc i} absorption components which tend to be  grouped around $\sim$--9 km/s, $\sim$--15/--23 km/s, and {\mbox{$\sim$--28 km/s}} (Table \ref{tab:narrow}), in agreement with \cite{welty94}. Most of the stars, with few exceptions, have two features with similar velocities in both Ca {\sc ii} and Na {\sc i} lines; depending on the star those two features are distributed among the three mentioned velocity ranges. The more blue-shifted components at $\sim$--28km/s are close to the G cloud velocity vector; the least blue-shifted components at $\sim$--9 km/s are often close to the radial velocities of the stars. The fact that similar features, including those in the range $\sim$--15/--23 km/s, are shared in one way or another by all stars strongly suggests their ISM origin, as already noticed by \cite{Rebollido18} for HD 138813 and HD 146897. In general, the strongest feature is the less blue-shifted one, i.e., the one close to the stars' v$_{\rm rad}$; this trend holds irrespectively of whether one or two absorption features are detected in any of the Ca {\sc ii} and/or Na {\sc i}; exceptions to this general trend are HD 146606, and HD 145964 - we refer to this star again in section \ref{variable} as it has a non-photospheric event identified by \cite{welsh13}. The above results point evidently out that the ISM towards Upper Scorpius is not homogeneous, but with a notorious complexity likely structured in clumps or relatively small cloudlets, with different properties and located at different distances, discernibles along very nearby lines of sight. Finally, we note that a faint emission feature at $\sim$--1.2 km/s is detected in both lines of the Na {\sc i} doublet towards HD 138813. As shown in Fig. \ref{fig:tell_hip76310}, the emission is not related to the telluric subtraction. The emission feature is at the radial velocity of the star, and is similar to the emission feature detected towards HD 156623, a star with Ca {\sc ii} variable events \citep{Rebollido18,rebollido19a}. 
These emissions are most likely originated in the CS medium, as they are observed in every spectra, regardless of the observing campaign or  atmospheric conditions. 

\begin{figure}[h!]
  \centering
 \includegraphics[width=0.5\textwidth]{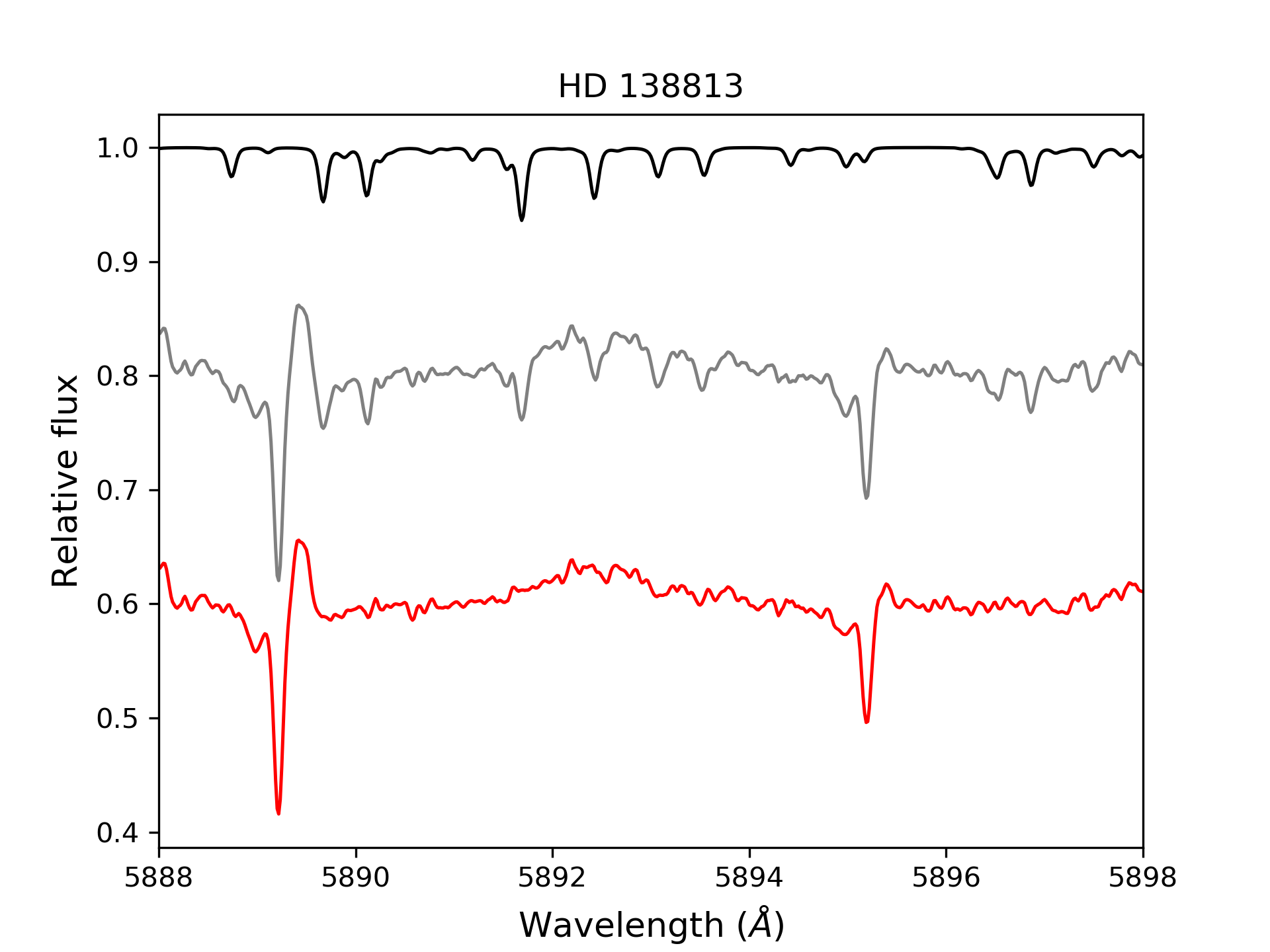} 
 \caption{Na {\sc i} D lines of HD 138813, where the emission near the radial velocity of the star is easily recognisable and clearly present in the uncorrected spectra (grey line), and not originated as an over subtraction of the atmospheric model (black line). Red line shows the telluric corrected spectrum.}
\label{fig:tell_hip76310}
\end{figure}

\end{itemize}

\subsection{Variable gas detection}
\label{variable}

Sixteen  stars  in our  sample were selected because of variable, $\beta$ Pic-like events of their Ca {\sc ii} H\&K and/or Na {\sc i} D lines. The stars HD 56537 ($\lambda$ Gem), HD 108767 ($\delta$ Crv), HD 109573 (HR 4796), and HD 148283 (HR  6123) also present  optical (or  UV) events \citep[e.g.][]{grady96b,welsh15,Iglesias18} although they were included on the basis of other criteria (Table \ref{tab:sample}). These stars are listed in Table \ref{tab:feblist} together with five new stars showing variability in non-photospheric features found in the  frame of this  work -  HD 36546, HD 37306, HD 39182 (HR 2025), HD  98058 ($\phi$ Leo), and HD 156623 (HIP 84881). We also include in Table \ref{tab:feblist} the star HD 132200; although this star was not included in our sample and has not directly been observed by us, a variable Ca {\sc ii} K absorption feature was found by \cite{Rebollido18}. Our results on $\phi$ Leo, HD 156623 and HR 10 have  already been discussed in \cite{eiroa16},  \cite{Rebollido18, rebollido19a},and \cite{montesinos19}, respectively.

\begin{center}
\begin{table}
  \caption{Stars with variable non-photospheric absorption features. References are within brackets.}
\begin{center}
\begin{tabular}{lll}
Name       & Prev. Detected    & Det. in this work  \\
\hline                    
HD 256 (HR 10)$^*$        &  Yes (1,12,15)    &          \bf{Yes}           \\
HD 9672 (49 Cet)          &  Yes (2)          &            No           \\
HD 21620                  &  Yes (3)          &            \bf{Yes}     \\
HD 32297                  &  Yes (4)          &            No     \\
HD 36546                  &  No               &            \bf{Yes}     \\
HD 37306                  &  No               &   \bf{Yes}     \\
HD 39182 (HR 2025)        &  No               &            \bf{Yes}     \\
HD 42111                  &  Yes (5,12)       &        \bf{Yes}          \\
HD 50241                  &  Yes (5,11)       &            No           \\
HD 56537 ($\lambda$  Gem) &  Yes (6)          &            No           \\
HD 64145 ($\phi$ Gem)     &  Yes (6)          &            No           \\
HD 80007 (HR 3685)        &  Yes (11,15)      &            \bf{Yes}    \\
HD 85905                  &  Yes (7,15)       &            \bf{Yes}     \\
HD 98058 ($\phi$ Leo)$^*$ &  No               &            \bf{Yes}    \\
HD 108767 ($\delta$ Crv)  &  Yes (6)          &            No           \\
HD 109573 (HR 4796)       &  Yes (6,16)       &            \bf{Yes}     \\
HD 110411                 &  Yes (3)          &            \bf{Yes}     \\
HD 138629 (HR 5774)       &  Yes (8)          &            No           \\
HD 132200$^*$             &  No               &            \bf{Yes}     \\ 
HD 145964                 &  Yes (3)          &            \bf{Yes}     \\
HD 148283 (HR 6123)       &  Yes (5,13)       &             No     \\
HD 156623 (HIP 84881)$^*$ &  No               &            \bf{Yes}     \\
HD 172555                 &  Yes (9)          &            \bf{Yes}     \\
HD 182919 (5 Vul)         &  Yes (2)          &            \bf{Yes}     \\
HD 183324                 &  Yes (10,16)      &            \bf{Yes}     \\
HD 217782                 &  Yes (2,5,14)     &            \bf{Yes}     \\
\hline
\end{tabular}
 \end{center}
(*) Results have been presented by \cite{eiroa16}, \cite{Rebollido18}, and Montesinos et al. (2019, A\&A in press).\\
References: (1) \cite{lagrangeHenri1990}; (2) \cite{montgomery12};
(3) \cite{welsh13}; (4) \cite{redfield07a}; (5) \cite{roberge08};
(6) \cite{welsh15}; (7) \cite{welsh98}; (8) \cite{Lagrange-Henri90b};
(9) \cite{kiefer14a}; (10) \cite{montgomery17}; (11) \cite{hempel03}; (12) \cite{lecavelier97};
(13) \cite{grady96b}; (14)  \cite{cheng03}; (15) \cite{redfield07b};
(16) \cite{Iglesias18}
\label{tab:feblist}
\end{table}
\end{center}

\subsubsection*{Comments on individual stars}
\label{sect:febs_indstars}

 \begin{itemize}

\item HD 9672 (49 Cet), HD 32297, HD 50241, HD 56537 ($\lambda$ Gem), HD 64145 ($\phi$ Gem), HD  108767 ($\delta$ Crv), and HD 148283 (HR  6123) do not present any apparent transient event in our time series spectra (Table \ref{tab:feblist}). Also, while we do not see any variability in HD 138629 (HR 5774) our spectra differ from previous ones (see below). Further, we note that the stars HD 56537, HD 64145, HD 110411, and HD 183324 do not present any stable, narrow absorption at the core of the photospheric line, although HD 110411 and HD 183324 seem to present variability at  the bottom of the  Ca {\sc ii} K line (Fig. \ref{fig:110411_183324},  \citep[see  also][]{Iglesias18}).

\begin{figure*}
\centering
\includegraphics[scale=0.55]{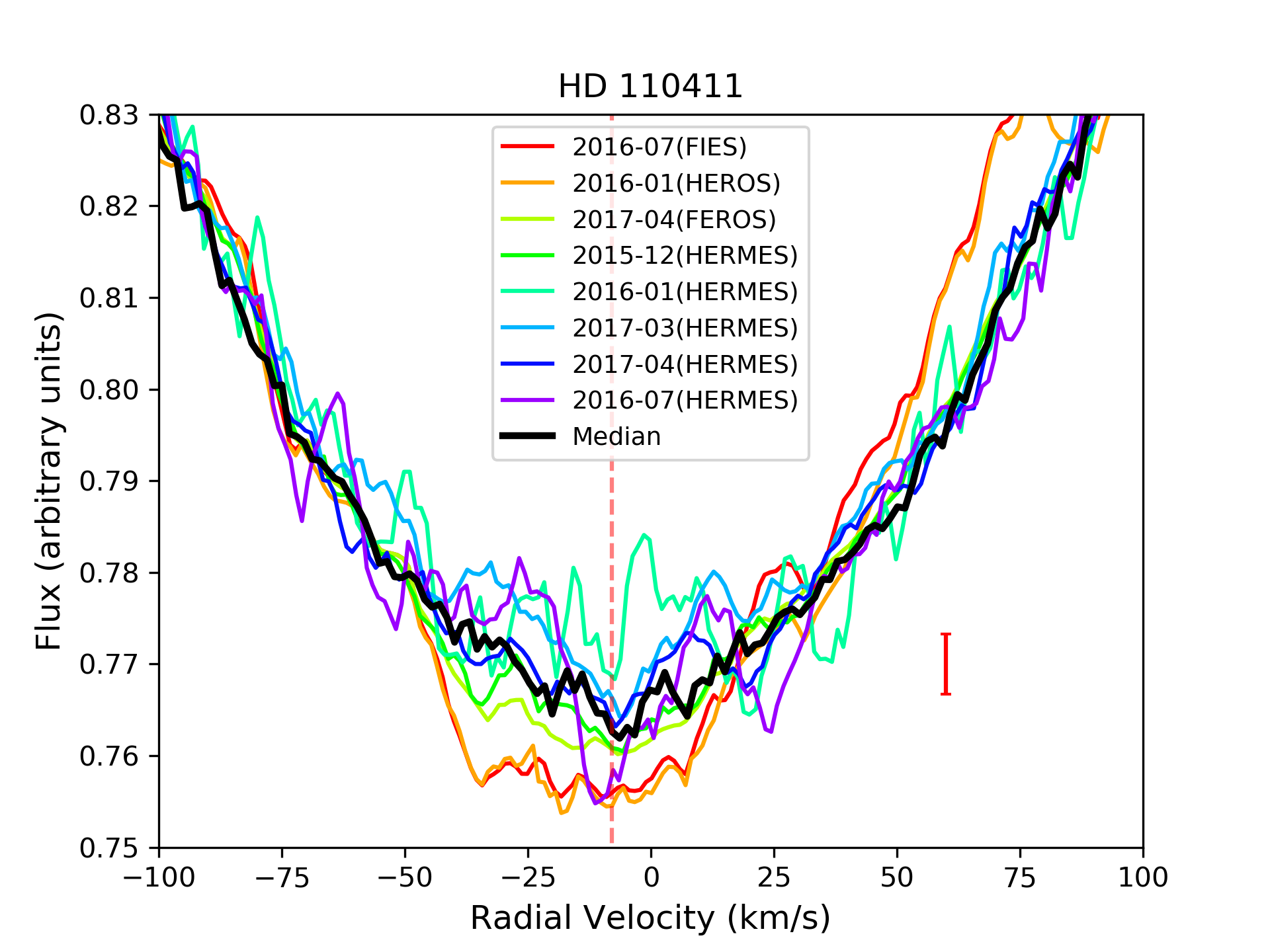}
\includegraphics[scale=0.55]{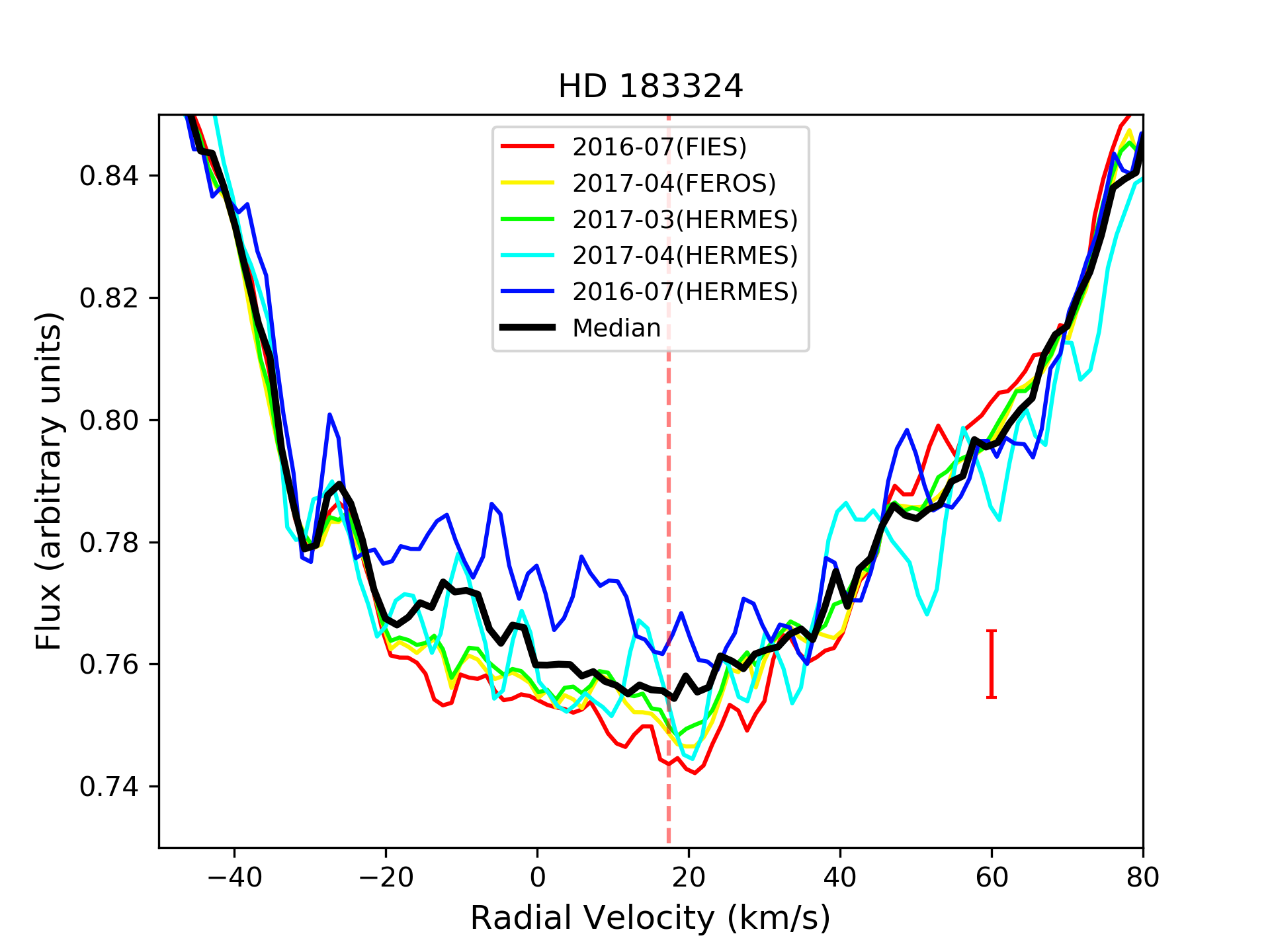}
\caption{Photospheric profile of Ca {\sc ii} K spectra of HD 110411 and HD 183324. Dates and instruments are colour-coded as indicated in the legend. The core of the line varies although neither a narrow feature nor transient events are clearly distinguished. The red dashed vertical line marks the radial velocity of the star. The red error bar shows the 3-$\sigma$ standard deviation in the region close to the bottom of the line. This applies to all upcoming figures.}
\label{fig:110411_183324}
\end{figure*}

\item 
HD 21620 presents one stable non-photospheric absorption in both Ca {\sc ii} and Na {\sc i} at $\sim$ 4 km/s (Fig. \ref{fig:na_absorptions} and Table \ref{tab:narrow}). This feature does not coincide either with the radial velocity of the star or any of the Colorado ISM clouds; it has, however, a plausible ISM origin as several neighbouring stars have similar Na {\sc i} absorptions \citep{genova03, welsh13}. A weak absorption detected in Ca {\sc ii} K at $\sim$ 16 km/s is close to the ISM velocity of the LIC cloud. Nonetheless the variability of this feature, also noted by \cite{welsh13} (see their Fig. 1 and 2), is remarkable and we attribute it a CS origin. Fig. \ref{fig:feb_hd21620} shows the Ca {\sc ii} K line of HD 21620 during the campaigns of  2015 and 2016. We detect mostly red-shifted variations in the range of $\sim$ 10-30 km/s, with a tentative, weak blue-shifted event on 04-05/09/2015 at $\sim$ 0 km/s (top left panel of Fig. \ref{fig:feb_hd21620}), and a potential red-shifted event at $\sim$ 50 km/s in the NOT median spectra of 07/2016 (bottom left panel of Fig. \ref{fig:feb_hd21620}). We note that \cite{welsh13} also detected a feature at a velocity close to this last event together with several blue-shifted ones. Some dynamical evolution might be traced by the $\sim$ 16 km/s events observed in January 2016 as suggested by their changes in velocity and depth  along three consecutive nights. All Ca {\sc ii} K events we detect are very weak with no detectable counterpart in the Ca {\sc ii} H line, which suggests that the gas is optically thin. As an example, Fig. \ref{fig:feb_hd21620_test} shows both Ca {\sc ii} lines as observed on Dec 23rd, 2015. The K line can be fitted with two Gaussians at velocities 4.3 and and 16.8 km/s, and equivalent widths of 14.2 and 3.3 m\AA, respectively; at the same time, the 4.3 km/s H absorption, which is the stable one, has an EW = 7.2  m\AA, i.e., this feature is optically thin, while the 16.8 km/s variable one is embedded in the noise, also suggesting optically thin gas.

\begin{figure}
\centering
\includegraphics[width=0.5\textwidth]{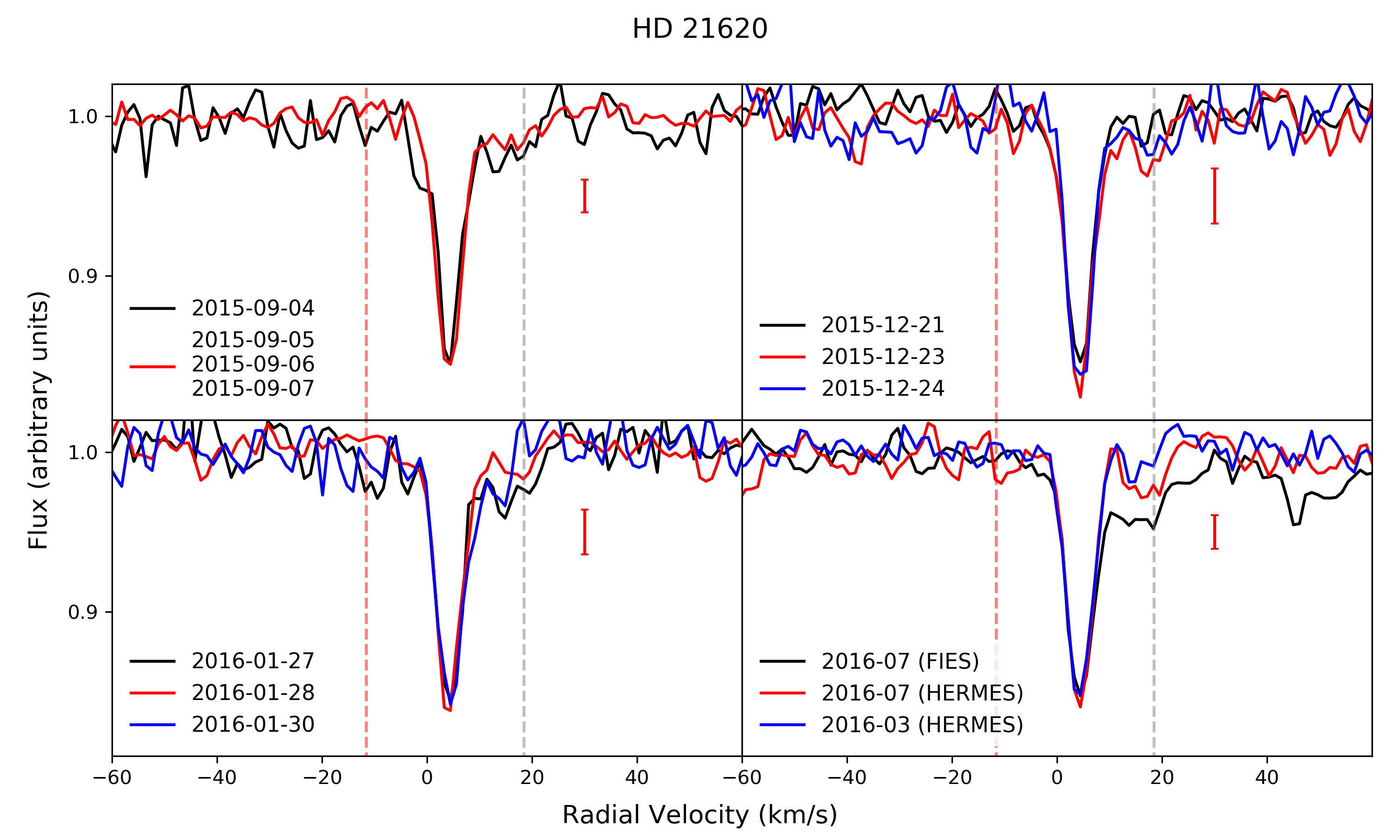}
\caption{Ca {\sc ii} K line of different epochs of HD 21620. Spectra plotted in the panels at the top and lower left were obtained with HERMES. Lower right panel shows median spectra of indicated period and telescope. FEB-like events appear at $\sim$ 16 km/s and tentatively at 0 (blue-shifted, top left panel), and 50 km/s (red-shifted, bottom right). The events of January 2016 appears to present some dynamical evolution. Vertical red and grey dashed lines show the stellar and ISM radial velocities respectively.}
\label{fig:feb_hd21620}
\end{figure}

\begin{figure}
\centering
\includegraphics[width=0.5\textwidth]{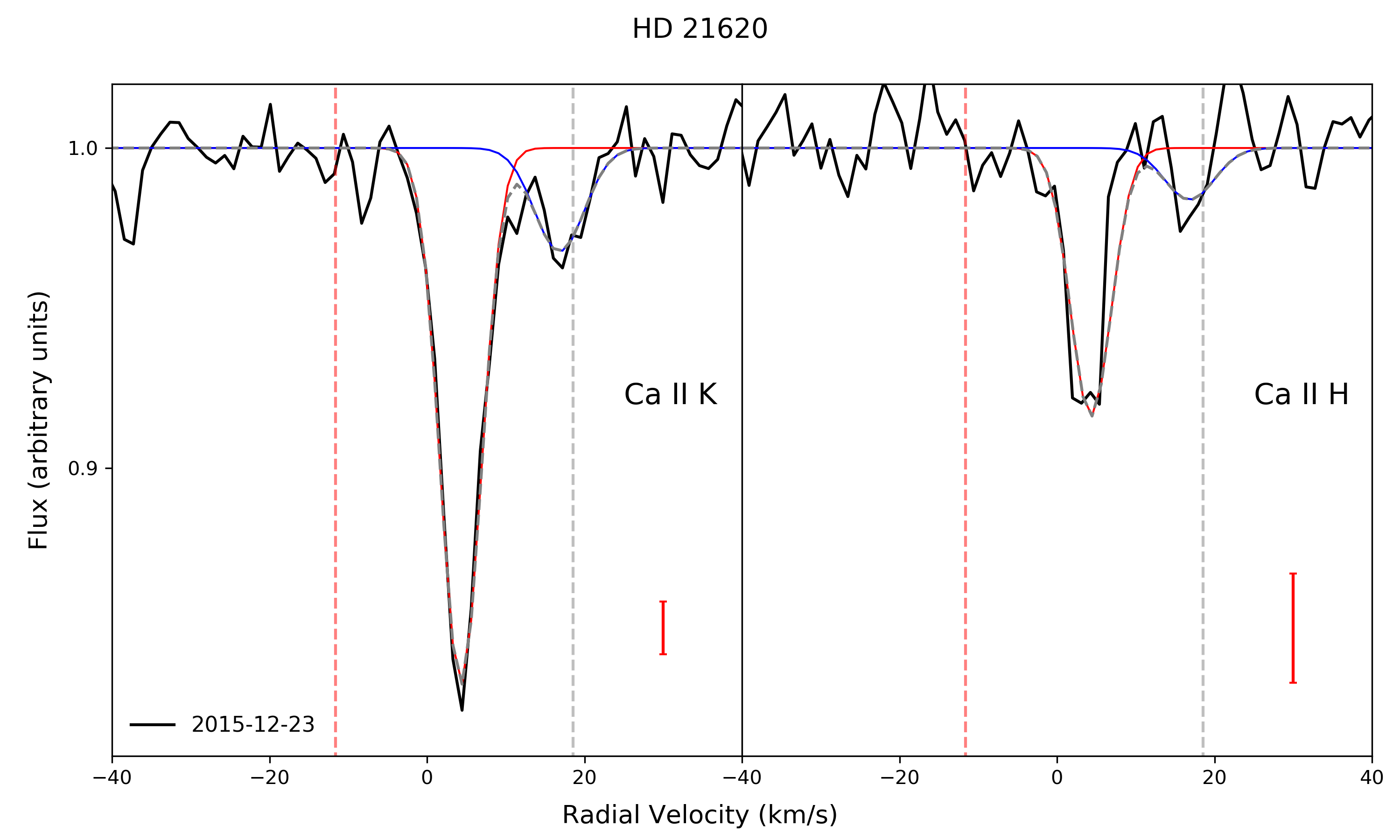}
\caption{The Ca {\sc ii} K and H lines as observed on 23/12/2015 using HERMES. Two Gaussians are fitted to the K non-photospheric feature with velocities 4.3 km/s(red continuous line) and 16.8 km/s (blue continuous line). The strongest, stable 4.3 km/s absorption is clearly detected in the H line, but the weakest, variable absorption at 16.8 km/s is embedded in the noise of the H spectrum. Vertical red and grey dashed lines show the stellar and ISM radial velocities respectively.}
\label{fig:feb_hd21620_test}
\end{figure}

\item HD 36546 presents a narrow feature at a velocity of $\sim$ 15 km/s visible in both Ca {\sc ii} and Na {\sc i} lines (Fig. \ref{fig:na_absorptions}). The origin of the feature is most likely CS as it coincides with the radial velocity of the star and is far from the ISM cloud along the line of sight (Table \ref{tab:narrow} and Fig. \ref{fig:na_absorptions}). A red-shifted event at a velocity of $\sim$ 20 km/s is detected in the Ca {\sc ii} K spectrum of 6/3/2017, apparently evolving in the following dates and practically disappearing in the spectrum of 8/3/2017 (Fig. ~\ref{fig:hd36546_feb}, left panel). A Gaussian deconvolution of the  6/3/2017 Ca {\sc ii} K absorption (Fig.\ref{fig:hd36546_feb_6_8}) gives an EW of 8.4$\pm$0.9 m\AA ~for the 20 km/s event, and 24.8$\pm$2.5 m\AA ~for the narrow stronger feature centered at 15 km/s, which is similar to the EW of 27.1$\pm$2.7 m\AA ~of this narrow feature in the spectrum of 8/3/2017. We note that the weak feature at 20 km/s is not discernible from the noise in the Ca {\sc ii} H line (not shown) suggesting that the gas is optically thin, somehow similar to the case of HD 21620. We also point out that the Na {\sc i} D2 line presents as well an asymmetry in the red wing, with a small change of slope when comparing the different dates (Fig. \ref{fig:hd36546_feb}, right panel). 

HD 36546 hosts a bright debris disc (Table \ref{tab:sample}) seen near edge-on with an inclination angle $i \sim 70 - 75^\circ$ \citep{currie17}, following the trend suggested by \cite{Rebollido18} between the disc inclination and the presence of narrow non-photospheric absorptions at the radial velocity of the star. \cite{lisse17} found evidence of a C-rich CS environment which makes HD 36546 similar to $\beta$ Pic and 49 Cet \citep{roberge06,roberge14}. Thus, it can be another example of an enhanced carbon abundance acting as a braking mechanism of the hot inner (<1 AU) CS gas released by evaporation of exocomets, dust grains or grain-grain collisions \citep{fernandez06,brandeker11}.

\begin{figure}
\centering
\includegraphics[width=0.5\textwidth]{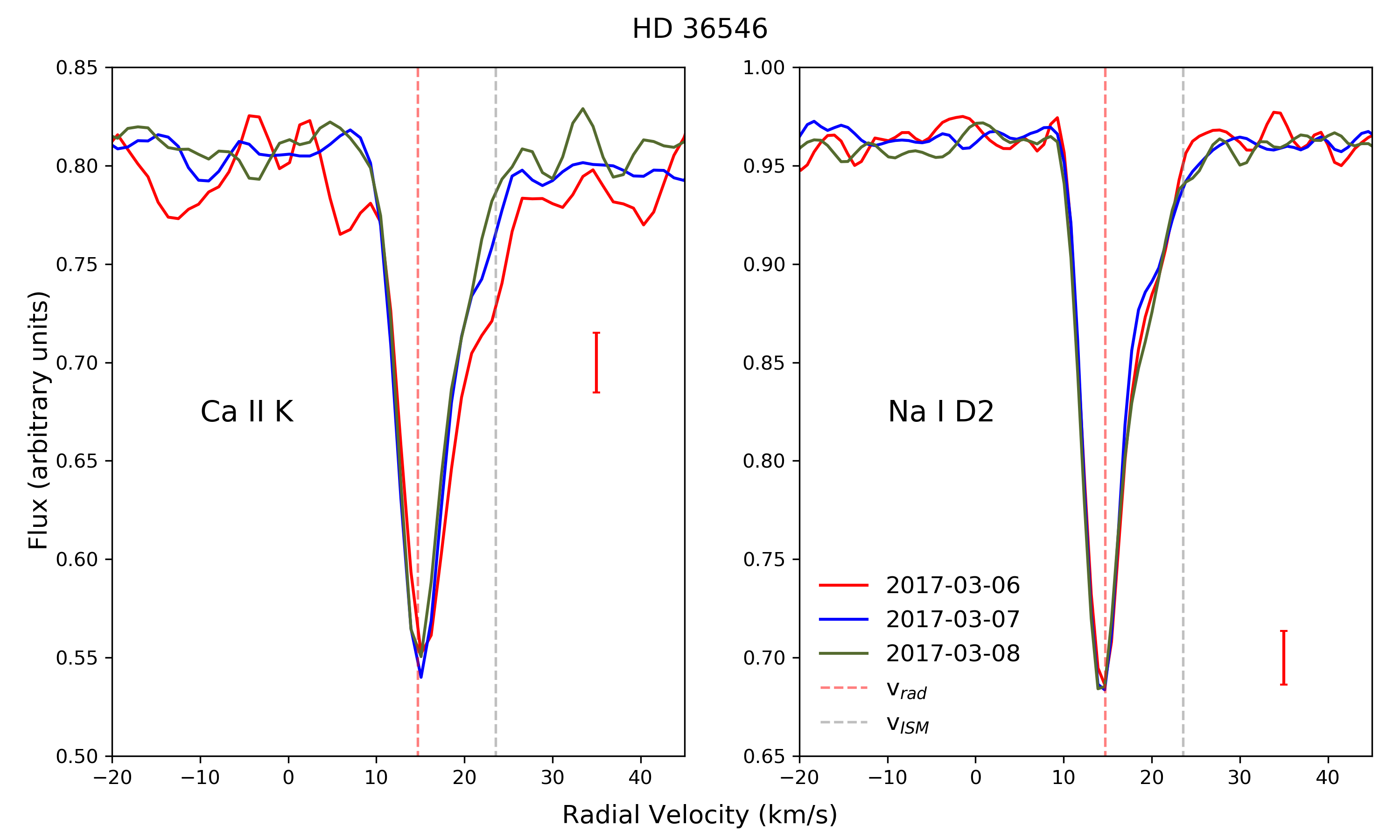}
\caption{Left panel: Ca {\sc ii} K line. A transient event is seen at $\sim$ 20 km/s superimposed on the red wing of the narrow non-photospheric absorption at $\sim$ 15km/s. No obvious event is seen in 08/03/2017. Right panel: Na {\sc i} D2 line. Dates are as indicated, and all spectra were obtained using HERMES. Red and grey vertical lines mark the radial velocity of the star and of the ISM respectively.}
\label{fig:hd36546_feb}
\end{figure}

\begin{figure}
\centering
\includegraphics[width=0.5\textwidth]{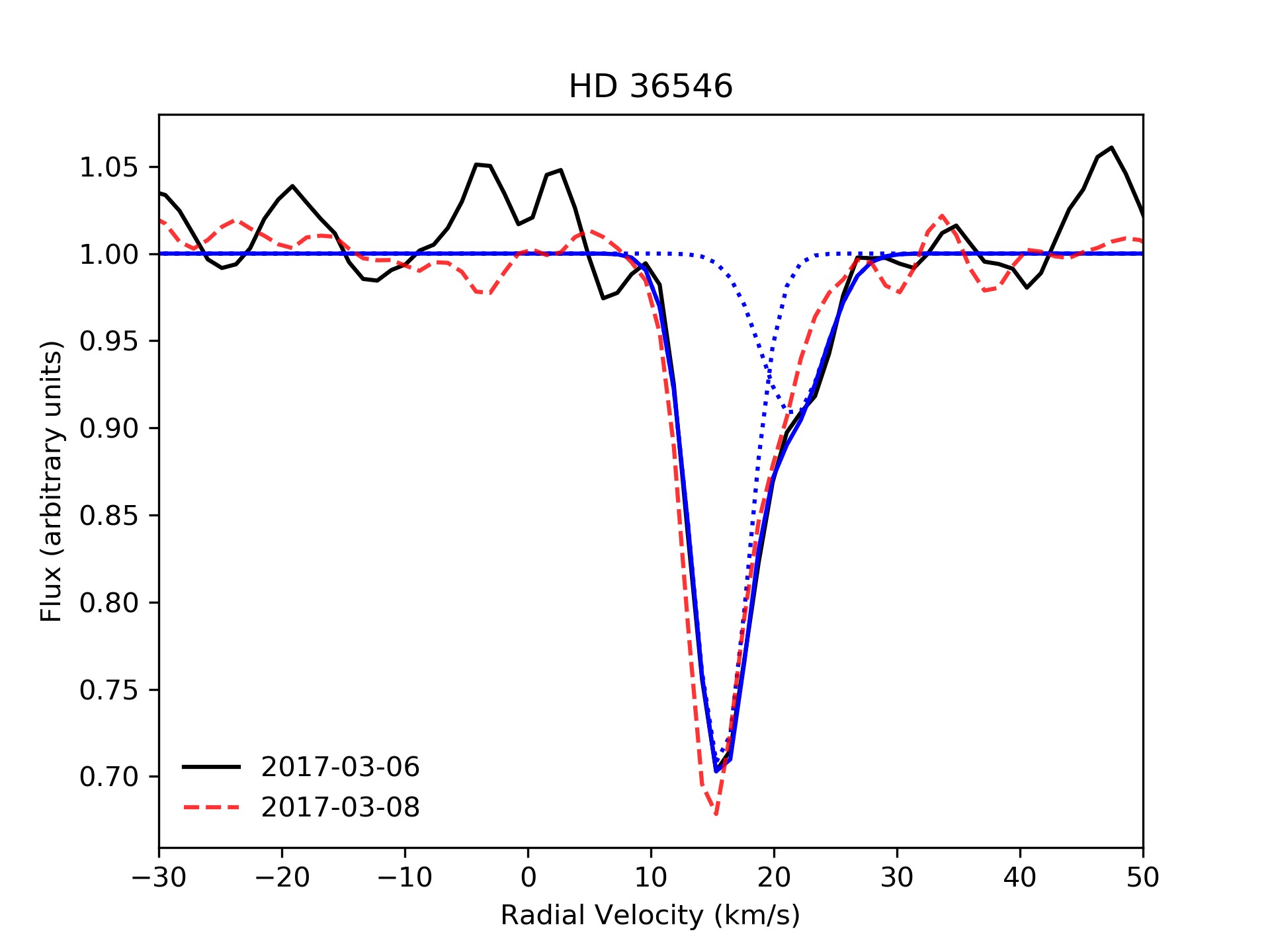}
\caption{The Ca {\sc ii} K spectra of the HD 36546 non-photospheric absorption for the day where the event in the  red wing is most conspicuous (06/03/2016, black line) and when it is practically  undetectable (08/03/2016, red dashed line). As in the previous figure, both spectra were obtained with HERMES. Blue solid line shows a fit of the 06/03/2016 spectrum with two Gaussians, each one plotted as dotted blue lines.}
\label{fig:hd36546_feb_6_8}
\end{figure}

\item HD 37306 presents two stable narrow Ca {\sc ii} and Na {\sc i} absorptions (Fig. \ref{fig:na_absorptions}) at velocities $\sim$11 km/s and $\sim$32 km/s, while the stellar v$_{\rm rad}$ is 25.1 km/s. Nonetheless, the most remarkable and striking behaviour is the strong shell-like spectrum that appeared on the spectra of September 2017. The Ca {\sc ii} H\&K lines developed a strong, symmetric, triangular profile superimposed on the photospheric lines and the two narrow interstellar features, together with narrow shell-like absorptions in the Ca {\sc ii} triplet or in several Fe {\sc ii} and Ti {\sc ii} lines (see Fig. \ref{fig:hd_37306}). At the same time, photospheric lines as e.g. the Mg {\sc ii} 4481 \AA ~or the O {\sc i} triplet at 7750 \AA ~remained constant, as well as the Na {\sc i} D lines. Further, the strong shell spectrum fully vanished in additional spectra taken in November 28, 2018 with the CARMENES  spectrograph in Calar Alto Observatory \citep{CARMENES}, and from December 14 to 18, 2018 with HERMES. This behaviour has also been observed by \cite{Iglesias19} partly using the same spectra. Due to the in principle unusual nature of this phenomenon, we have checked possible sources of contamination, such as instrumentation issues or additional sources in the fiber, but we have discarded these scenarios. Thus, the appearance/disappearance of the shell-like profiles does point out to the presence of CS gas, but no blue-/red-shifted FEB-like events are detected in any of our spectra. We note that variability of shell spectra and even its appearance/disappearance in some stars is well known \citep[e.g.][]{jaschek88}. At the same time, and despite the lack of any identified ISM clouds in the line of sight, the fact that no remarkable changes are seen in the mentioned $\sim$11 km/s and $\sim$32 km/s narrow features during the 8 days of observations in September 2017, while drastic changes are seen in the CS (shell-type) environment, suggests an ISM origin as the more plausible alternative for those two absorptions, in agreement with \cite{Iglesias18}.

\begin{figure}[h!]
  \centering
 \includegraphics[width=0.5\textwidth]{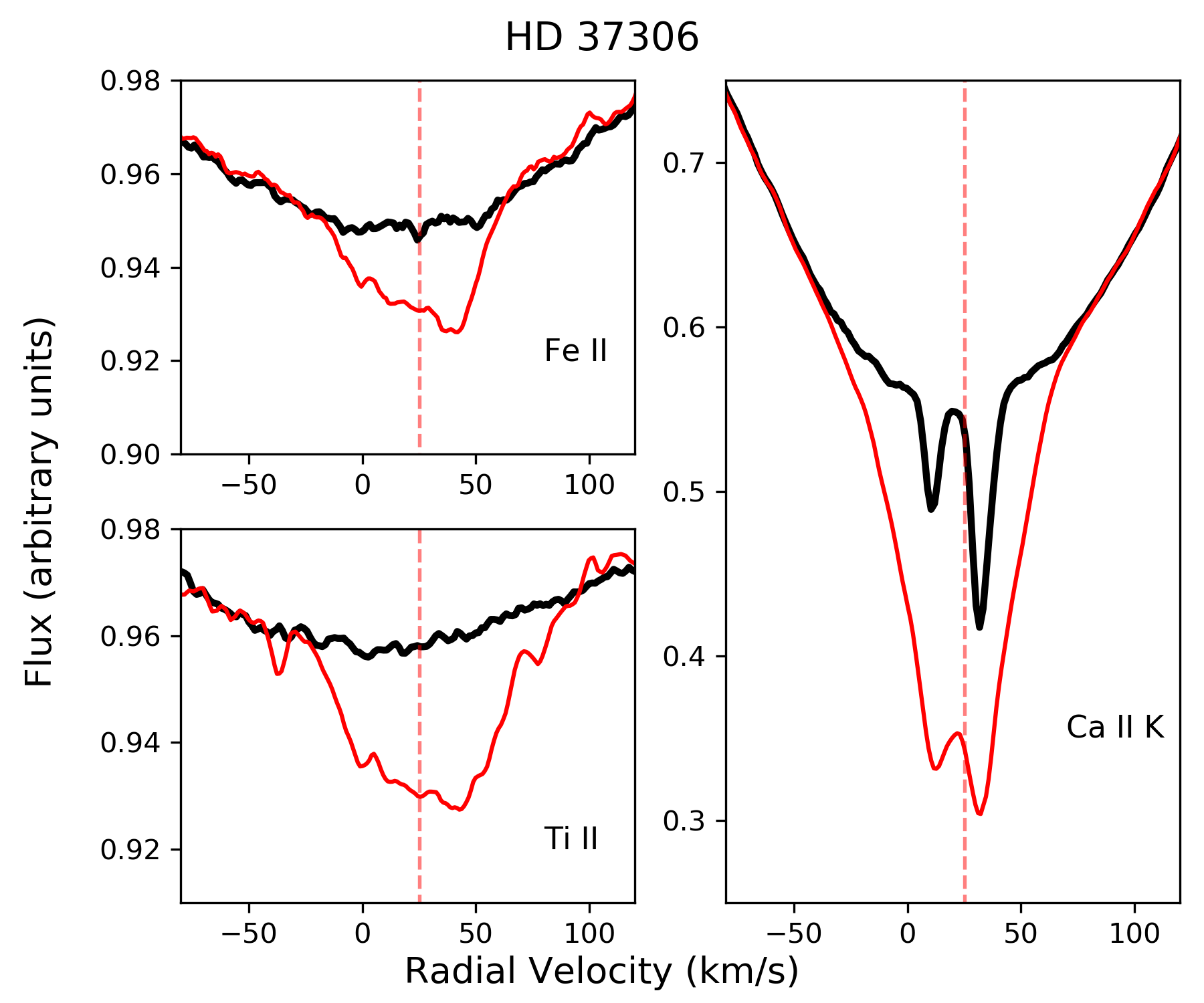}
 \caption{Fe {\sc ii} (4583.83 \AA), Ti {\sc ii} (4443.80 \AA) and Ca {\sc ii} K lines of HD 37306. While most of the spectra, represented by the median spectrum in the plot (black line), do no vary and show two Ca {\sc ii} non-photospheric absorptions, a strong shell-like profile appeared in September 2017 (red line, FEROS). It was not observed again in later spectra of November and December 2018, not shown in the figure. Red vertical line marks the radial velocity of the star.}
\label{fig:hd_37306}
\end{figure} 

\begin{figure}[h!]
  \centering
  \includegraphics[width=0.5\textwidth]{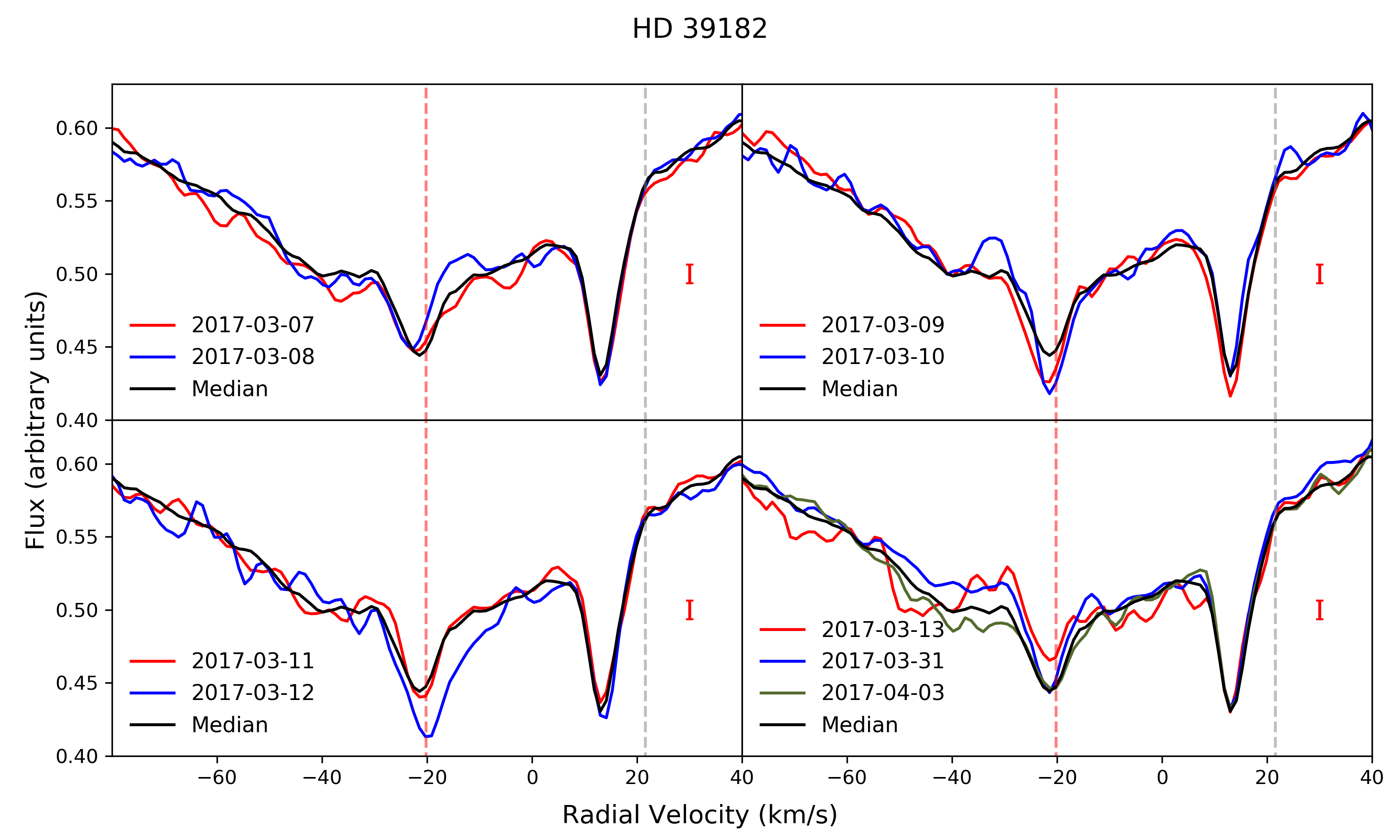}
 \caption{Ca {\sc ii} K spectra  of HD 39182 grouped by observing dates obtained with HERMES. In all panels,  the median spectra is also  plotted. The red and grey vertical lines correspond to the  radial velocity  of the star and the velocity vector of the LIC Colorado cloud, respectively.}
\label{fig:HR2025}
\end{figure} 

\item HD 39182 (HR 2025), one of the selected Ti {\sc ii} stars, has a sharp triangular-like absorption (Fig. \ref{fig:na_absorptions}) with two narrow  Ca {\sc ii} components; one is found at $\sim$ --22 km/s, centred in the photospheric line coinciding with the stellar v$_{\rm rad}$, and varies significantly; the second, strongest one at $\sim$ 13 km/s is clearly displaced from the line center (Fig. \ref{fig:HR2025}). A weak extra absorption at $\sim$ --41 km/s is present in some  spectra, i.e., in the blue wing of the --22 km/s component (Figs. \ref{fig:na_absorptions} and \ref{fig:HR2025}). The Na {\sc i} D lines only show a strong narrow absorption at $\sim$  14 km/s, coinciding with the strongest Ca {\sc ii} component. None of the velocity components coincides with the velocity vector (v$_{\rm ISM}$ = 21.62 km/s) of the LIC cloud, which is seen along the line of sight to the star. As an example of the observed Ca {\sc ii} variability Fig. \ref{fig:HR2025_hk} shows both H and K lines taken in two consecutive nights, where remarkable variations of the depth and profile of both non-photospheric features are observed. We also note that our spectra differ from the one reported by \cite{lagrangehenri90}. Their spectrum shows a strong Ca {\sc ii} feature at the bottom of the stellar line, i.e., like our --22 km/s feature, but the strong 13 km/s absorption in our spectra, if present, is much weaker. Further, \cite{lagrangehenri90} do not report any Na  {\sc i} component. Thus, all these results suggest that the origin of all absorptions are CS.

\begin{figure}[h!]
  \centering
  \includegraphics[width=.5\textwidth]{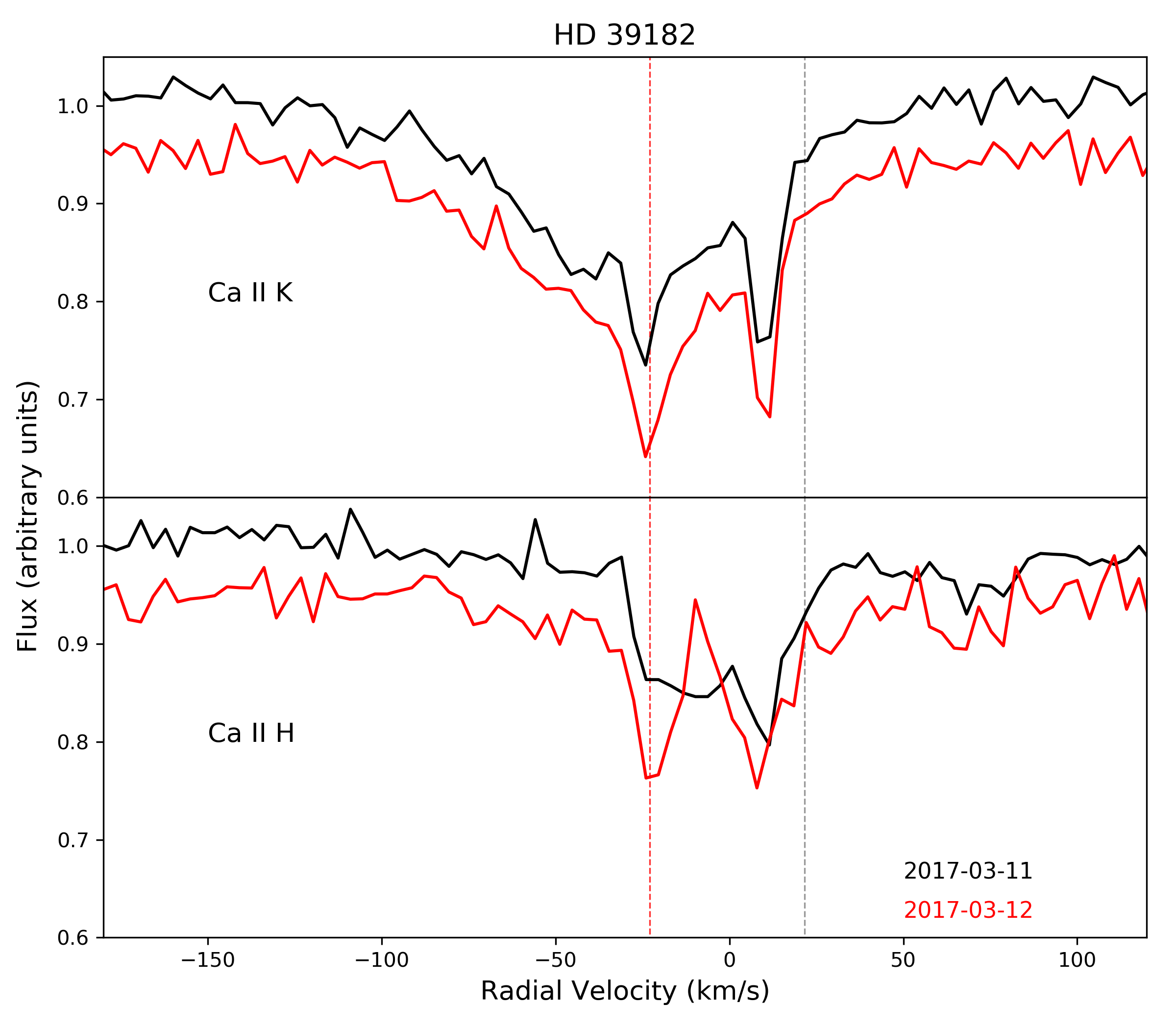}
 \caption{ Ca {\sc ii} K and H lines of HD 39182 colour-coded for two different observing dates obtained with HERMES. Spectra have been shifted 0.05 units in the Y-axis aiming to facilitate the visualisation of the variability. The red and grey vertical lines correspond to the  radial velocity  of the star and the velocity vector of the LIC Colorado cloud, respectively.}
\label{fig:HR2025_hk}
\end{figure} 

\begin{figure}[h!]
  \centering
 \includegraphics[width=0.5\textwidth]{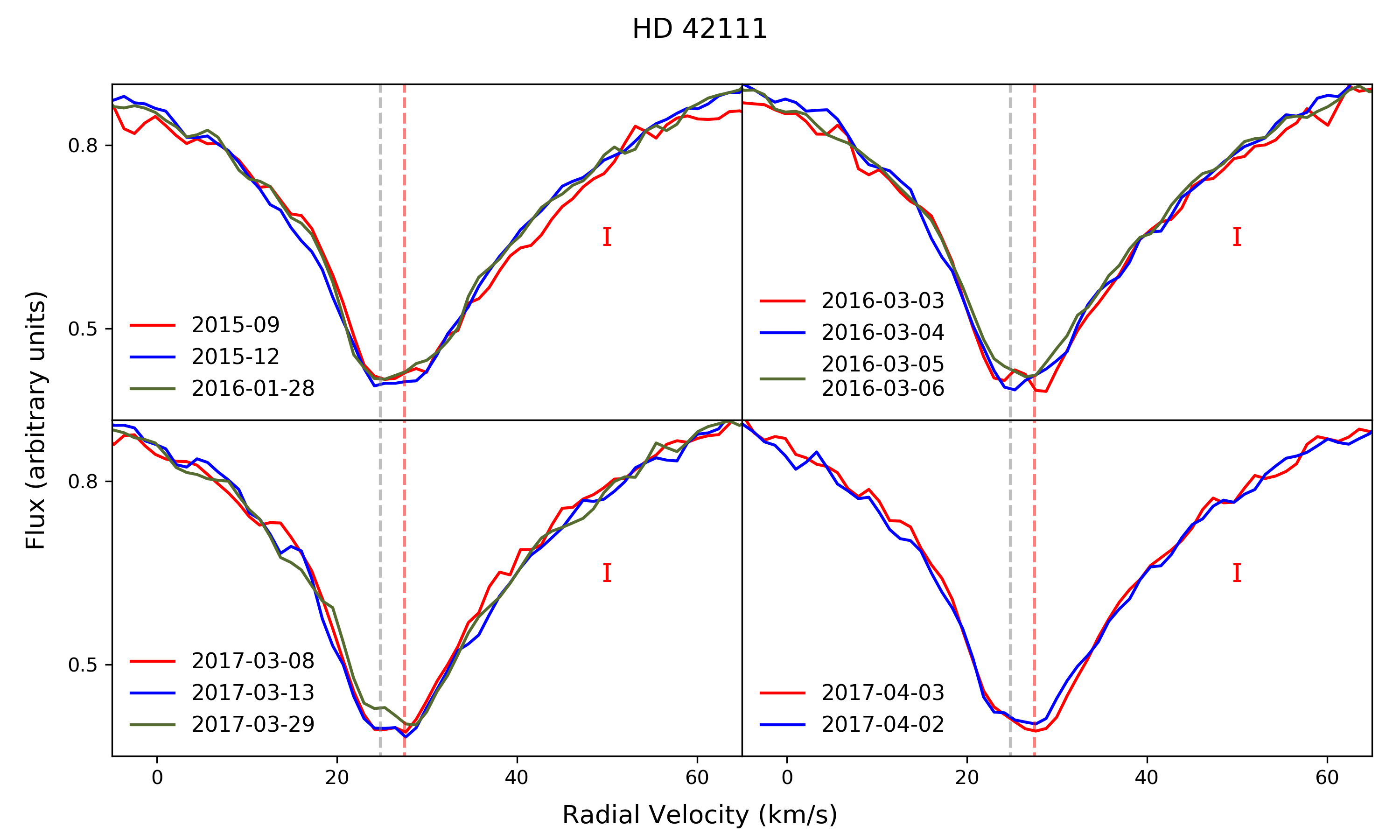} 
 \caption{Ca  {\sc ii}  K spectra  of  HD 42111 obtained with HERMES and grouped by  observing dates. Spectra have been shifted 0.05 units in the Y-axis. The red and grey vertical lines correspond  to the radial  velocity of the  star  and the  velocity  vector of  the ISM in the line of sight.}
\label{fig:HD42111}
\end{figure}

\begin{figure}[h!]
  \centering
 \includegraphics[width=0.5\textwidth]{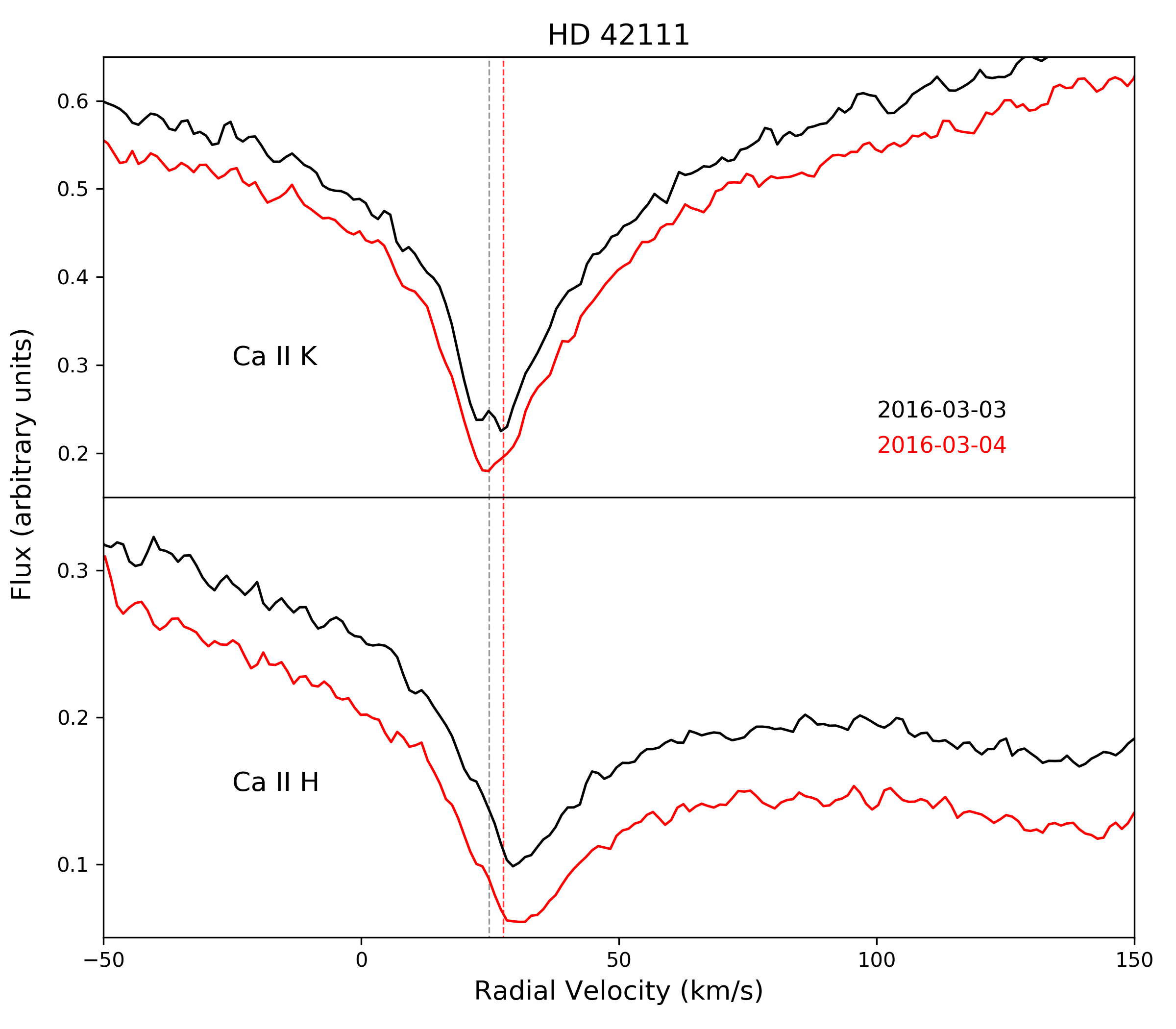} 
 \caption{Ca  {\sc ii} H\&K spectra  of  HD 42111 obtained with HERMES and grouped as observed in the indicated consecutive dates. The photospheric contribution has not been removed in this plot.The vertical lines correspond  to the radial  velocity of the  star  and the  velocity  vector of  the ISM in the line of sight.}
\label{fig:HD42111_CaiiHK}
\end{figure}

\item
HD 42111 (HR 2174) is a shell  star with strong {\mbox{Ca {\sc ii}}} and Na {\sc i} absorptions in the median spectrum close to the radial velocity of the star and of the Aur cloud velocity (Fig. \ref{fig:na_absorptions} and  Table  \ref{tab:narrow}), which have previously been reported by \cite{lagrangehenri90}. Those authors attributed a CS origin (at least partially) to the Ca {\sc ii} absorption based on a comparison of the dispersion velocities (FWHMs) of Ca {\sc ii} K and Na {\sc i} lines; this result was later confirmed by \cite{lagrangehenri91} by comparing the non-photospheric features  of HD 42111 and of the nearby star HD 42092. Further, the (at least partly) Ca {\sc ii} CS origin is corroborated by the fact that the EW and FWHM of the Ca {\sc ii} K absorption, as estimated with our spectra, are much larger that the ones reported in those works, while at the same time the EWs and FWHM of the strong Na I absorptions are similar - a small change of the Ca {\sc ii} K absorption was also pointed out by \cite{welsh13}.  Individual spectra of HD 42111 show that the Ca {\sc ii} K feature is formed by two components at velocities $\sim$25 km/s and 27.5 km/s, supporting the suggestion made by \cite{lagrangehenri90} concerning the plausible blend of two distinct features. Both Ca {\sc ii} K components have very similar strength (Fig. \ref{fig:HD42111}), although the feature  at $\sim$25 km/s varies in depth while the one at $\sim$27.5 km/s appears distinctly only in some selected dates as a kind of $\beta$ Pic-like events, e.g. 03/03/2016 or 08/03/2017 (Fig. \ref{fig:HD42111}), likely with a small dynamical evolution, at least in the March 2016 spectra, (see top-right panel of Fig. \ref{fig:HD42111}). In this respect, we note that \cite{welsh13} reported a FEB-like Ca {\sc ii} K event at 75 km/s, and that \cite{grady96b} and  \cite{lecavelier97} detected gas in UV lines of Fe {\sc ii}, Mn {\sc  ii}, and Mg {\sc  ii}, interpreted as CS clumps falling onto the star. With respect to the Ca {\sc ii} H feature,  our individual spectra do not resolve both K components, and are all well represented by their median profile; also, the peak of the H line feature is slightly red-shifted with respect to the K absorption. As an example, Fig. \ref{fig:HD42111_CaiiHK} plots the Ca {\sc ii} H and K lines of the two consecutive nights where that behaviour can be seen.
This could be due to the fact that the broad Ca {\sc ii} H feature is severely blended with the strong triangular-like profile at the core of the Balmer H$\epsilon$ line - such strong triangular-like profiles are clearly present in all Balmer lines. Obviously, additional higher resolution spectra  are needed in order to attempt to resolve the Ca {\sc ii} H absorption without the interference of the H$\epsilon$ line, and to study its plausible variability. We further note that a weak, but very broad, absorption is observed in both Na {\i} D lines producing the observed secondary peak (Fig. \ref{fig:na_absorptions}. That absorption,  present in all individual and the median spectrum of HD 42111, is not evident in the spectra by \cite{lagrangehenri90}, \cite{lagrangehenri91}, or the high-resolution, unpublished spectrum obtained by EXPORT \citep{mora01}.

\item HD 80007 has very weak Ca {\sc ii} K and Na {\sc  i} D2 absorptions at the stellar  v$_{\rm rad}$ (Table \ref{tab:narrow} and Fig. \ref{fig:na_absorptions}). The corresponding Ca {\sc ii} H and Na {\sc i} D1  might be present in our median spectrum but at the noise level (Fig. \ref{fig:na_absorptions}); and new spectra are required before a sound confirmation can be made. \cite{hempel03} noticed a change in the equivalent width, shape, and velocity of the Ca {\sc ii} absorption. \cite{redfield07b} also found variability in the velocity while the column density of the Ca {\sc ii} absorption remains relatively constant; in contrast, those authors found more remarkable variability in the velocity (two absorptions at $\sim$ --7 km/s and $\sim$ 7 km/s) and column densities of the Na {\sc i} feature. Further, \cite{welsh15} found a quasi two-component Ca {\sc ii} K feature in two consecutive nights and one single-component absorption in two  other nights, with changes in the equivalent width. In addition, \cite{wood71} found a quasi-periodic oscillation in the strength of the H$\beta$ Balmer line, and  suggested it could be due to flares generated  by acoustic oscillations of the stellar atmosphere.

\begin{figure}[h!]
  \centering
  \includegraphics[width=0.45\textwidth]{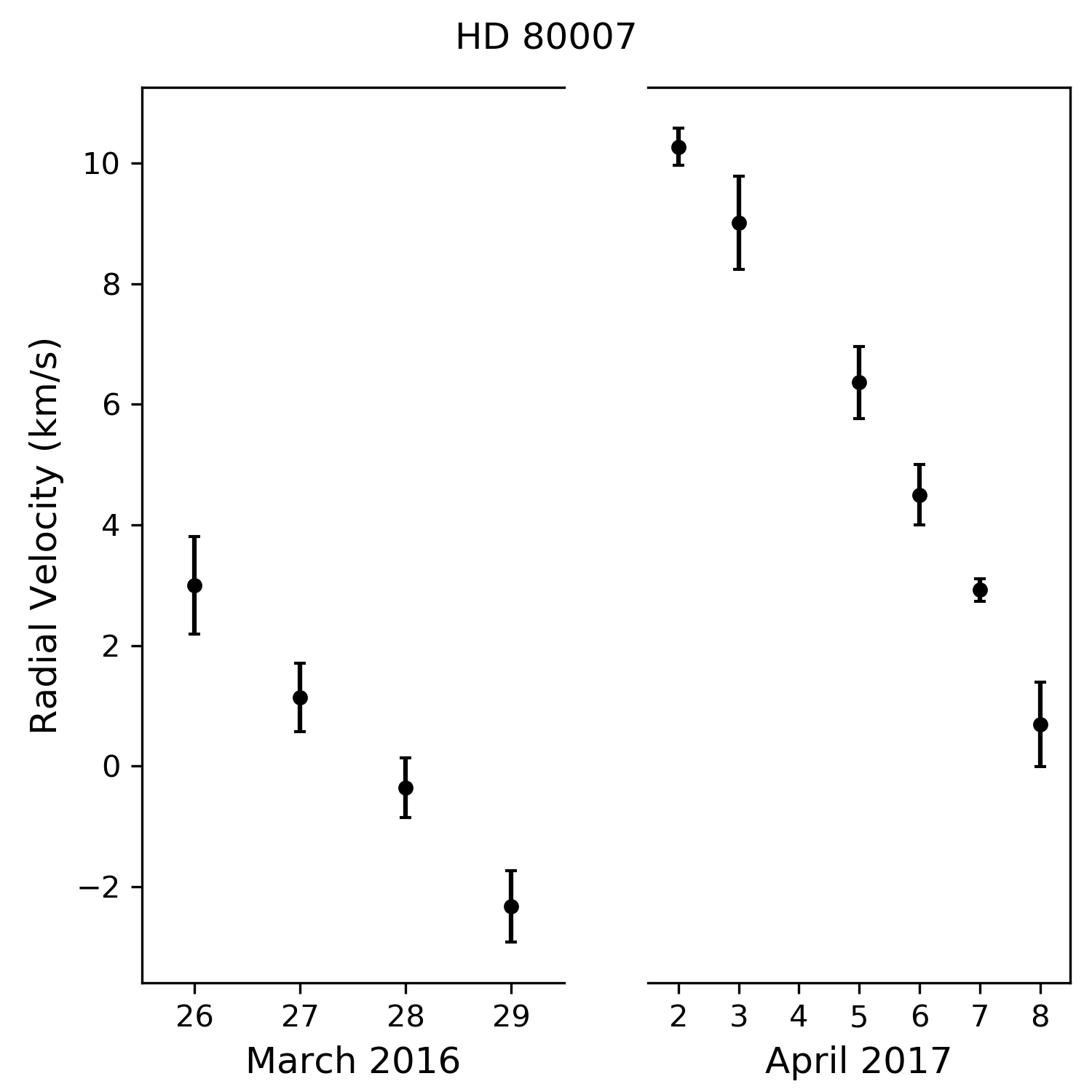}
 \caption{Radial velocity variation of HD 80007 in both observing periods. The observations suggest the presence of an unseen companion. }
\label{fig:hd80007_vrad}
\end{figure}

\begin{figure}[h!]
  \centering
 \includegraphics[width=0.5\textwidth]{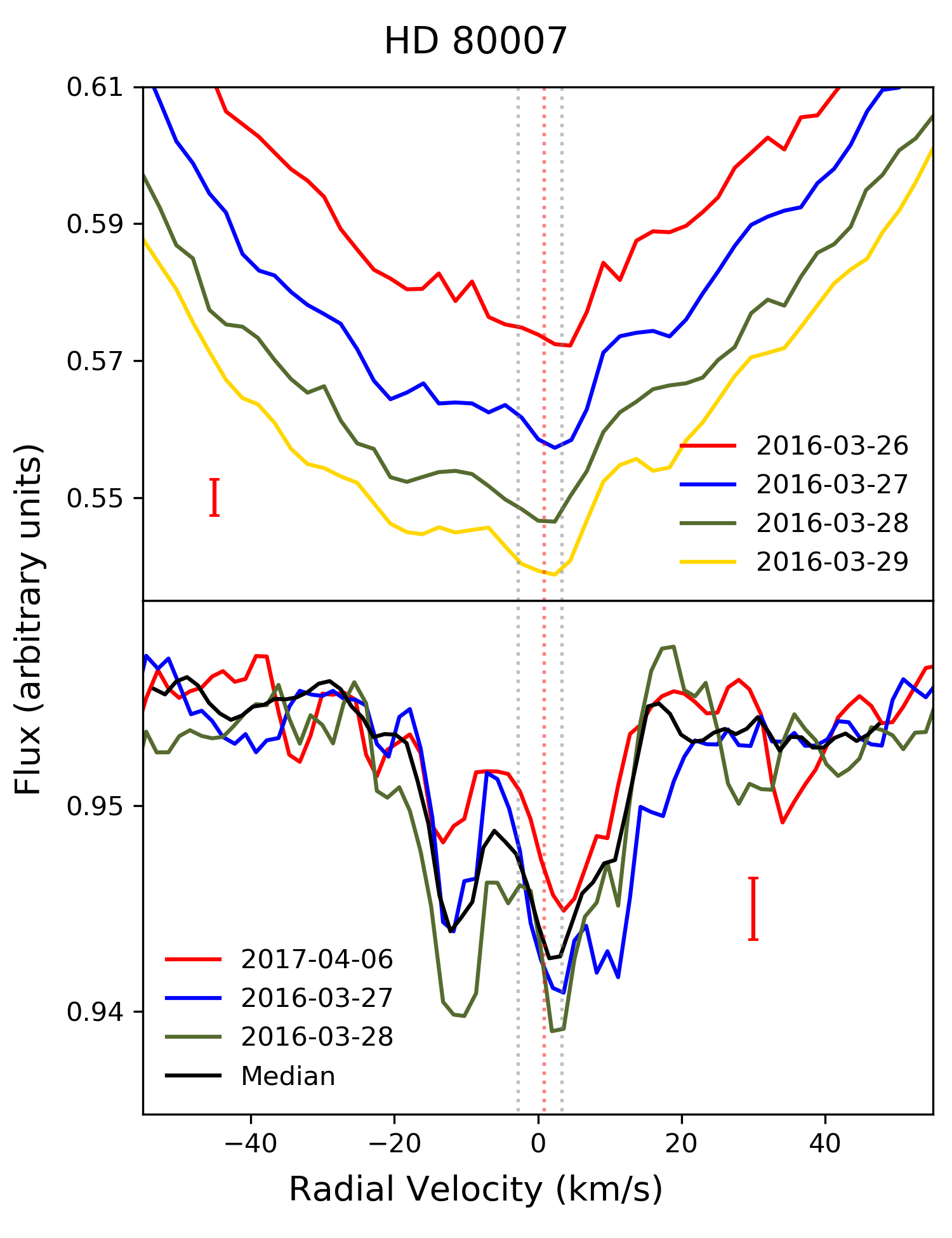}
 \caption{Top panel: Ca {\sc ii} K line of HD 80007 for the selected days, observed with FEROS. Spectra have been shifted {0.01 units} in the Y-axis aiming to facilitate the visualisation of the variability. The central absorption is seen at $\sim$ 2 km/s, as well as blue- and red-shifted variable absorptions. Bottom panel: Na {\sc i} D2  line of HD 80007 for the selected days. Variability is seen at $\sim$ -10 and  $\sim$ 2 km/s. In both panels it is noticeable the slight shift  at the bottom of the narrow absorptions presumably produced by a companion. }
\label{fig:HD80007_febs}
\end{figure} 

Our spectra show new aspects of both Ca {\sc ii} and Na {\i} absorptions as well as in the stellar radial velocity. Firstly, the radial velocity  of the star shows a regular variation of the order of $\sim$ 1.5 km/s per day in both 2016 and 2017 observing periods, Fig. \ref{fig:hd80007_vrad}. A possibility is that HD 80007 is a binary system and that the radial velocity variability is induced by an unseen companion. Secondly, the Ca {\sc ii} K absorption shows a "central" feature with small changes in its strength accompanied in some spectra with blue- and red-shifted components (Fig. \ref{fig:HD80007_febs}, top panel). At the same time the Na {\sc i} D2 feature presents two components. One is a broad, variable feature centred at $\sim$ 2 km/s, i.e., the velocity of the star and the Ca {\sc ii} feature, with a red-shifted wing up to $\sim$ 18.5 km/s even discernible in the median spectrum (Fig. \ref{fig:HD80007_febs}, bottom panel). Further, one red-shifted event at $\sim$10 km/s and extending up to $\sim$22 km/s might tentatively be present in the spectrum of 2016 March 27. The second Na {\sc i} D2 component at a velocity $\sim$ --11.5 km/s appears in all spectra but it varies its depth. We note that the velocity difference between both features is approximately the same as the ones sporadically observed by \cite{redfield07b}. 

\begin{figure}[h!]
  \centering
  \includegraphics[width=0.5\textwidth]{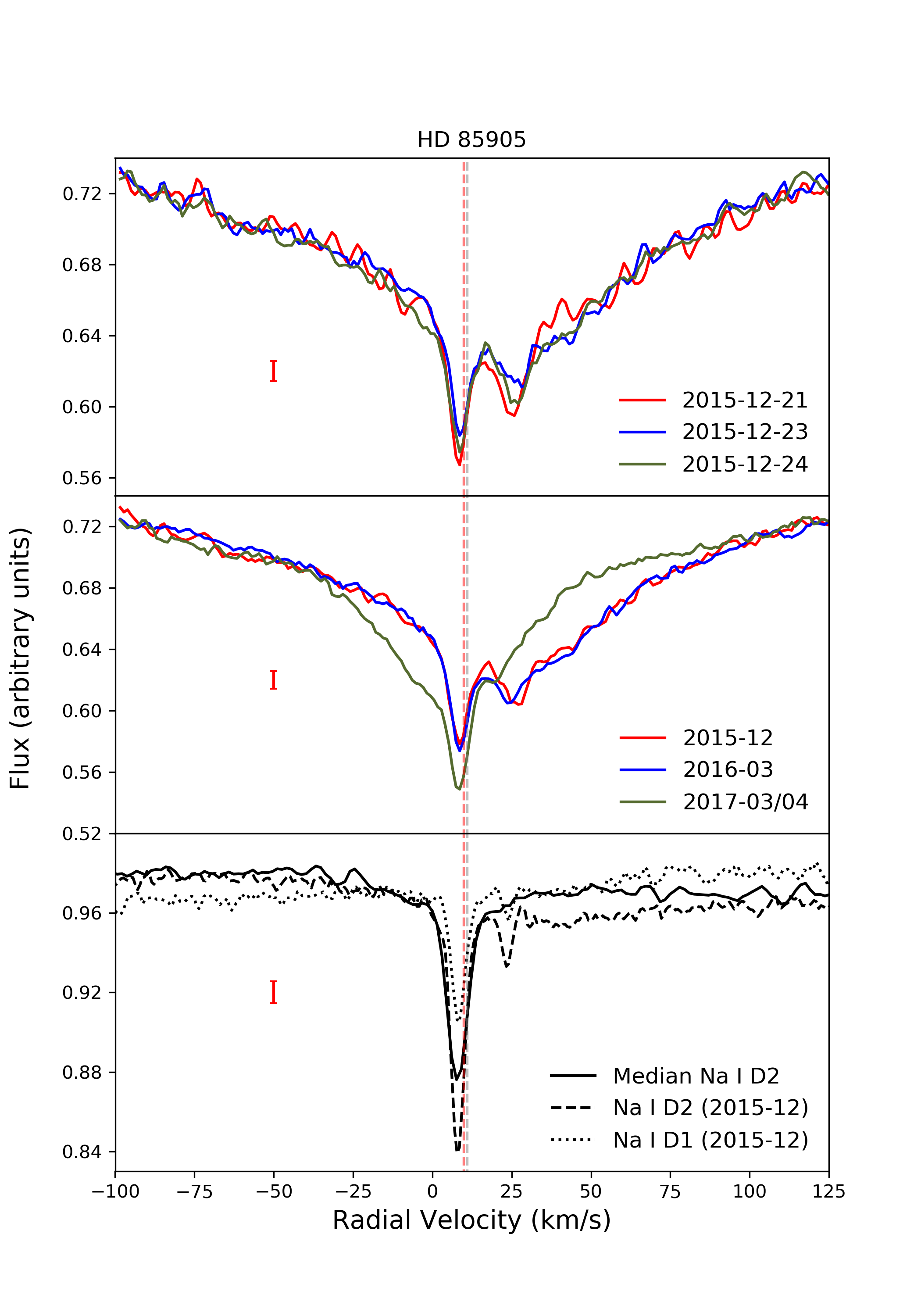}
 \caption{Top panel: Days 21, 23 and 24 of December 2015  where variability of the  $\sim$ 25 km/s Ca  {\sc ii} K feature can be seen. Middle panel: Ca  {\sc ii} K median spectra of three different campaigns using HERMES where the variability in the triangular profile is seen. The absorption at $\sim$ 25 km/s disappears in March 2017. Lower panel: Na {\sc i} D lines of December 2015 where the absorption at $\sim$ 23 km/s is visible. The median of all spectra in the Na {\sc i} D2 line where the absorption is no longer present is also shown. The vertical lines correspond to the radial velocities of the star and of the ISM.}
\label{fig:hd85905}
\end{figure}

\begin{figure}[h!]
  \centering
  \includegraphics[width=0.5\textwidth]{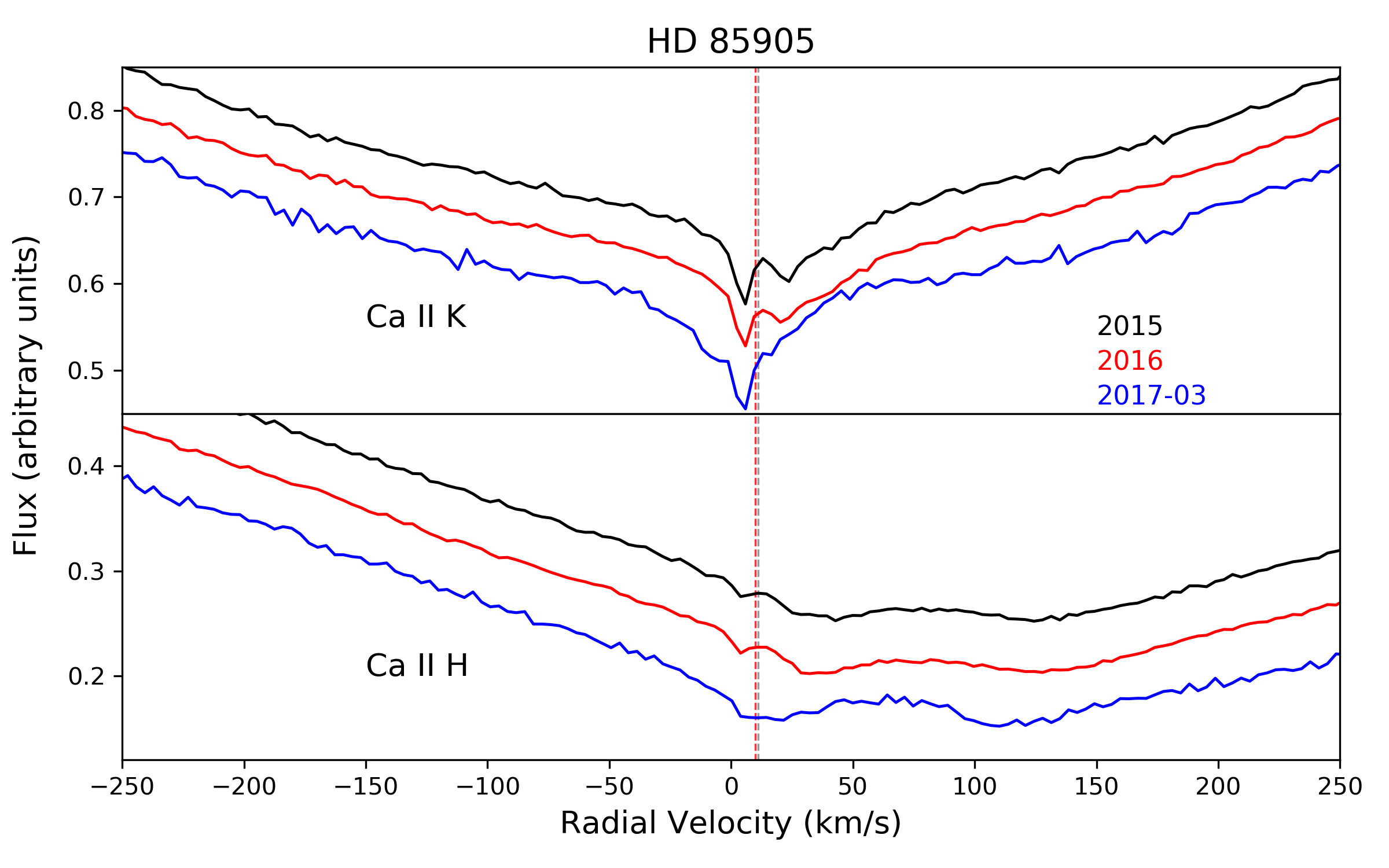}
 \caption{Ca {\sc ii} H and K median spectra of the campaigns indicated in the labels obtained using HERMES, where the variability of both lines can be appreciated. Spectra have been shifted 0.05 units in the Y-axis The photospheric contribution has not been removed in this plot. The vertical lines correspond to the radial velocities of the star and of the ISM.}
\label{fig:HD85905_CaiiHK}
\end{figure}

\item HD 85905 is a shell star whose median spectrum shows a sharp triangular-like Ca {\sc ii} absorption with two components at velocities at $\sim$8.4 km/s and $\sim$25.0 km/s, and one Na {\sc i} feature at $\sim$8.1 km/s. The feature at $\sim$8 km/s coincides with the radial velocity of the star and is also close to the velocity vector of the G cloud (Fig. \ref{fig:na_absorptions} and Table \ref{tab:narrow}). Nonetheless, individual spectra from the different epochs show remarkable variability; Fig. \ref{fig:hd85905} shows some examples. During December 2015 both Ca {\sc ii} K components experienced noticeable variations, while the corresponding Na {\sc i} 8.0 km/s feature remained constant. However, there is a relatively strong feature at a velocity of $\sim$23.7 km/s in both Na {\sc i} D lines, (i.e. close to the Ca {\sc ii} 25.0 km/s component) visible in all dates of that period (December 2015) but not in any other of our observing epochs. The lower panel of Fig. \ref{fig:hd85905} shows the median of December 2015 spectra of both Na {\sc i} D lines where this result can be appreciated. Further, while during the periods of December 2015, January and March 2016 both Ca {\sc ii} components were present, only the component at $\sim$8 km/s was visible during the two different campaigns of 2017 (March 8 to 11, and March 29 to April 8). During these campaigns, a strong absorption appeared as well in the blue wing of the Ca {\sc ii} K, while the red-shifted 25.0 km/s feature practically disappears. Fig. \ref{fig:HD85905_CaiiHK} shows the profiles of both the Ca {\sc ii} H and K lines of the median spectra of the three previously mentioned periods where changes of the H line profile can also be appreciated. The above results clearly suggest a CS origin of the non-photospheric absorptions observed in HD 85905, maybe related to a variability of the CS shell as suggested by the variations observed in other shell lines of e.g. Fe {\sc ii}, but not in photospheric lines as Mg {\sc ii} 4481 \AA ~or the O {\sc i} triplet at 7775 \AA ~(not shown). A detailed analysis will be published elsewhere. \cite{welsh98} and \cite{redfield07b} also attributed a CS origin to the Ca {\sc ii} and Na {\sc i} absorptions they detected.

\begin{figure}[h!]
  \centering
  \includegraphics[width=0.5\textwidth]{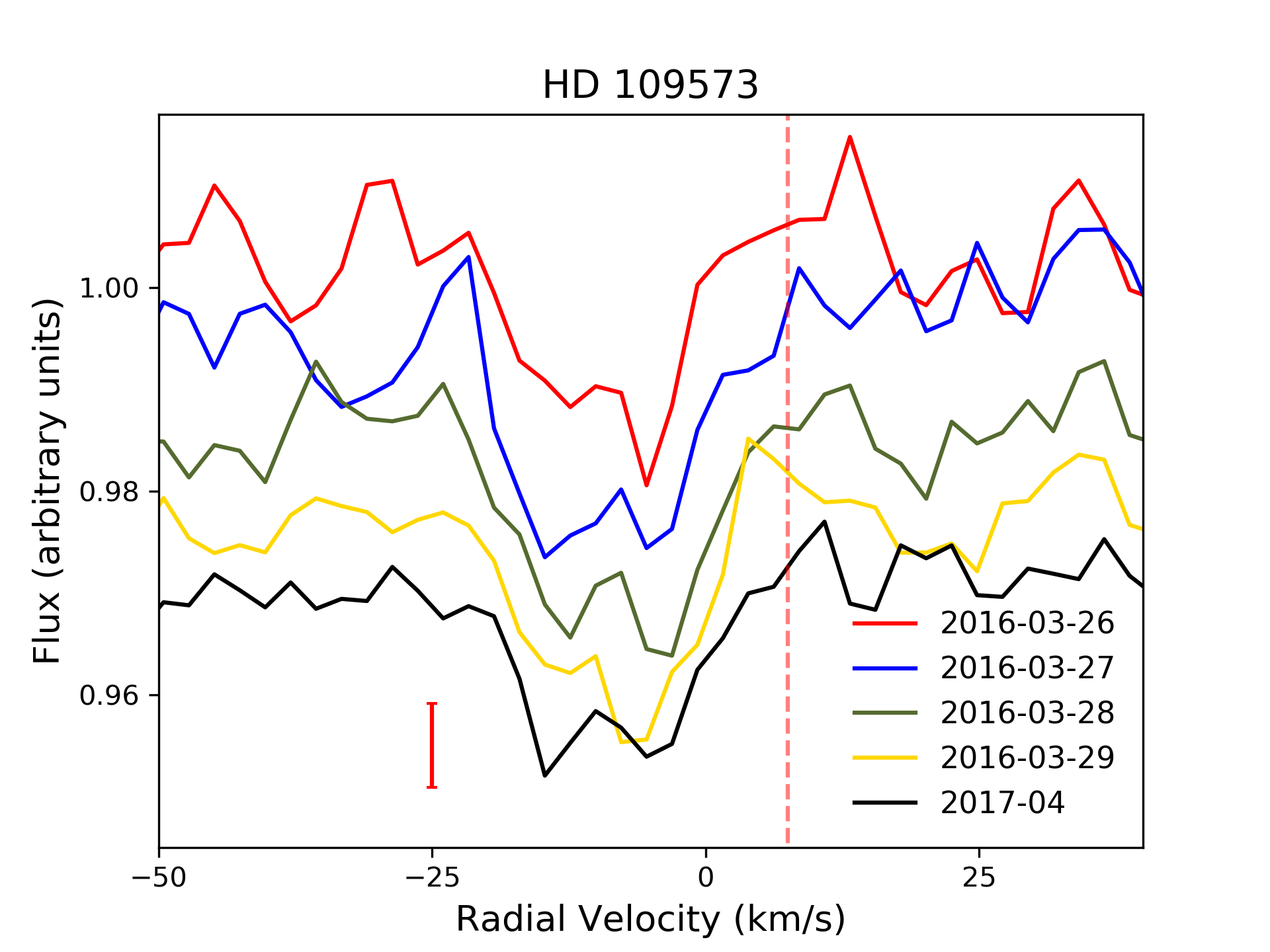}
 \caption{Ca  {\sc ii}  K  line of HD 109573. A shift of 0.005 has been added to the Y-axis in order to help differentiate the variations. All spectra were obtained using FEROS. The red vertical line corresponds to the radial velocity of the star. Ca {\sc ii} K line shows two absorptions  not coincident with the radial velocity of the star but with variations in their strength.}
\label{fig:HR4796}
\end{figure}

\item HD  109573 (HR  4796) has two very weak Ca {\sc  ii} K absorptions at $\sim$ --14.2 and --4.7  km/s in the median spectrum (Fig. \ref{fig:na_absorptions}). The one at --14.2 km/s is not detected in the Ca {\sc ii} H line, suggesting the gas is optically thin, as it is the --4.7 km/s feature. This latter absorption  is also present in the Na {\sc i} D lines. None of these features coincides  with the  radial velocity of the  star. Fig. \ref{fig:HR4796} shows details of the Ca {\sc ii} K line on different dates (the Ca {\sc ii} H line is only revealed with the median spectra of all 23 individual spectra of HR 4796). Both Ca {\sc ii} K components vary in depth and shape, in some cases close to the noise level. Nonetheless, a discernible variability is seen, e.g. the -14.2 km/s component of March 2016 26th and 29th  is distinctively weaker than of March 2016 27th, 28th. or the median of April 2017.  
Our spectra and the variability of the Ca {\sc ii} K features are quite similar to those in \cite{welsh15}, strongly suggesting a CS origin. Nonetheless, \cite{Iglesias18} attribute the --5  km/s feature an ISM origin, as the field star 1 Cen (HD 110073; 30 pc behind HR 4796) also has a similar absorption feature. \cite{welsh15} detected a faint FEB-like event at $\sim$ 60 km/s in two spectra of a single night; and \cite{Iglesias18} report a faint variable feature at the velocity of the star. None of these are apparent in our spectra.

\item HD 138629 (HR 5774) has three non stellar Ca {\sc ii} features at velocities  --31.8, --22.9, and --13.8 km/s with no sign of variability, none of them clearly coincident with v$_{\rm rad}$  or with v$_{\rm ISM}$. The  Na {\sc i} D lines also present three absorptions, two of them coinciding with two Ca {\sc  ii} ones (Fig. \ref{fig:HR5774}). Our spectra differ from the two and four Ca {\sc ii} components, and from one Na {\sc i} feature, reported by \cite{Lagrange-Henri90b}. 
We attribute at least partially a CS origin due to the apparent changes with previous works, but an ISM origin can not be excluded at least for the features at $\sim$ --22 km/s and $\sim$ --12 km/s.

\begin{figure}[h!]
  \centering
 \includegraphics[width=0.5\textwidth]{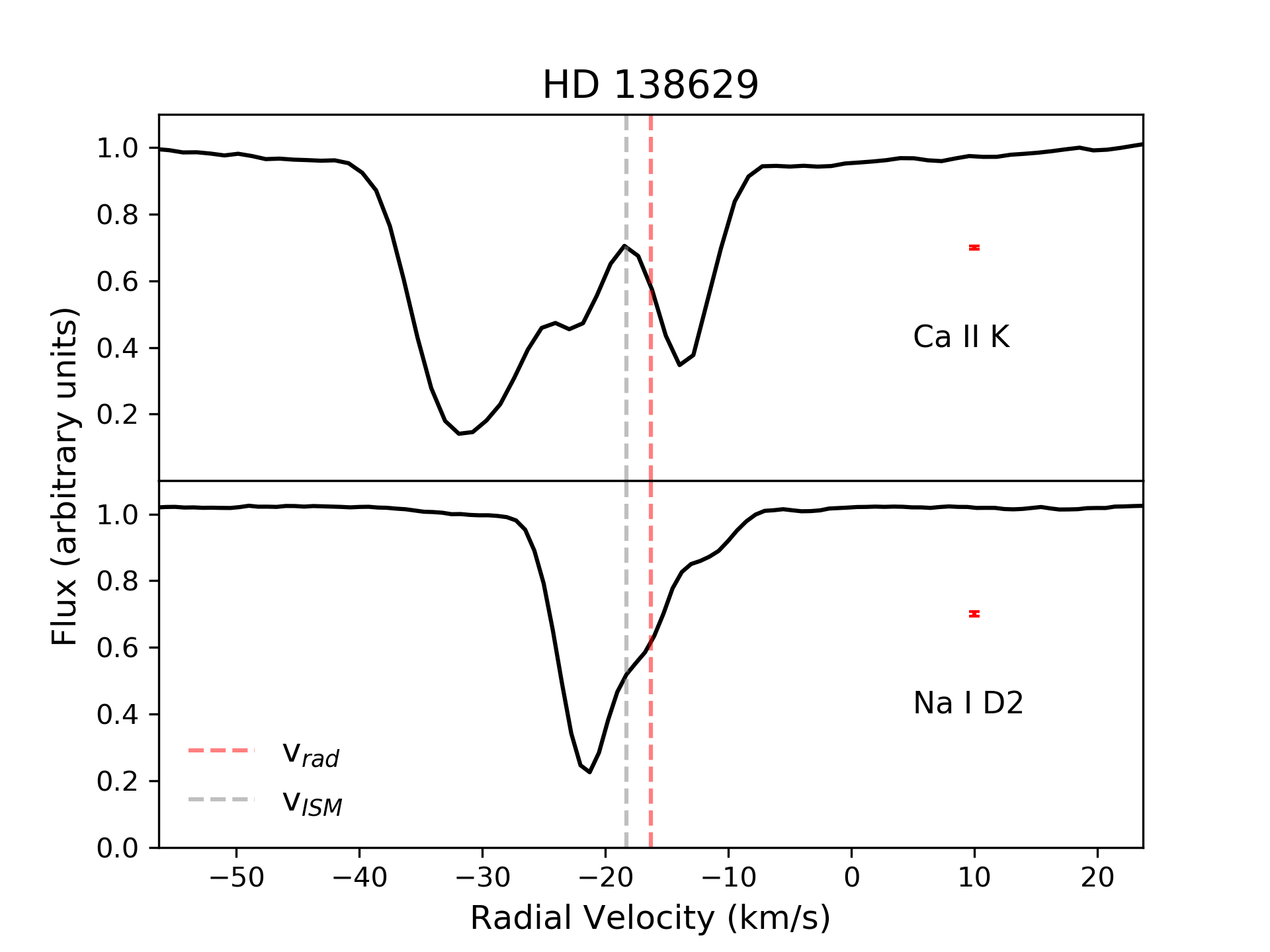}
\caption{ Median HERMES spectra of Ca {\sc ii} K and Na {\i} D2 non-photospheric absorption features observed towards HR 5774. Dashed lines correspond to the radial velocity of the star (red) and the velocity vector of the interstellar NGP cloud (blue).}
\label{fig:HR5774}
\end{figure}

\begin{figure}[b!]
 \centering
 \includegraphics[width=0.5\textwidth]{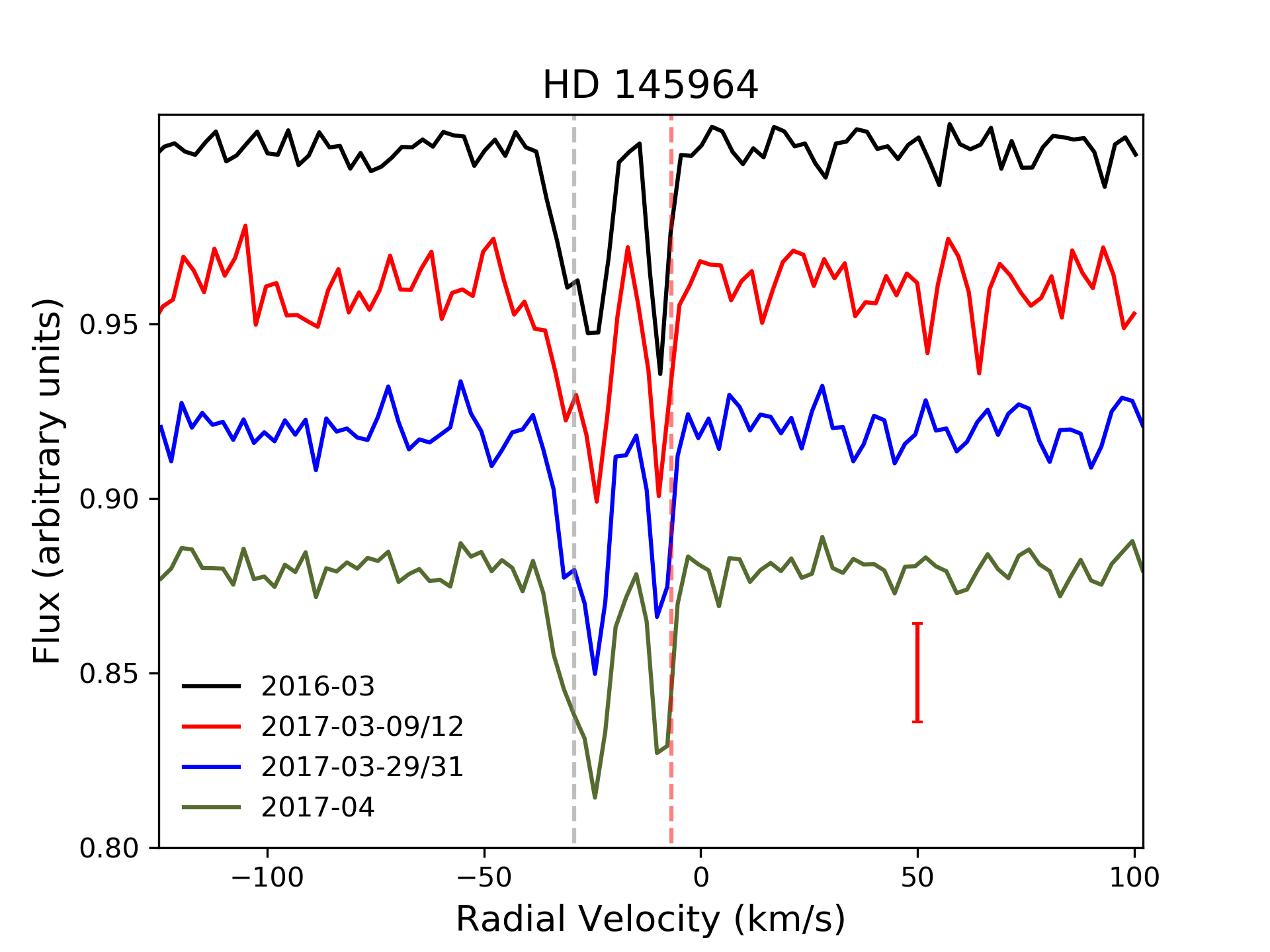}
 \caption{Ca {\sc ii} K spectra of HD 145964 as indicated in the labels, obtained using HERMES. An offset of 0.04 was introduced in the Y-axis to better perceive the variations. The red and grey vertical lines  correspond to  the radial velocity of the star  and the velocity vector  of the  G Colorado cloud, respectively}
\label{fig:HD145964}
\end{figure}

\item HD 145964 is one of the Upper Scorpius stars.  Two absorption components at velocities $\sim$ --9.5 km/s and $\sim$--25  km/s  are detected in Ca {\sc  ii} and Na {\sc i} lines (Fig. \ref{fig:na_absorptions} and Table \ref{tab:narrow}). Those components can in principle be attributed to the ISM medium as they are also detected in many other UpSco stars (see Sect. \ref{sect:individual_stars}). However, while the profile of Ca {\sc ii} K feature at --9.0 km/s, which is the velocity of the star, remains practically constant, the asymmetric blue wing of the --25 km/s absorption shows a  weak $\sim$ --30 km/s component in most spectra of all observing runs, but not e.g. in both the Hermes and FEROS spectra taken during the same dates in April 2017 (Fig. \ref{fig:HD145964}). We also note that a marginal variation in the relative depth of the --9.0 km/s and --25 km/s features might be present. The same behaviour is observed in the Ca {\sc ii} H line (Fig. \ref{fig:HD145964_febs}).
 
Thus, a CS contribution is suggested, in particular for the --25 km/s and --30 km/s absorptions. Similar Ca {\sc ii} K absorptions were detected by \cite{welsh13}, who also reported a weak FEB-like event at a velocity of $\sim$ 50 km/s that is not detectable in our data.

\begin{figure}[b!]
 \centering
 \includegraphics[width=0.5\textwidth]{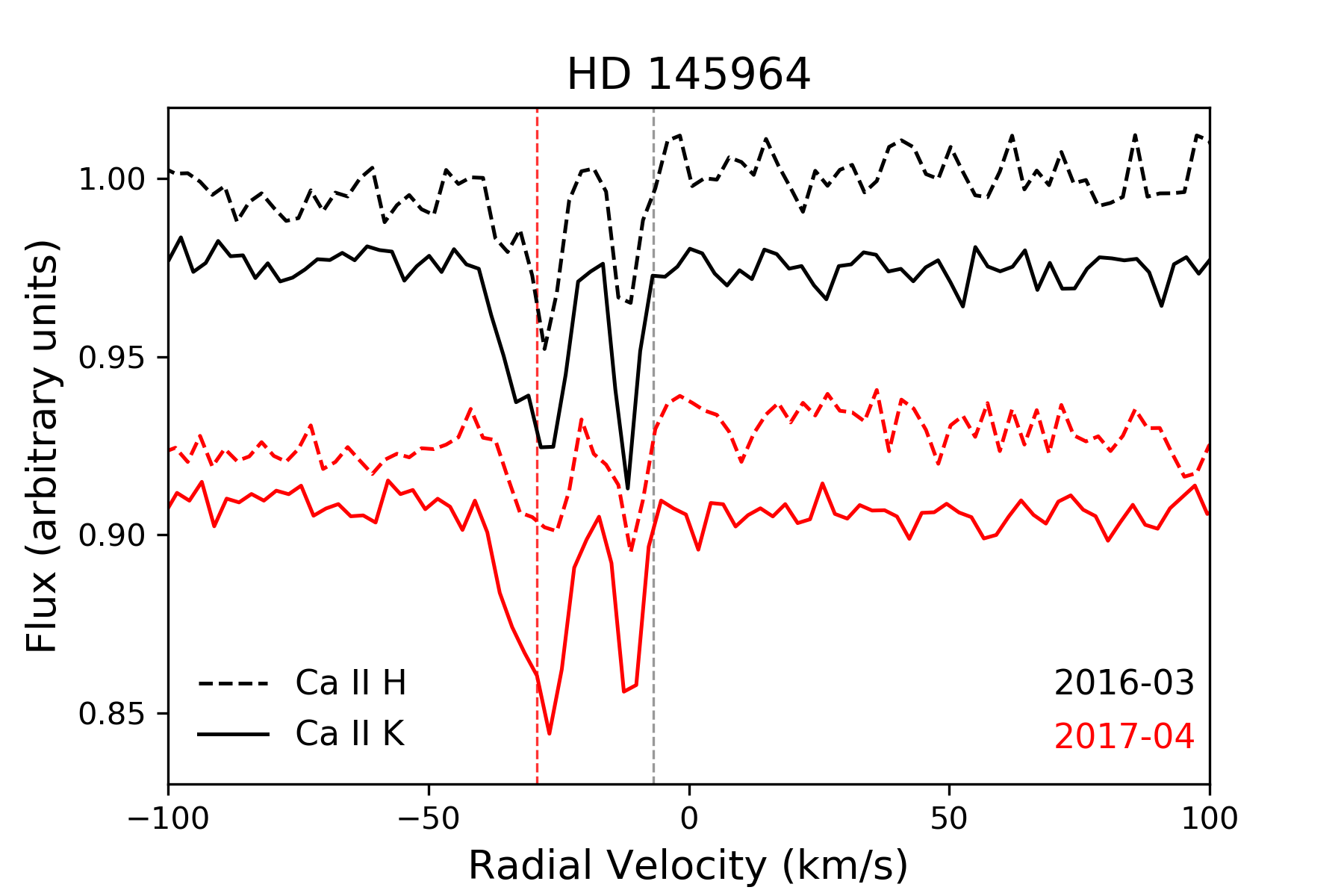}
 \caption{Spectra of HD 145964 colour coded as indicated in the labels for Ca {\sc ii} K \& H lines (continuous and dashed lines, respectively), and obtained with HERMES. An offset {of 0.05} was introduced in the Y-axis to better perceive the variations between both dates, and 0.02 between the K and H lines. The red and grey vertical lines  correspond to  the radial velocity of the star  and the velocity vector  of the  G Colorado cloud.}
\label{fig:HD145964_febs}
\end{figure}

\item HD 172555 was already discussed by \cite{Rebollido18} as one of the debris disc stars with both cold and hot gas in its circumstellar environment  (see also the first paragraph of Sect. \ref{sect:individual_stars}). Here we want to stress that a weak ISM  feature is detected in both Ca {\sc ii} lines at a velocity of $\sim$--20 km/s, in good agreement with \cite{kiefer14a}. In addition, our spectra reveal a weak, broad Na {\sc i} D2 feature centred at $\sim$15.3km/s and extending from $\sim$--5 km/s up to $\sim$35 km/s. Although the individual spectra are relatively noisy we are confident on this detection as it appears in all spectra. The top panel of Fig. \ref{fig:HD172555} shows the Na {\sc i} D2 median spectrum, and spectra  of 08/03/2016, 08/04/2017. The broad Na {\sc i} D2 feature presents a clear variability, denoting its CS nature. The bottom panel of Fig. \ref{fig:HD172555} shows the telluric subtraction for the indicated dates, where it is clear the variability is not related to telluric lines. Fig. \ref{fig:HD172555} (top panel) also shows the median of all Na {\sc i} D1 line spectra, where the ISM feature is clearly visible but not the broad CS one. We note that \cite{grady18} detected some UV broad, variable absorptions of ions like e.g. C{\sc ii}.
  
\begin{figure}[h!]
  \centering
 \includegraphics[width=0.5\textwidth]{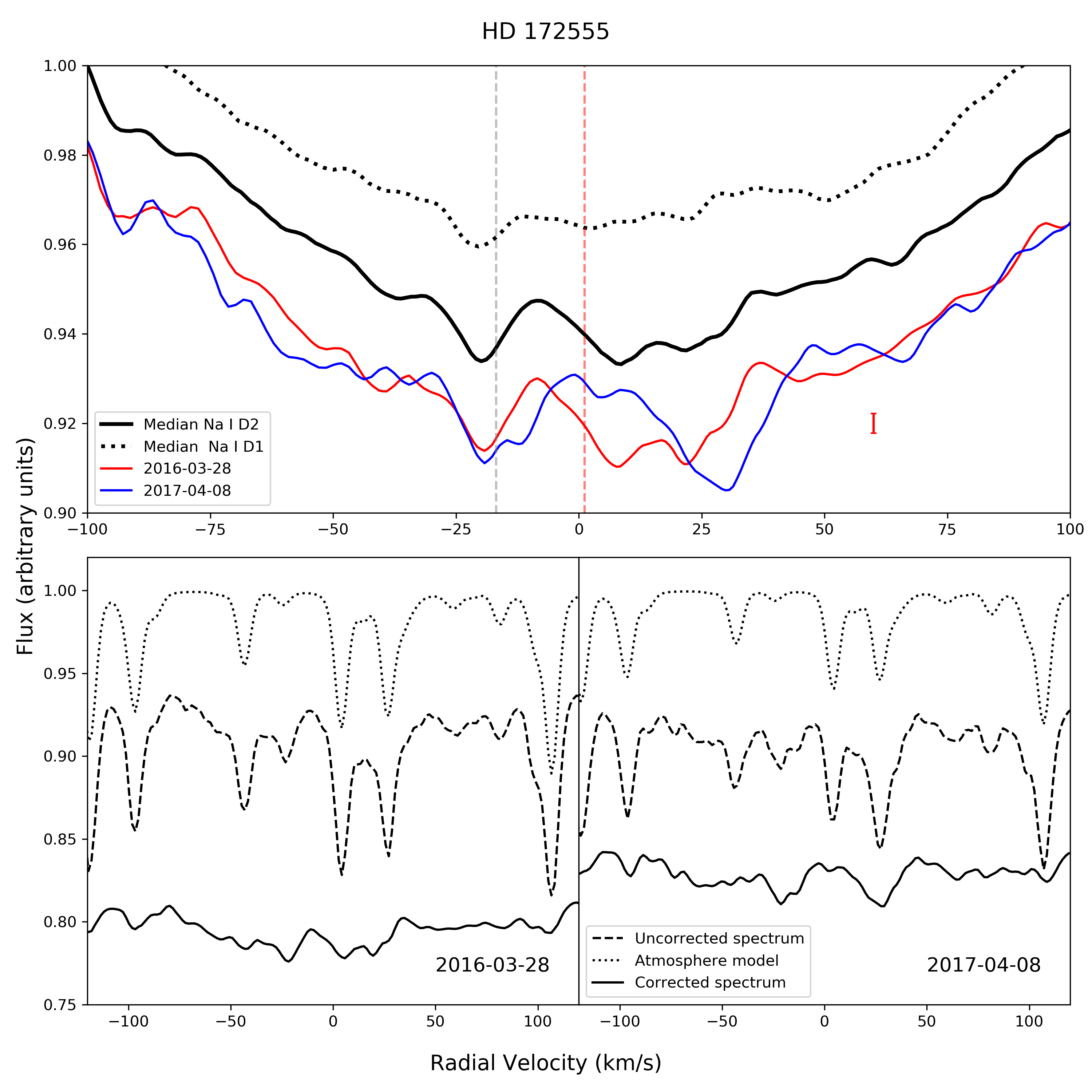}
 \caption{ Top panel: Absorptions detected in the Na {\sc i} D2 line of HD 172555. Black lines corresponds to the median of all spectra for Na {\sc i} D2 (solid line) and D1 (dotted line). Blue and red lines correspond to two different dates, where the variations in the $\sim$15 km/s Na {\sc i} D2 component can be perceived. Bottom panel: Examples of the telluric subtraction are plotted, in order to show that this process it is not the source of the variability. In both cases, spectra were obtained using FEROS. Red and grey vertical lines mark the stellar and ISM radial velocities respectively.}
\label{fig:HD172555}
\end{figure}

\item HD 182919 (5 Vul)  has a  weak, narrow  absorption  feature at {\mbox{$\sim$--18.5 km/s,}} close  the velocity of the star and  to the G, Mic, and Aql clouds (Table \ref{tab:narrow} and Fig. \ref{fig:na_absorptions}). The feature varies its depth ($\sim$ 6.0$\sigma$) when analysing  the median spectra of different epochs (Fig. \ref{fig:HD182919}); thus,  it most likely has a CS origin, at least partly. During the 2016 July observing run at Mercator a very weak Ca {\sc ii} K blue-shifted absorption with an EW of 1.3 m\AA ~is apparent at a velocity of $\sim$ -35 km/s; in addition, the spectrum of 2016-07-14 shows a Ca {\sc ii} K red-shifted event at $\sim$ 25 km/s and EW 1.6 m\AA. This absorption is not detected in the Ca {\sc ii} H line. Since its value is consistent with a 3.1 $\sigma$ detection, we consider the detection {\it tentative}. We note that \cite{montgomery12} also  noticed  the  variability of  the narrow absorption as well as the presence of a FEB-like event with a velocity range 15--60 km/s in  one of their spectra.
\begin{figure}[h!]
  \centering
 \includegraphics[width=0.5\textwidth]{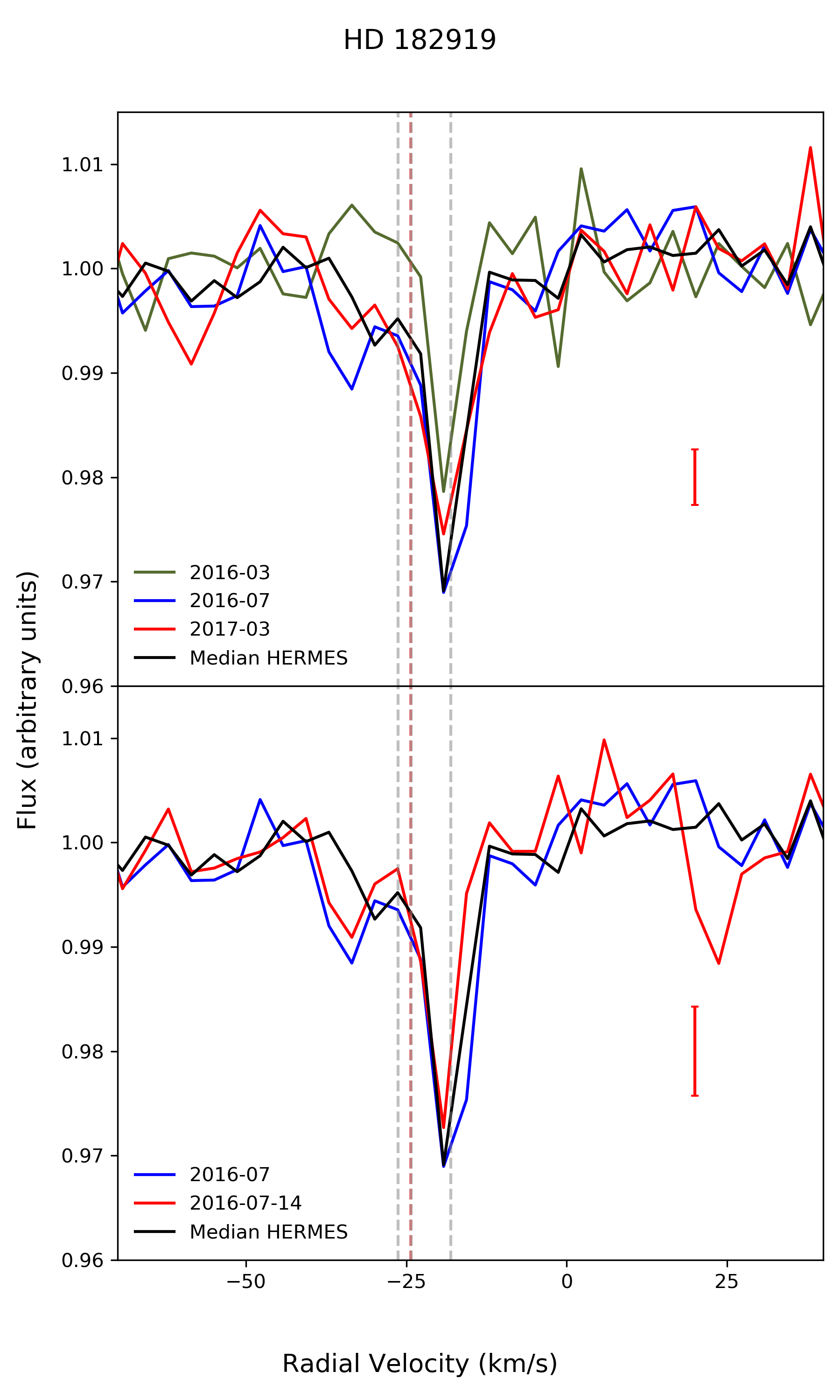}
 \caption{Ca {\sc ii} K spectra of HD  182919 (5 Vul) taken with HERMES in the dates indicated in the labels. Top panel shows the variation of the features at $\sim$ --18 and $\sim$ --35 km/s. In the bottom panel it is visible a possible FEB-like event at $\sim$ 25 km/s. The red and grey lines correspond to the radial velocity of the star and the velocity vector of Colorado clouds, respectively.}
\label{fig:HD182919}
\end{figure}

\item HD 217782 (2 And) has three Ca {\sc ii} absorptions at velocities $\sim$ --17.1 km/s, --9.2 km/s, and 5.2 km/s (see Fig. \ref{fig:na_absorptions}). Two of them, blue-shifted with respect to the radial velocity of the star, are also detected in Na {\sc i}. The weakest feature at 5.1 km/s is at the stellar v$_{\rm rad}$ and the velocity vector of the Hyades ISM cloud (Table \ref{tab:narrow}). While the stronger and narrower {\mbox{-9.3}} km/s Ca {\sc ii} K feature remains unchanged (variations below 1$\sigma$), the -16.5 and likely the 5.1 km/s components present some variability (Fig. \ref{fig:HD217782}). Particularly,  the -16.5 km/s feature seems to experience some dynamical evolution changing its depth and velocity within hours, e.g. up to 3 $\sigma$ EW variation along the night  6$^{th}$/7$^{th}$ September 2015, as well as a shape and depth change (Fig. \ref{fig:HD217782}, bottom left panel). Similar changes are also seen on other nights. Those changes, although much less remarkable, can tentatively be present in the weaker Ca {\sc ii} H feature, as seen in Fig. \ref{fig:HD217782_febs} where both Ca {\sc ii} non-photospheric lines of the mentioned night are shown. These results  suggest the presence of CS gas around 2 And. \cite{cheng03,montgomery12,welty96} found similar UV/optical results.

\begin{figure}[h!]
  \centering
  \includegraphics[width=0.5\textwidth]{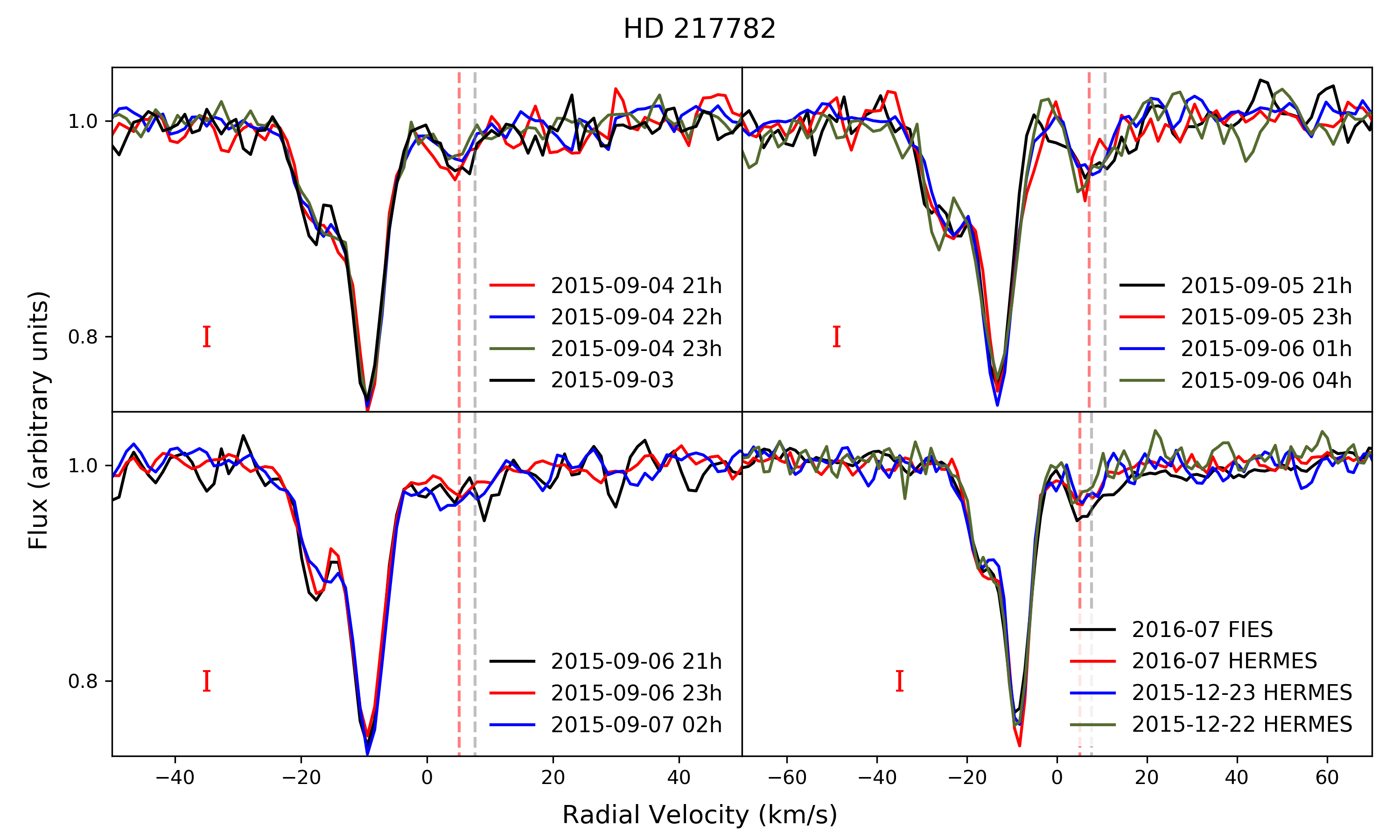}
 \caption{Ca {\sc ii} K spectra of HD  217782 as indicated in  the labels. In the bottom right panel the labels indicate the instrument used in each case. The rest of the spectra were obtained with HERMES. The dashed red and grey lines correspond to the radial velocity of the star and the velocity vector of Colorado clouds, respectively.}
\label{fig:HD217782}
\end{figure}

\begin{figure}[h!]
  \centering
  \includegraphics[width=0.5\textwidth]{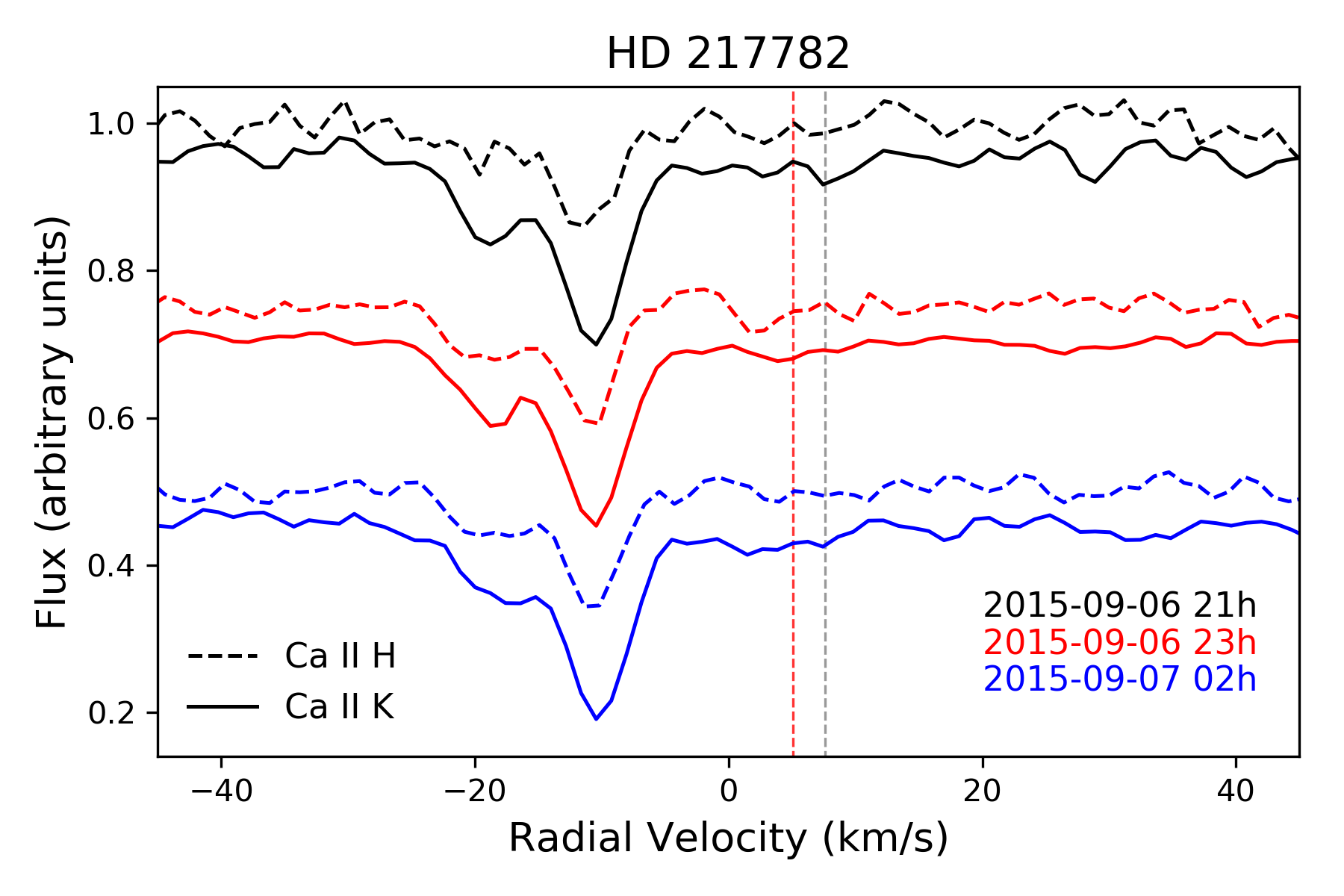}
 \caption{Ca {\sc ii} H and K non-photospheric features of HD  217782 taken during the night indicated in the labels. All spectra were taken using HERMES. A shift of 0.2 was introduced in the Y-axis to help differentiate the variations.  The dashed red and grey lines correspond to the radial velocity of the star and the velocity vector of Colorado clouds, respectively.}
\label{fig:HD217782_febs}
\end{figure}

\end{itemize}

\subsection{Summary of CS gas detections}

Fig. \ref{fig:na_absorptions} shows the observed median Ca {\sc ii} K\&H  and Na {\sc i} D lines and the non-photospheric residuals, once the stellar contribution has been subtracted, of the 60 stars that have narrow stable absorptions; radial velocities of the stars and the velocity vector of the ISM Colorado clouds velocities are also plotted. Table \ref{tab:narrow} gives radial velocities, FWHM, equivalent widths, and column densities of the narrow features. Table \ref{tab:feblist} lists the stars showing variable absorptions detected in this work, and also the stars reported in the literature as hosting sporadic events but not detected by us.

We find evidence of hot CS gas in 32 objects out of the initial sample of 117 stars, being 30 in the form of stable non-photospheric components, and variable absorptions in the other two cases. Variable red- and/or blue-shifted events with respect to the radial velocity of the stars have been detected in 18 objects, including the serendipitous detection of HD 132200, which was not in the initial sample \citep{Rebollido18}. Among those 18 objects, all but HD 110411 and HD 183324, also have stable narrow features. These figures mean we have found evidence of a close-in gaseous CS environment in $\sim$27\% of the sample. We note, however, that it is not statistically significant as the selection criteria were highly biased, for instance including stars for which  the presence of hot or cold CS gas was already known. Nonetheless, the figure does indicate that a non-negligible amount of stars, particularly A-type (see below), are surrounded by CS gas which can be detected by means of high resolution optical spectroscopy. We note that we are not considering the 8 stars where the detected narrow feature has a dubious origin as we are unable to soundly attribute it either to a CS or/and ISM environment. 

The detected CS gas clearly has distinct origins. Red- and blue-shifted events are plausibly linked to the presence of FEBs, as in the well known case of $\beta$ Pic; even in some cases our spectra likely trace the dynamical evolution of those exocomets, i.e., a change in depth and velocity. In none of the cases the exocomet activity is as rich as in $\beta$ Pic, which remains unique in this context. Stable absorption features in some stars are also likely related to exocomets and/or evaporation of grains in the immediate CS environment. In the case of shell stars, the non-photospheric stable features, including many metallic shell lines of species like Fe {\sc ii} or Ti {\sc ii}, are related to the shell itself, and likely arising from mass loss phenomena experienced by the central star. Nonetheless, some shell stars also present sporadic red- or blue-shifted absorption events in Ca {\sc ii} reminiscent of exocomets. HD 37306 represents an extreme case, where we detected the appearance and disappearance of a strong shell spectrum but no trace of any exocomet-like event.

\section{Discussion}
\label{discussion}

While our stellar sample is heterogeneous and highly biased we can still try to find some trends among the incidence of CS gas and some general properties of the stars, and the different groups of stars according to the selection criteria. 

\begin{figure}[h!]
  \centering
 \includegraphics[width=0.5\textwidth]{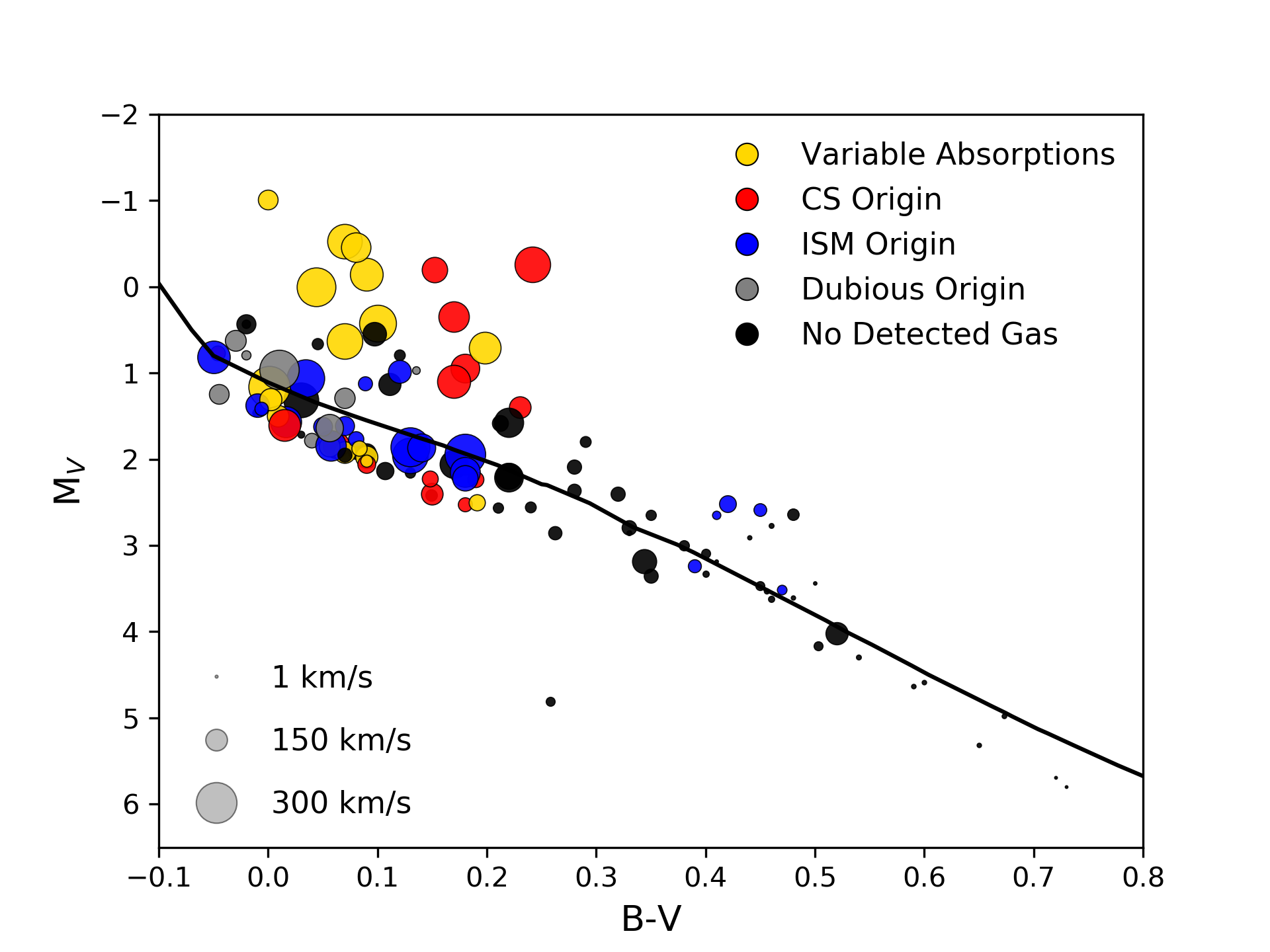}
\caption{Colour-Magnitude diagram of the whole sample. Colours represent stars with different non-photospheric features while the size of the symbols is proportional to  $v \sin i$.}
\label{fig:hrdiagram}
\end{figure}

Fig. \ref{fig:hrdiagram} shows the HR diagram of the sample where the absolute magnitude M$_V$ and colour index B-V are estimated from the magnitudes and parallaxes given in SIMBAD; the MS track is taken from \cite{pecaut13}. To estimate M$_V$ we have not taken into account the potential extinction towards the stars; nonetheless, the true loci of the individual stars in the HR diagram would not significantly alter the conclusions. Objects with evidence of hot CS gas (variable or stable) are A-type stars, and are therefore located in the upper-left region of the diagram. This is in line with previous works, although as far as we know they have only been concentrated in the study of A-type stars \citep[e.g.][and references therein]{holweger99,welsh18}. We only know of one later spectral type star, HD 109085 ($\eta$ Crv, F2 V) for which one exocometary-like event has recently been reported although it requires confirmation \citep{welsh19}; while cold CO is most likely present in this system \citep{marino17}, we do not find any trace of CS gas  associated with this star \citep[][this work]{Rebollido18}. In this respect, photometric transits are more efficient than spectroscopy to detect exocomets around later type stars \citep[e.g.][]{rappaport18,ansdell19}. The inability, at least up to now, of spectroscopy to detect exocomet signatures in late-type stars might be due to the concurrence of several causes, e.g. stellar activity that makes it extremely difficult to detect faint variable events superimposed on the profiles of the chromospheric active Ca {\sc ii} and Na {\sc i} lines; also, late type stars usually have small rotational velocities so that narrow stable absorptions are practically indiscernible from the core of the narrow stellar lines. 

\begin{figure}[h!]
  \centering
 \includegraphics[width=0.5\textwidth]{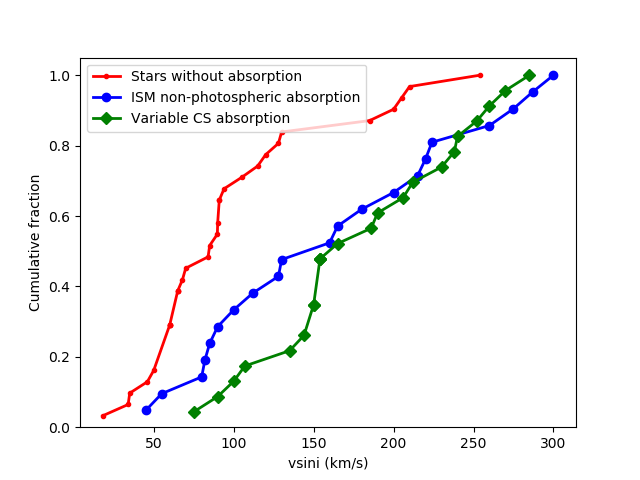}
\caption{Cumulative distribution functions of the subsamples labelled in the plot.}
\label{fig:ks_sample}
\end{figure}

It is obvious from Fig. \ref{fig:hrdiagram} that stars with both stable and/or sporadic CS features tend to be in many cases located above the main sequence, in the $\delta$ Scuti instability strip of the HR diagram \citep{breger79,rodriguez94}. A few of the hot-gas-bearing stars in our sample are identified as $\delta$ Scuti stars in SIMBAD - HD 110411, HD 183324, and HD 192518; recently, \cite{mellon19} have found that HD 156623 is also a $\delta$ Scuti star; we also point out that $\beta$ Pic itself  has $\delta$ Scuti pulsations \citep{koen03,mekarnia17}. In addition, the CS gas stars have distinctly larger $v \sin i$. As expected, \citep[e.g.][]{nielsen13}, the highest $v \sin i$ values are found for the earlier spectral types (symbol sizes in Fig. \ref{fig:hrdiagram} are proportional to v $\sin i$). Excluding stars later than F2, Fig. \ref{fig:ks_sample} shows the cumulative distribution functions (CDF) of the projected rotational velocity of the early type stars without non-photospheric features, stars with features identified as ISM absorptions, and stars with variable events. It is evident from Fig. \ref{fig:ks_sample} that stars with non-photospheric ISM features or with variable CS events, have larger $v \sin i$ values than stars without non-photosperic features. Table \ref{tab:kstest} shows the results of a Kolmogorov-Smirnov test separating those three subsamples. We refer to \cite{maldonado12} for the meaning of the parameters D, p-value, and n$_{\rm eff}$ in that table. In particular, the p-value indicates that the subsample of stars with ISM absorptions differ with a probability of $\sim$97\%, of the stars without non-photospheric absorptions; for the case of stars with variable events the probability is practically 100\%. At the same time, the results of the Kolmogorov-Smirnov test indicates that the ISM and variable events subsamples do not differ significantly. Nonetheless, a visual inspection of Fig. \ref{fig:ks_sample} suggests that there might be a paucity of the variable hot-gas-bearing stars with $v \sin i$ up to  $\sim$150 km/s with respect to the sample of stars with ISM absorptions, which is lost in the statistical test when comparing the whole range of $v \sin i$.

\begin{table}
\caption{Kolmogorov-Smirnov test comparing the projected rotational velocity distribution of the subsample of stars earlier than F2 without non-photospheric features, stars with absorptions with an ISM origin, and stars with variable events.}
\begin{small}
\begin{tabular}{lllll}
\hline
\hline
Sample 1             &   Sample 2           &   D        &   p-value     &   n$_{\rm eff}$      \\
\hline  
ISM absorp.         &   No absorp.		    & 0.39     &   0.030	&	12.52 \\
CS Var. absorp.	    &   No absorp. 	        & 0.66     & 6.376e-06	&   12.20 \\
ISM absorp.         &   CS Var. absorp.		& 0.30     & 0.222		&   10.98 \\ 
\hline
\end{tabular}
\end{small}    
    \label{tab:kstest}
\end{table}

\begin{figure}[h!]
  \centering
  \includegraphics[width=0.5\textwidth]{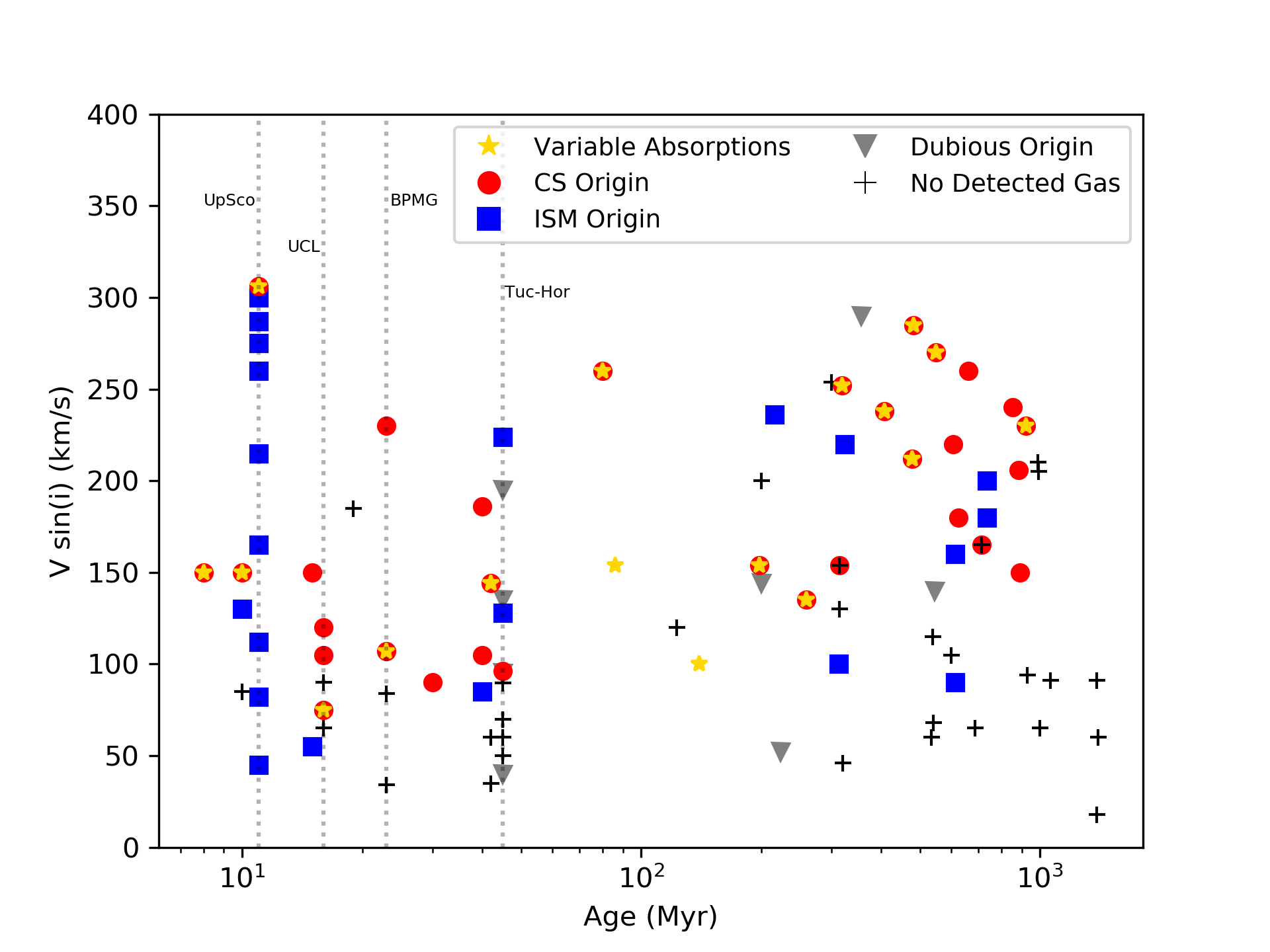}
 \caption{Age versus $v \sin i$ of the stars in the sample. Symbols mark the type of absorptions detected for each object. Vertical dotted lines mark the location of some of the young associations: Upper Scorpious, Upper Centaurus Lupus, Beta Pic Moving Group and Tucana-Horologium. }
\label{fig:age_vsini}
 \end{figure}

When considering the age of all stars with non-photospheric CS and ISM absorptions, $\sim$58\%  have ages below 100 Myr (Table \ref{tab:sample}). 
If we just consider stars plausibly hosting CS features that figure is $\sim$51\%, and reduces $\sim$42\%  (11 out of the 26 stars) when the stars with variable features are considered. Thus, although ages of field stars might be highly uncertain and the nature of the variability of the features is not always due to exocomet-like events, e.g. the case of HD 37306, stars with FEB-like events do not tend to be young objects but they are distributed in a wide range of ages, from $\sim$10 Myr to $\sim$1 Gyr. All this is clearly recognized in Figure \ref{fig:age_vsini} where a plot of the rotational velocity of the stars (up to F2) versus age is shown. Stars with non-photospheric features are all younger than 1000 Myr, and its distribution is clearly modulated by stars in young clusters - mainly UpSco, UCL, BPMG and Tuc-Hor. Stars with CS features are found among the whole range of ages, and have higher rotational velocities than stars without non-photospheric features. They also appear to have higher rotational rotational velocities than stars with ISM features, in fact reflecting the results of the Kolgomorov-Smirnov test above, and the mentioned paucity of CS-feature stars with low $v\sin i$.

\begin{figure}[h!]
  \centering
  \includegraphics[width=0.5\textwidth]{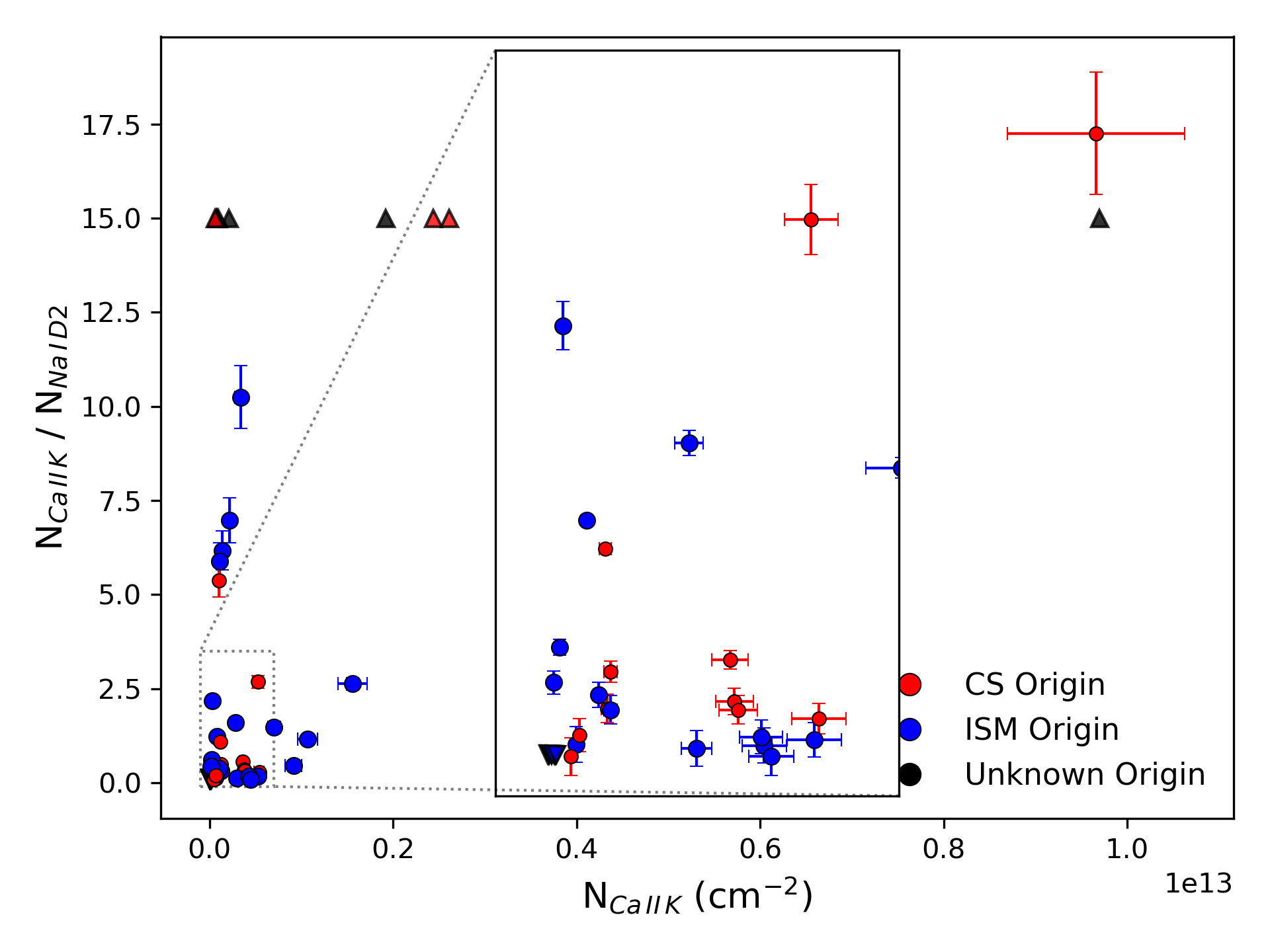}
 \caption{Column densities of Ca {\sc ii} K and ratio of column density of Ca {\sc ii} K and Na {\sc i} D2 of those absorptions of similar radial velocities. For the case where an absorption was detected in only one of the lines, triangles pointing up are upper limits and triangles pointing down are lower limits. The outlier in the upper-left corner of the plot is HD 42111. Colour denotes the suggested origin as in the legend.} 
\label{fig:col_den}
 \end{figure}

Out of the 32 stars in our sample with non-photospheric absorptions, we find in many cases coincident radial velocities (within the errors reported in Sect. 4.1) between the features observed in Ca and in Na. 
Fig. \ref{fig:col_den} shows the ratio of Ca {\sc ii} K and Na {\sc i} D2 column densities against the Ca {\sc ii} K column density, colour coded for the attributed origin. In those cases where only one of the lines was detected, an upper limit was calculated using the EW uncertainty to compute the column density. The distribution of the stars in this diagram is consistent with other works \citep[e.g.][]{welsh10,gudennavar12}. There is no clear separation between CS or ISM absorptions, i.e., the ratios of their column densities do not show any significant trend when comparing them regarding their origin, in agreement with previous results (see section 4.1.1, first paragraph).  It is worth noticing though, that it seems that the presence of both Ca and Na components is more common in ISM absorptions. HD 42111 seems to behave as an extreme outlier in our sample as it has a large N$_{\rm Ca {\sc ii} ~K}$. However, this shell star has two practically blended narrow, absorptions (not discernible in the median spectrum), which might explain the large N$_{\rm Ca {\sc ii} ~K}$. 

While objects in the sample are distributed in the sky without preferential locations (Fig. \ref{fig:allsky}), we find a possible trend when examining distances. Stars without narrow absorptions (either CS or ISM) are located at $<$ 50 pc, but there is no clear differentiation between stars with CS and ISM narrow absorptions. This could be due to the lower frequency of interstellar clouds at shorter distances, and the exponential growth of the number of stars with distance.

\begin{table}
\caption{Debris disc stars in the sample with narrow non-photospheric features. Bold-faced stars have features of CS origin, and those in italics also have variable absorptions. HD 110411 and HD 183324 do not have narrow absorptions but have CS gas.}
 \begin{center}
\begin{tabular}{lll}
\hline
\hline
HD 3003             &   HD 118232           &  HD 146897            \\
HD 5267             &   HD 125162           &  HD 147137            \\
{\bf \em HD 9672}   &   {\bf HD 131488}     &  {\bf \em HD 156623}  \\
{\bf \em HD 21620}  &   HD 131835           &  {\bf HD 158352}      \\
{\bf \em HD 32297}  &   HD 138813           &  {\bf \em HD 172555}  \\
{\bf \em HD 36546}  &   HD 142315           &  {\bf HD 181296}      \\
{\bf \em HD 37306}  &   HD 144587           &  {\bf  \em HD 182919} \\
HD 71043  	        &   HD 144981           &  {\bf \em  HD 183324} \\
HD 71722  	        &   HD 145554           &  HD 188228            \\
HD 105850  	        &   HD 145631           &  HD 198160            \\
{\bf \em HD 109573} &   {\bf HD 145689}     &  HD 221756            \\
{\bf HD 110058}     &   {\bf \em HD 145964} &				        \\
{\bf \em HD 110411} &   HD 146606           &                       \\
\hline
\end{tabular}
\end{center}
\label{tab:debris}
\end{table}

\subsection{Stars with debris discs}
\label{sect:debris}

To our knowledge, 76 out of the 117 stars in the sample are associated with a known debris disc (Table \ref{tab:sample}). We find that 35 out of those 76 debris discs  have  at least a  non-photospheric narrow feature (Table \ref{tab:narrow} and Table \ref{tab:debris}), and 15 among these 35 debris disc stars plausibly have at least one CS component - this figure does not include stars with an ambiguous origin of the detected non-photospheric  absorptions. Stable CS features at the velocity of the star have been interpreted as proof of a  CS gas disc, and its persistence requires the presence of a braking mechanism that prevents the hot gas from being blown away by the strong radiation pressure from the star \citep[e.g.][]{lagrange98}. \cite{fernandez06} suggested that in $\beta$ Pic such mechanism could be exerted by the observed enhanced carbon abundance \citep[see also][]{roberge06,brandeker11}, a fact also suggested for HD 9672 \citep{roberge14}. Among the stars in our sample with stable features, in addition to HD 9672 and HD 36546 (see section 4.2), that scenario might be at work for HD 32297, where a carbon overabundance is suggested by the the 3.7$\sigma$ Herschel detection of [C {\sc ii}] emission at 158 $\mu$m \citep{donaldson13}.

Concerning variable absorptions, 12 out of the total 18 stars showing variability either   in one or both of the Ca {\sc ii} H\&K lines or  in one or both of the Na {\sc i} D lines, have as well excesses associated with the presence of a debris discs (Table \ref{tab:debris} and Table \ref{tab:feblist}). All these stars are younger than 200 Myr. 

In all cases, but HD 37306, transient red- or blue-shifted events attributed to exocomets have been observed. The observed variability in HD 37306 is due to the appearance, and later disappearance, of a strong shell spectrum (see above). Complementary to these results, the rest of the stars with variable features listed in Table \ref{tab:feblist}, i.e. 14 stars, do not host a debris disc, and are older than 200 Myr. We point out, however,  that in four cases - HD 39182, HD 64145, HD 132200, and HD 138629 - there is no observational information concerning the potential presence of a debris disc, and in two cases - HD 132200 and HD 138629 - we do not have information about their age.
 
In parallel, to our knowledge 36 out of the 76 debris discs have been spatially resolved after the compilations by N. Pawellek and A. Krivov\footnote{https://www.astro.uni-jena.de/index.php/theory/catalog-of-resolved-debris-disks.html}, and C. McCabe, I.H. Jansen, and K. Stapelfeldt\footnote{https://www.circumstellardisks.org/}. Further,  28 out of those 36 resolved debris disc stars have spectral types  earlier than F2 and, therefore, sensitive to show the presence of exocometary signals. Also, \cite{moor17} reported inclinations for the A1 stars HD 121617 and HD 131488. Table \ref{tab:resolved} lists the early-type 30 resolved debris discs together with their corresponding inclination angles, taken from the mentioned catalogues.  
Among the resolved discs, 17 have non-photospheric features (Table \ref{tab:resolved}). In 6 cases the absorption features are more plausibly interstellar or ambiguous - HD 71722, HD 125162, HD 131835, HD 138813, HD 146897, and HD 188228.  In 11 stars, a CS origin seems to be the most reasonable one, for at least one of the observed non-photospheric absorption features; 8 out of these 11 stars have variable absorptions. Although based on a relatively low number of objects, an inspection of the inclination angles in  Table \ref{tab:resolved} reveals: i)  there is no a preferential distinction between discs seen edge- or polar-on for those debris discs without non-photospheric absorptions; ii) similar result is seen for those stars with ISM features; and iii) the trend is clearly different when we inspect debris discs with CS absorptions, i.e., most of them are clearly seen at $i > 70^\circ$ ($\sim$ 72\%), and this trend is reinforced when only those discs with variable features are considered. In other words, debris discs associated with stars hosting CS absorptions  tend to be seen edge-on, while debris discs associated with stars without non-photospheric absorptions or stars with ISM features do not show a preferential inclination. This result is consistent with the fact that the detection of CS features, i.e. hot gas, is favoured when the systems are seen close to edge-on, as well as with the large projected rotational velocities shown by the stars with CS gas compared to those without such gas. We note that the observed trend, which is just a geometrical effect, does not exclude the existence of hot gas, i.e. the eventual existence of comet-like bodies around the debris discs systems seen away from edge-on. A similar trend was already pointed out by \cite{Rebollido18} in their analysis of debris discs with measurable amounts of cold gas seen in emission in the far-IR and (sub-) mm wavelength regimes.

\begin{table}
  \caption{Resolved debris discs and inclination angles. Bold-faced are those with narrow non-photospheric absorption features while those in italics have at least one component attributed to a CS origin. HD 110411 and HD 183324 do not have narrow absorptions but have CS gas.}
\begin{tabular}{llllll}
  \hline
  \hline
 Star             &  $i^o$ &  star              & $i^o$ & star                 &$i^o$\\
\hline
{\bf \em HD 9672}  & 79  & {\bf HD 71722}     & 78  & {\bf \em HD 131488}$^{2}$& 82   \\
HD 14055           & 83  & HD 74873           & 27  & {\bf HD 131835}     & 75   \\
HD 15115           & 80  & HD 95418           & 84  & HD 139006           & 80   \\
HD 21997           & 26  & HD 102647          & 30  & {\bf \em HD 138813} & 28   \\
HD 27290           & 69  & HD 109085          & 35  & {\bf HD 146897}     & 84   \\
HD 28355           & 76  & {\bf \em HD 109573}& 76  & {\bf \em HD 156623} & 30   \\
HD 31295$^{*,1}$   &     & {\bf \em HD 110058{$\dagger$}}& 50  &  {\bf \em HD 172555} & 75  \\
{\bf \em HD 32297} & 90  & {\em HD 110411}    & 70  & {\bf \em HD 181296} & 90   \\
{\bf \em HD 36546} & 75  & HD 121617$^{2}$    & 37  & {\em HD 183324}     & 2    \\
HD 38206           & 60  & {\bf HD 125162}    & 48   & {\bf HD 188228}     & 49   \\
\hline 
\end{tabular}\\
(*) resolved debris disc with unreported inclination angle; ({$\dagger$}) \cite{kasper15} report an edge-on inclination

(1) \cite{draper16}; (2) \cite{moor17}
\label{tab:resolved}
\end{table}

\subsection{Near infra-red excesses}
 
There are 22 stars (Table \ref{tab:sample}) in our sample explicitly taken from the literature searching for hot excesses in the H (1.6 $\mu$m) and K(2.2 $\mu$m) bands (Table \ref{tab:sample}). In addition, HD 172555 also presents such excess \citep{absil13,ertel14,ertel16,nunhez17}. 

Within the stars in our sample with near-IR excesses plausibly due to hot dust (i.e., excluding stars where binarity is the cause of the excess), 11 stars have spectral types earlier than F2, and 6 out of those 11 stars present non-photospheric features, either detected in this work or in previous ones (Table \ref{tab:hotexcesses}).

To assess the significance of the incidence of CS absorption around hot dust stars we consider all reported 31 near-IR excess stars \citep{absil13,ertel14,nunhez17}. Excluding the binaries, 14 stars - the 11 studied by us plus Vega, $\beta$ Pic, and HD 210049 ($\mu$ PsA) - have spectral types earlier than F2. 
Among those 14 stars, 5 have variable absorptions features (HD 2262, HD 56573, HD 108767, HD 172555, and $\beta$ Pic); 2 stars have an ambiguous CS/ISM narrow feature (HD 177724, and HD 210418); 2 stars are associated with pole-on debris discs (HD 102647 and Vega), i.e. an unfavourable orientation to detect CS features; 1 star (HD 28355) has a resolved debris disc with an inclination angle $ i = 70^\circ$, but our spectra do no reveal any non-photospheric absorption; for the last 4 stars (HD 40136, HD 187642, HD 203280, and HD 210049) the orientation of the system is unknown. If we assume random orientations, the probability to observe a system with an inclination larger than 65$^\circ$--70$^\circ$, i.e., edge-on or close to it, lies between $\sim$42\% and 34\%, respectively, which is similar to the percentage of stars with hot CS gas among the hot dust stars. We also note that it agrees  with the percentage of edge-on discs, either debris or other type of disc, that can be found in the Catalogue of Resolved Discs compiled by C. McCabe, I.H. Jansen, and K. Stapelfeldt. Given those figures, a relationship between hot dust and hot CS gas is suggested, which should be confirmed by increasing the sample of stars with the appropriate near-IR interferometric and UV/optical spectroscopic data.

\begin{center}
\begin{table}
  \caption{Stars showing near-IR excesses.}
\begin{center}
\begin{tabular}{lcr}
\hline
\hline
Name      &   Non-photospheric &  H/K excess   \\
          &       absorption   &  \\
\hline
HD 2262    &   Yes*(1)      &      (2)     \\
HD 28355   &   No           &      (2)     \\
HD 40136   &   No           &      (1)     \\
HD 56537   &   Yes(2)       &      (1)     \\
HD 102647  &   No           &      (1)     \\
HD 108767  &   Yes(2,3)      &      (2)     \\
HD 172555   &   Yes(3,4)      &      (2)        \\
HD 177724  &   Yes(3)       &      (1,3)   \\
HD 187642  &   No           &      (1,3)   \\
HD 203280  &   No           &      (1)     \\
HD 210418  &   Yes(3)       &      (3)     \\
\hline
\end{tabular}
\end{center}
*Not clear if exocomet-like or stellar mass loss events.

References for non-photospheric absorptions: (1): \cite{welsh18}; (2): \cite{welsh15}; (3) This work ; (4) \cite{kiefer14a}

References for H/K excess: (1): \cite{absil13}, (2): \cite{ertel14}, (3) \cite{nunhez17}
\label{tab:hotexcesses}
\end{table}
\end{center}

\subsection{Ti {\sc ii} / Shell stars}
\label{sec:tiii}
In addition to the stars selected on the basis of their Ti {\sc ii} lines, there is a number of stars which have been classified as shell stars \citep{abt08,lagrangehenri90,hauck00,roberge08}. Table \ref{tab:shell} lists, according to our knowledge, the shell stars in our sample. No Ca {\sc ii} or Na {\sc i} non-photospheric absorptions are observed towards HD 39283 and HD 77190; in the cases of HD 118232 and HD 196724 the origin of the Ca {\sc ii} absorption is likely ISM or ambiguous. As a matter of fact, the spectra of these four stars do not reveal any prominent shell-like characteristics. The rest of the stars show triangular-like CS absorptions (bold-faced in the table) - those stars with variable features are also emphasized.  Not all absorptions can be attributed to FEB-like events; for example, the stable absorption seen towards HD 192518 is clearly formed in a gaseous shell, while as already mentioned the variability in HD 37306 is due to the appearance/disappearance of a shell around this star. Nonetheless, 11 stars do present variable events, most of them attributable to exocomets; even, our spectra might be tracing their dynamical evolution  of some events by changing their depths and velocities.

\begin{table}
\caption{Shell stars. Stars with non-photospheric CS features are bold-faced while those with variability are 
emphasized.}
\begin{tabular}{lll}
\hline
\hline
Star   &  Star &  Star \\
\hline
{\bf \em HD 256}      & {\bf \em HD 50241}   & {\bf \em HD 148283}  \\
{\bf HD 21688}        & HD 77190             & {\bf  HD 158352}  \\
{\bf \em HD 37306}    & {\bf \em HD 85905}   & {\bf HD 168646}      \\
{\bf \em HD 39182}    & {\bf \em HD 98058}   & {\bf HD 192518}      \\
HD 39283              &  HD 118232           & HD 196724      \\
{\bf HD \em 42111}    & {\bf \em HD 138629}  & {\bf \em HD 217782}  \\
\hline 
\end{tabular}\\
\label{tab:shell}

\end{table}

\subsection{ $\lambda$ Boo stars}

The sample contains 12 objects previously classified as $\lambda$ Boo type stars (Table \ref{tab:lambda_boo}) although three of them - HD 39283, HD 210418, and HD 217782 - have recently been considered as non-members of this stellar class \citep[][and references therein]{murphy15}. Several criteria have been used to classify  $\lambda$ Boo stars \citep[e.g.][]{murphy15,gray17}; among them, optical line ratios between volatiles, like CNO, and heavier elements (e.g. Mg) are useful to ascribe stars to this stellar class \citep{cheng17}. This criterion is based on the basic definition  of the $\lambda$  Boo stars, i.e. stars with  a remarkable low abundance of heavy (e.g. Fe, Al, Mg) elements while volatiles  as CNO have near solar abundances \citep[e.g.]{baschek88}. In order to eventually find new $\lambda$ Boo candidates, we have measured in all stars earlier than F2 in our sample the ratio of the Mg {\sc ii} 4481 \AA ~and the O {\sc i} 7774 \AA ~triplet, which are strong and easy to measure in our spectra. 

\begin{figure}
\includegraphics[width=0.5\textwidth]{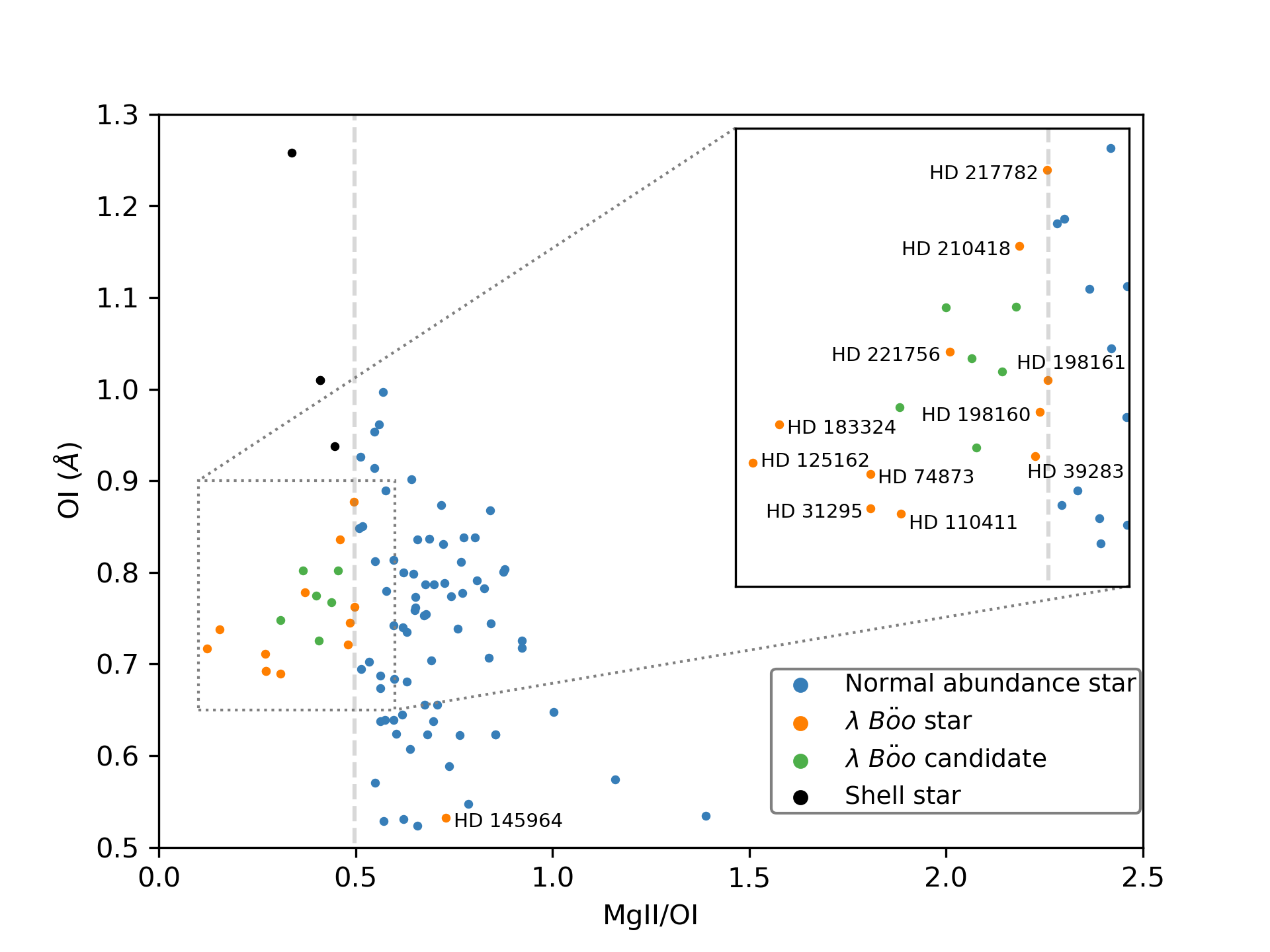}
\caption{Ratio O {\sc i} (7744 ~\AA) and Mg {\sc ii} (4481 ~\AA) versus the EW of O {\sc i}. Orange dots mark stars previously classified as $\lambda$ Boo stars; green dots mark new $\lambda$ Boo candidates; and blue dots mark the stars with normal abundances. The vertical dashed line delimits the locus of bona fide  $\lambda$ Boo stars (see text).}
\label{fig:lambda_boo}
\end{figure}

Fig. \ref{fig:lambda_boo} is a plot of the EW of the O {\sc i} line versus the ratio of Mg {\sc ii}  to O {\sc i} lines. All identified $\lambda$ Boo stars in the sample are clearly located to the left in that plot, well separated from the bulk of the stars. The exception is HD 145964 classified as \textit{weak} $\lambda$ Boo by \cite{welsh13} based on the measurements of \cite{abt95}. The identified $\lambda$ Boo stars have low metallicities ([Fe/H] < --0.25) excluding HD 145964 and HD 217782.  A vertical dashed line at \mbox{Mg{\sc ii }/O {\sc i}< 0.49} marks the limit for the identified $\lambda$ Boo objects in our sample. This figure is the one for the $\lambda$ Boo stars HD 198160 and HD 198161, and approximately the one for those  removed  from the class by \cite{murphy15} stars. We note that 6 stars  are additionally located at similar locations in Fig. \ref{fig:lambda_boo} as the  $\lambda$ Boo stars, and they also have low metallicities (Table \ref{tab:parameters}). We consider that those stars are new candidates, Table \ref{tab:lambda_boo}.  In addition, 3 shell stars -HD 39182, HD 42111, and HD 168646 - have low Mg {\sc ii}/O {\sc i} ratio but a very high EW (O {\sc i}), and  metallicities >0.0. 

With respect to the presence of non-photospheric narrow and/or variable features, 15 out of the 18 $\lambda$ Boo stars (including the new candidates) listed in Table \ref{tab:lambda_boo} present evidences of non-photospheric gas in their spectrum; in 5 cases the feature is interstellar, in 2 cases the origin is ambiguous; and the remaining 8 stars (44\% of the $\lambda$ Boo stars), have exocometary-like events. Comparing this figure to the 26\% incidence of exocometary-like events of the whole sample, there seems to be an enhanced probability for $\lambda$ Boo stars to have such events. A connection between metal-poor stars and the presence of debris discs or remainings of planet formation processes has been suggested before in the literature \citep{jura15,murphy17}, as heavier elements are blown away by radiation pressure, while volatile elements are accreted onto the star \citep[see][and references therein]{draper16}. Therefore, exocomets represent a rather likely scenario, that could replenish the atmosphere of $\lambda$ Boo stars of C, N, O and S elements.

\begin{table}
\caption{EW of the Mg {\sc ii} 4481 ~\AA and O {\sc i} 7744 ~\AA lines of previously known $\lambda$ Boo stars, together with stars of the sample with similar characteristics. The new $\lambda$ Boo candidates are indicated.}
\begin{tabular}{lrrc}
\hline
\hline
Star &  EW Mg {\sc ii} & EW O {\sc i} & $\lambda$ B$\ddot o$o \\  
     &  (\AA)         &     (\AA)   & \\
\hline
HD 31295     &   0.188     &   0.692     &   Yes        \\
HD 32297     &   0.336     &   0.767     &   New        \\
HD 36546     &   0.231     &   0.748     &   New        \\
HD 39182     &   0.413     &   1.010     &   Shell      \\
HD 39283     &   0.347     &   0.720     &   Yes        \\
HD 42111     &   0.424     &   1.258     &   Shell      \\
HD 71722     &   0.294     &   0.802     &   New        \\
HD 74873     &   0.192     &   0.711     &   Yes        \\
HD 110058    &   0.310     &   0.774     &   New        \\
HD 110411    &   0.213     &   0.689     &   Yes        \\
HD 125162    &   0.087     &   0.717     &   Yes        \\
HD 145964    &   0.388     &   0.532     &   Yes        \\
HD 156623    &   0.366     &   0.802     &   New        \\
HD 168646    &   0.418     &   0.937     &   Shell      \\
HD 177724    &   0.294     &   0.726     &   New        \\
HD 183324    &   0.114     &   0.738     &   Yes        \\
HD 198160    &   0.362     &   0.745     &   Yes        \\
HD 198161    &   0.378     &   0.762     &   Yes        \\
HD 210418    &   0.385     &   0.836     &   Yes        \\
HD 217782    &   0.435     &   0.877     &   Yes        \\
HD 221756    &   0.289     &   0.778     &   Yes        \\

\end{tabular}
\label{tab:lambda_boo}
\end{table}

\section{Conclusions}

We have presented the observational results of a systematic study aiming at detecting and monitoring hot gas attributed to the presence of exocomets in the CS environment. The study has been based on the analysis of  the Ca {\sc ii} K\&H and Na {\sc i} D lines in a heterogeneous and biased sample of 117 main-sequence late B to G type stars. This is the largest systematic study searching for exocomets carried out so far. The main results are the following.

Narrow non-photospheric ISM or CS absorptions have been detected towards $\sim$50\% of the sample (60 stars). Among the stars with those absorptions, at least one of the detected narrow features can be attributed to CS gas in 30 objects, i.e., 26\% of the whole sample. This figure is not statistically significant as the studied sample included stars with previously detected CS gas, but it does show that gas in the CS environment of A-type stars is relatively common. In some stars, the gas is originated in a CS shell surrounding the stars; in some other cases, the narrow absorptions can be attributed to the evaporation of exocomets or to gas released by dust grain collisions or evaporation. 

Sporadic red-shifted and  blue-shifted events that in some cases (but not all) might be due to $\beta$ Pic-like exocomets have been found in the CS environment of 16 stars, out of which 6 are new to the literature (including HD 132200, not contained in the original sample). In a few cases, our spectra seem to trace the dynamical evolution of such events as suggested by changes in their velocity and depth. Nonetheless, the exocometary-like activity in our stars, maybe with the exception of $\phi$ Leo \citep[see][]{eiroa16}, is very poor compared with $\beta$ Pic, which remains a unique object. 

The variations observed towards the two other stars, namely HD 256 (HR 10) and HD 37306 (HR 1919) do not require the presence of exocomets for their interpretation. The variability of the CS features in HR 10 is mainly due to the binary character of this star. A detailed analysis has  being carried out in a separated paper \citep{montesinos19}. In the particular case of HD 37306, our spectra have witnessed the appearance and disappearance of a strong shell around this star. There are no hints in our data of exocometary events in this star, although we point out that some shell stars in our sample do show such events. 

Hot CS gas is only detected towards stars earlier than A9, in line with previous works. The F2 V star $\eta$ Crv is the only star for which $\beta$ Pic-like events with a 2.9$\sigma$ detection have been claimed \citep{welsh19}. This paucity in detecting exocometary events around late type stars might be due to the inability of spectroscopy to detect faint non-photospheric absorptions on top of cool photospheres. In this respect, photometric transits have demonstrated to be competitive and successful in detecting exocomets around late type stars.

Hot gas-bearing main sequence stars have distinctly higher projected rotational velocities, and spread over a large range of ages, from $\sim$10 Myr up to at least $\sim$1 Gyr. Some of them are also known to present $\delta$ Scuti pulsations, as $\beta$ Pic itself.

Exocometary-like events are often associated with edge-on debris disc stars, in particular towards those with cold gas \citep[see also][]{Rebollido18}. This result is interpreted as a geometrical effect, but it does suggest that debris disc stars with non-photospheric absorptions (i.e. hot gas) are excellent targets to search in the far-IR and (sub-)mm spectral range for the presence of cold gas released by the evaporation of solid bodies at distances relatively far from the central star. We note that not all stars with FEB-events are associated with a debris disc. It cannot be excluded that this is due to the current, limited sensitivity to detect debris discs, ${\rm L_{dust}}/{\rm L_*}\!\sim \!10^{-6}$.

We find that FEB-like events are detected towards 17\% of stars with near-IR excesses denoting the presence of hot exozodies. Both hot dust and gas might be related phenomena, although more observations are needed to confirm or deny the trend pointed out in our study.

Our sample includes 18 $\lambda$ Bootis stars, with 6 new candidates found in this work. A relevant fraction of them, 8 out the 18 stars, have FEB-like events, suggesting again that both phenomena could be related.

\begin{acknowledgements} The authors wish to thank the careful reading of the manuscript and comments by the referee which have helped to improve the content of this work. Based on observations made with the Mercator Telescope, operated on the island of La Palma by the Flemmish Community, and the Nordic Optical Telescope, operated by the Nordic Optical Telescope Scientific Association, at the Spanish Observatorio del Roque de los Muchachos of the Instituto de Astrofísica de Canarias. Based on observations made with ESO Telescopes at the La Silla Observatory under programmes 099.A-9029(A),099.A-9004(A) and 094.A-9012. Also based on observations made with the TIGRE telescope funded and operated by the Universities of Hamburg, Guanajuato and Liège. We thank David Montes for providing the December 2018 Mercator spectra and to Santos Pedraz and Ana Guijarro for the CARMENES spectra of HD 37306. We thank Sam Kim and Markus Rabus for the FEROS spectra obtained in September 2017. I.R. thanks Angela Hempel for her support with FEROS observations in March 2016 and April 2017, and the Calar Alto Observatory staff, where a large part of this paper was written. Partially based on data obtained from the ESO Science Archive Facility, programs 096.C-0876, 076.B-0055, 185.D-0056, 082.B-0610, 072.C-0488, 0100.C-0090, 093.C-0409, 081.D-0610, 080.A-9006, 184.C-0815, 077.C-0295. C.E., G.M., B.M., I.R., and E.V. are supported by Spanish grant AYA 2014-55840-P. H.C. acknowledges funding from the ESA Research Fellowship Programme. J.~O., A.~B., and D.~I. acknowledge support from the ICM (Iniciativa Cient\'ifica Milenio) via the Nucleo Milenio de Formación planetaria grant. J.~O. acknowledges support from the Universidad de Valpara\'iso and from Fondecyt (grant 1180395). A.~B. acknowledges support from Fondecyt (grant 1190748). A.~M. acknowledges the support of the Hungarian National Research, Development and Innovation Office NKFIH Grant KH-130526. This research has made use of the SIMBAD database, operated at CDS, Strasbourg, France. This work has used Seth Redfield's Colorado model of interstellar clouds and his online tool "LISM Kinematic Calculator". This work has also used the debris disc catalogues "Resolved debris discs" from Jena University compiled by N. Pawellek and A. Krivov; and the one in circumstellardisks.org compiled and mantained by C. McCabe, I. H. Jansen and K. Stapelfeldt.

\end{acknowledgements}
\bibliographystyle{aa} 
\bibliography{biblio} 

\newpage
\begin{appendix}
\onecolumn
\section{Figures}
\subsection{Narrow Stable Absorptions}

\onecolumn

\begin{figure}
\centering
\includegraphics[width=1.\textwidth]{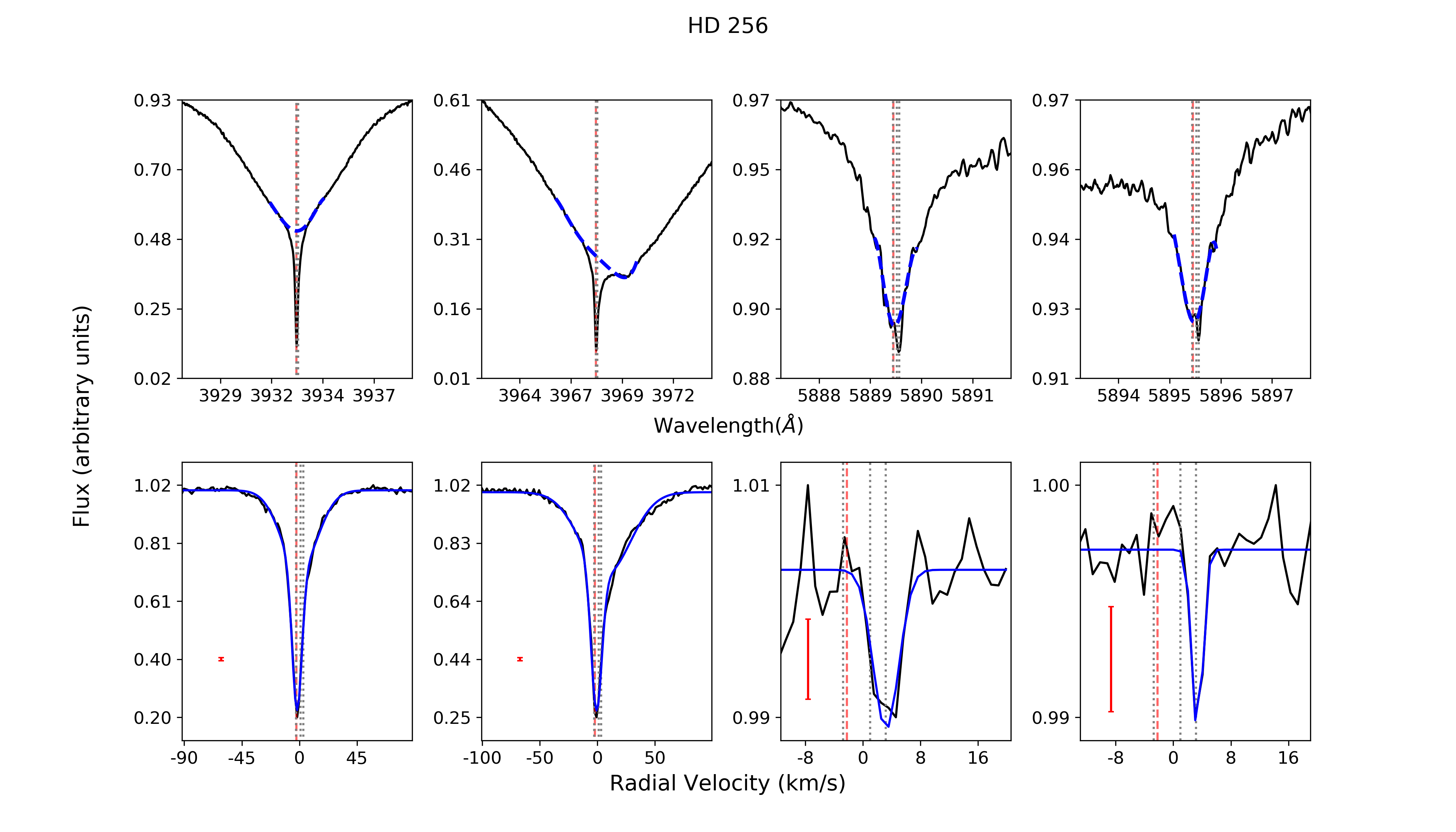}      
\includegraphics[width=1.\textwidth]{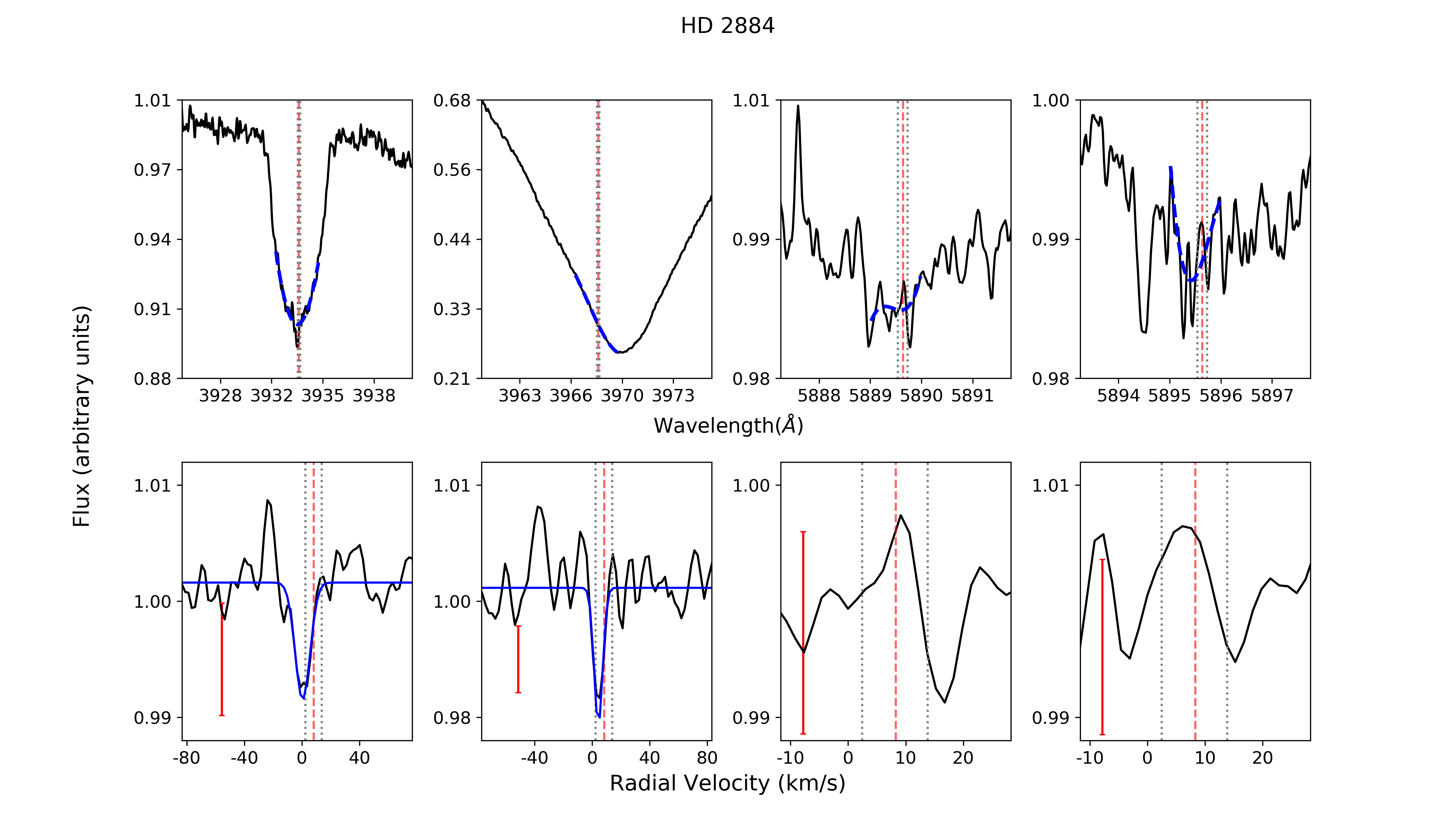}    
\caption{Stars showing narrow non-photospheric absorptions. Top panels: Photospheric Ca {\sc ii} H \& K and Na {\sc i} D lines with fitted modeling dashed blue line, x-axis shows the wavelength. Bottom panels: Residuals once the spectrum is divided by the photosphere, x-axis in velocity. Blue lines mark the fits to the non-photospheric absorptions. Vertical red dashed and grey dotted lines represent the stellar radial velocity and the ISM velocities respectively.  Red error bars show three sigma value measured in the continuum adjacent to the photospheric line.}
\label{fig:na_absorptions}
\end{figure}

\addtocounter{figure}{-1}

\begin{figure}
\centering
\includegraphics[width=1.\textwidth]{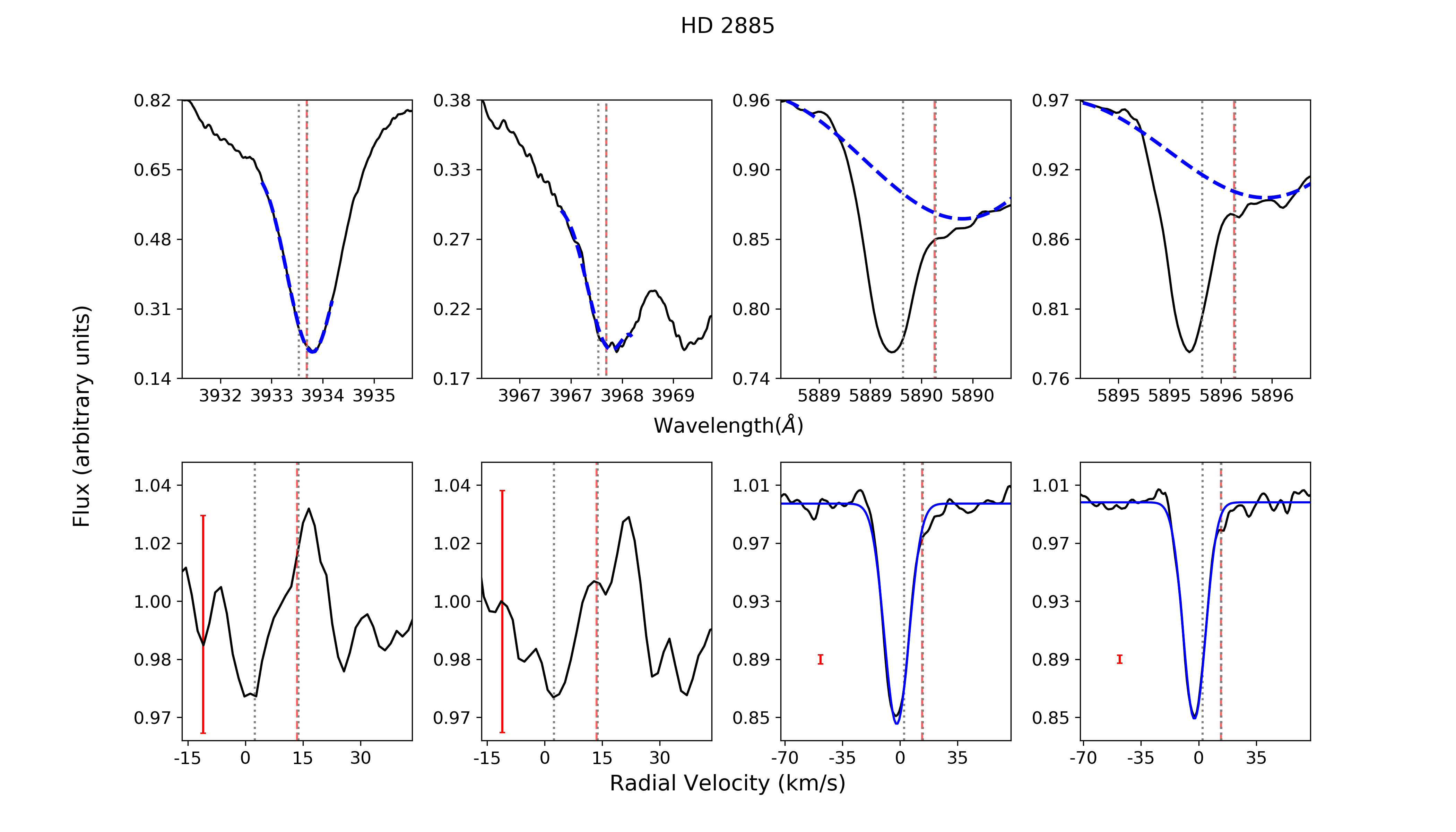}
\includegraphics[width=1.\textwidth]{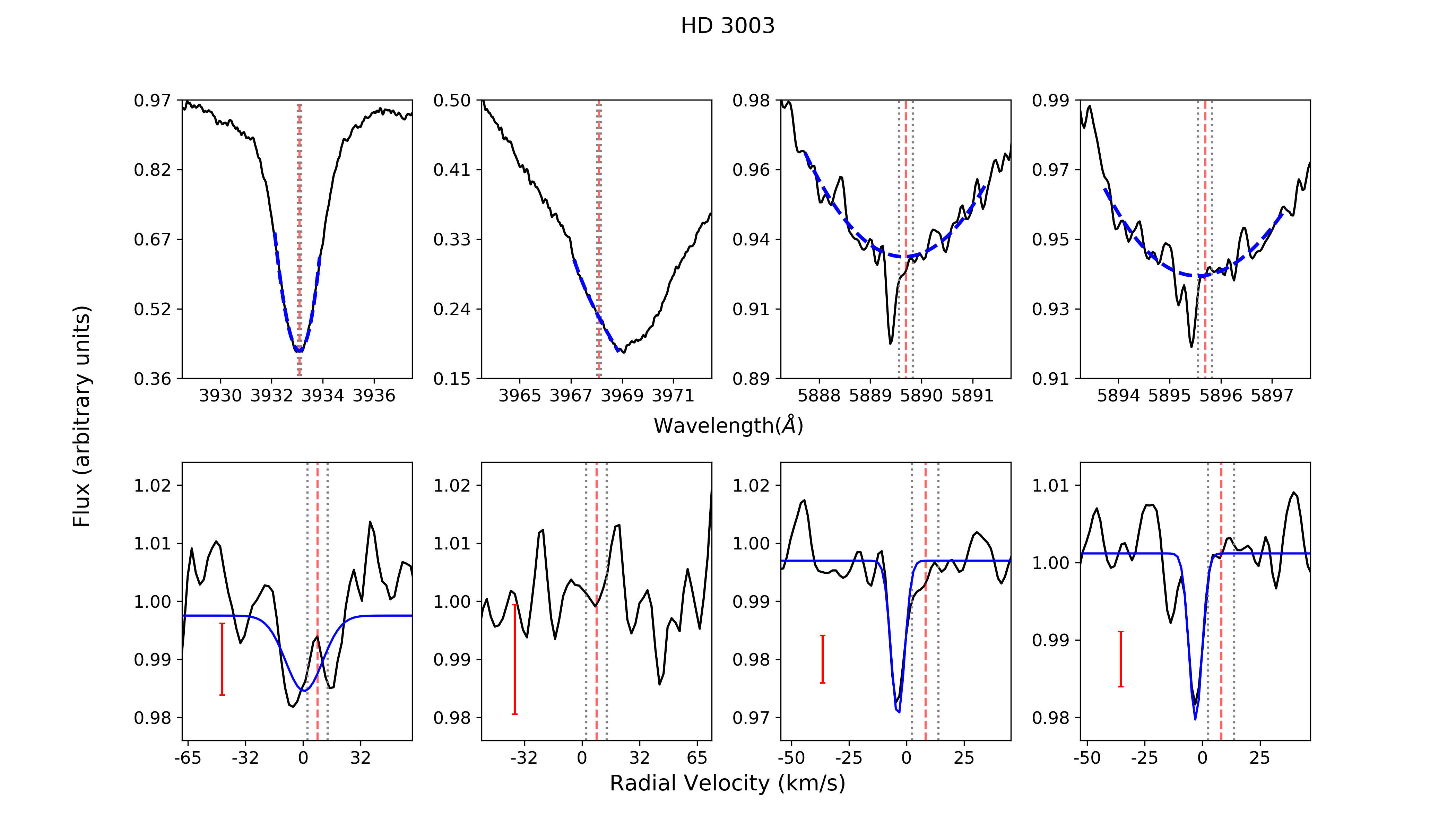}
\caption{Stars showing narrow non-photospheric absorptions. Top panels: Photospheric Ca {\sc ii} H \& K and Na {\sc i} D lines with fitted modeling dashed blue line, x-axis shows the wavelength. Bottom panels: Residuals once the spectrum is divided by the photosphere, x-axis in velocity. Blue lines mark the fits to the non-photospheric absorptions. Vertical red dashed and grey dotted lines represent the stellar radial velocity and the ISM velocities respectively.  Red error bars show three sigma value measured in the continuum adjacent to the photospheric line.}
\end{figure}

\renewcommand{\thefigure}{\arabic{figure} (Cont.)}
\addtocounter{figure}{-1}

\begin{figure}
\centering
\includegraphics[width=1.\textwidth]{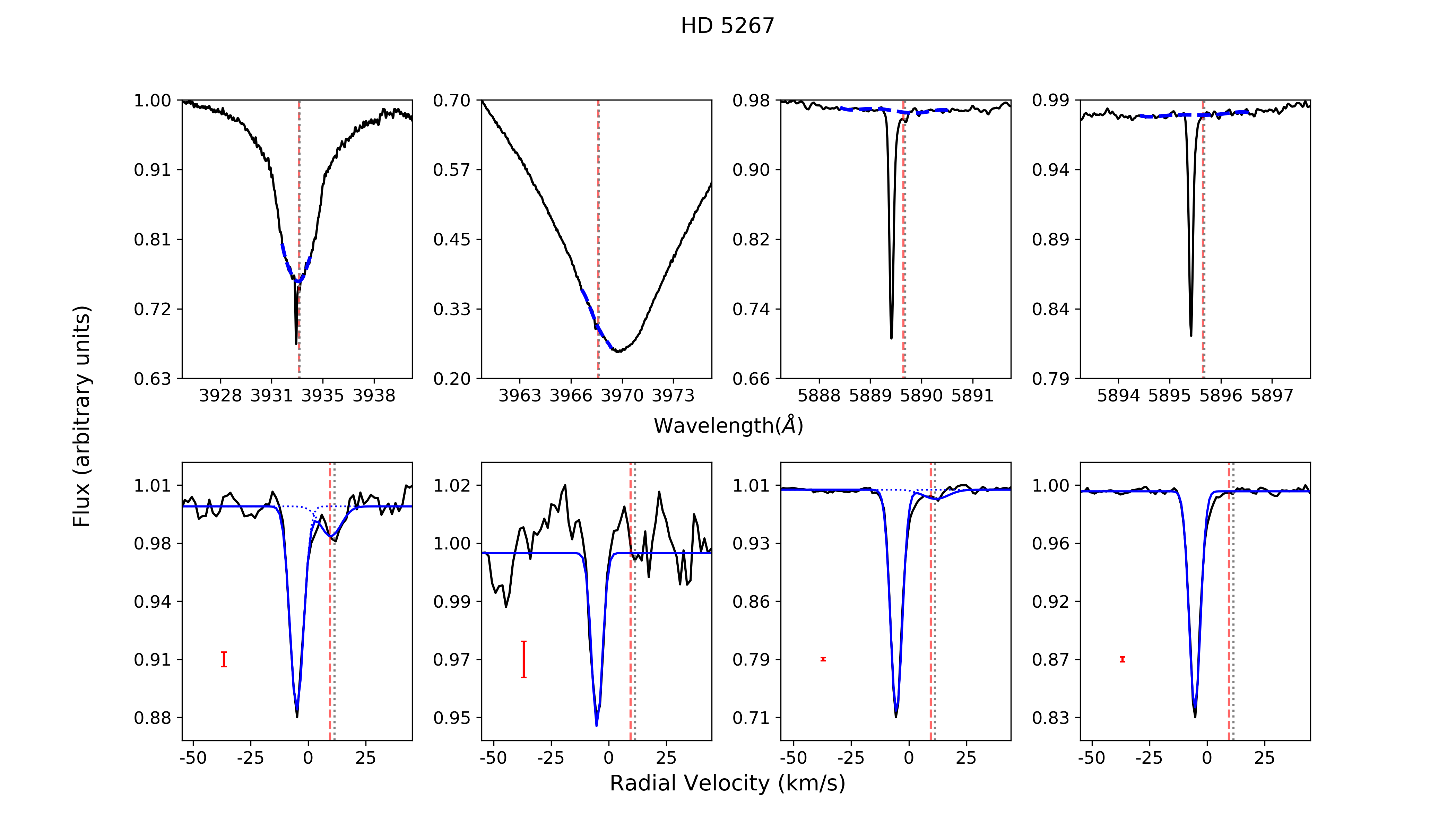}    
\includegraphics[width=1.\textwidth]{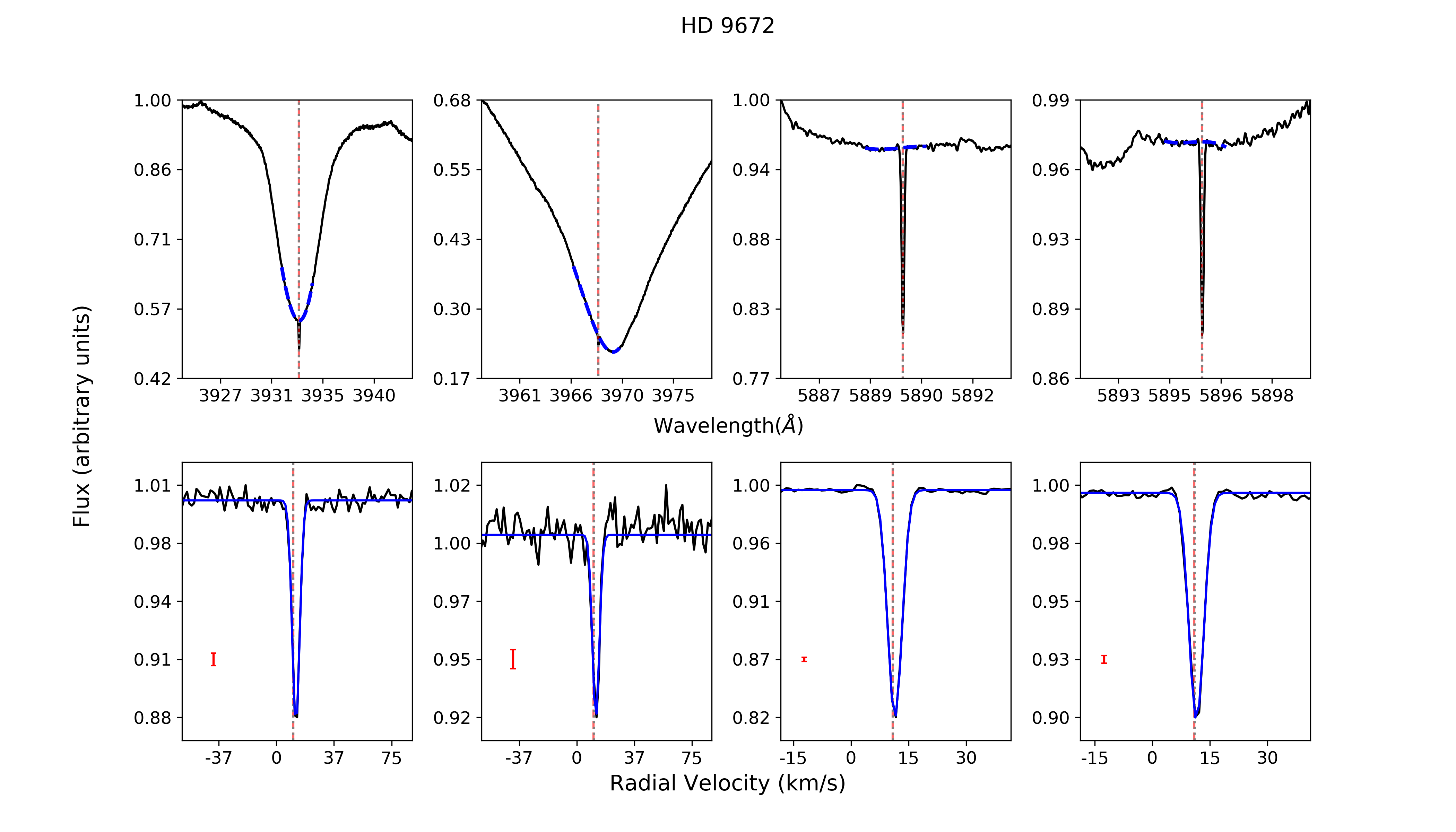}
\caption{Stars showing narrow non-photospheric absorptions. Top panels: Photospheric Ca {\sc ii} H \& K and Na {\sc i} D lines with fitted modeling dashed blue line, x-axis shows the wavelength. Bottom panels: Residuals once the spectrum is divided by the photosphere, x-axis in velocity. Blue lines mark the fits to the non-photospheric absorptions. Vertical red dashed and grey dotted lines represent the stellar radial velocity and the ISM velocities respectively.  Red error bars show three sigma value measured in the continuum adjacent to the photospheric line. }
\end{figure}

\renewcommand{\thefigure}{\arabic{figure} (Cont.)}
\addtocounter{figure}{-1}

\begin{figure}
\centering
\includegraphics[width=1.\textwidth]{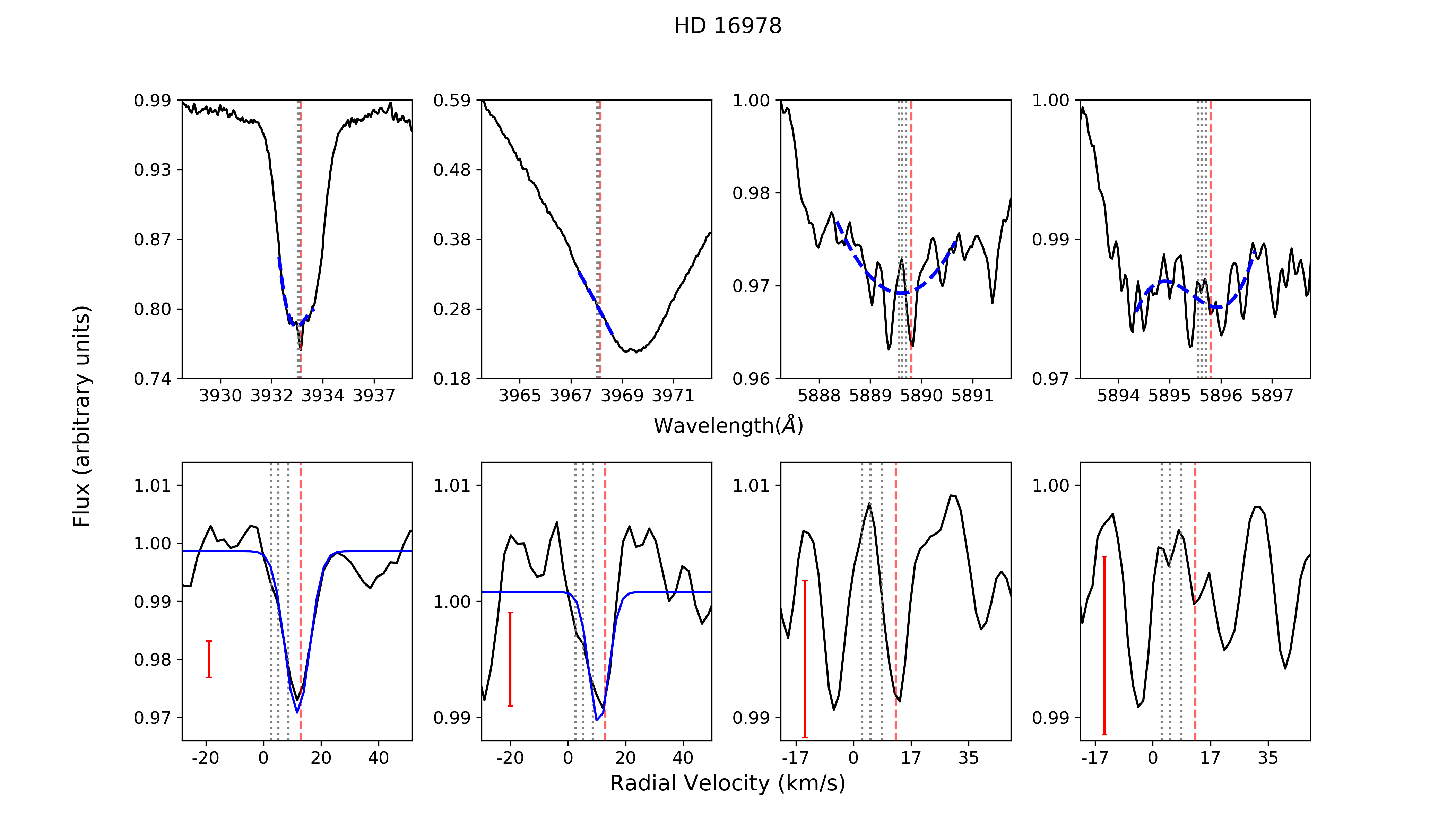}   
\includegraphics[width=1.\textwidth]{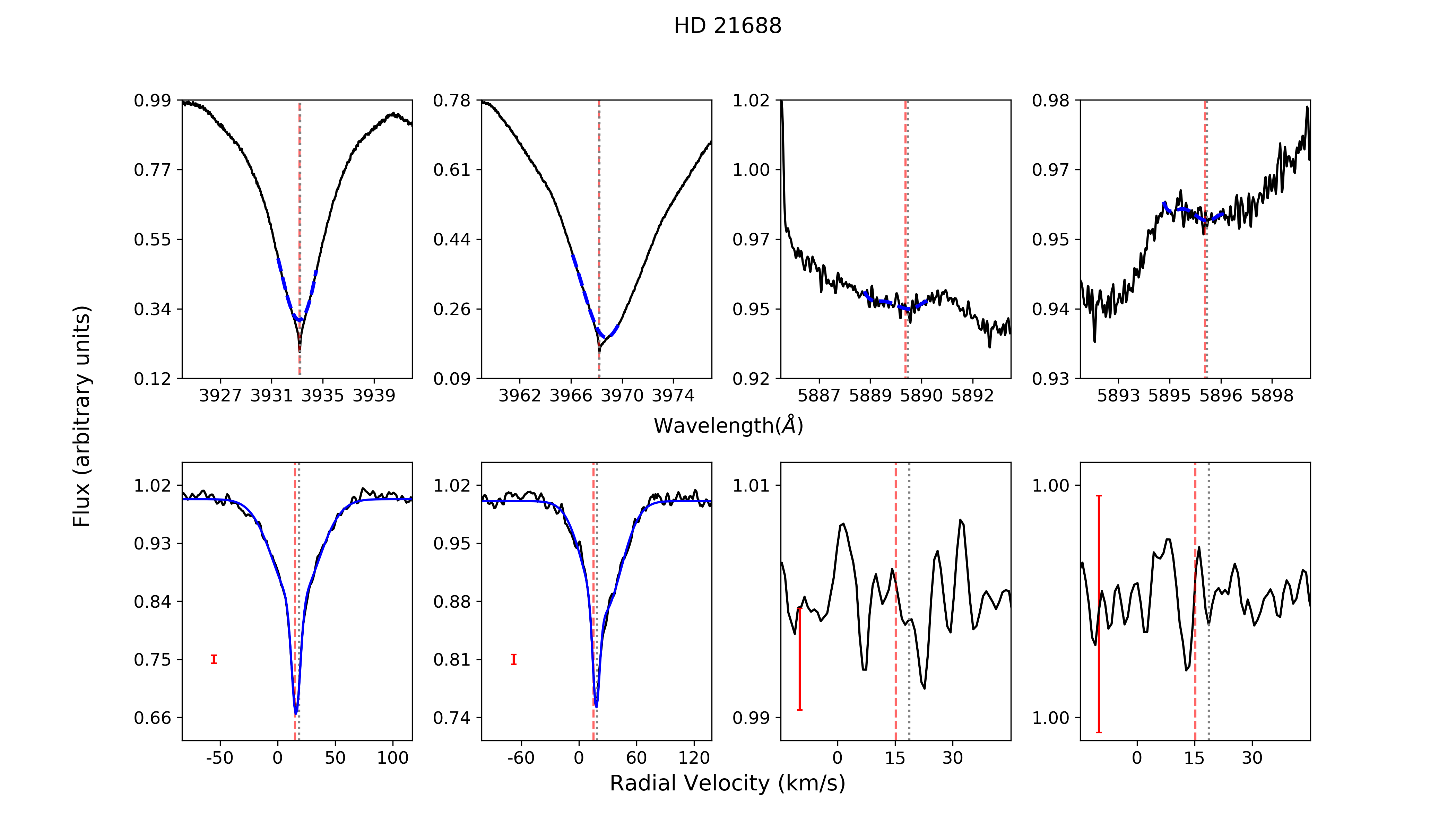}    
\caption{Stars showing narrow non-photospheric absorptions. Top panels: Photospheric Ca {\sc ii} H \& K and Na {\sc i} D lines with fitted modeling dashed blue line, x-axis shows the wavelength. Bottom panels: Residuals once the spectrum is divided by the photosphere, x-axis in velocity. Blue lines mark the fits to the non-photospheric absorptions. Vertical red dashed and grey dotted lines represent the stellar radial velocity and the ISM velocities respectively.  Red error bars show three sigma value measured in the continuum adjacent to the photospheric line. }
\end{figure}

\renewcommand{\thefigure}{\arabic{figure} (Cont.)}
\addtocounter{figure}{-1}

\begin{figure}
\centering
\includegraphics[width=1.\textwidth]{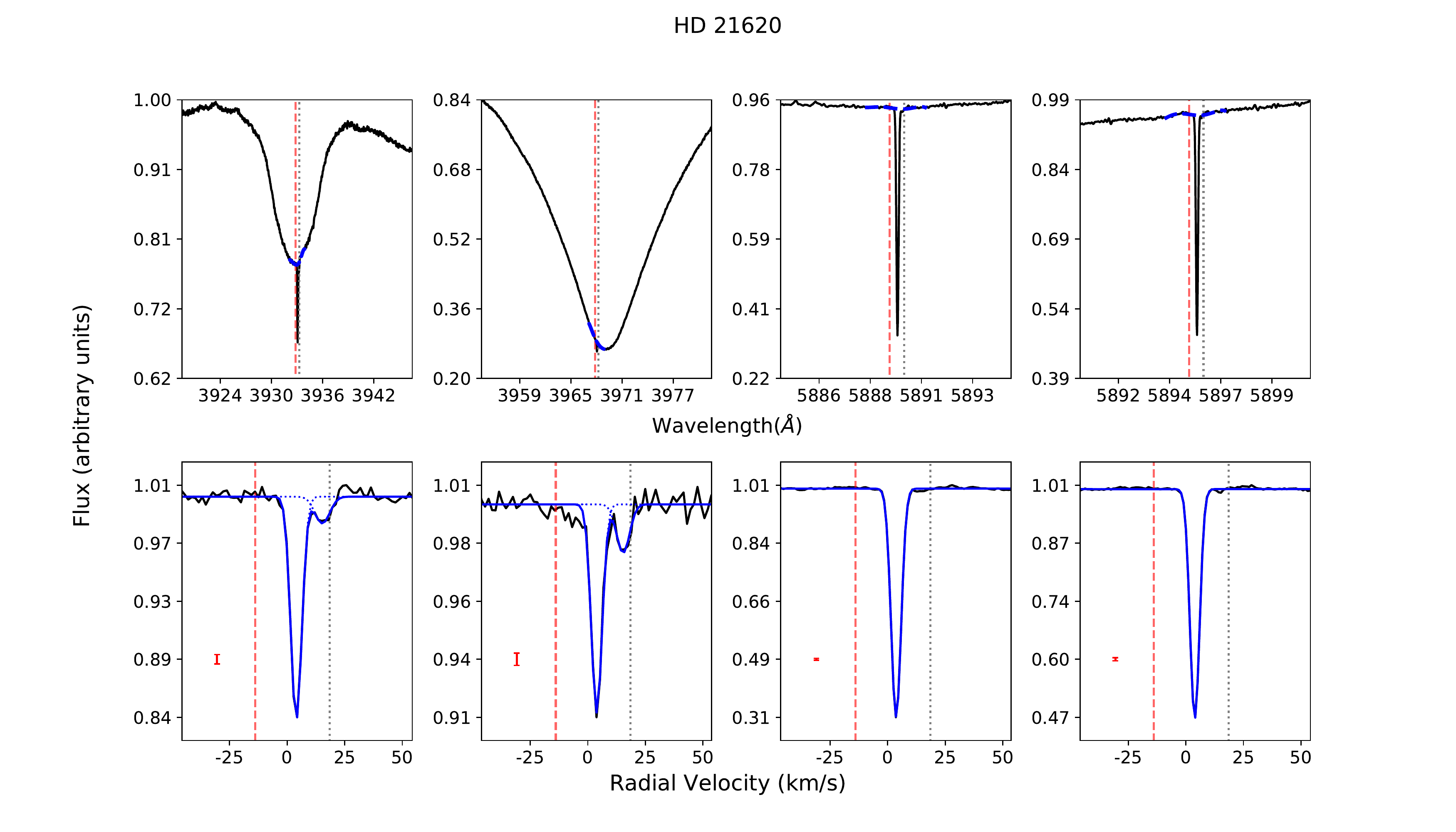}
\includegraphics[width=1.\textwidth]{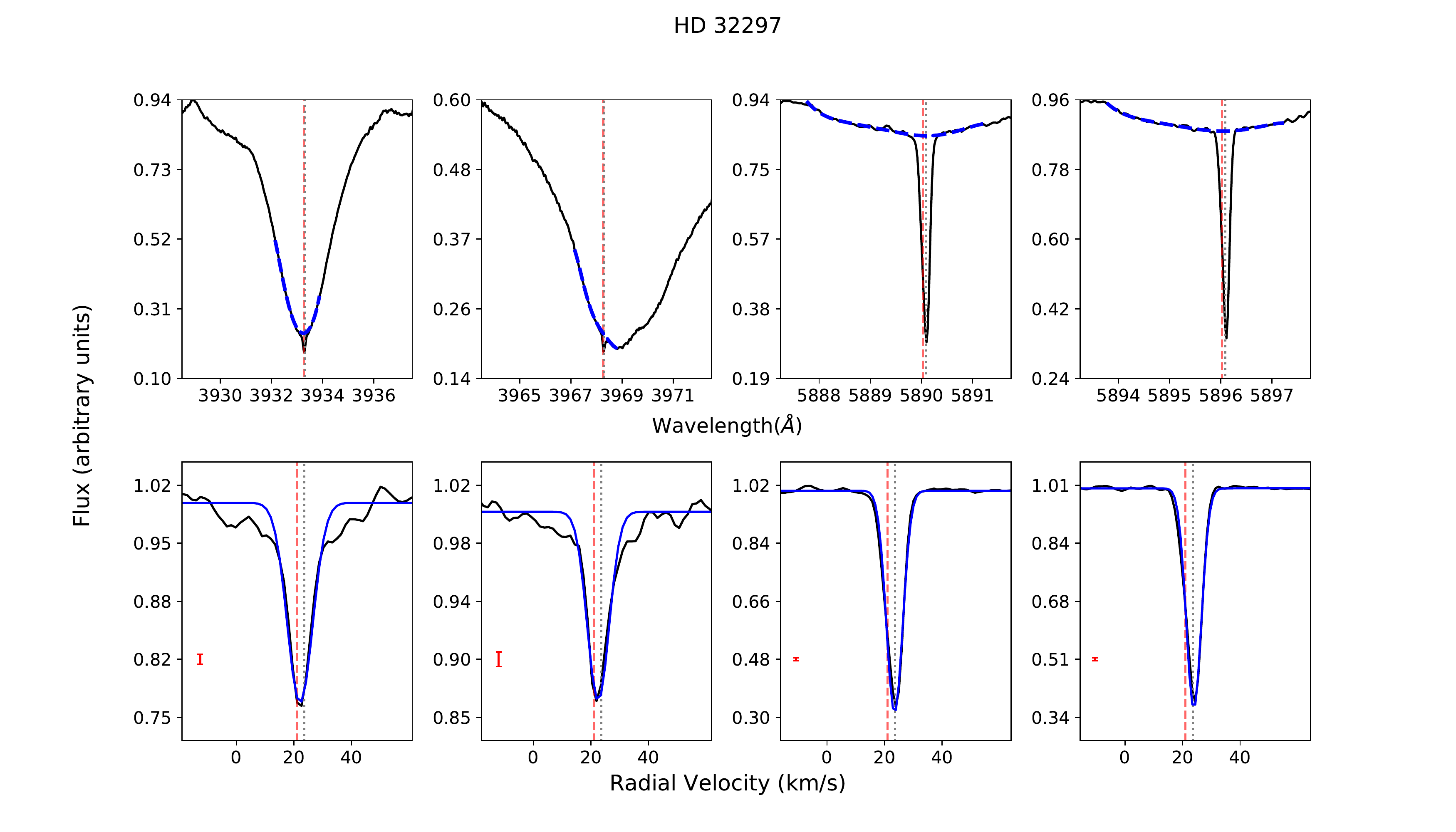}   
\caption{Stars showing narrow non-photospheric absorptions. Top panels: Photospheric Ca {\sc ii} H \& K and Na {\sc i} D lines with fitted modeling dashed blue line, x-axis shows the wavelength. Bottom panels: Residuals once the spectrum is divided by the photosphere, x-axis in velocity. Blue lines mark the fits to the non-photospheric absorptions. Vertical red dashed and grey dotted lines represent the stellar radial velocity and the ISM velocities respectively.  Red error bars show three sigma value measured in the continuum adjacent to the photospheric line. }
\end{figure}

\renewcommand{\thefigure}{\arabic{figure} (Cont.)}
\addtocounter{figure}{-1}

\begin{figure}
\centering   
\includegraphics[width=1.\textwidth]{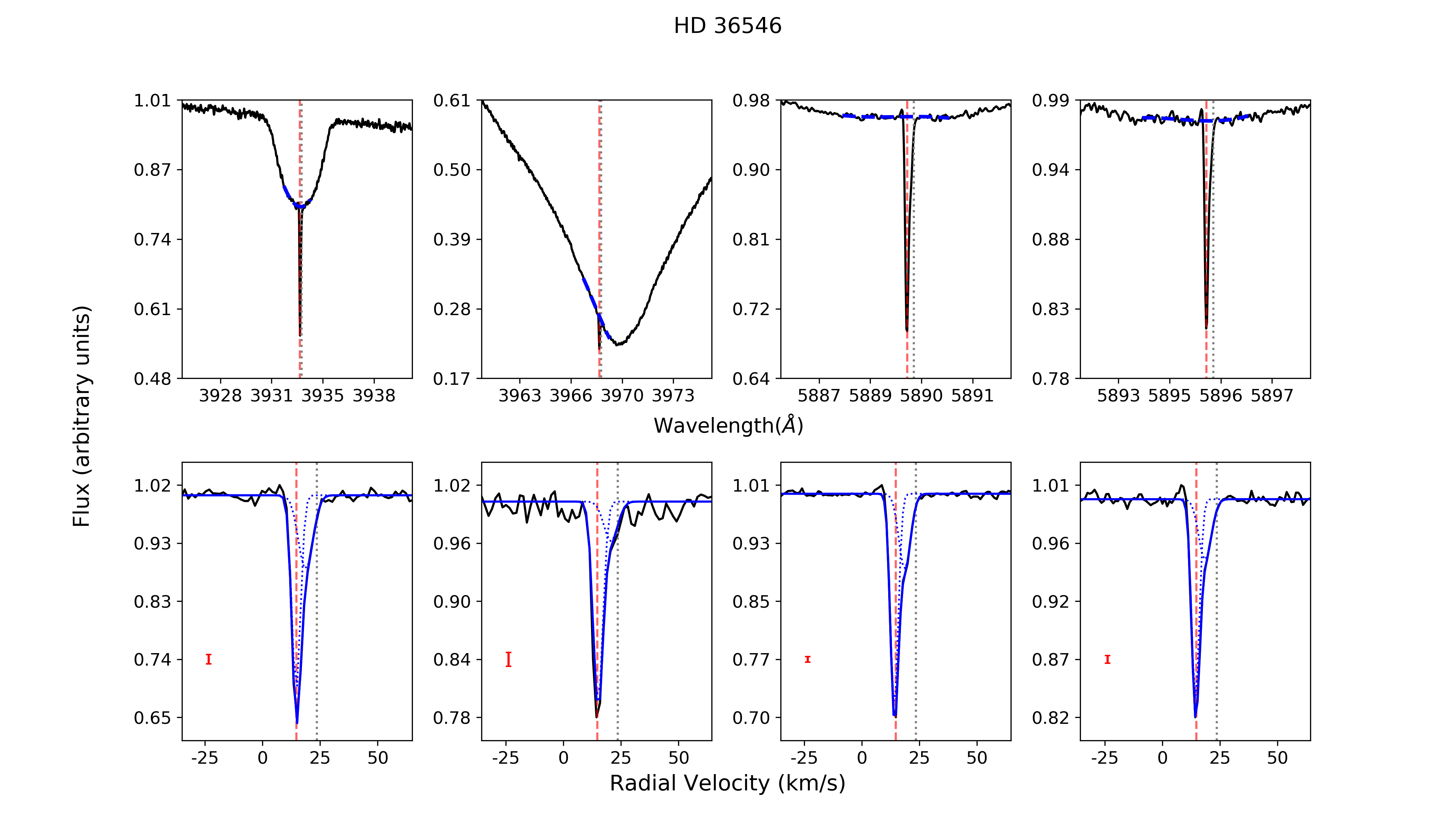}   
\includegraphics[width=1.\textwidth]{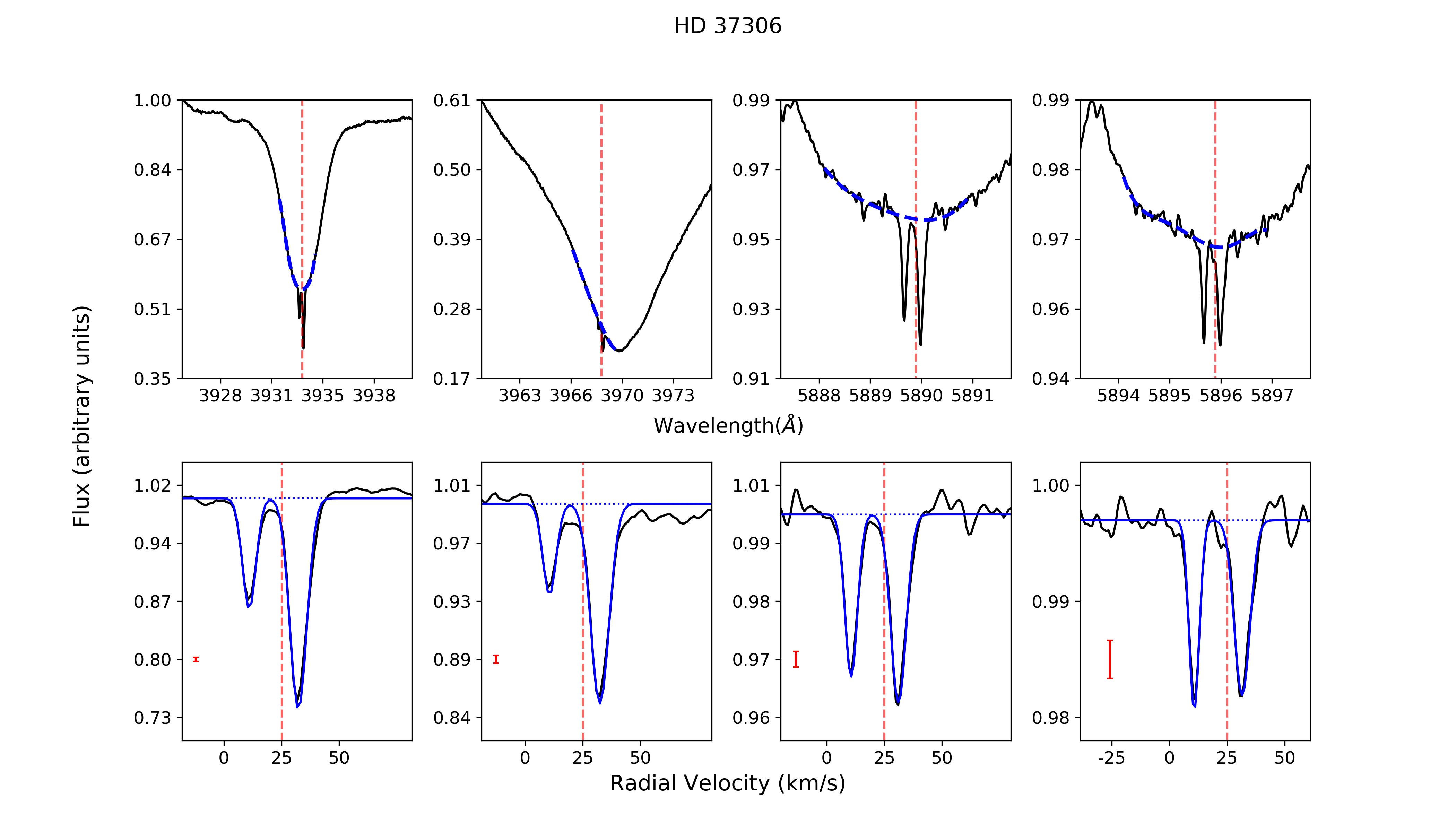}
\caption{Stars showing narrow non-photospheric absorptions. Top panels: Photospheric Ca {\sc ii} H \& K and Na {\sc i} D lines with fitted modeling dashed blue line, x-axis shows the wavelength. Bottom panels: Residuals once the spectrum is divided by the photosphere, x-axis in velocity. Blue lines mark the fits to the non-photospheric absorptions. Vertical red dashed and grey dotted lines represent the stellar radial velocity and the ISM velocities respectively.  Red error bars show three sigma value measured in the continuum adjacent to the photospheric line. }
\end{figure}

\renewcommand{\thefigure}{\arabic{figure} (Cont.)}
\addtocounter{figure}{-1}

\begin{figure}
\centering
\includegraphics[width=1.\textwidth]{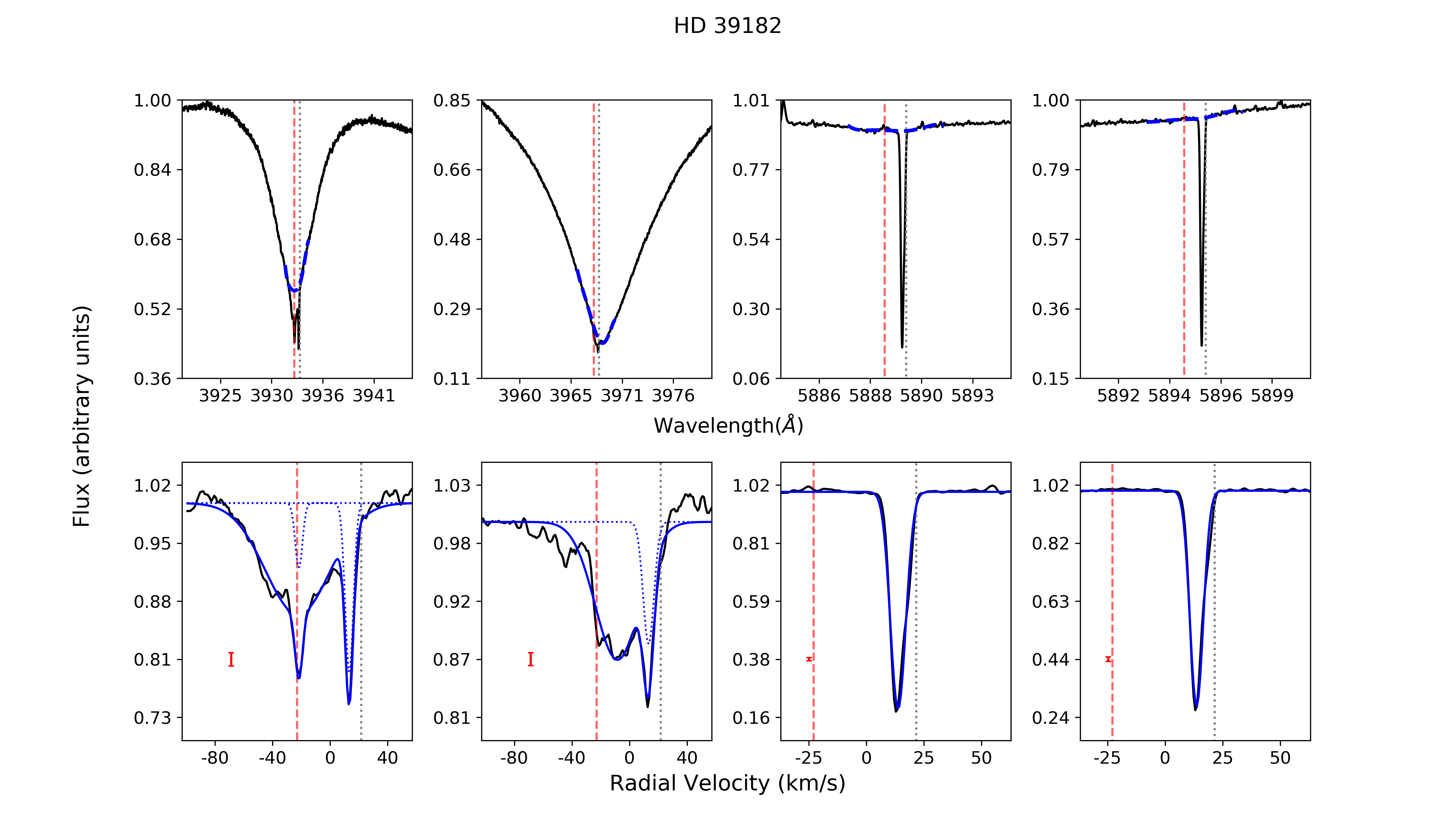}    
\includegraphics[width=1.\textwidth]{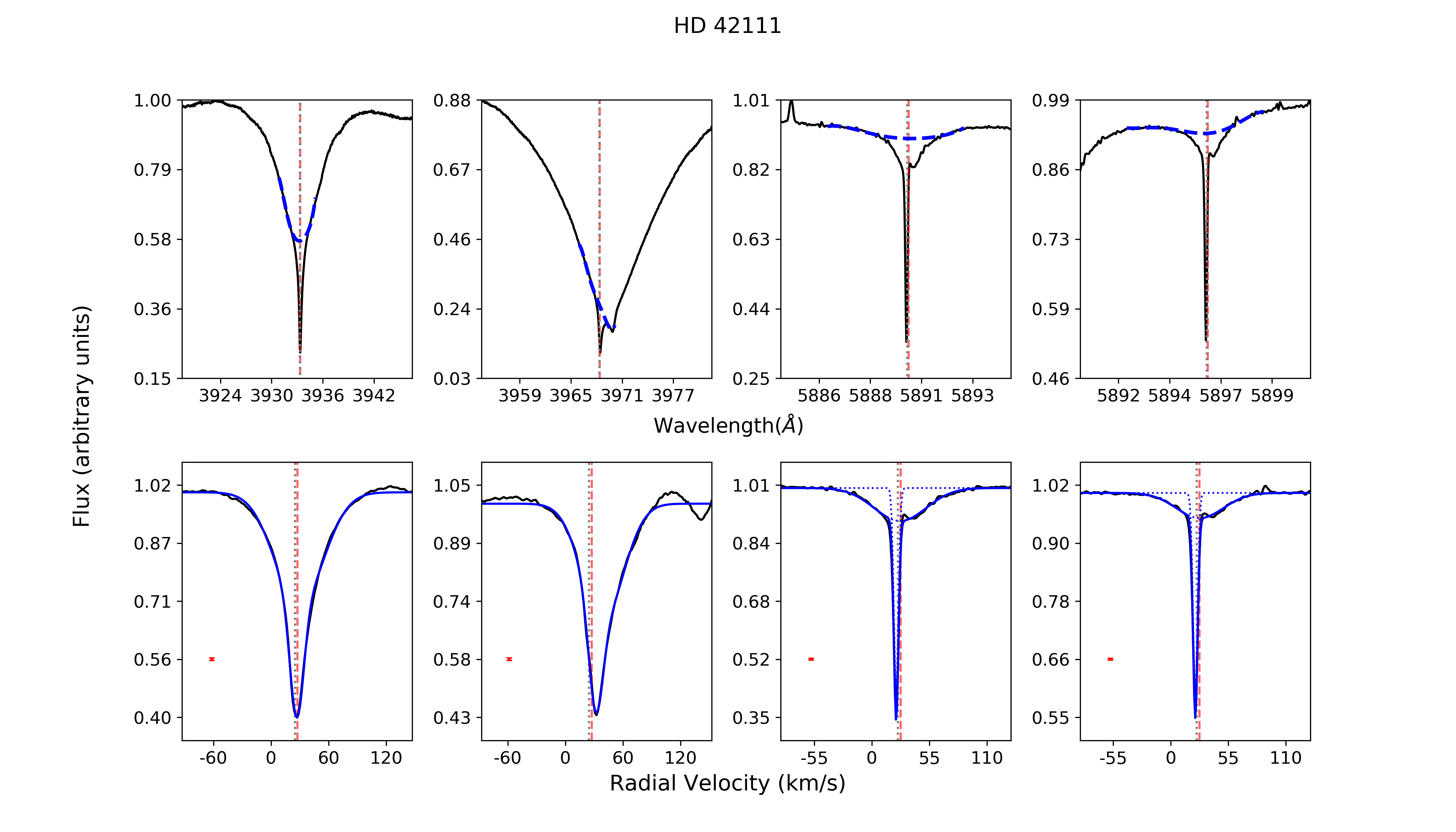}   
\caption{Stars showing narrow non-photospheric absorptions. Top panels: Photospheric Ca {\sc ii} H \& K and Na {\sc i} D lines with fitted modeling dashed blue line, x-axis shows the wavelength. Bottom panels: Residuals once the spectrum is divided by the photosphere, x-axis in velocity. Blue lines mark the fits to the non-photospheric absorptions. Vertical red dashed and grey dotted lines represent the stellar radial velocity and the ISM velocities respectively.  Red error bars show three sigma value measured in the continuum adjacent to the photospheric line. }
\end{figure}

\renewcommand{\thefigure}{\arabic{figure} (Cont.)}
\addtocounter{figure}{-1}

\begin{figure}
\centering
\includegraphics[width=1.\textwidth]{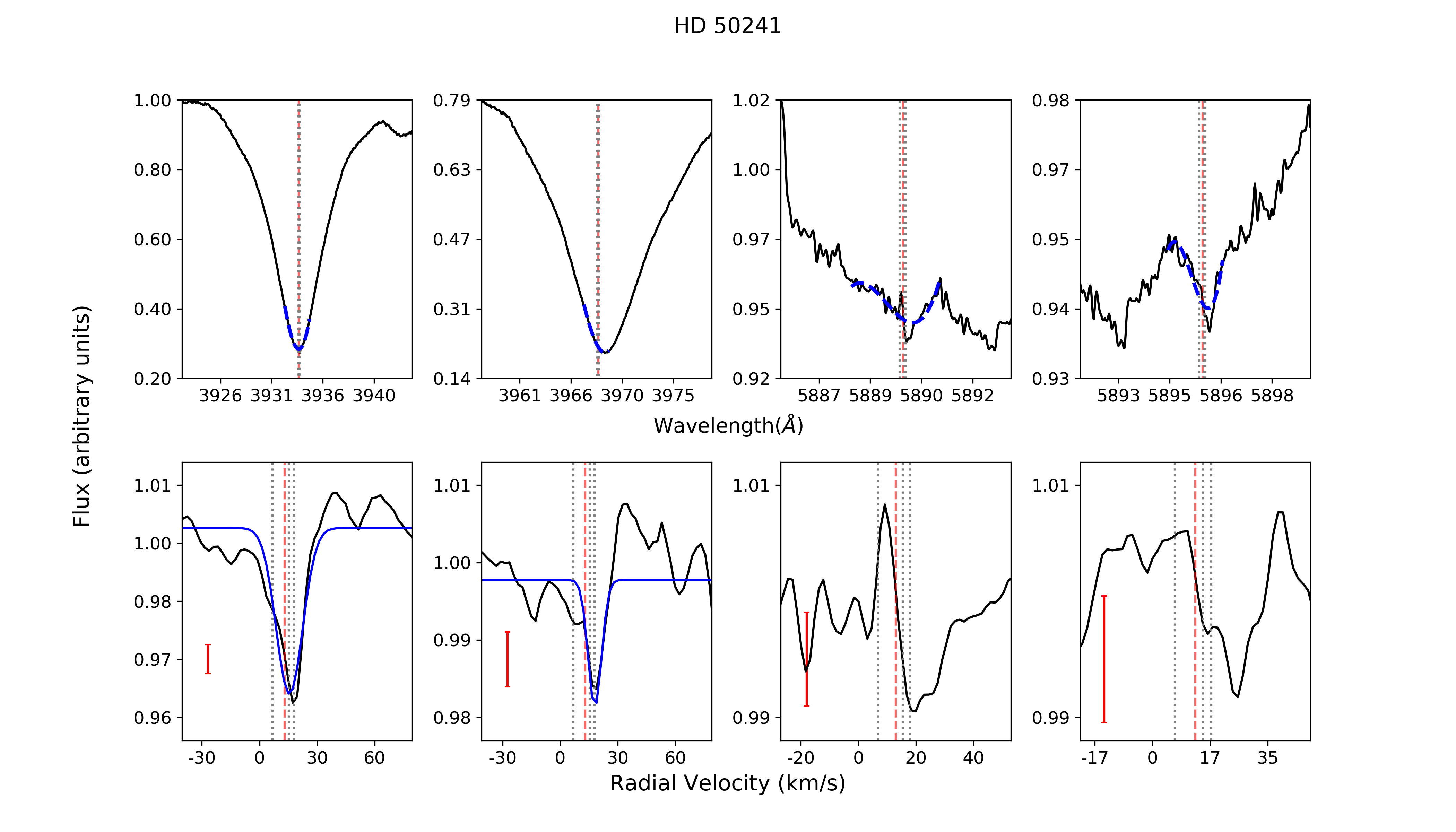}
\includegraphics[width=1.\textwidth]{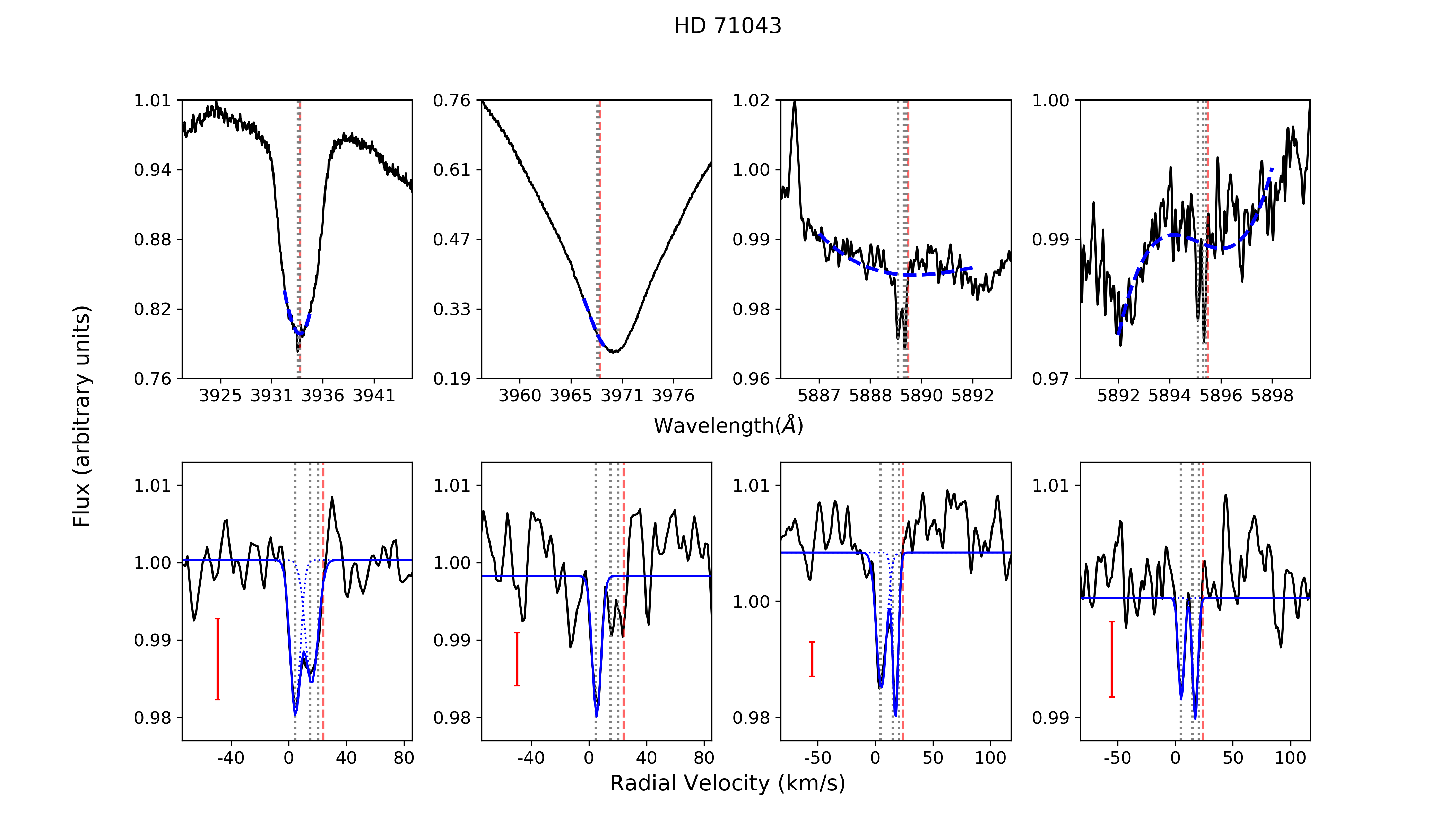}   
\caption{Stars showing narrow non-photospheric absorptions. Top panels: Photospheric Ca {\sc ii} H \& K and Na {\sc i} D lines with fitted modeling dashed blue line, x-axis shows the wavelength. Bottom panels: Residuals once the spectrum is divided by the photosphere, x-axis in velocity. Blue lines mark the fits to the non-photospheric absorptions. Vertical red dashed and grey dotted lines represent the stellar radial velocity and the ISM velocities respectively.  Red error bars show three sigma value measured in the continuum adjacent to the photospheric line. }
\end{figure}

\renewcommand{\thefigure}{\arabic{figure} (Cont.)}
\addtocounter{figure}{-1}

\begin{figure}
\centering   
\includegraphics[width=1.\textwidth]{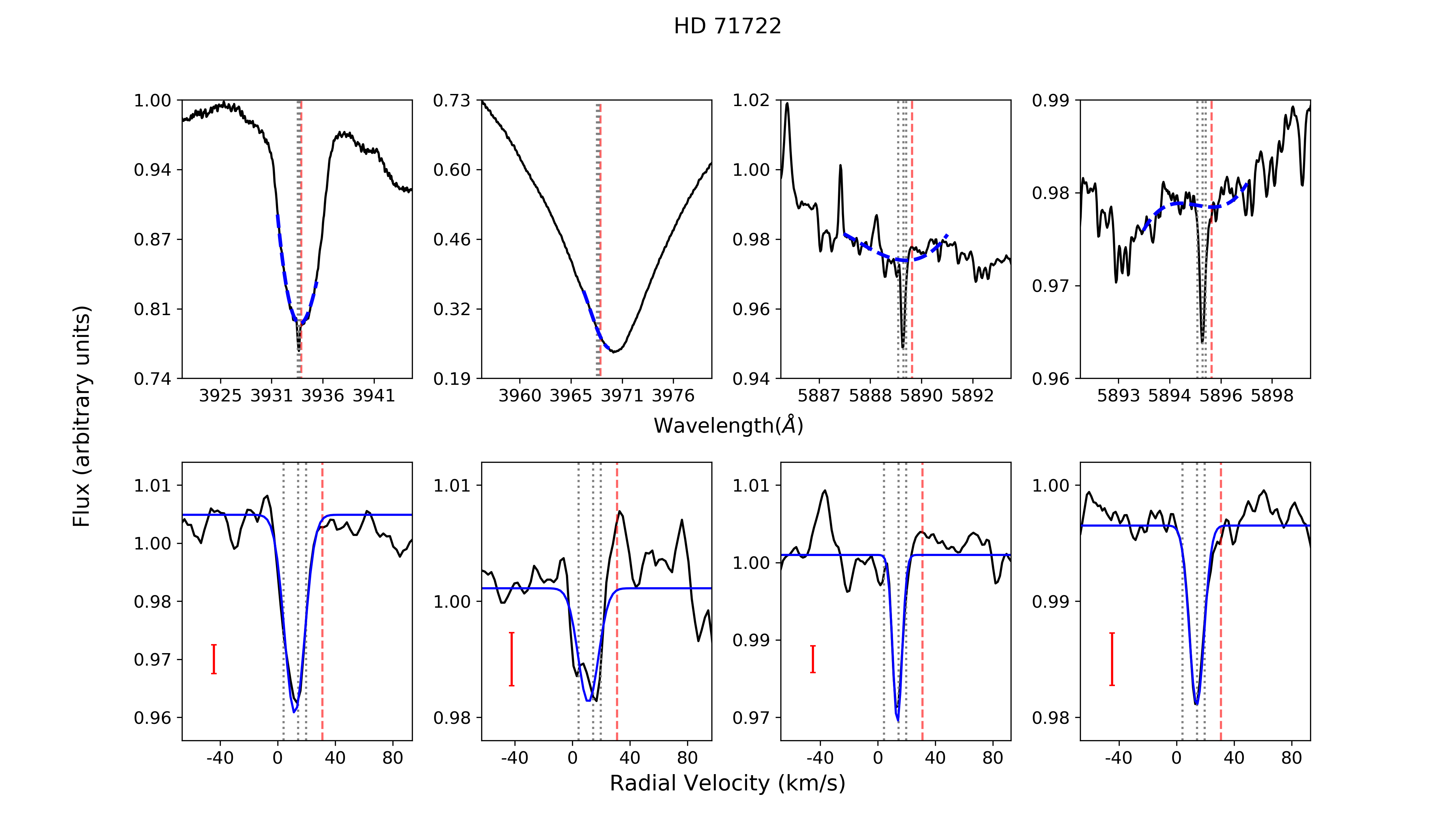}   
\includegraphics[width=1.\textwidth]{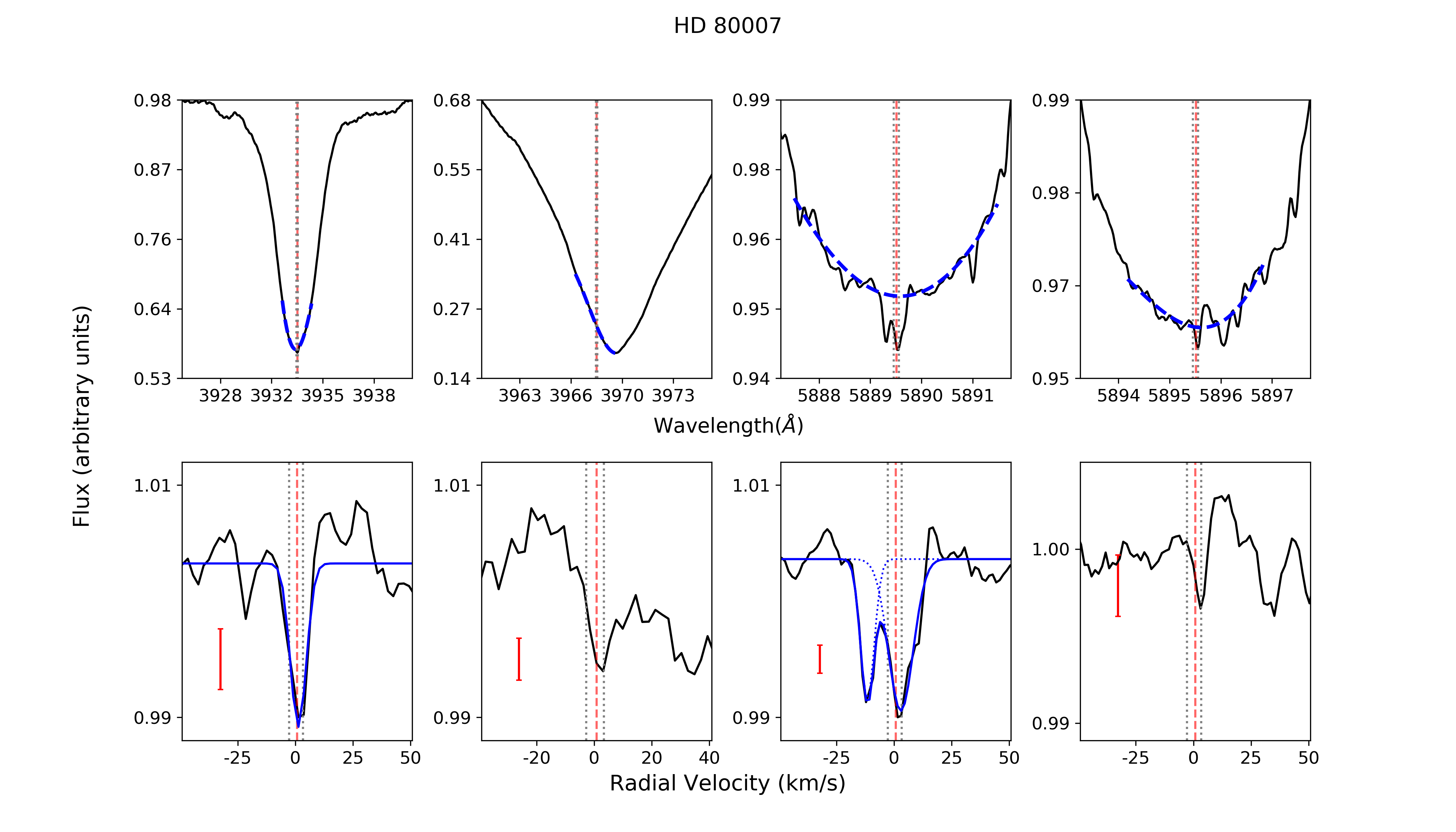} 
\caption{Stars showing narrow non-photospheric absorptions. Top panels: Photospheric Ca {\sc ii} H \& K and Na {\sc i} D lines with fitted modeling dashed blue line, x-axis shows the wavelength. Bottom panels: Residuals once the spectrum is divided by the photosphere, x-axis in velocity. Blue lines mark the fits to the non-photospheric absorptions. Vertical red dashed and grey dotted lines represent the stellar radial velocity and the ISM velocities respectively.  Red error bars show three sigma value measured in the continuum adjacent to the photospheric line. }
\end{figure}

\renewcommand{\thefigure}{\arabic{figure} (Cont.)}
\addtocounter{figure}{-1}

\begin{figure}
\centering
\includegraphics[width=1.\textwidth]{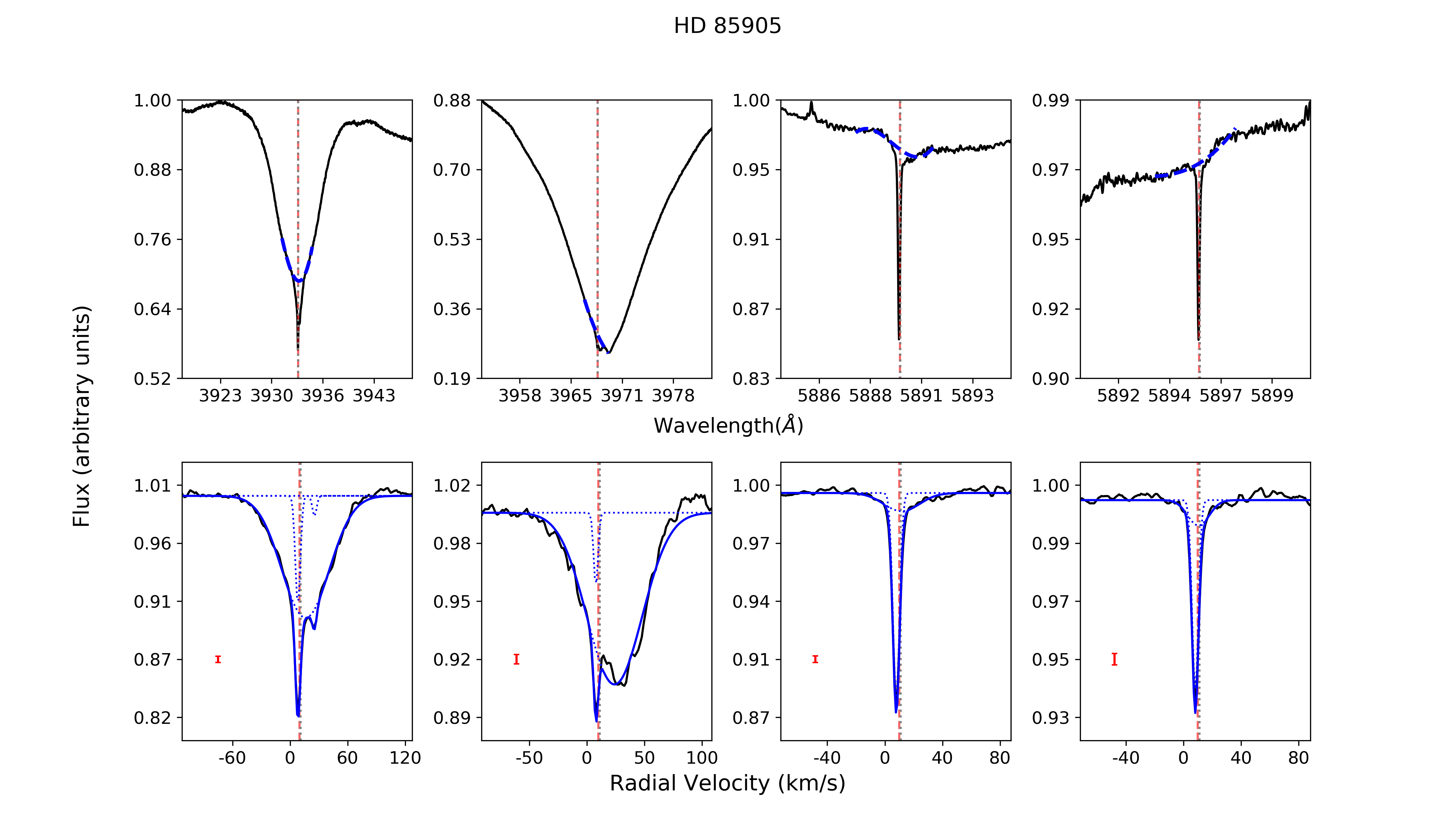}   
\includegraphics[width=1.\textwidth]{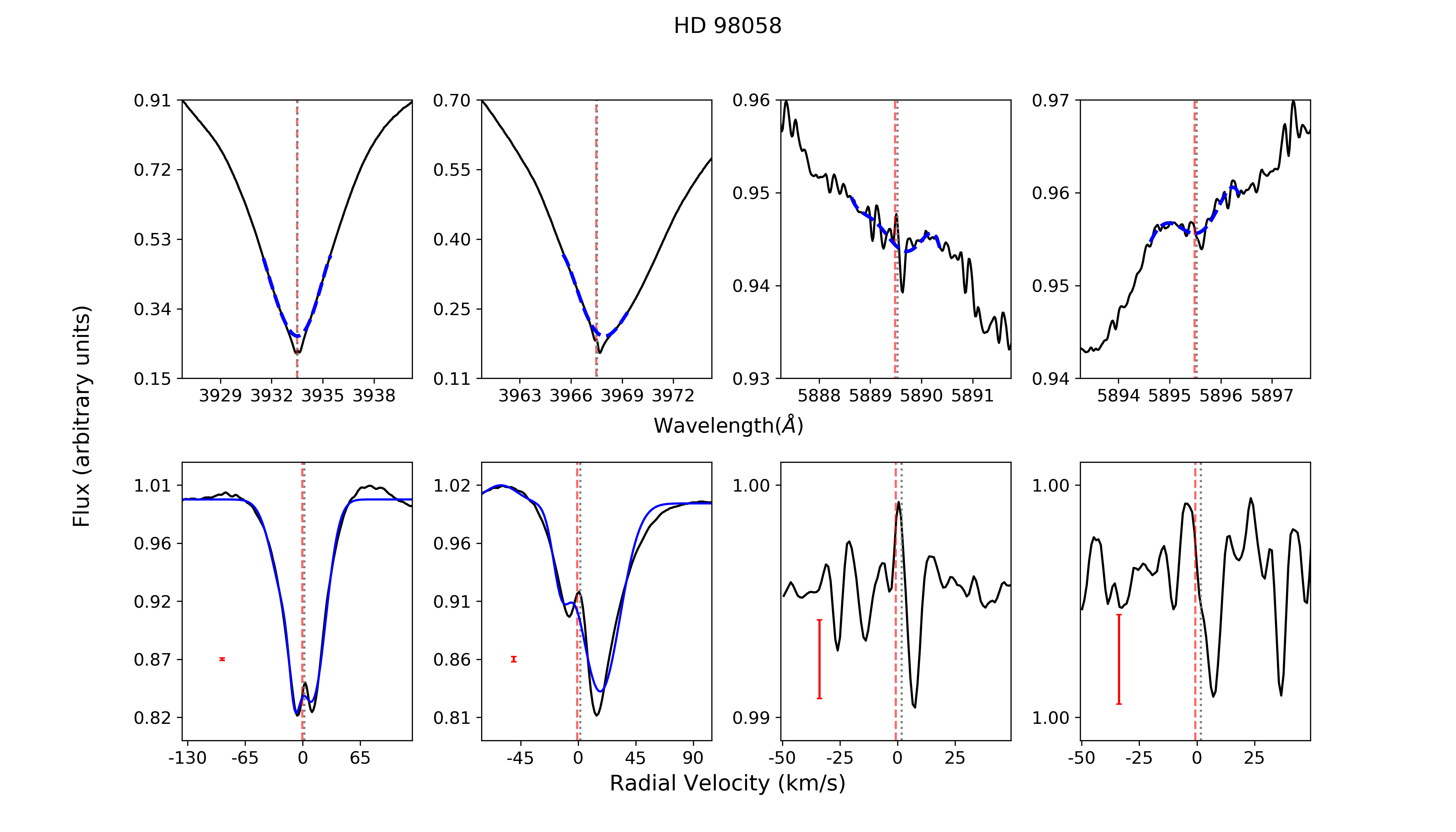}    
\caption{Stars showing narrow non-photospheric absorptions. Top panels: Photospheric Ca {\sc ii} H \& K and Na {\sc i} D lines with fitted modeling dashed blue line, x-axis shows the wavelength. Bottom panels: Residuals once the spectrum is divided by the photosphere, x-axis in velocity. Blue lines mark the fits to the non-photospheric absorptions. Vertical red dashed and grey dotted lines represent the stellar radial velocity and the ISM velocities respectively.  Red error bars show three sigma value measured in the continuum adjacent to the photospheric line. }
\end{figure}

\renewcommand{\thefigure}{\arabic{figure} (Cont.)}
\addtocounter{figure}{-1}

\begin{figure}
\centering
\includegraphics[width=1.\textwidth]{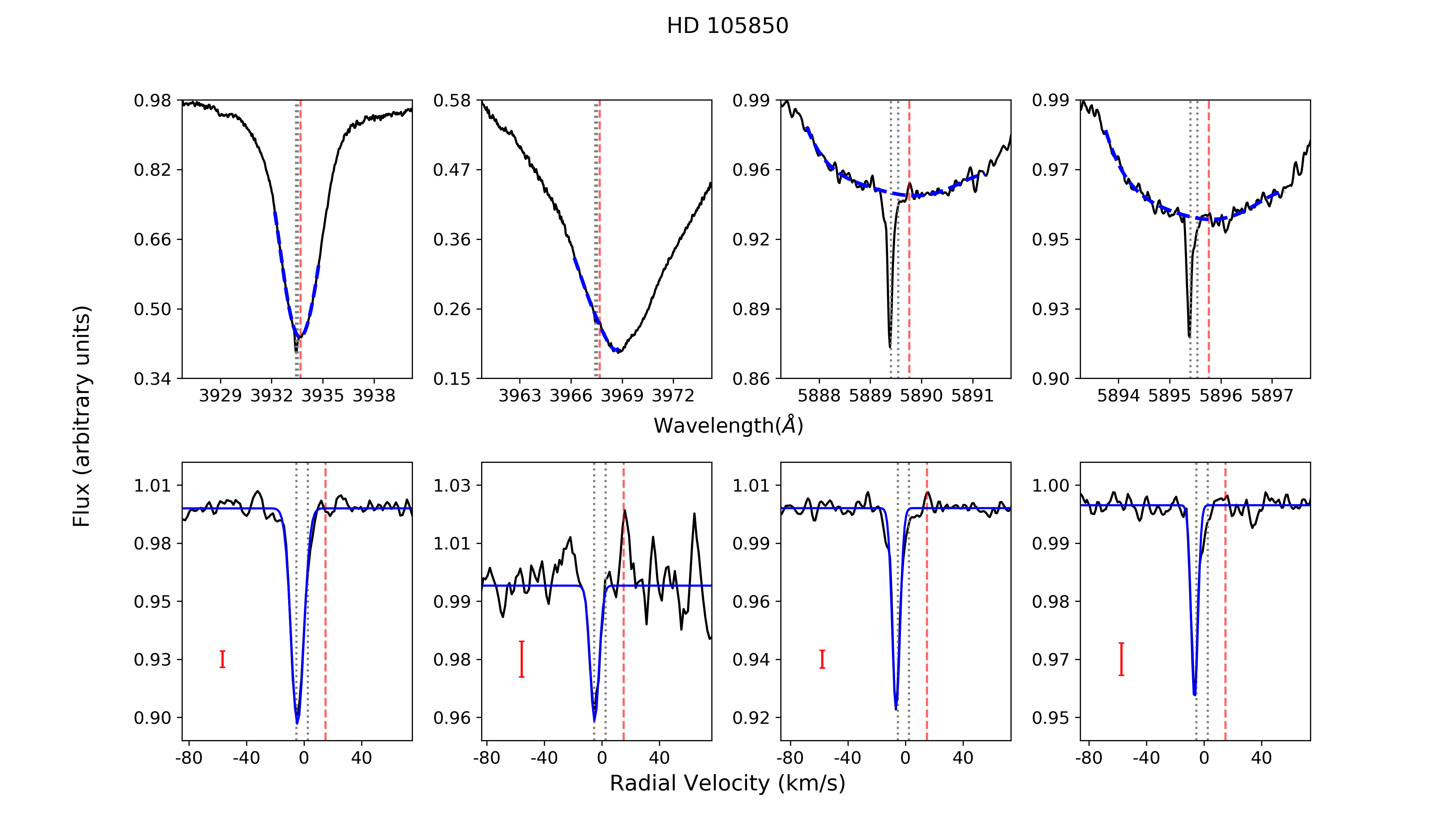} 
\includegraphics[width=1.\textwidth]{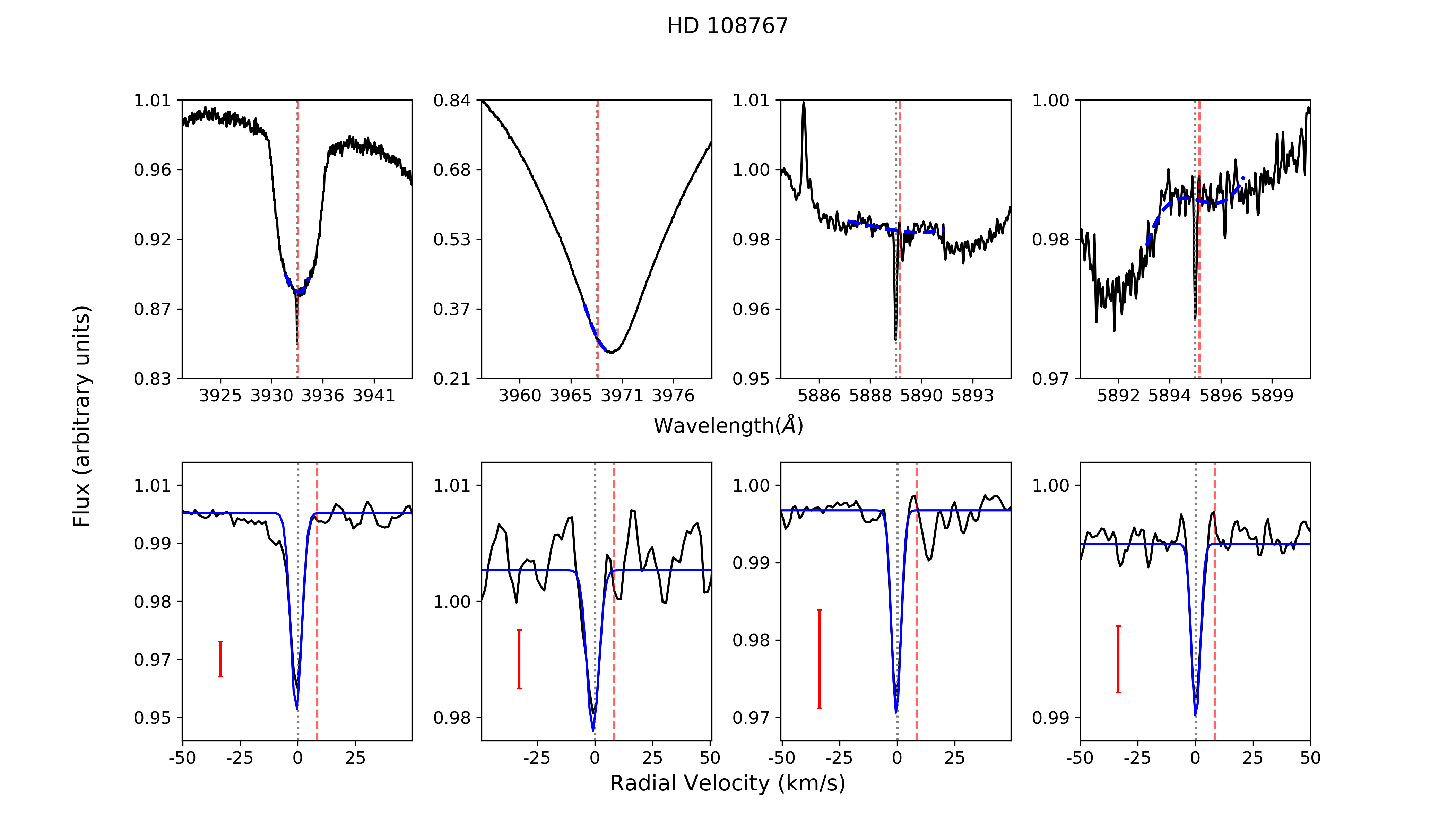}  
\caption{Stars showing narrow non-photospheric absorptions. Top panels: Photospheric Ca {\sc ii} H \& K and Na {\sc i} D lines with fitted modeling dashed blue line, x-axis shows the wavelength. Bottom panels: Residuals once the spectrum is divided by the photosphere, x-axis in velocity. Blue lines mark the fits to the non-photospheric absorptions. Vertical red dashed and grey dotted lines represent the stellar radial velocity and the ISM velocities respectively.  Red error bars show three sigma value measured in the continuum adjacent to the photospheric line. }
\end{figure}

\renewcommand{\thefigure}{\arabic{figure} (Cont.)}
\addtocounter{figure}{-1}

\begin{figure}
\centering
\includegraphics[width=1.\textwidth]{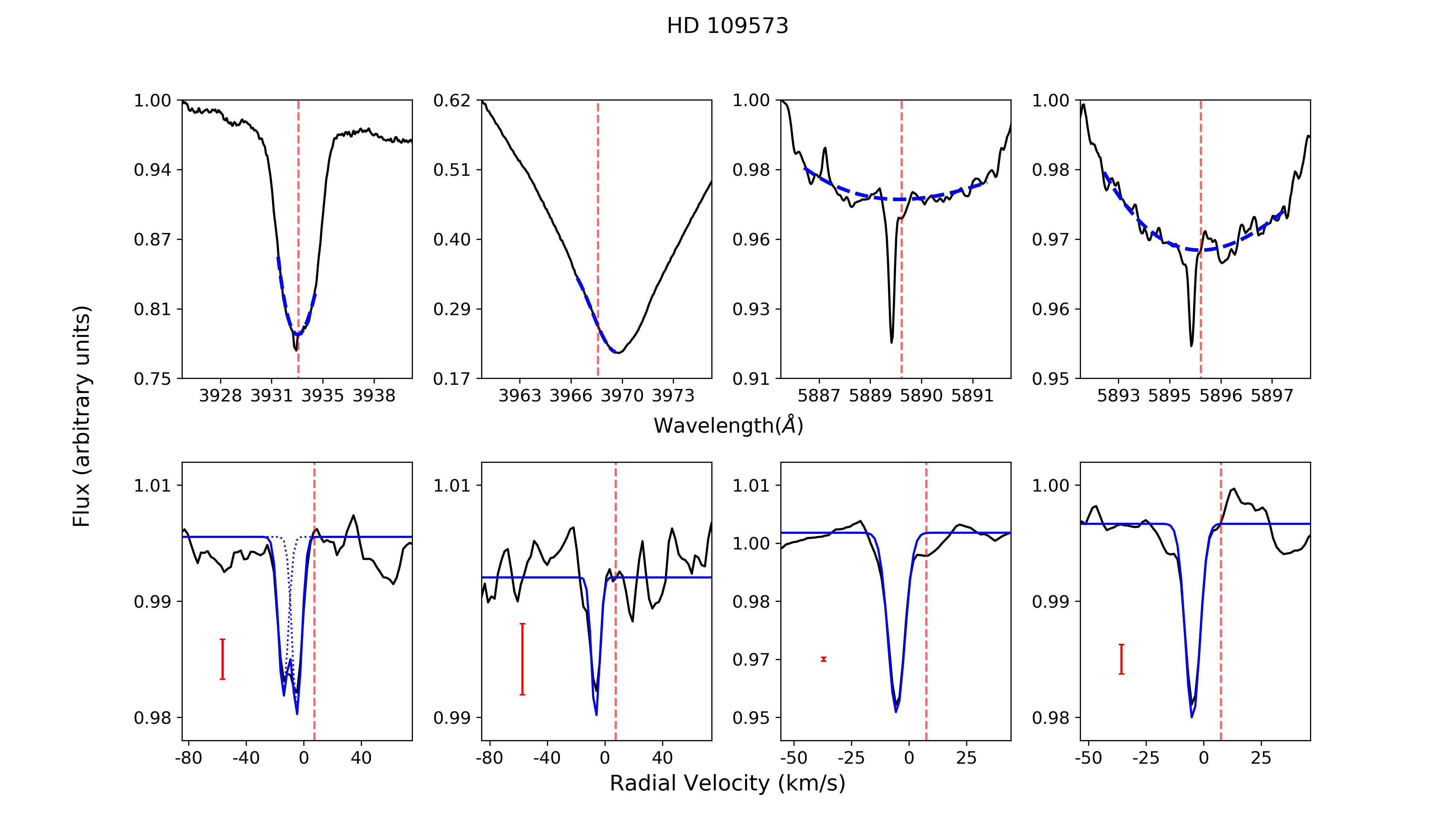}    
\includegraphics[width=1.\textwidth]{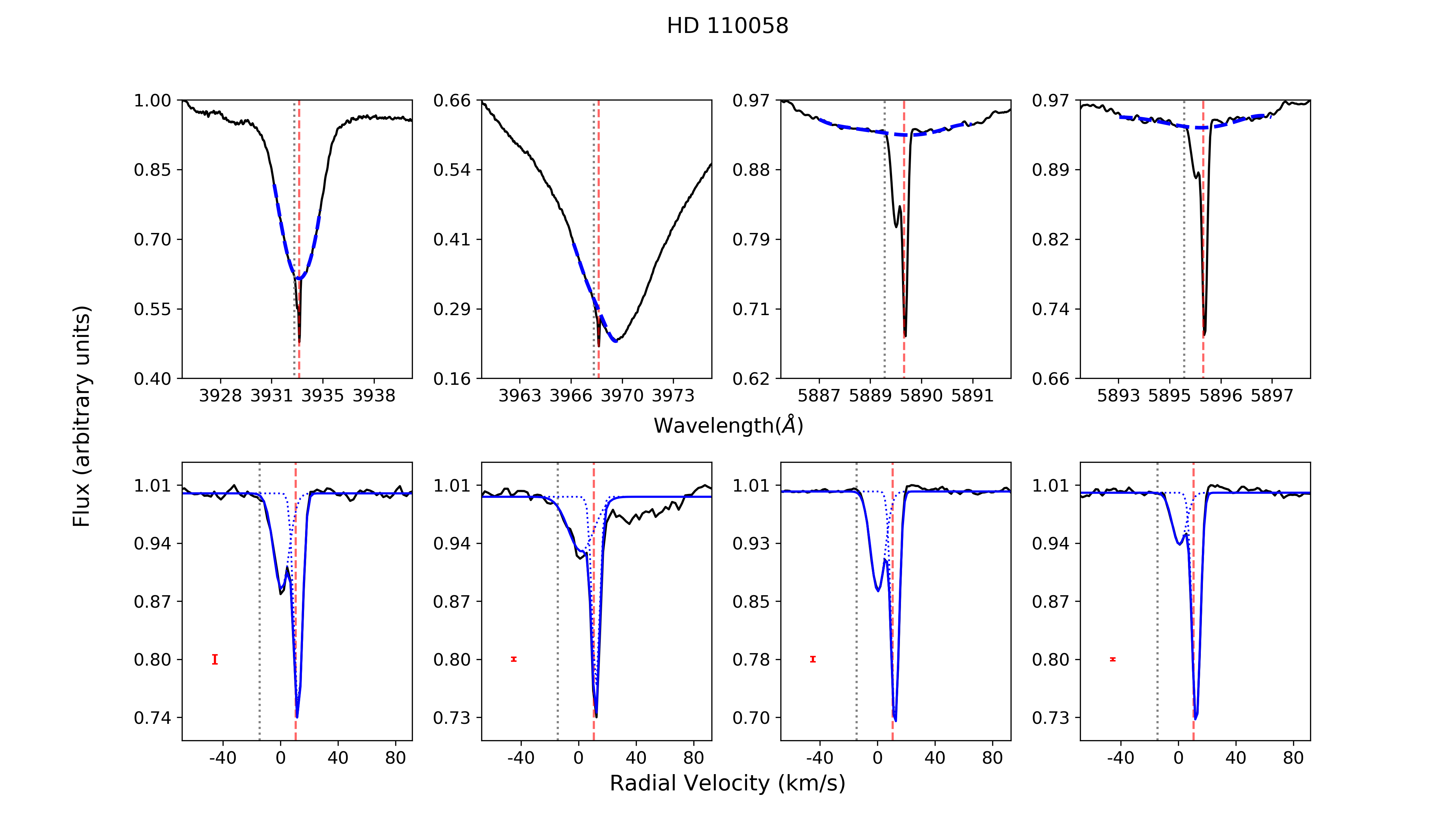}
\caption{Stars showing narrow non-photospheric absorptions. Top panels: Photospheric Ca {\sc ii} H \& K and Na {\sc i} D lines with fitted modeling dashed blue line, x-axis shows the wavelength. Bottom panels: Residuals once the spectrum is divided by the photosphere, x-axis in velocity. Blue lines mark the fits to the non-photospheric absorptions. Vertical red dashed and grey dotted lines represent the stellar radial velocity and the ISM velocities respectively.  Red error bars show three sigma value measured in the continuum adjacent to the photospheric line. }
\end{figure}

\renewcommand{\thefigure}{\arabic{figure} (Cont.)}
\addtocounter{figure}{-1}

\begin{figure}
\centering
\includegraphics[width=1.\textwidth]{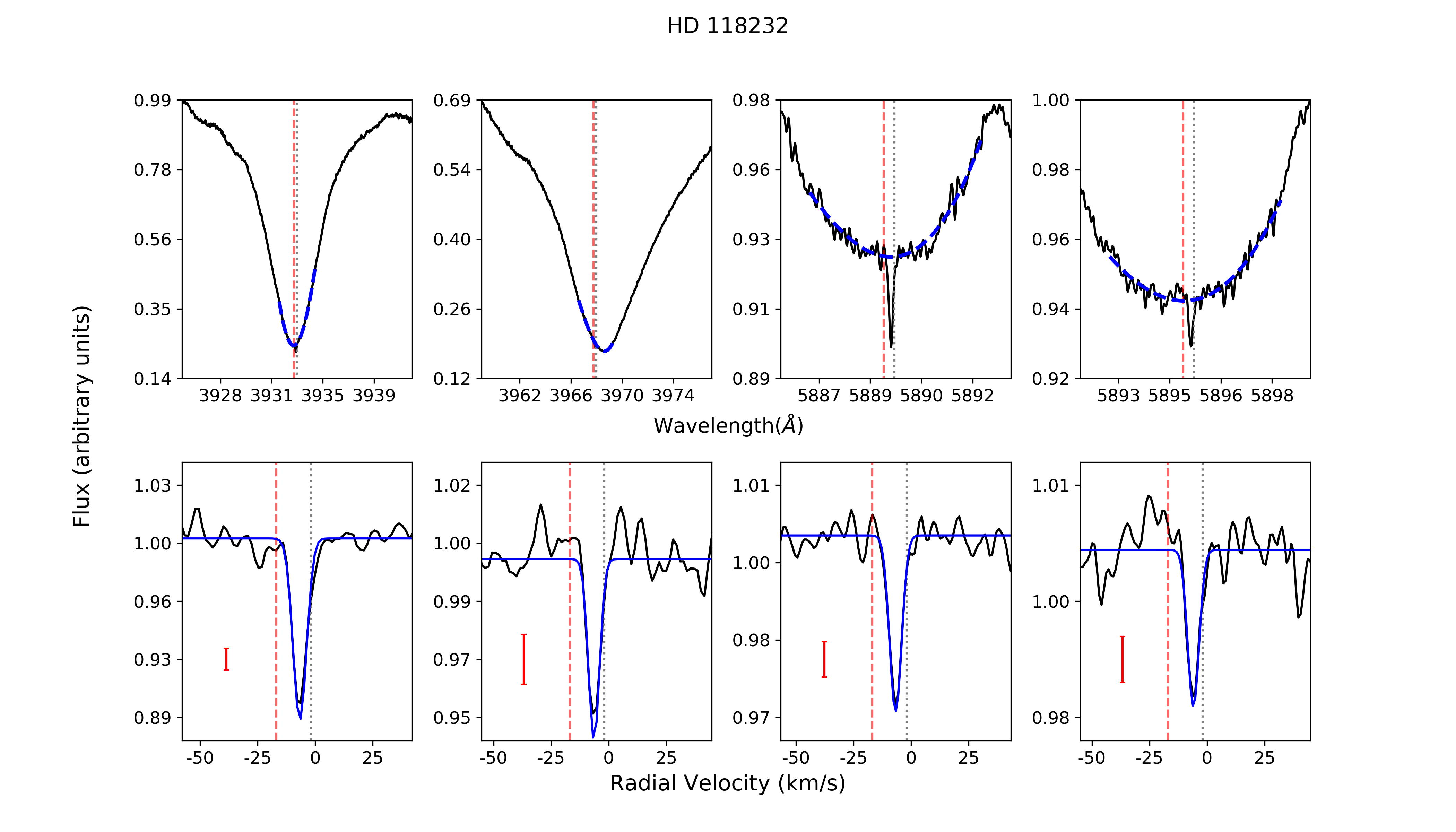}  
\includegraphics[width=1.\textwidth]{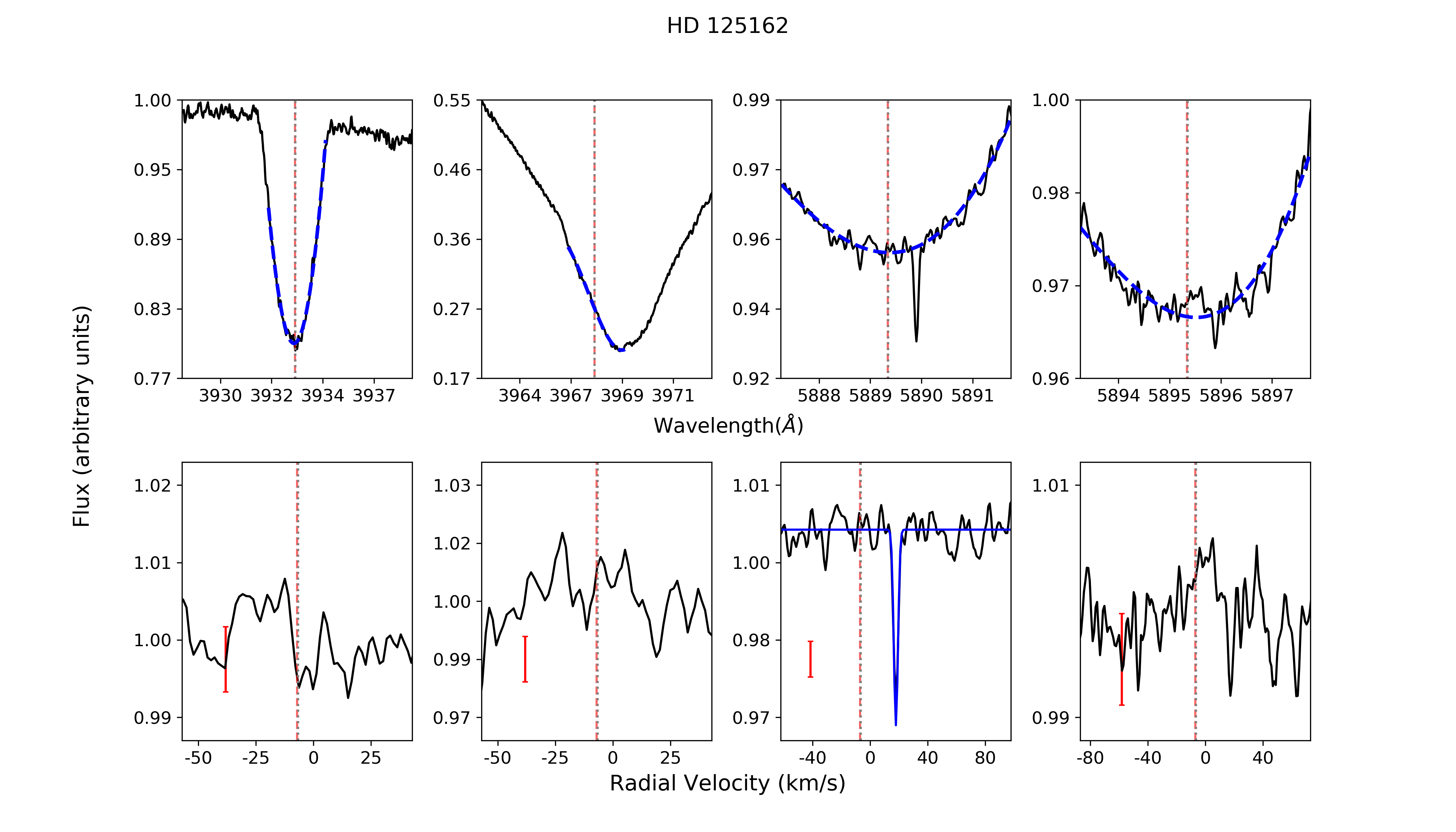} 
\caption{Stars showing narrow non-photospheric absorptions. Top panels: Photospheric Ca {\sc ii} H \& K and Na {\sc i} D lines with fitted modeling dashed blue line, x-axis shows the wavelength. Bottom panels: Residuals once the spectrum is divided by the photosphere, x-axis in velocity. Blue lines mark the fits to the non-photospheric absorptions. Vertical red dashed and grey dotted lines represent the stellar radial velocity and the ISM velocities respectively.  Red error bars show three sigma value measured in the continuum adjacent to the photospheric line. }
\end{figure}

\renewcommand{\thefigure}{\arabic{figure} (Cont.)}
\addtocounter{figure}{-1}

\begin{figure}
\centering
\includegraphics[width=1.\textwidth]{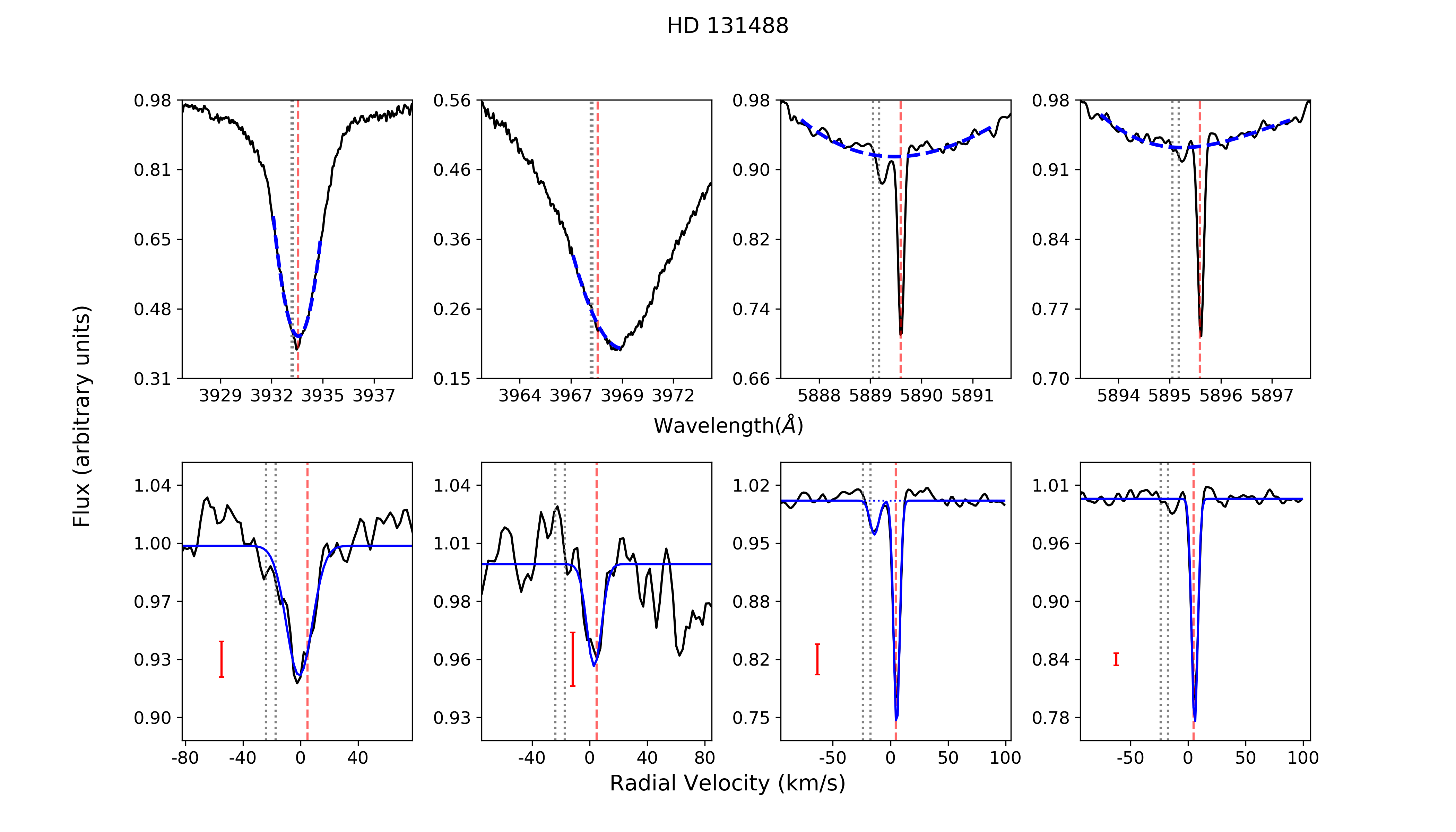}
\includegraphics[width=1.\textwidth]{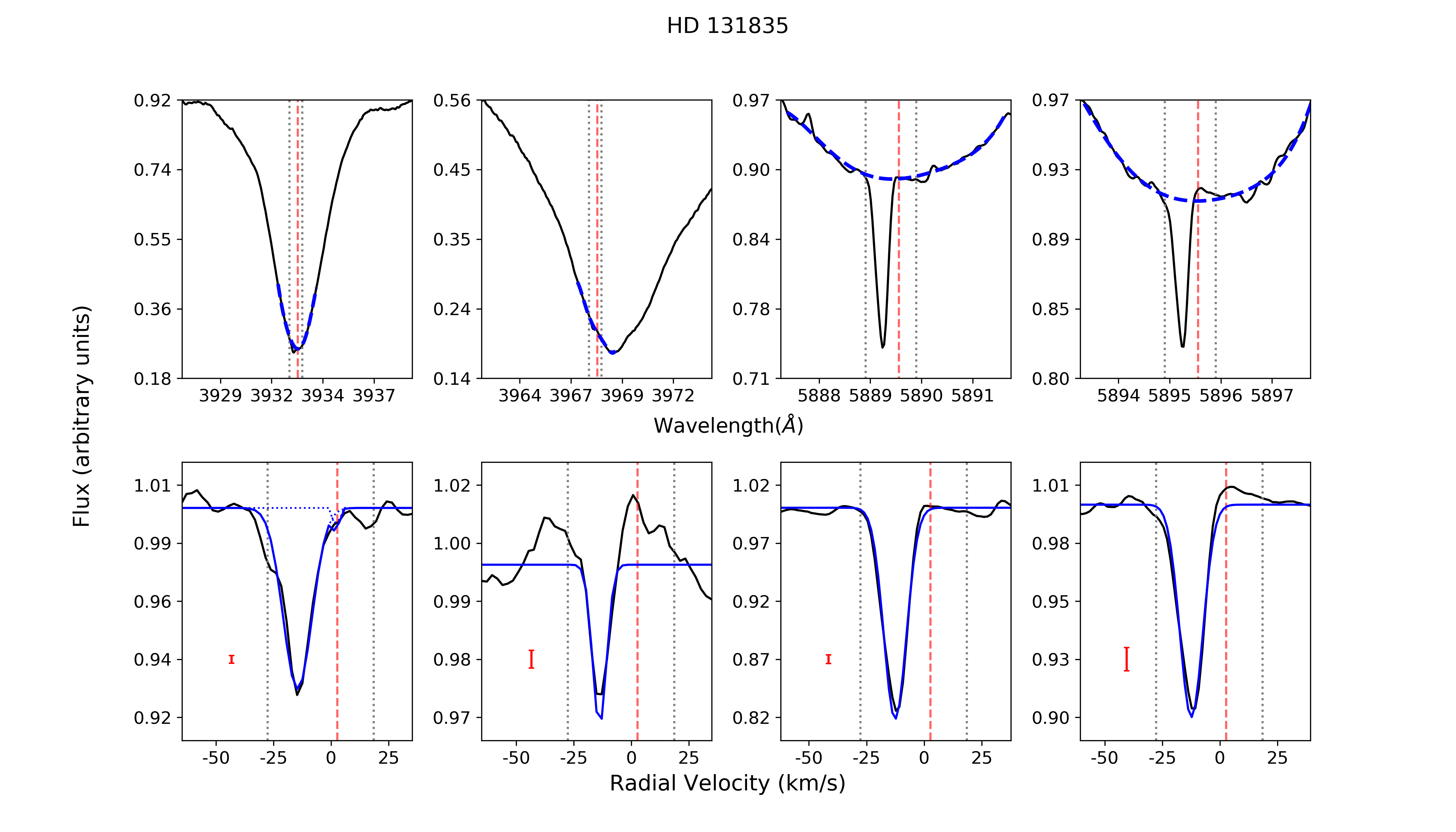}   
\caption{Stars showing narrow non-photospheric absorptions. Top panels: Photospheric Ca {\sc ii} H \& K and Na {\sc i} D lines with fitted modeling dashed blue line, x-axis shows the wavelength. Bottom panels: Residuals once the spectrum is divided by the photosphere, x-axis in velocity. Blue lines mark the fits to the non-photospheric absorptions. Vertical red dashed and grey dotted lines represent the stellar radial velocity and the ISM velocities respectively.  Red error bars show three sigma value measured in the continuum adjacent to the photospheric line. }
\end{figure}

\renewcommand{\thefigure}{\arabic{figure} (Cont.)}
\addtocounter{figure}{-1}

\begin{figure}
\centering  
\includegraphics[width=1.\textwidth]{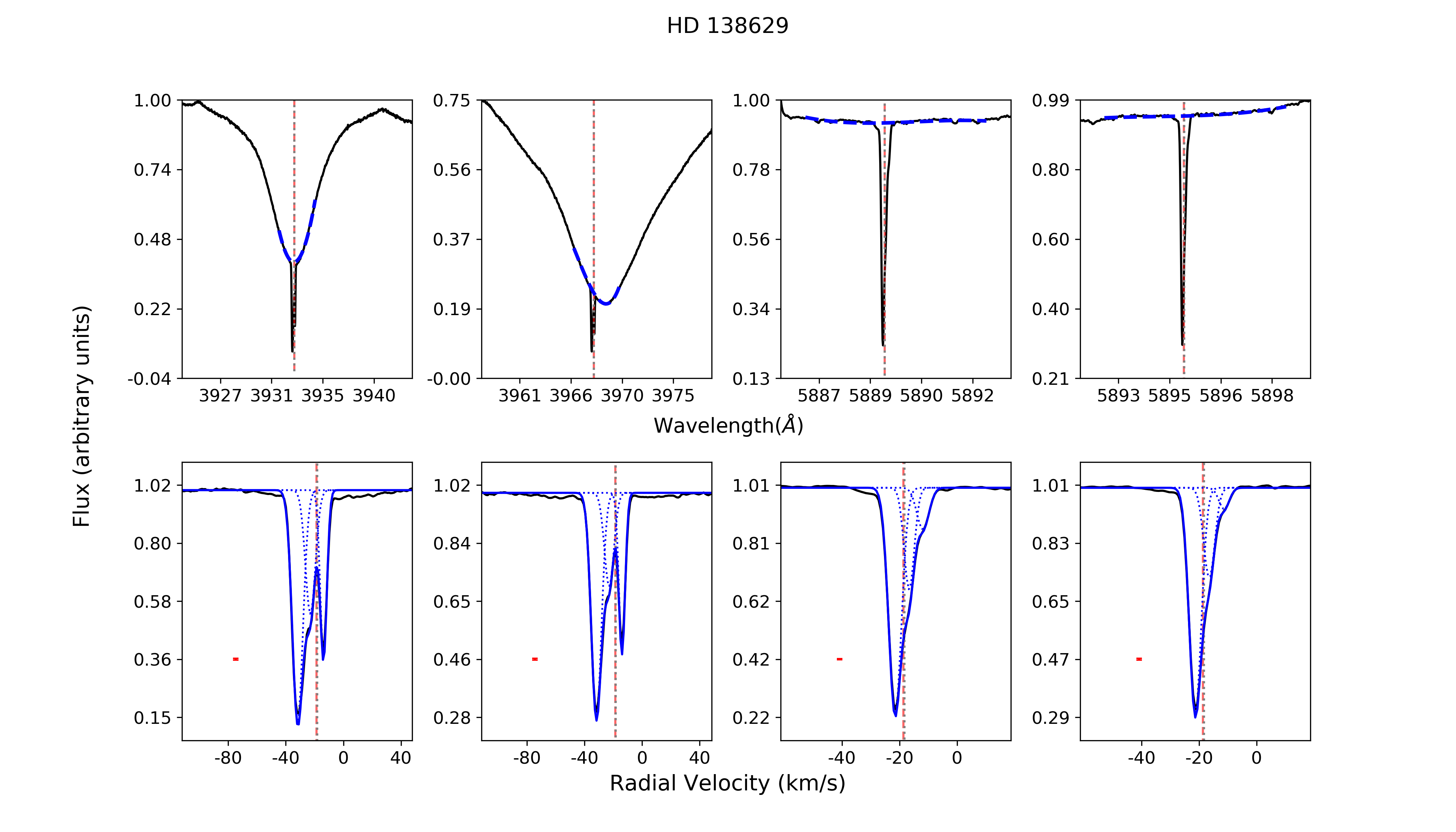}    
\includegraphics[width=1.\textwidth]{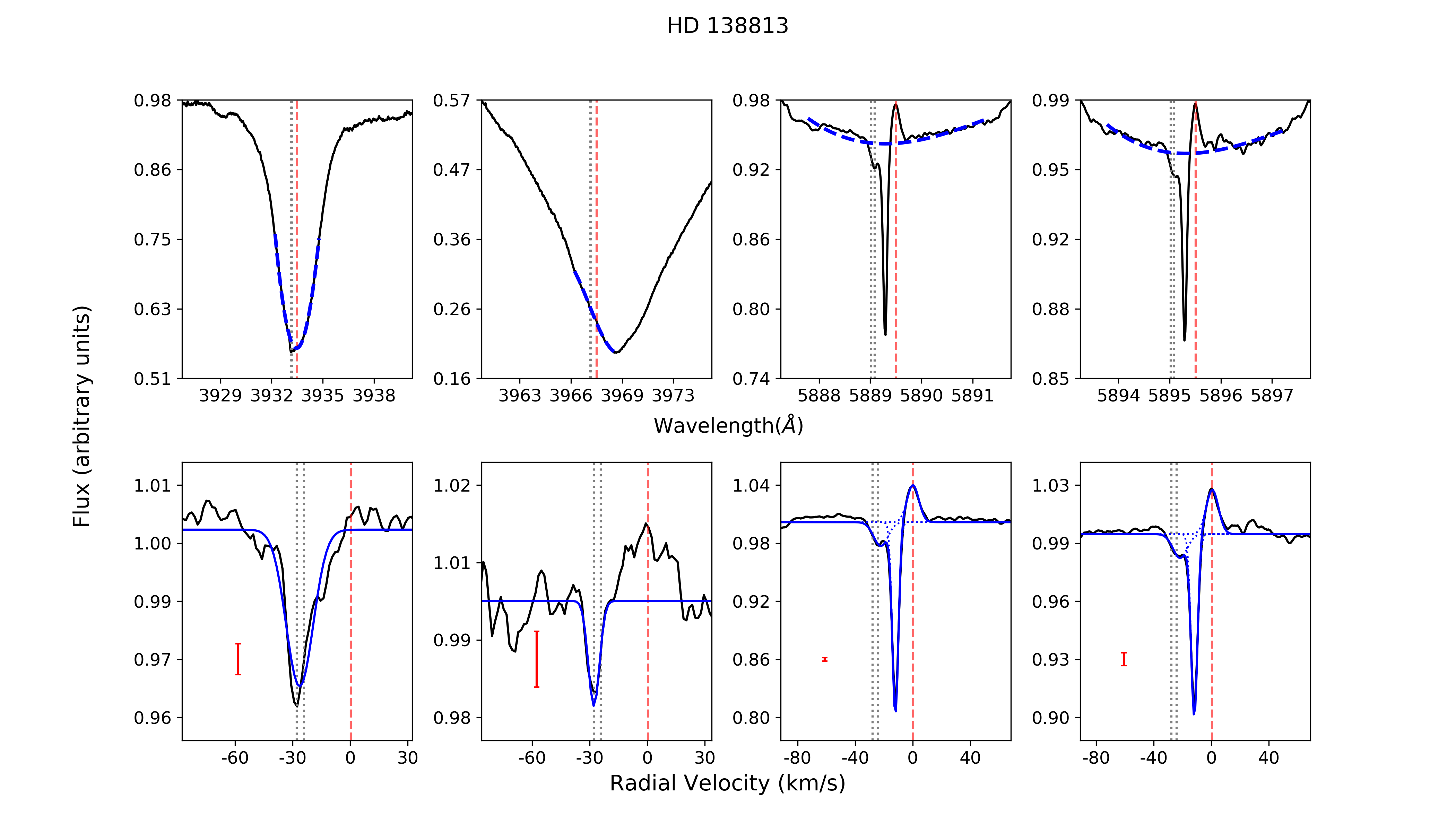} 
\caption{Stars showing narrow non-photospheric absorptions. Top panels: Photospheric Ca {\sc ii} H \& K and Na {\sc i} D lines with fitted modeling dashed blue line, x-axis shows the wavelength. Bottom panels: Residuals once the spectrum is divided by the photosphere, x-axis in velocity. Blue lines mark the fits to the non-photospheric absorptions. Vertical red dashed and grey dotted lines represent the stellar radial velocity and the ISM velocities respectively.  Red error bars show three sigma value measured in the continuum adjacent to the photospheric line. }
\end{figure}

\renewcommand{\thefigure}{\arabic{figure} (Cont.)}
\addtocounter{figure}{-1}

\begin{figure}
\centering
\includegraphics[width=1.\textwidth]{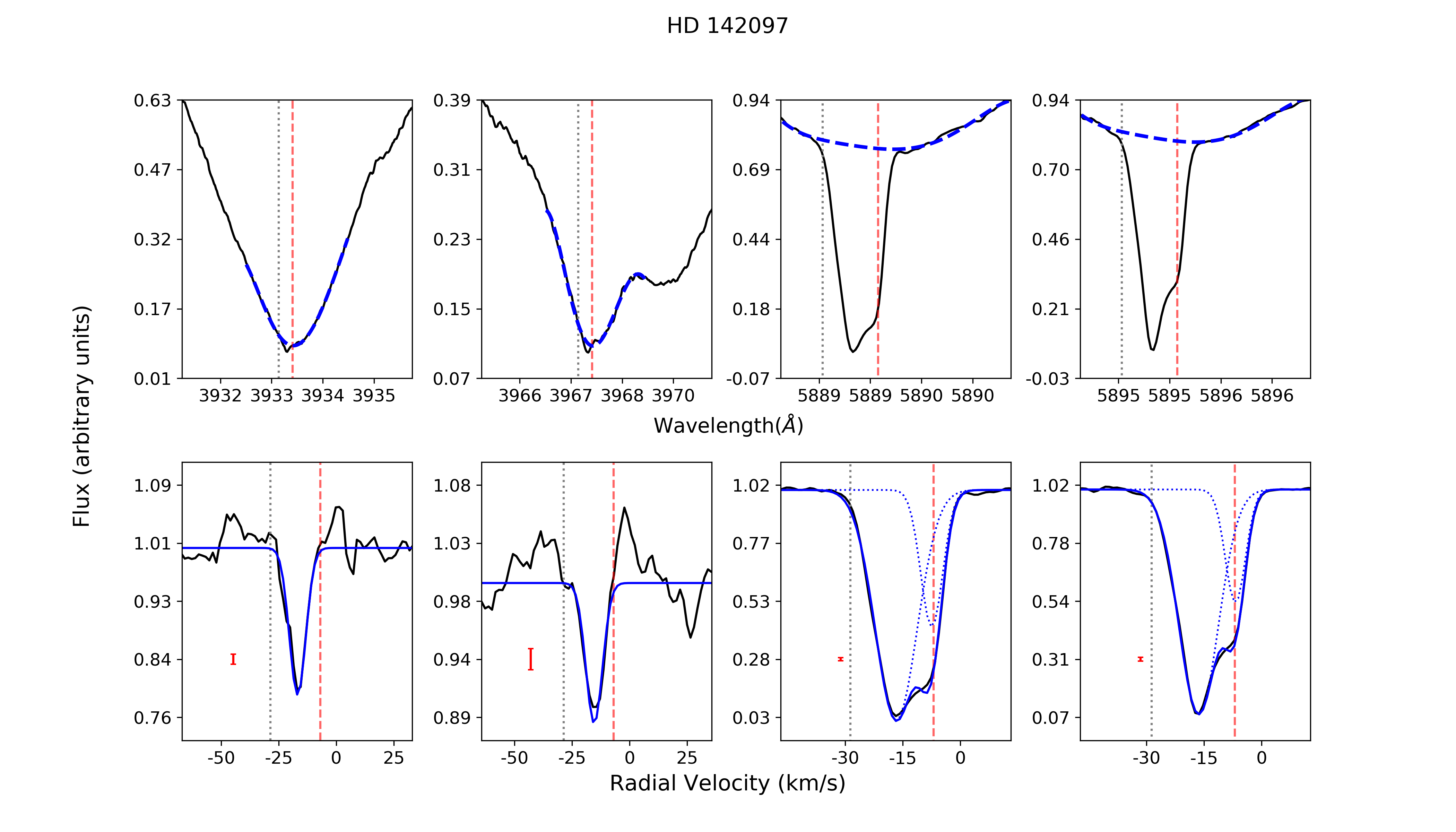}  
\includegraphics[width=1.\textwidth]{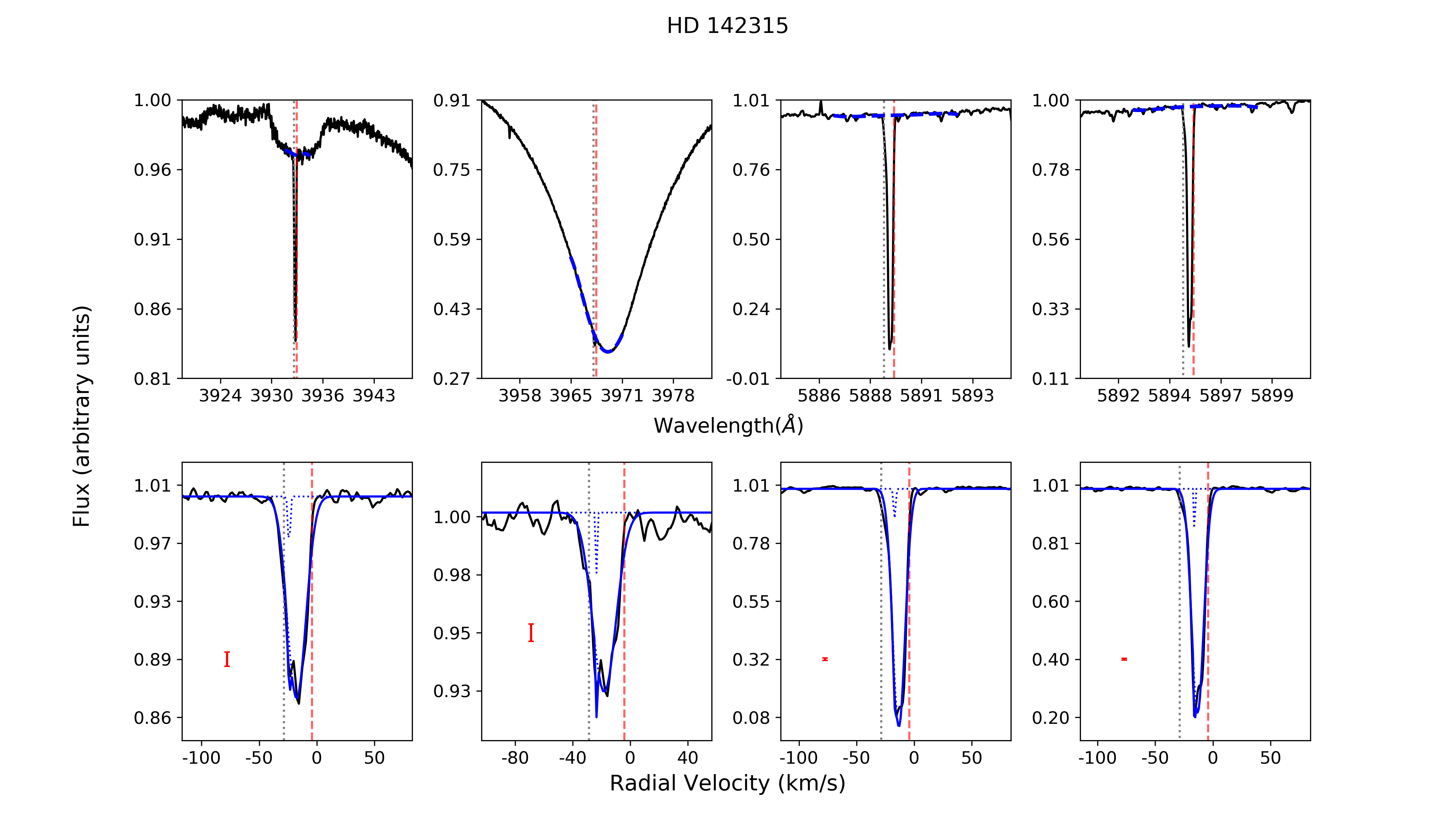}  
\caption{Stars showing narrow non-photospheric absorptions. Top panels: Photospheric Ca {\sc ii} H \& K and Na {\sc i} D lines with fitted modeling dashed blue line, x-axis shows the wavelength. Bottom panels: Residuals once the spectrum is divided by the photosphere, x-axis in velocity. Blue lines mark the fits to the non-photospheric absorptions. Vertical red dashed and grey dotted lines represent the stellar radial velocity and the ISM velocities respectively.  Red error bars show three sigma value measured in the continuum adjacent to the photospheric line. }
\end{figure}

\renewcommand{\thefigure}{\arabic{figure} (Cont.)}
\addtocounter{figure}{-1}

\begin{figure}
\centering
\includegraphics[width=1.\textwidth]{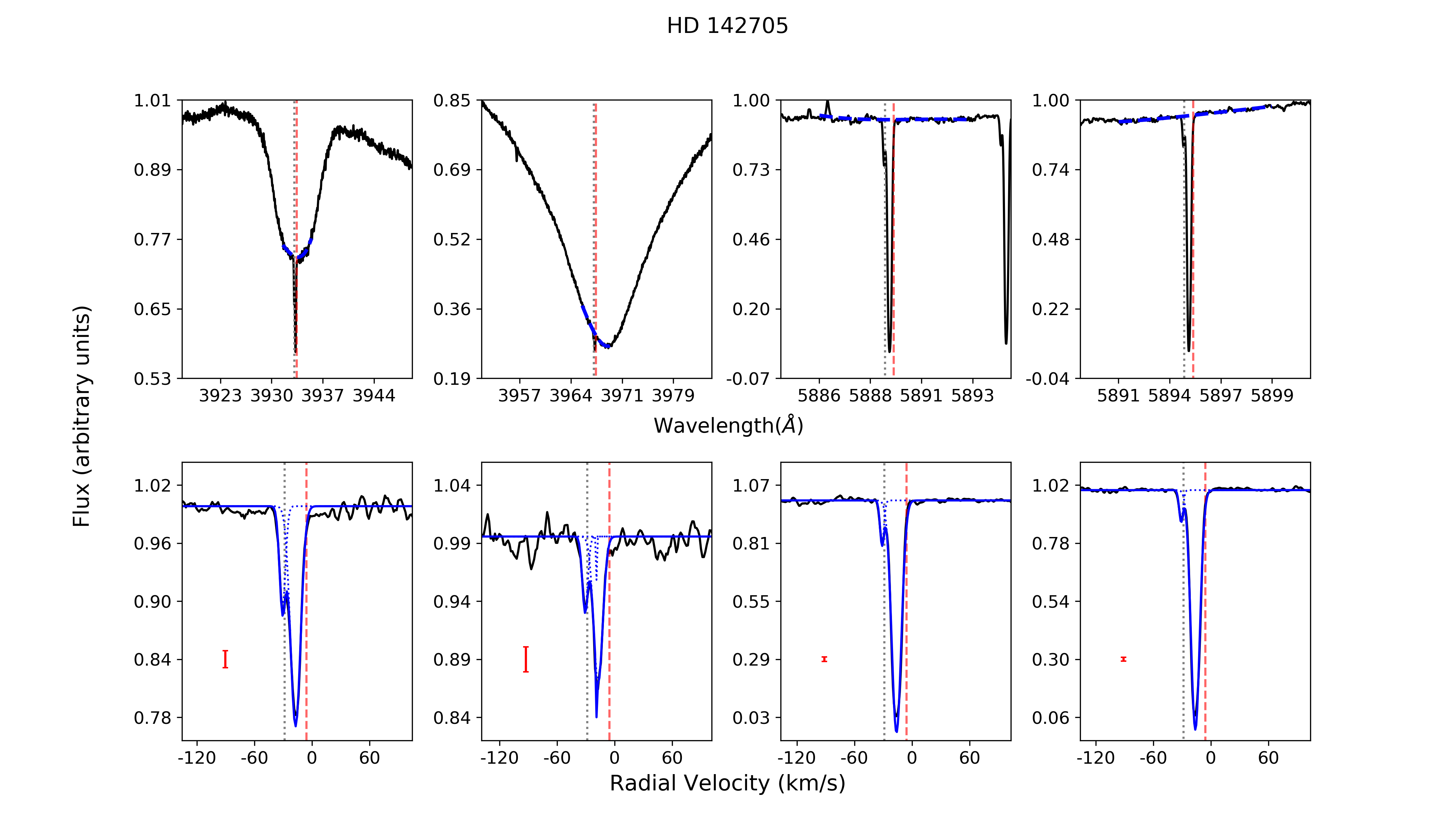}
\includegraphics[width=1.\textwidth]{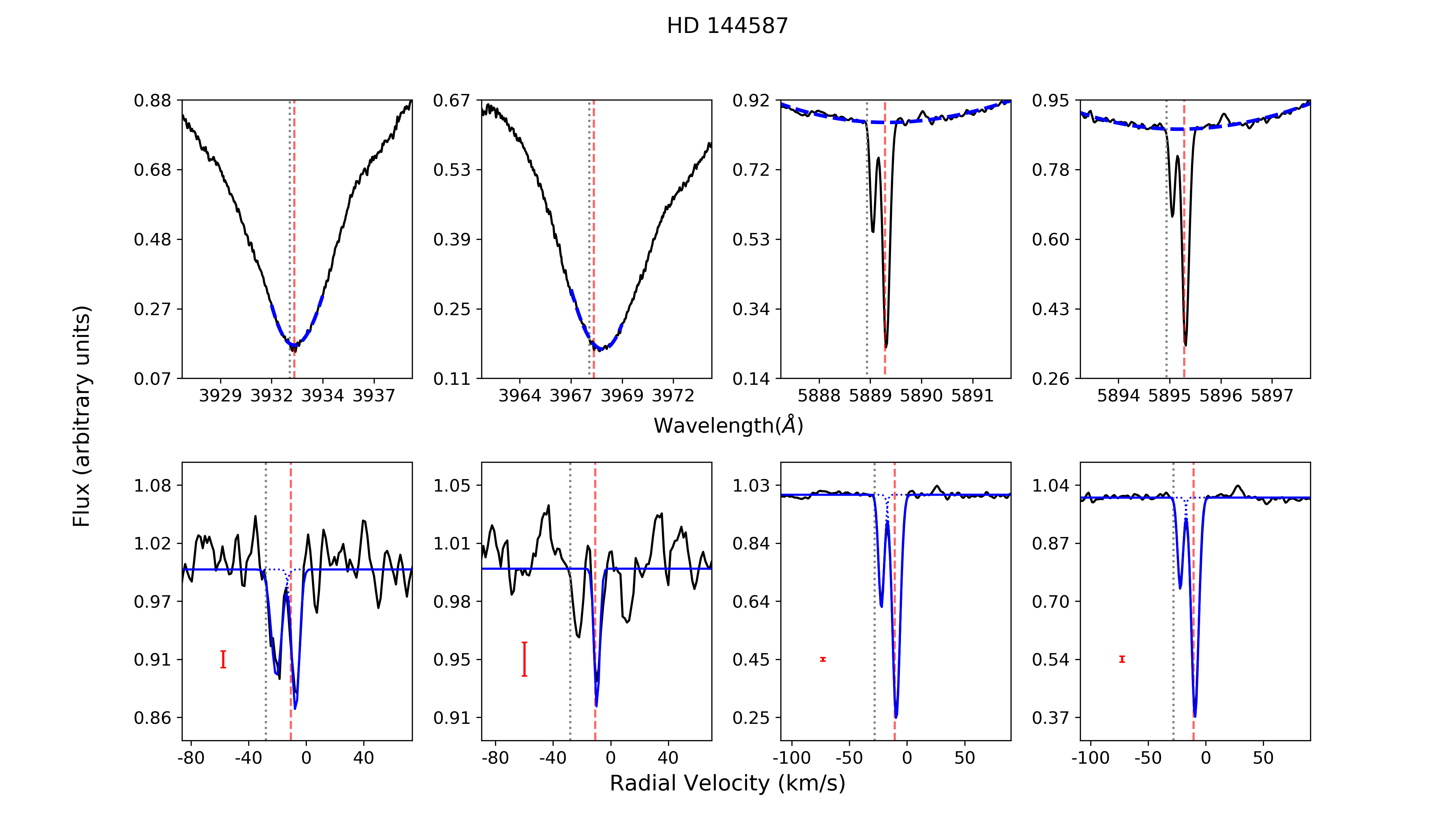}   
\caption{Stars showing narrow non-photospheric absorptions. Top panels: Photospheric Ca {\sc ii} H \& K and Na {\sc i} D lines with fitted modeling dashed blue line, x-axis shows the wavelength. Bottom panels: Residuals once the spectrum is divided by the photosphere, x-axis in velocity. Blue lines mark the fits to the non-photospheric absorptions. Vertical red dashed and grey dotted lines represent the stellar radial velocity and the ISM velocities respectively.  Red error bars show three sigma value measured in the continuum adjacent to the photospheric line. }
\end{figure}

\renewcommand{\thefigure}{\arabic{figure} (Cont.)}
\addtocounter{figure}{-1}

\begin{figure}
\centering  
\includegraphics[width=1.\textwidth]{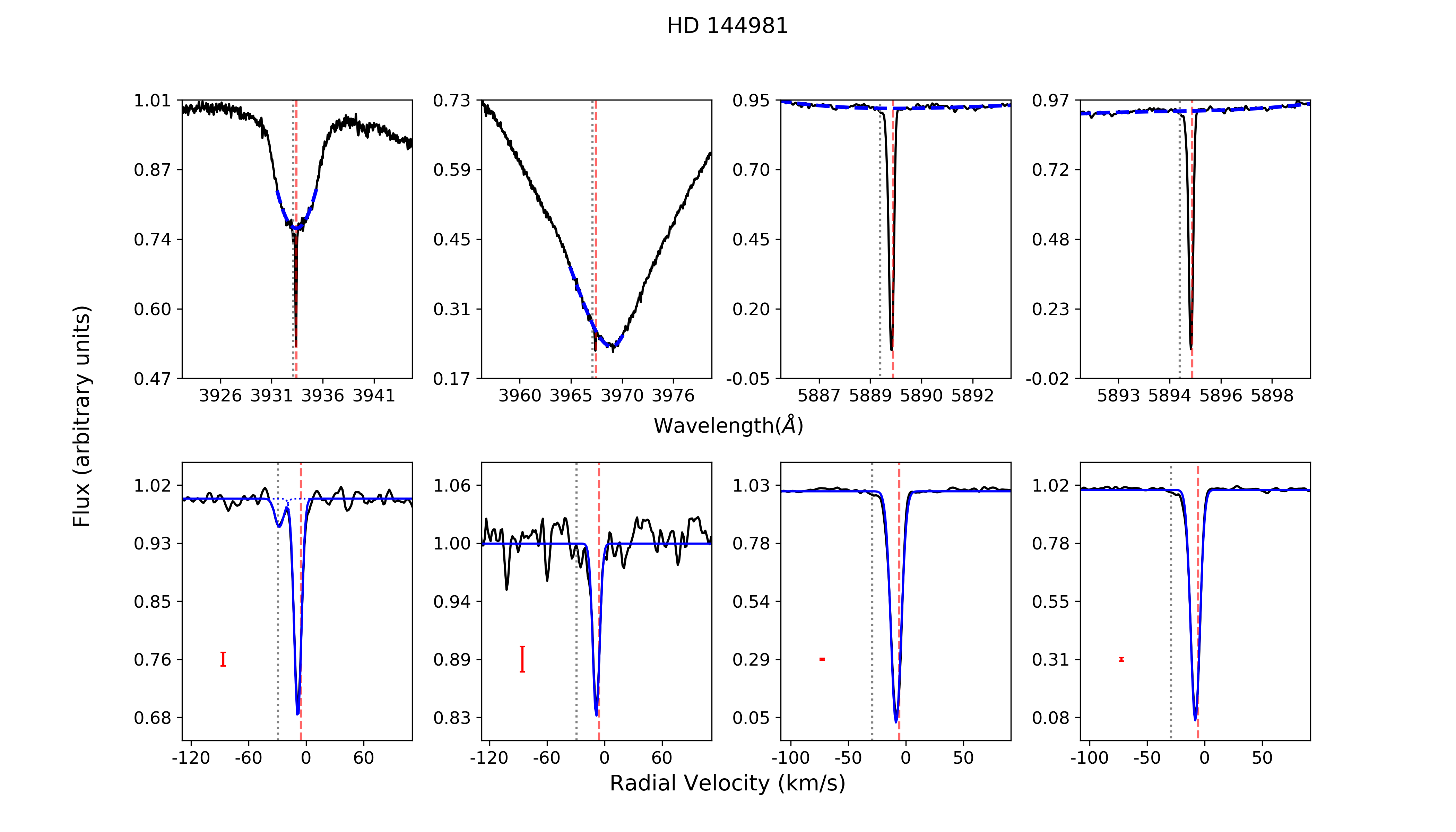}  
\includegraphics[width=1.\textwidth]{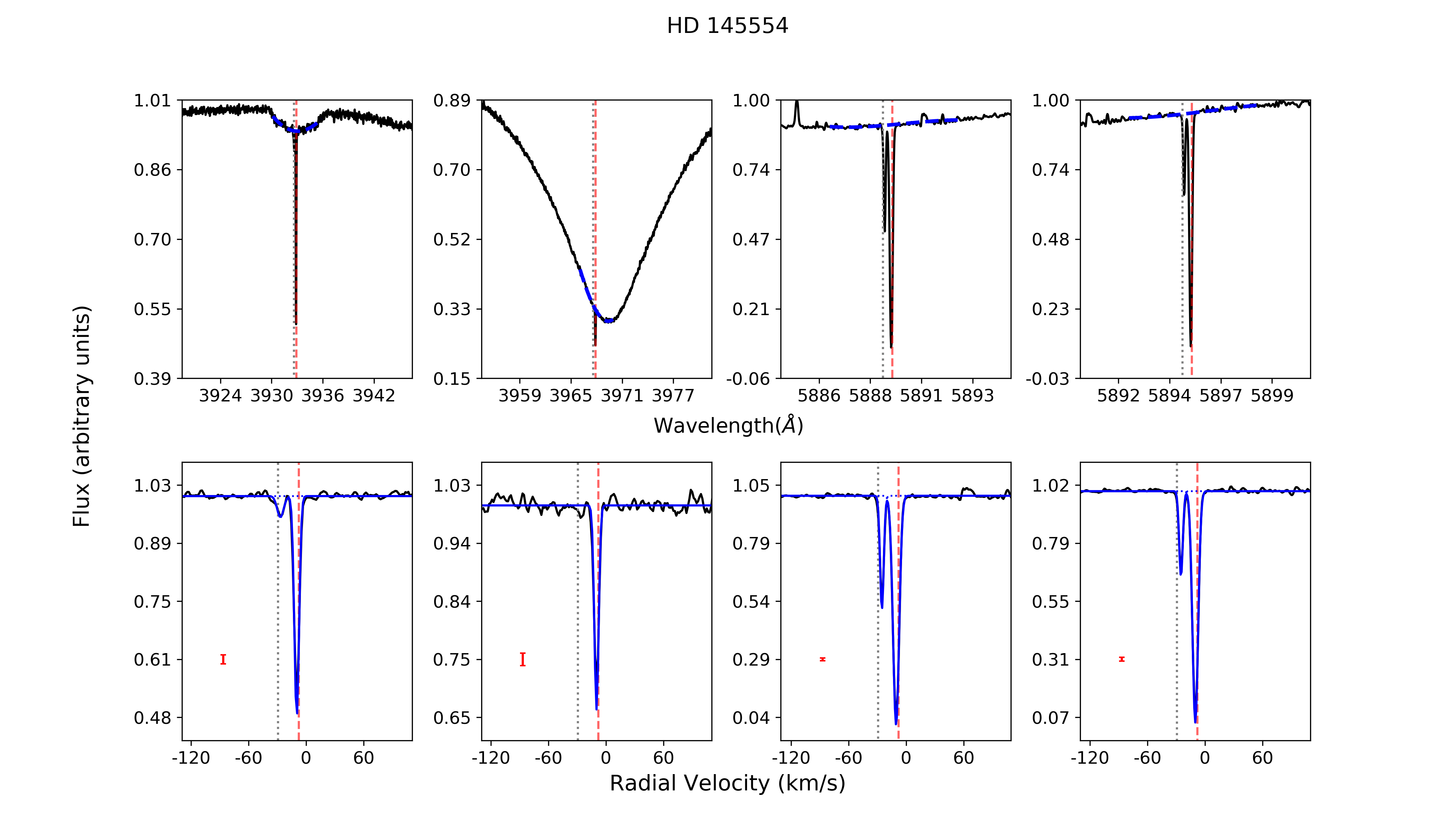} 
\caption{Stars showing narrow non-photospheric absorptions. Top panels: Photospheric Ca {\sc ii} H \& K and Na {\sc i} D lines with fitted modeling dashed blue line, x-axis shows the wavelength. Bottom panels: Residuals once the spectrum is divided by the photosphere, x-axis in velocity. Blue lines mark the fits to the non-photospheric absorptions. Vertical red dashed and grey dotted lines represent the stellar radial velocity and the ISM velocities respectively.  Red error bars show three sigma value measured in the continuum adjacent to the photospheric line. }
\end{figure}

\renewcommand{\thefigure}{\arabic{figure} (Cont.)}
\addtocounter{figure}{-1}

\begin{figure}
\centering
\includegraphics[width=1.\textwidth]{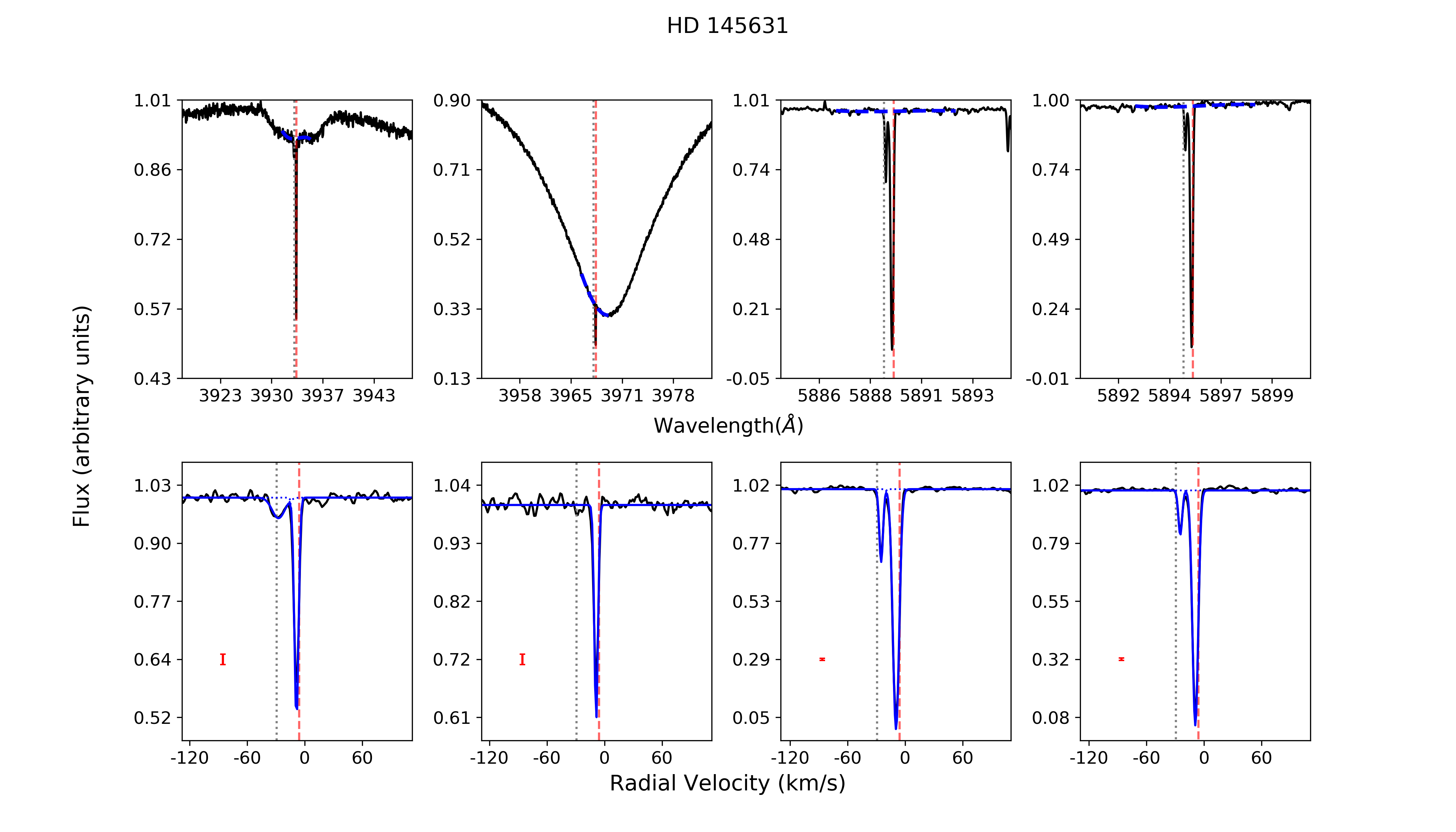}  
\includegraphics[width=1.\textwidth]{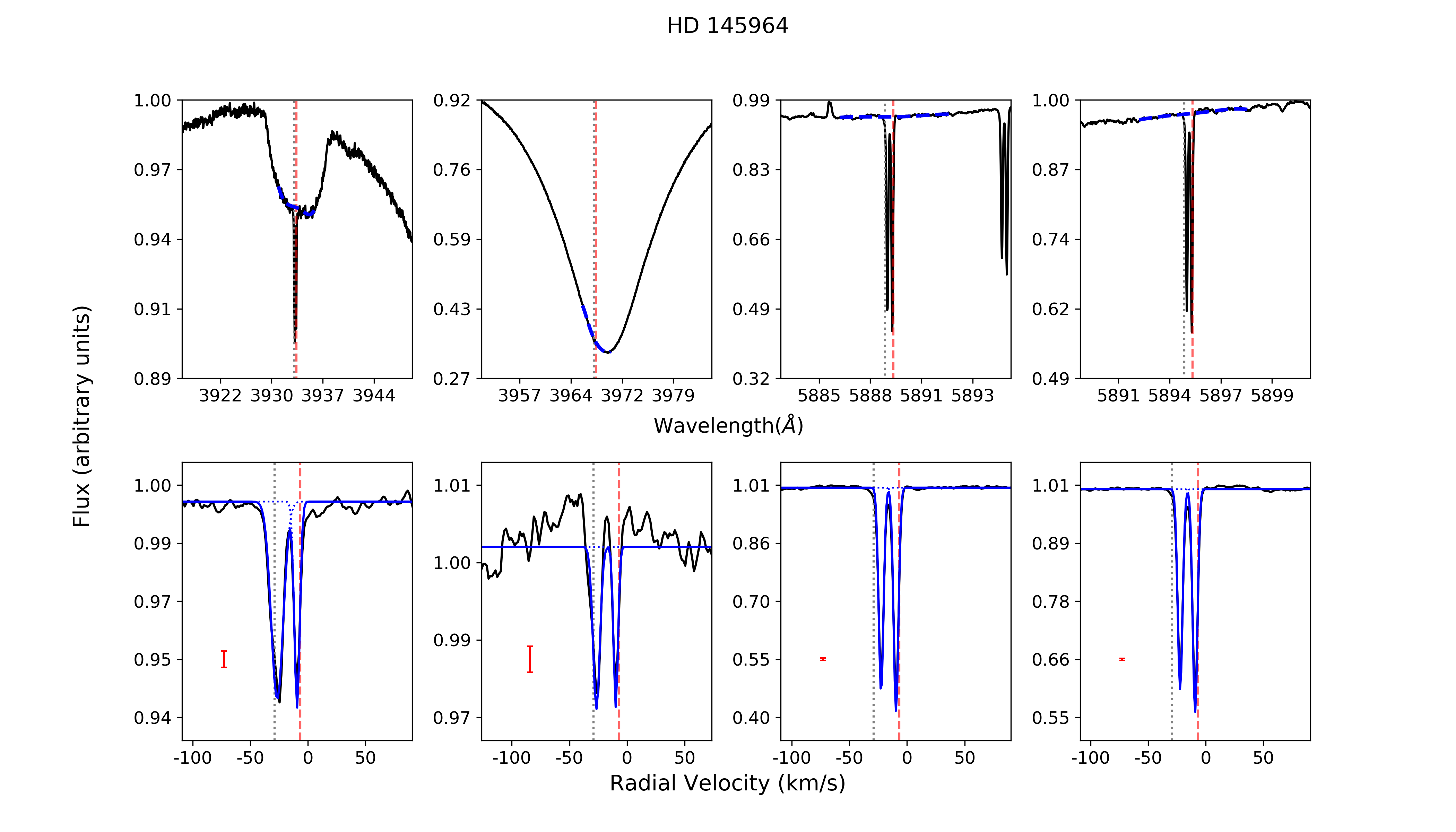}  
\caption{Stars showing narrow non-photospheric absorptions. Top panels: Photospheric Ca {\sc ii} H \& K and Na {\sc i} D lines with fitted modeling dashed blue line, x-axis shows the wavelength. Bottom panels: Residuals once the spectrum is divided by the photosphere, x-axis in velocity. Blue lines mark the fits to the non-photospheric absorptions. Vertical red dashed and grey dotted lines represent the stellar radial velocity and the ISM velocities respectively.  Red error bars show three sigma value measured in the continuum adjacent to the photospheric line. }
\end{figure}

\renewcommand{\thefigure}{\arabic{figure} (Cont.)}
\addtocounter{figure}{-1}

\begin{figure}
\centering
\includegraphics[width=1.\textwidth]{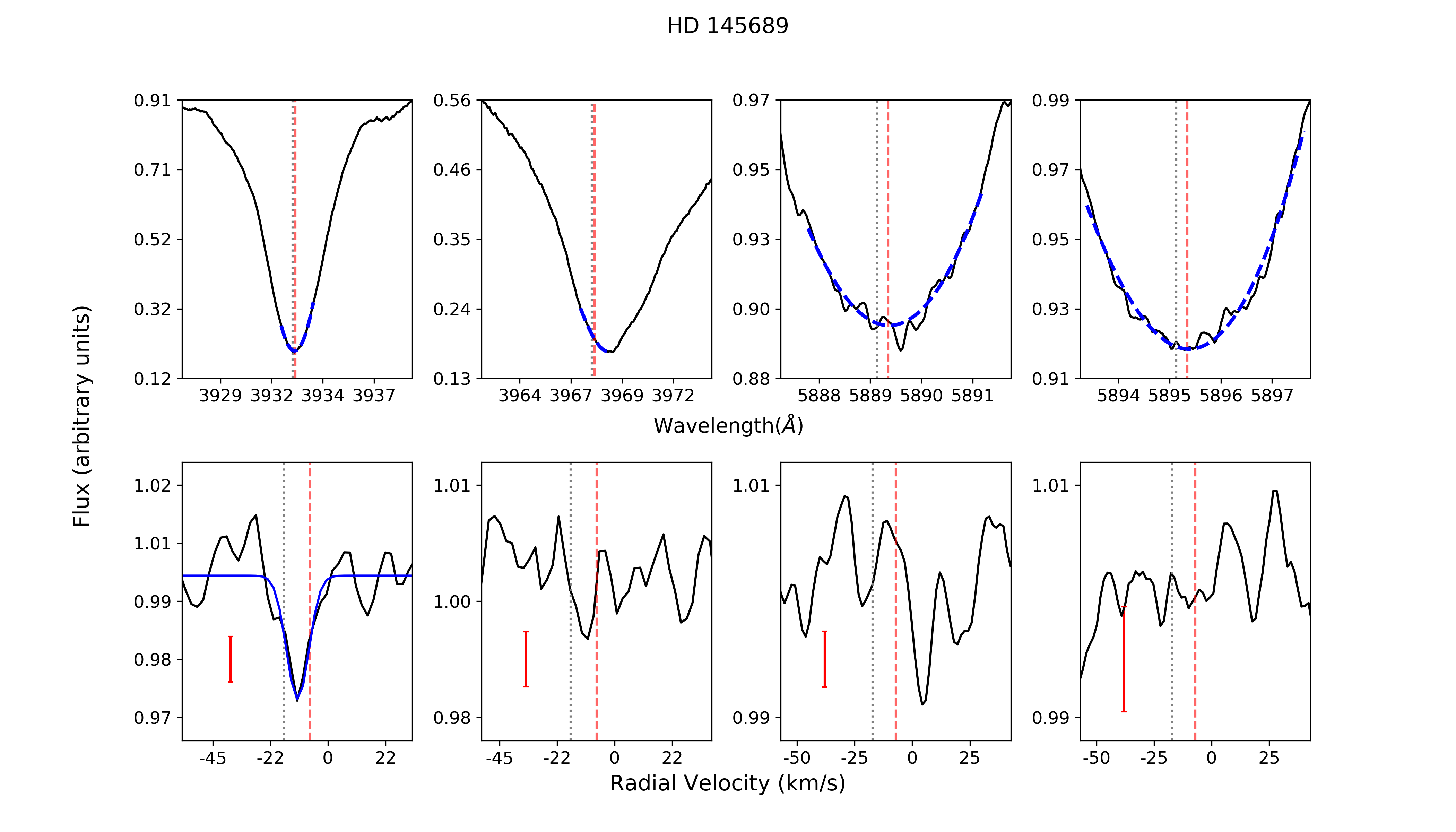}
\includegraphics[width=1.\textwidth]{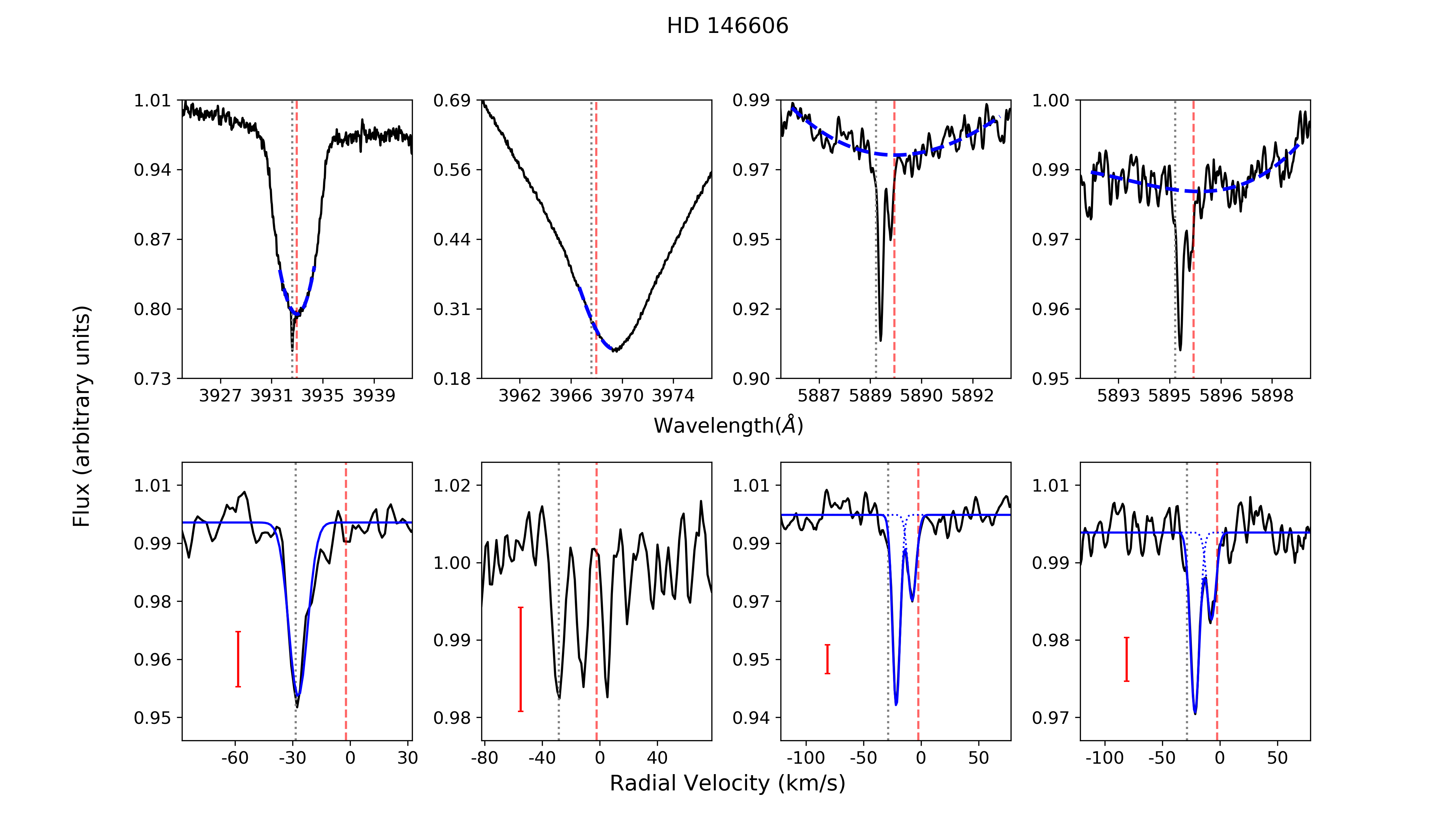}  
\caption{Stars showing narrow non-photospheric absorptions. Top panels: Photospheric Ca {\sc ii} H \& K and Na {\sc i} D lines with fitted modeling dashed blue line, x-axis shows the wavelength. Bottom panels: Residuals once the spectrum is divided by the photosphere, x-axis in velocity. Blue lines mark the fits to the non-photospheric absorptions. Vertical red dashed and grey dotted lines represent the stellar radial velocity and the ISM velocities respectively.  Red error bars show three sigma value measured in the continuum adjacent to the photospheric line. }
\end{figure}

\renewcommand{\thefigure}{\arabic{figure} (Cont.)}
\addtocounter{figure}{-1}

\begin{figure}
\centering  
\includegraphics[width=1.\textwidth]{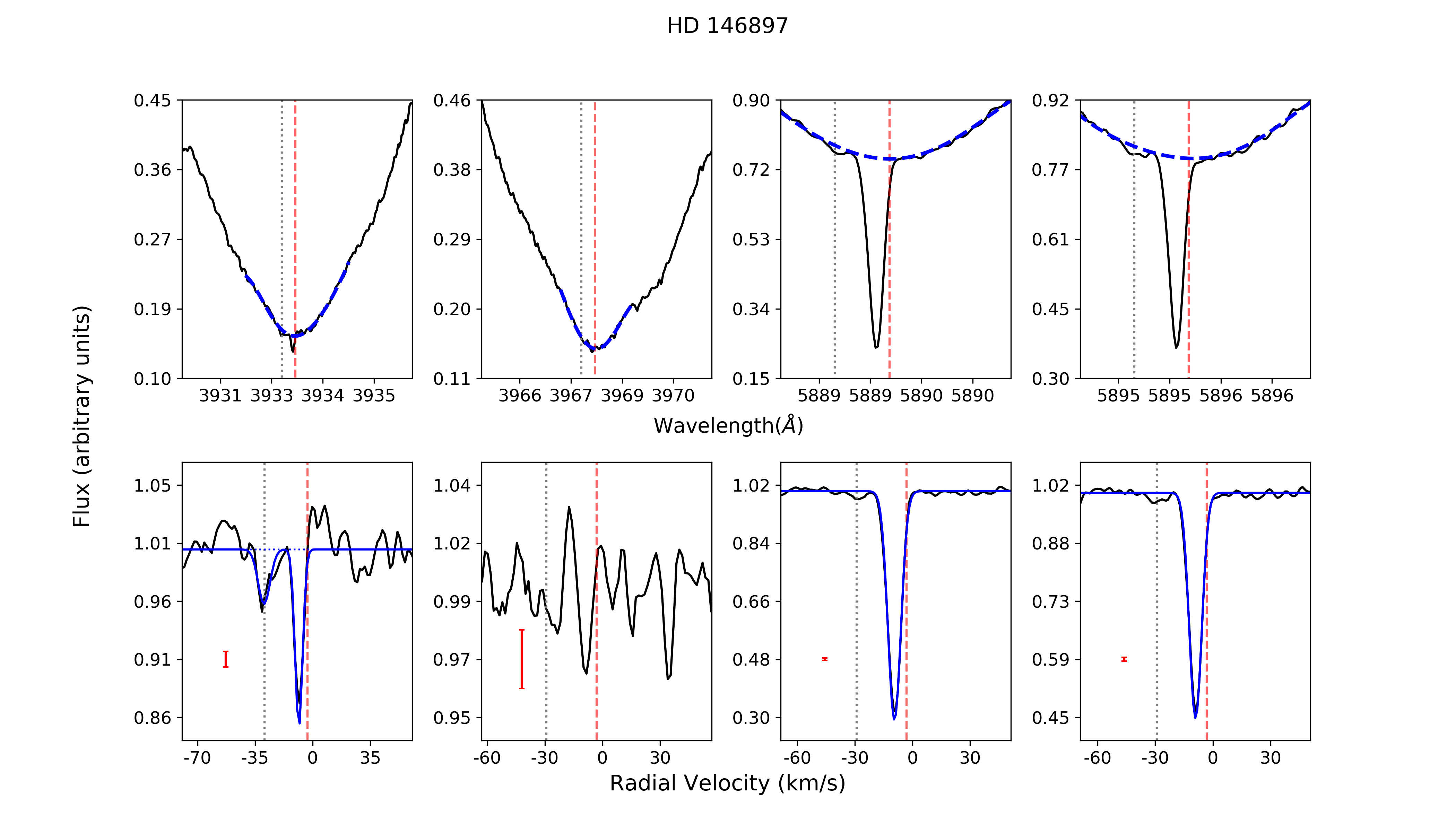}  
\includegraphics[width=1.\textwidth]{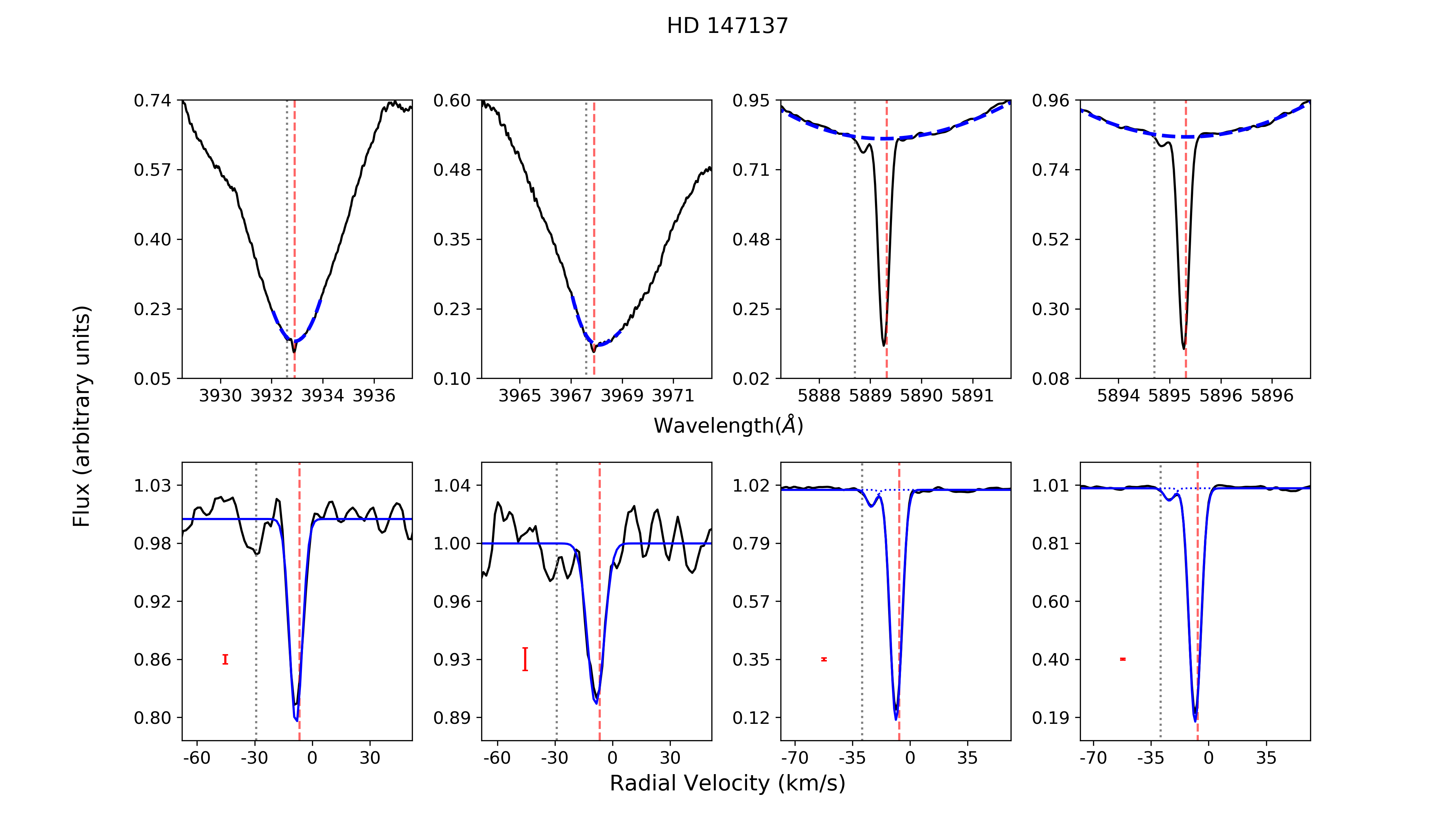}
\caption{Stars showing narrow non-photospheric absorptions. Top panels: Photospheric Ca {\sc ii} H \& K and Na {\sc i} D lines with fitted modeling dashed blue line, x-axis shows the wavelength. Bottom panels: Residuals once the spectrum is divided by the photosphere, x-axis in velocity. Blue lines mark the fits to the non-photospheric absorptions. Vertical red dashed and grey dotted lines represent the stellar radial velocity and the ISM velocities respectively.  Red error bars show three sigma value measured in the continuum adjacent to the photospheric line. }
\end{figure}

\renewcommand{\thefigure}{\arabic{figure} (Cont.)}
\addtocounter{figure}{-1}

\begin{figure}
\centering
\includegraphics[width=1.\textwidth]{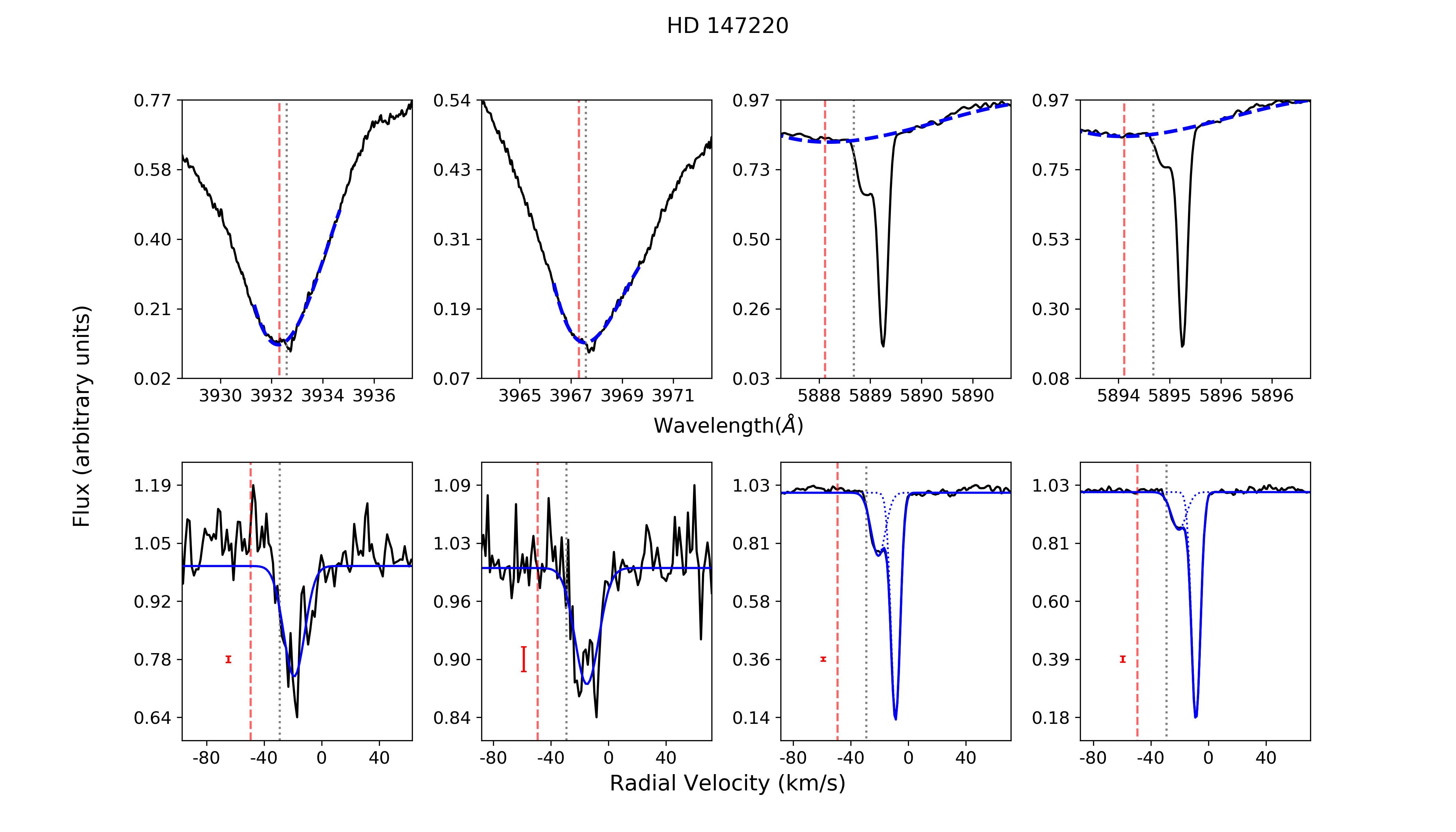}  
\includegraphics[width=1.\textwidth]{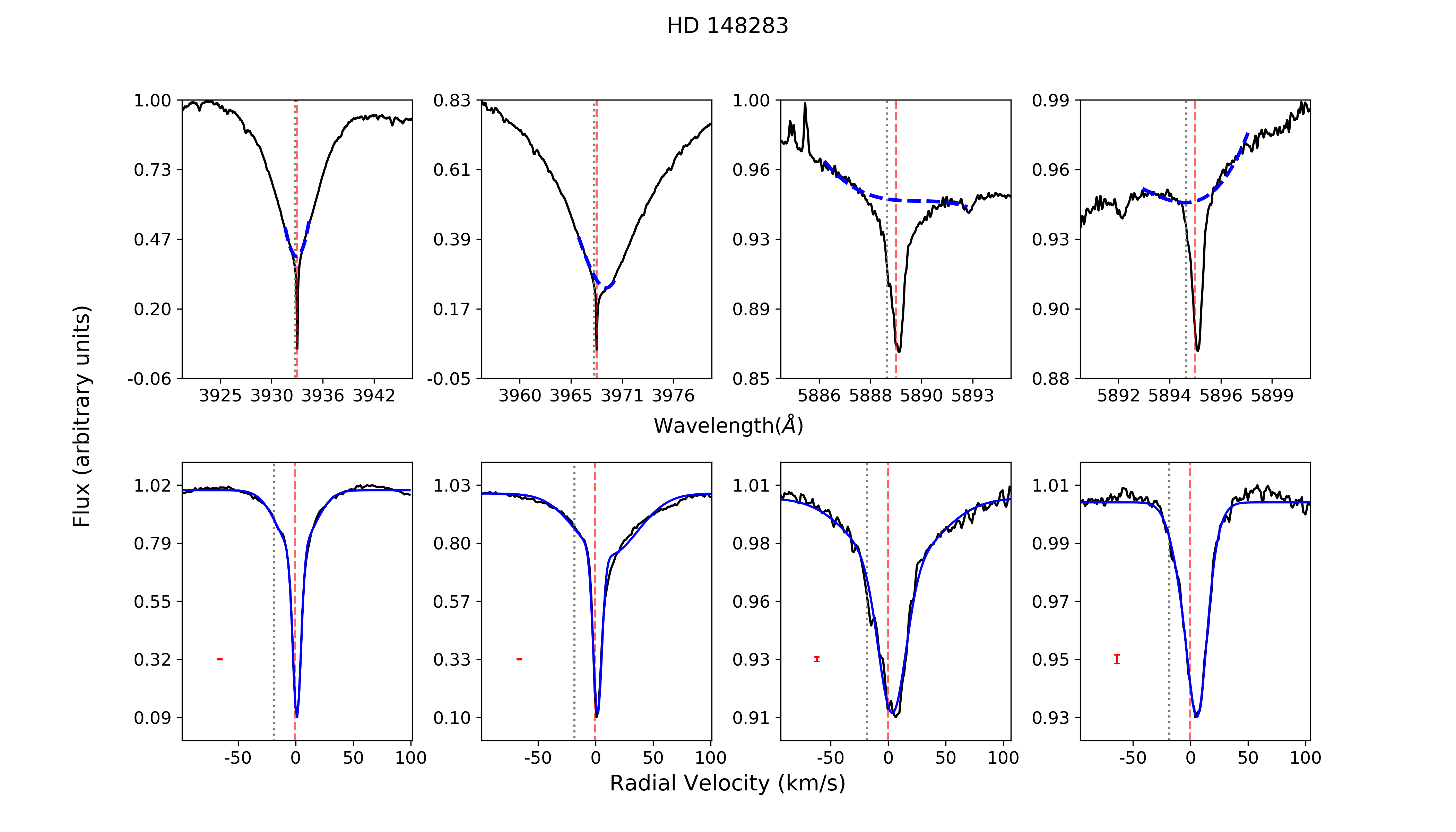}    
\caption{Stars showing narrow non-photospheric absorptions. Top panels: Photospheric Ca {\sc ii} H \& K and Na {\sc i} D lines with fitted modeling dashed blue line, x-axis shows the wavelength. Bottom panels: Residuals once the spectrum is divided by the photosphere, x-axis in velocity. Blue lines mark the fits to the non-photospheric absorptions. Vertical red dashed and grey dotted lines represent the stellar radial velocity and the ISM velocities respectively.  Red error bars show three sigma value measured in the continuum adjacent to the photospheric line. }
\end{figure}

\renewcommand{\thefigure}{\arabic{figure} (Cont.)}
\addtocounter{figure}{-1}

\begin{figure}
\centering
\includegraphics[width=1.\textwidth]{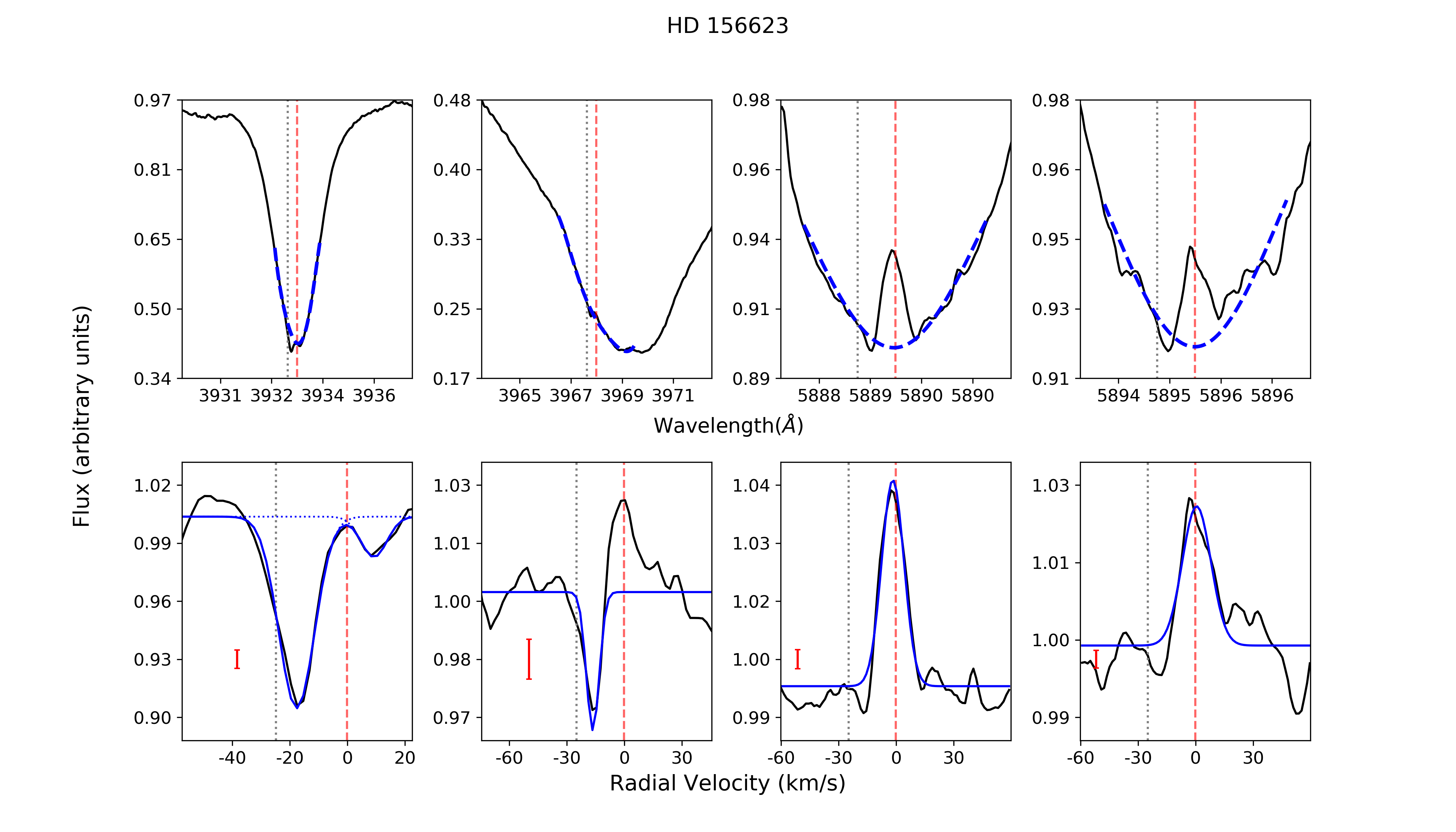}
\includegraphics[width=1.\textwidth]{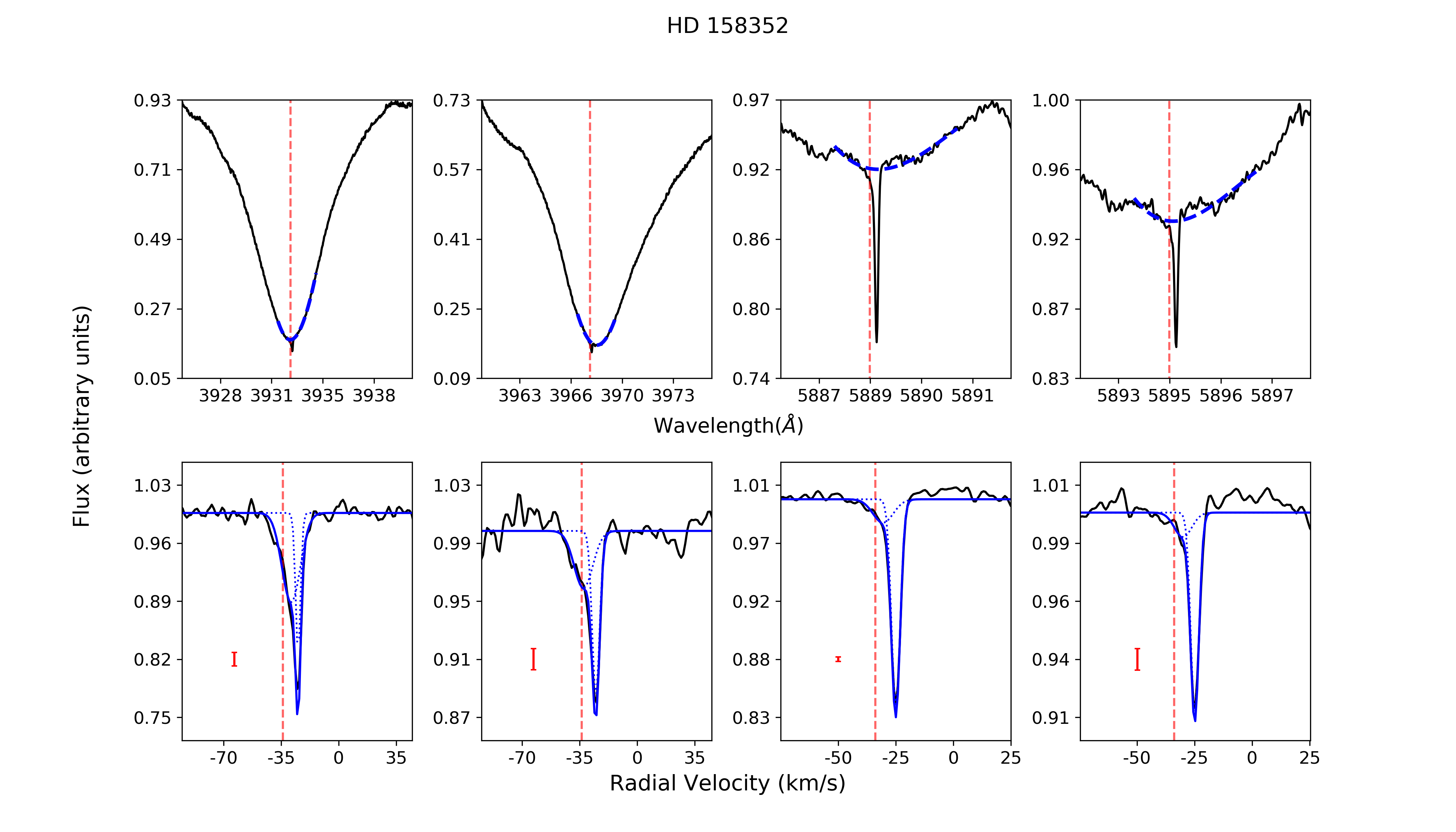}  
\caption{Stars showing narrow non-photospheric absorptions. Top panels: Photospheric Ca {\sc ii} H \& K and Na {\sc i} D lines with fitted modeling dashed blue line, x-axis shows the wavelength. Bottom panels: Residuals once the spectrum is divided by the photosphere, x-axis in velocity. Blue lines mark the fits to the non-photospheric absorptions. Vertical red dashed and grey dotted lines represent the stellar radial velocity and the ISM velocities respectively.  Red error bars show three sigma value measured in the continuum adjacent to the photospheric line. }
\end{figure}

\renewcommand{\thefigure}{\arabic{figure} (Cont.)}
\addtocounter{figure}{-1}

\begin{figure}
\centering  
\includegraphics[width=1.\textwidth]{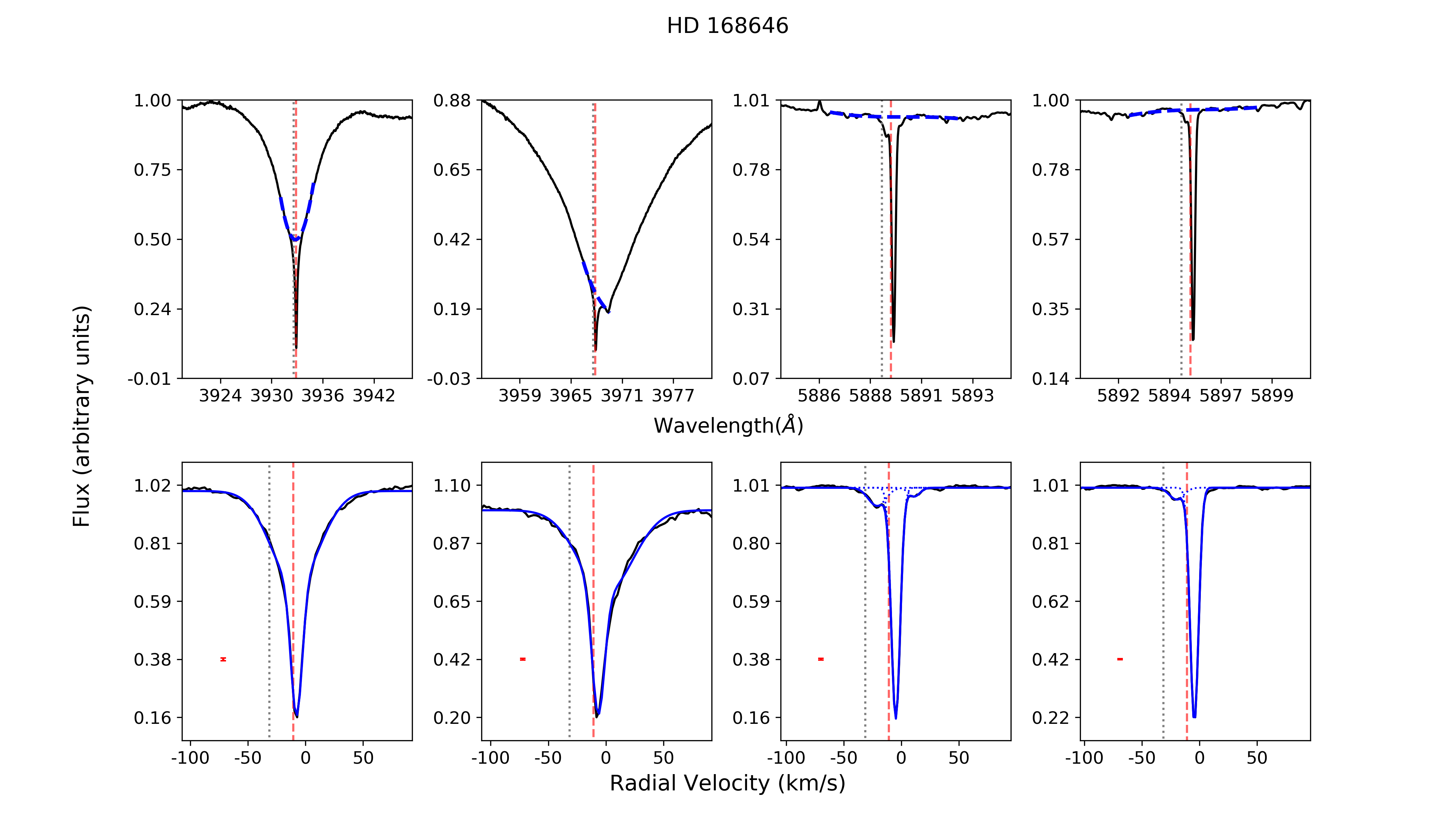}  
\includegraphics[width=1.\textwidth]{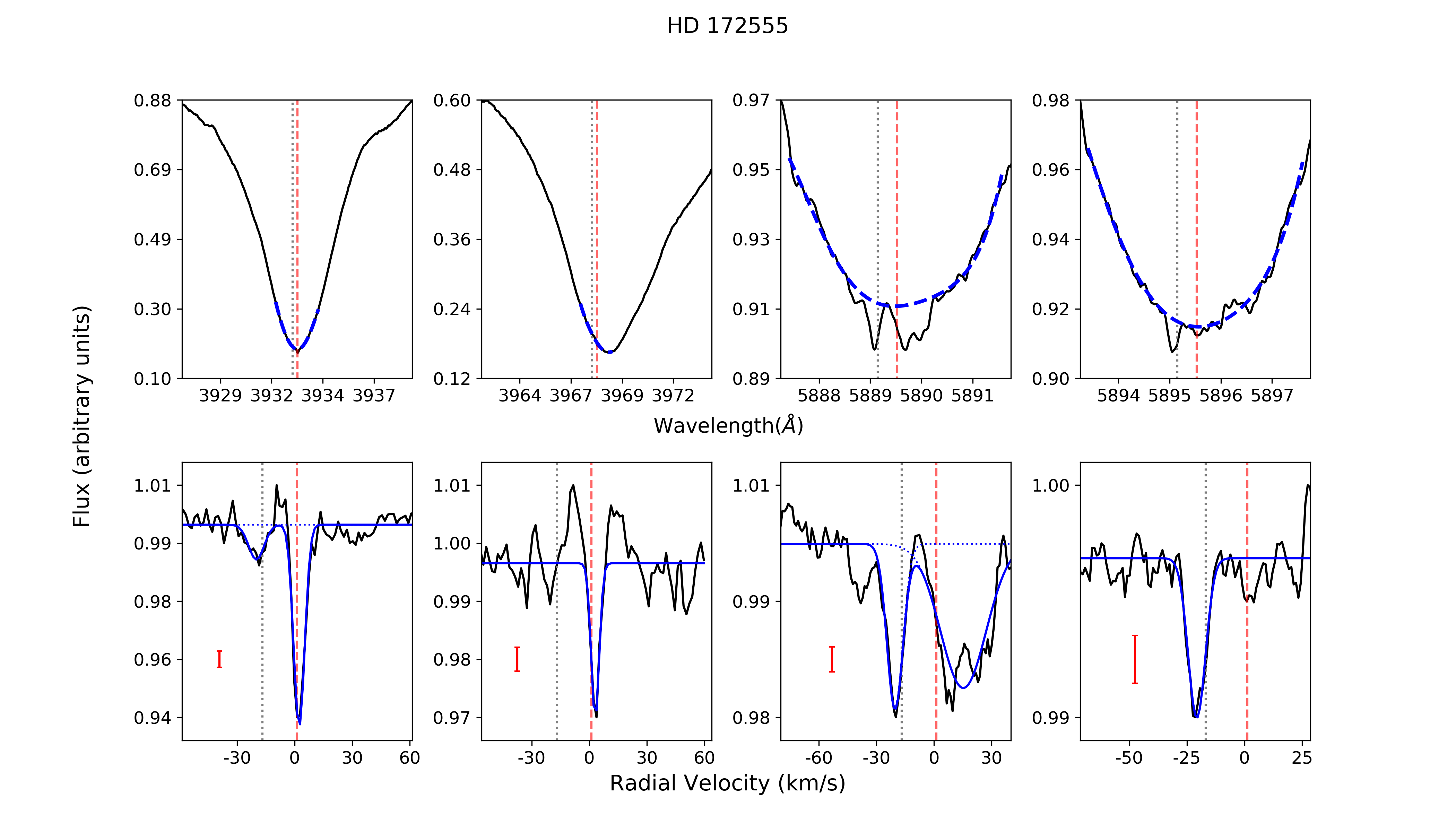} 
\caption{Stars showing narrow non-photospheric absorptions. Top panels: Photospheric Ca {\sc ii} H \& K and Na {\sc i} D lines with fitted modeling dashed blue line, x-axis shows the wavelength. Bottom panels: Residuals once the spectrum is divided by the photosphere, x-axis in velocity. Blue lines mark the fits to the non-photospheric absorptions. Vertical red dashed and grey dotted lines represent the stellar radial velocity and the ISM velocities respectively.  Red error bars show three sigma value measured in the continuum adjacent to the photospheric line. }
\end{figure}

\renewcommand{\thefigure}{\arabic{figure} (Cont.)}
\addtocounter{figure}{-1}

\begin{figure}
\centering
\includegraphics[width=1.\textwidth]{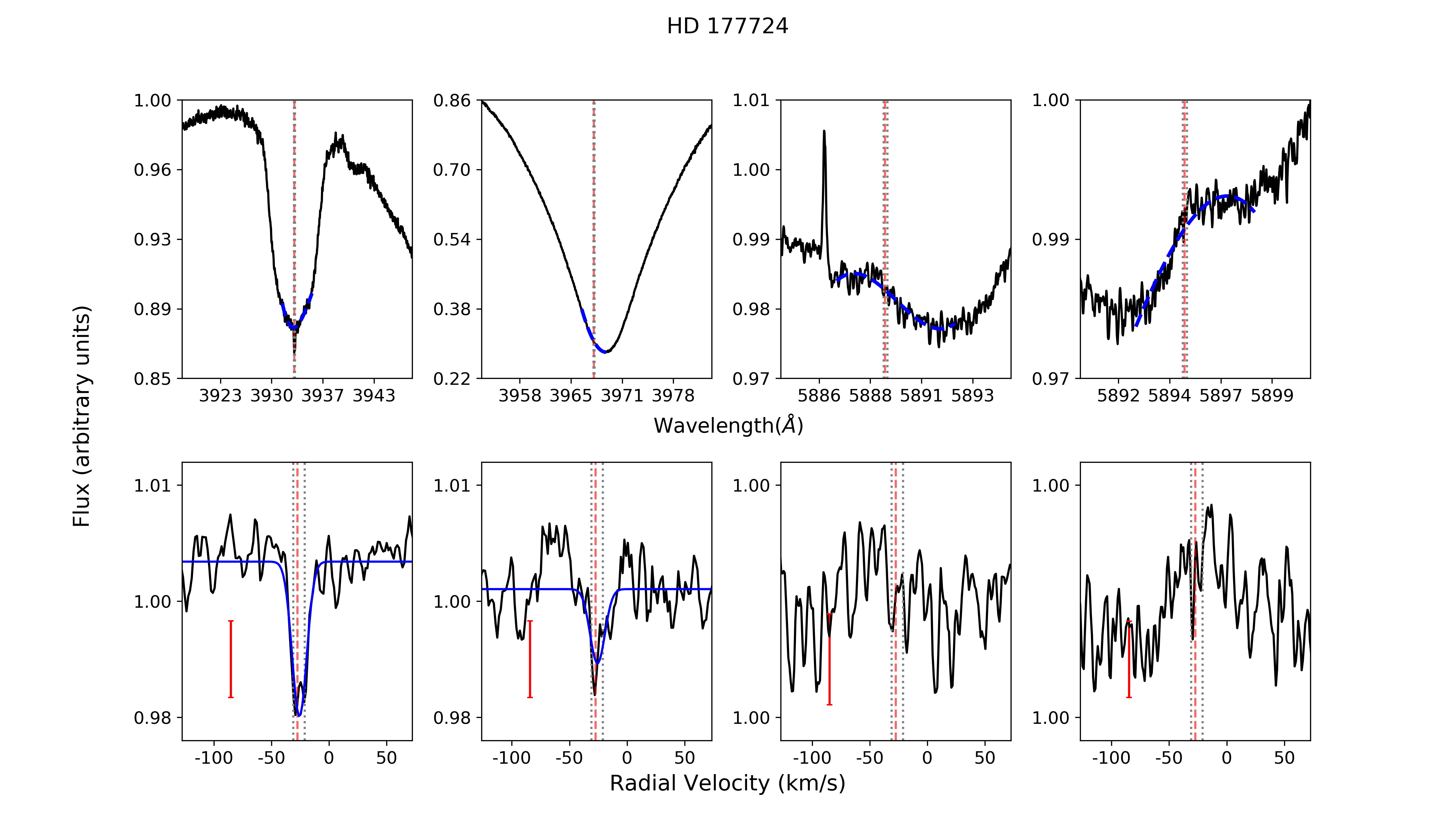}  
\includegraphics[width=1.\textwidth]{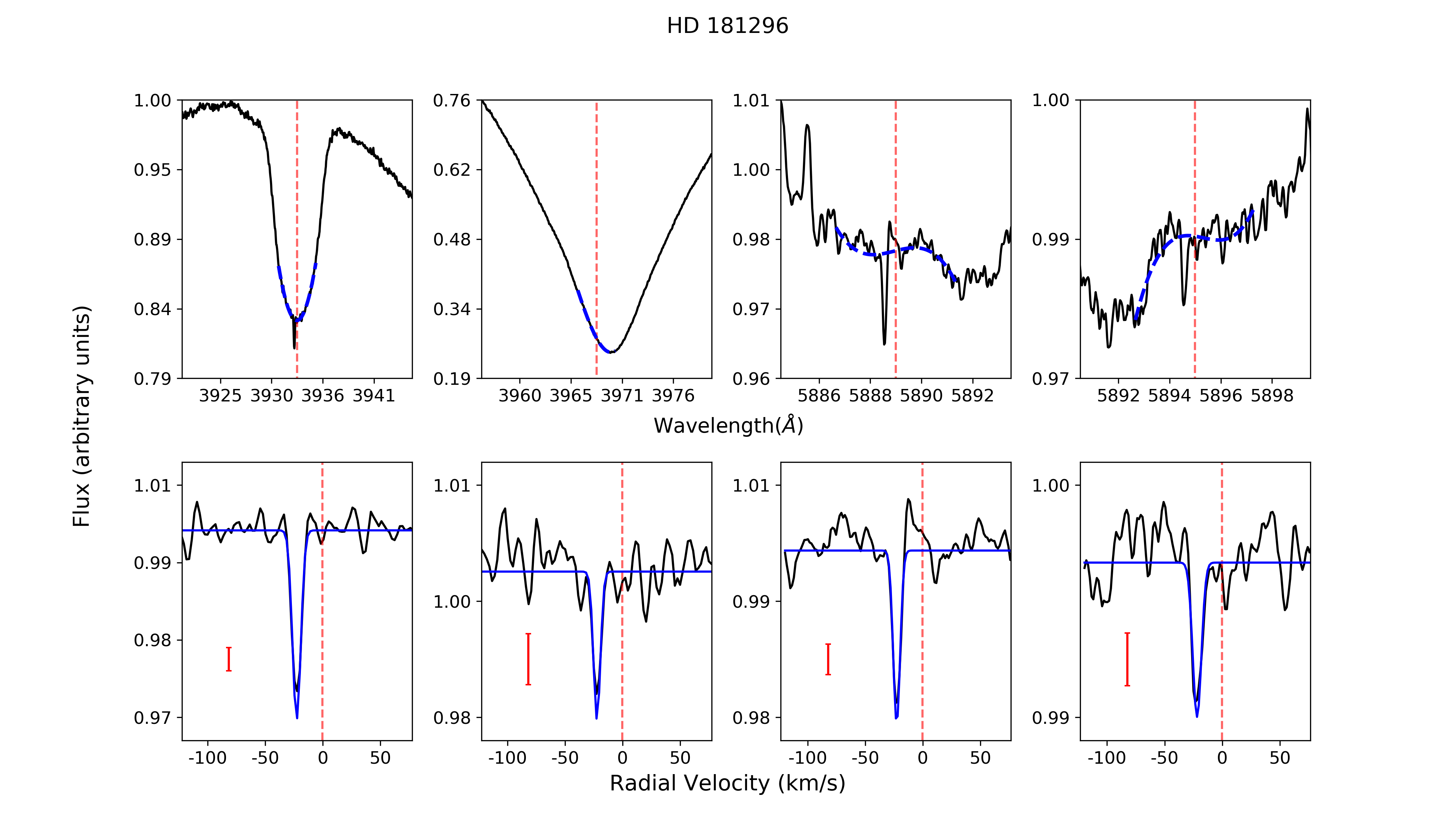}  
\caption{Stars showing narrow non-photospheric absorptions. Top panels: Photospheric Ca {\sc ii} H \& K and Na {\sc i} D lines with fitted modeling dashed blue line, x-axis shows the wavelength. Bottom panels: Residuals once the spectrum is divided by the photosphere, x-axis in velocity. Blue lines mark the fits to the non-photospheric absorptions. Vertical red dashed and grey dotted lines represent the stellar radial velocity and the ISM velocities respectively.  Red error bars show three sigma value measured in the continuum adjacent to the photospheric line. }
\end{figure}

\renewcommand{\thefigure}{\arabic{figure} (Cont.)}
\addtocounter{figure}{-1}

\begin{figure}
\centering
\includegraphics[width=1.\textwidth]{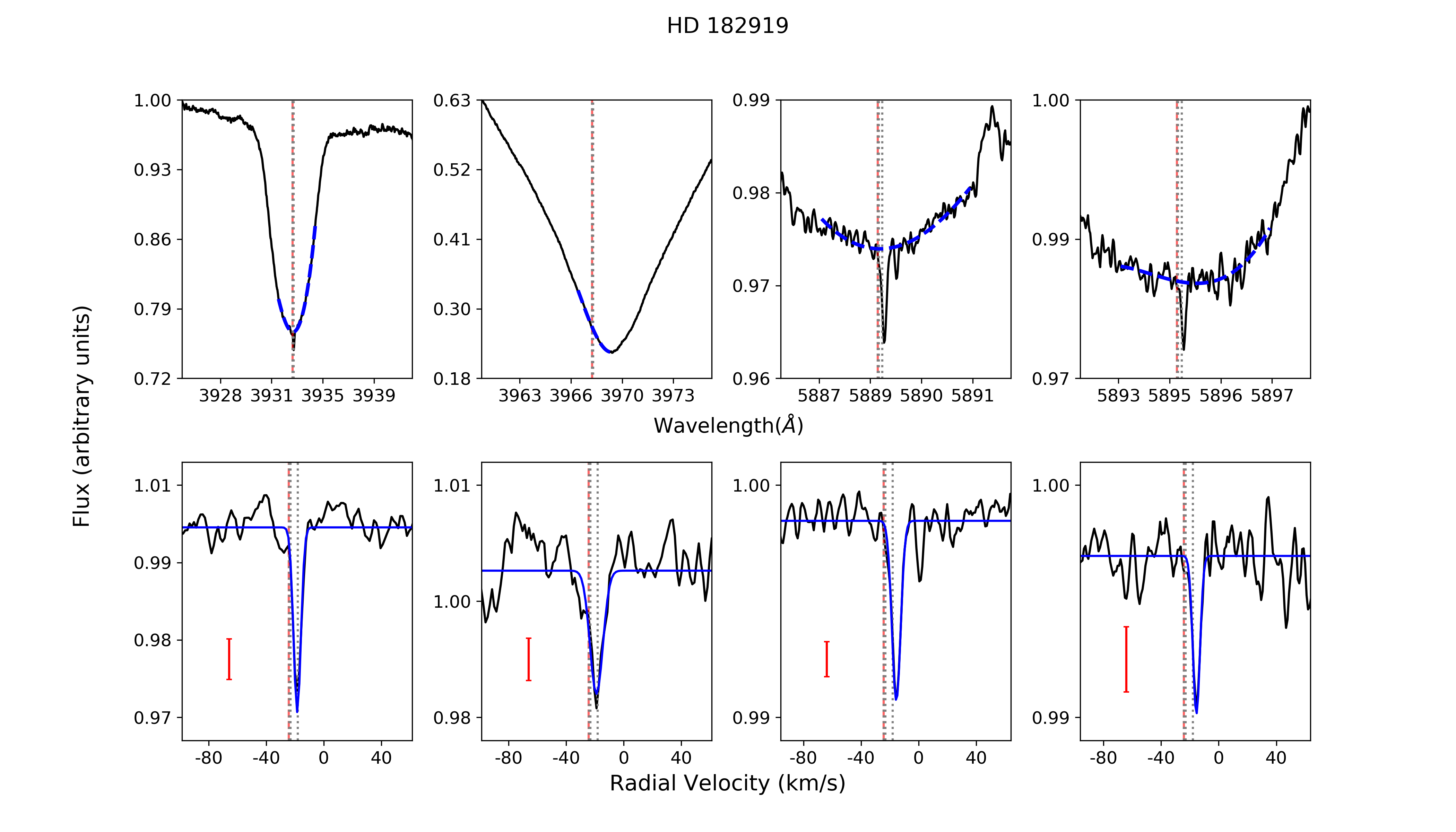} 
\includegraphics[width=1.\textwidth]{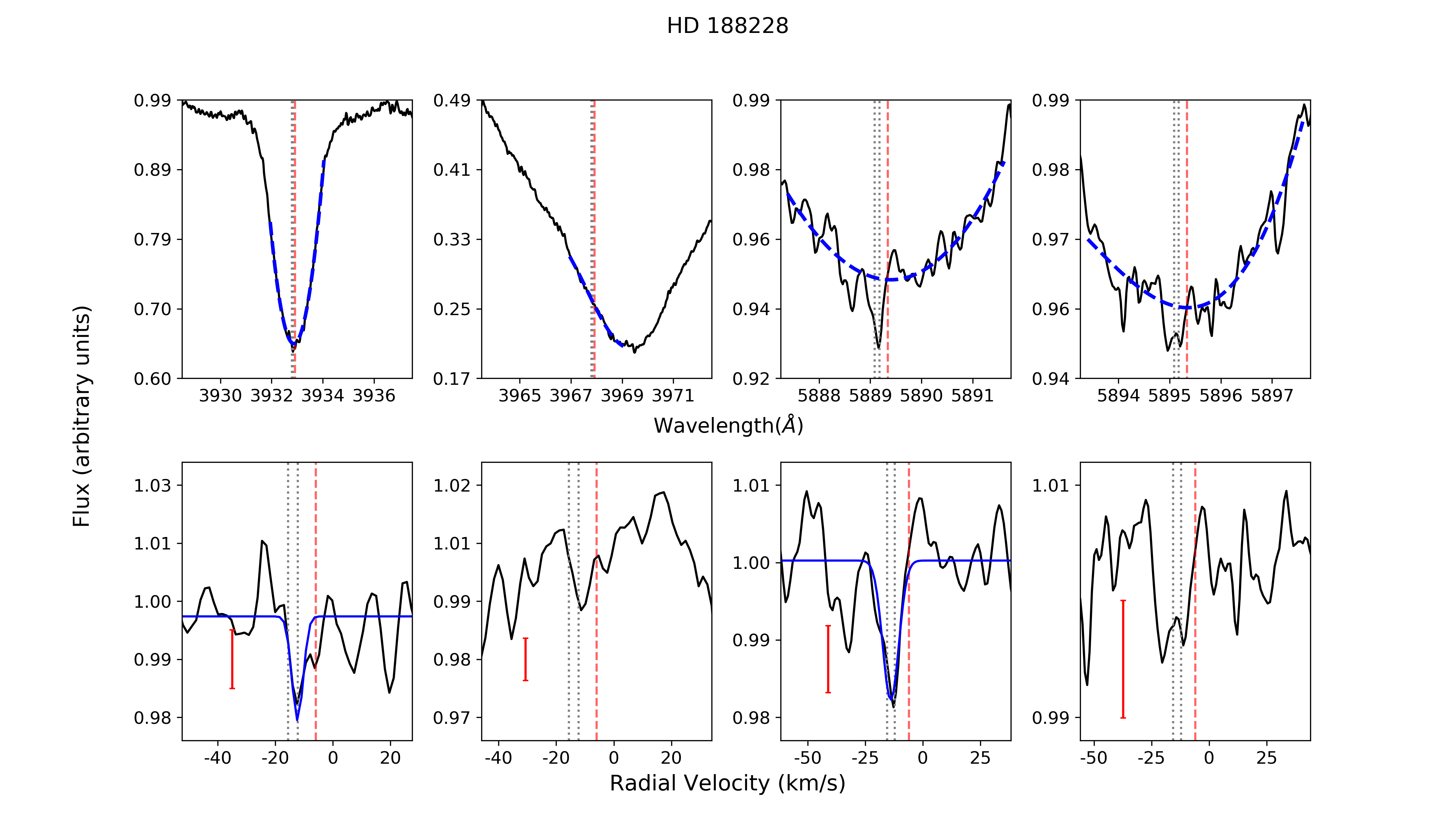}     
\caption{Stars showing narrow non-photospheric absorptions. Top panels: Photospheric Ca {\sc ii} H \& K and Na {\sc i} D lines with fitted modeling dashed blue line, x-axis shows the wavelength. Bottom panels: Residuals once the spectrum is divided by the photosphere, x-axis in velocity. Blue lines mark the fits to the non-photospheric absorptions. Vertical red dashed and grey dotted lines represent the stellar radial velocity and the ISM velocities respectively.  Red error bars show three sigma value measured in the continuum adjacent to the photospheric line. }
\end{figure}

\renewcommand{\thefigure}{\arabic{figure} (Cont.)}
\addtocounter{figure}{-1}

\begin{figure}
\centering   
\includegraphics[width=1.\textwidth]{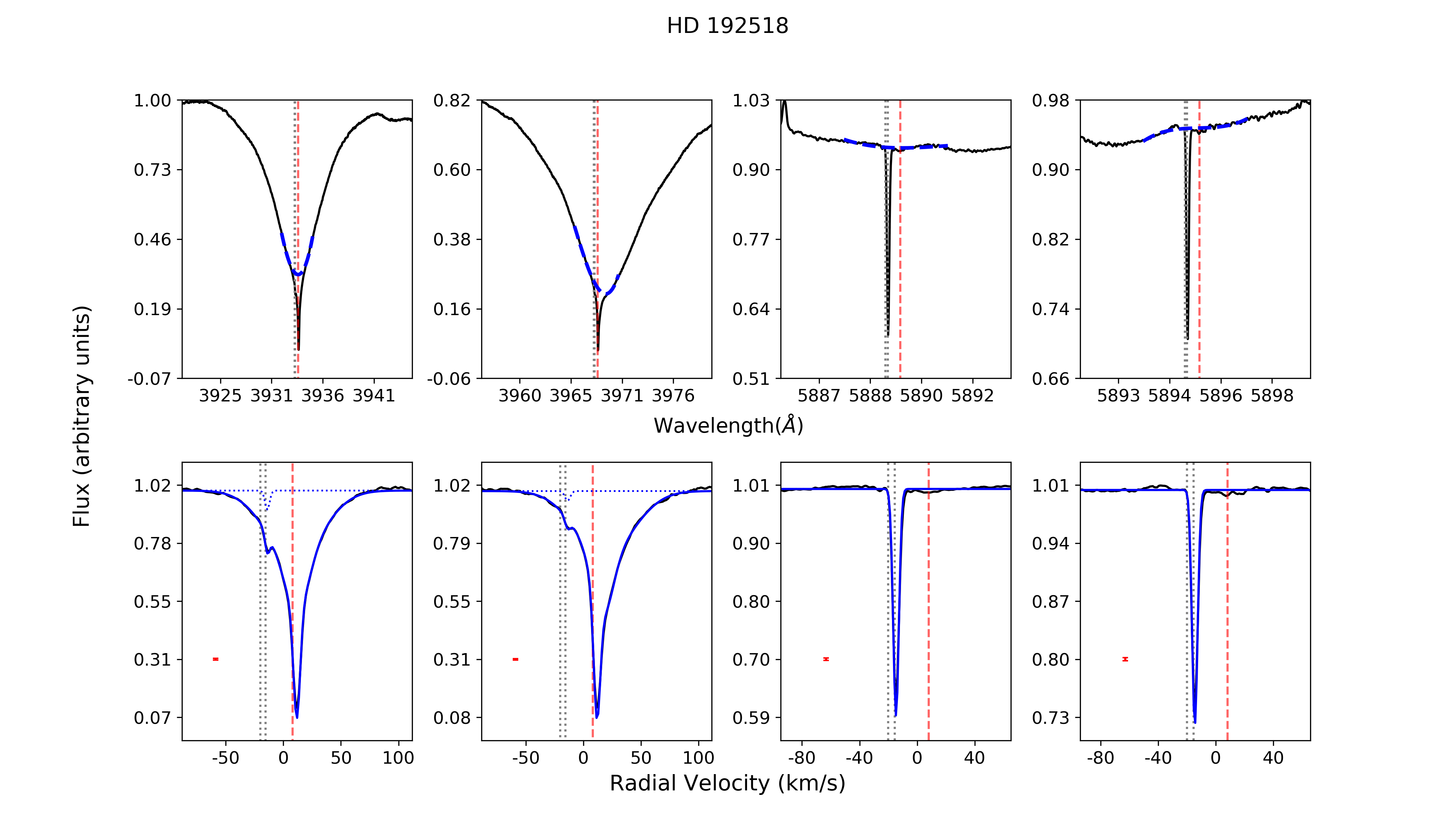}     
\includegraphics[width=1.\textwidth]{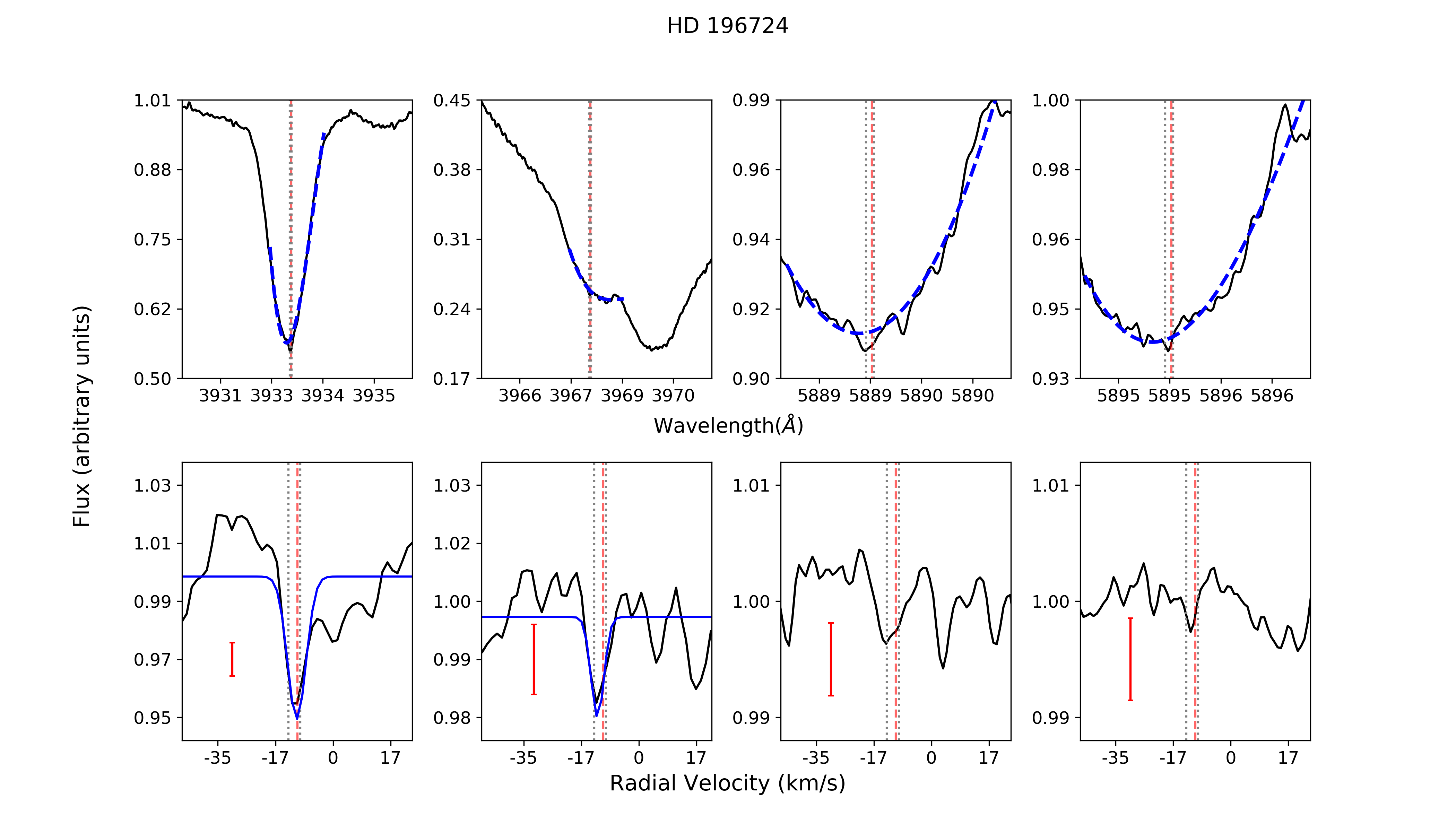}
\caption{Stars showing narrow non-photospheric absorptions. Top panels: Photospheric Ca {\sc ii} H \& K and Na {\sc i} D lines with fitted modeling dashed blue line, x-axis shows the wavelength. Bottom panels: Residuals once the spectrum is divided by the photosphere, x-axis in velocity. Blue lines mark the fits to the non-photospheric absorptions. Vertical red dashed and grey dotted lines represent the stellar radial velocity and the ISM velocities respectively.  Red error bars show three sigma value measured in the continuum adjacent to the photospheric line. }
\end{figure}

\renewcommand{\thefigure}{\arabic{figure} (Cont.)}
\addtocounter{figure}{-1}

\begin{figure}
\centering
\includegraphics[width=1.\textwidth]{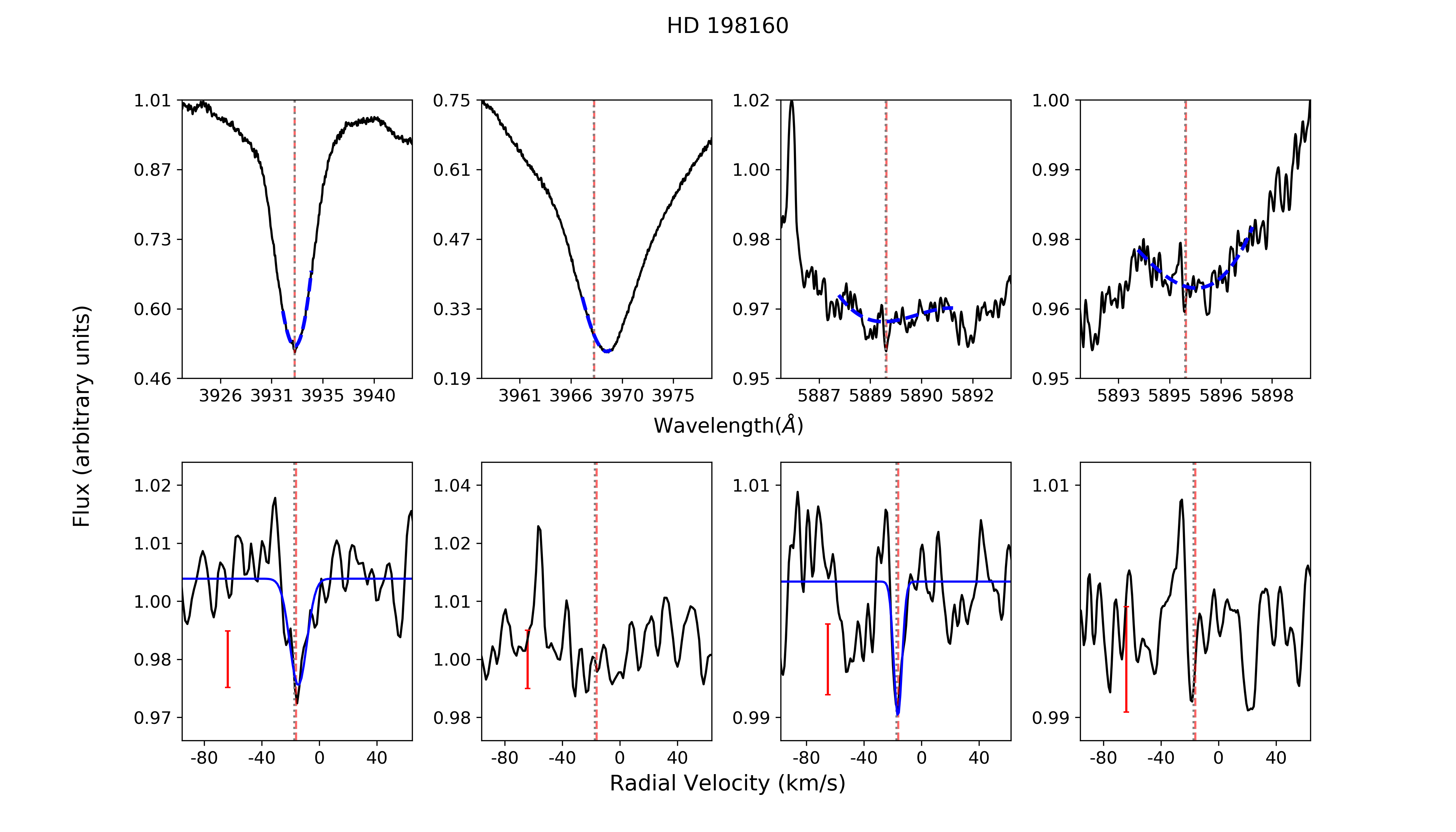} 
\includegraphics[width=1.\textwidth]{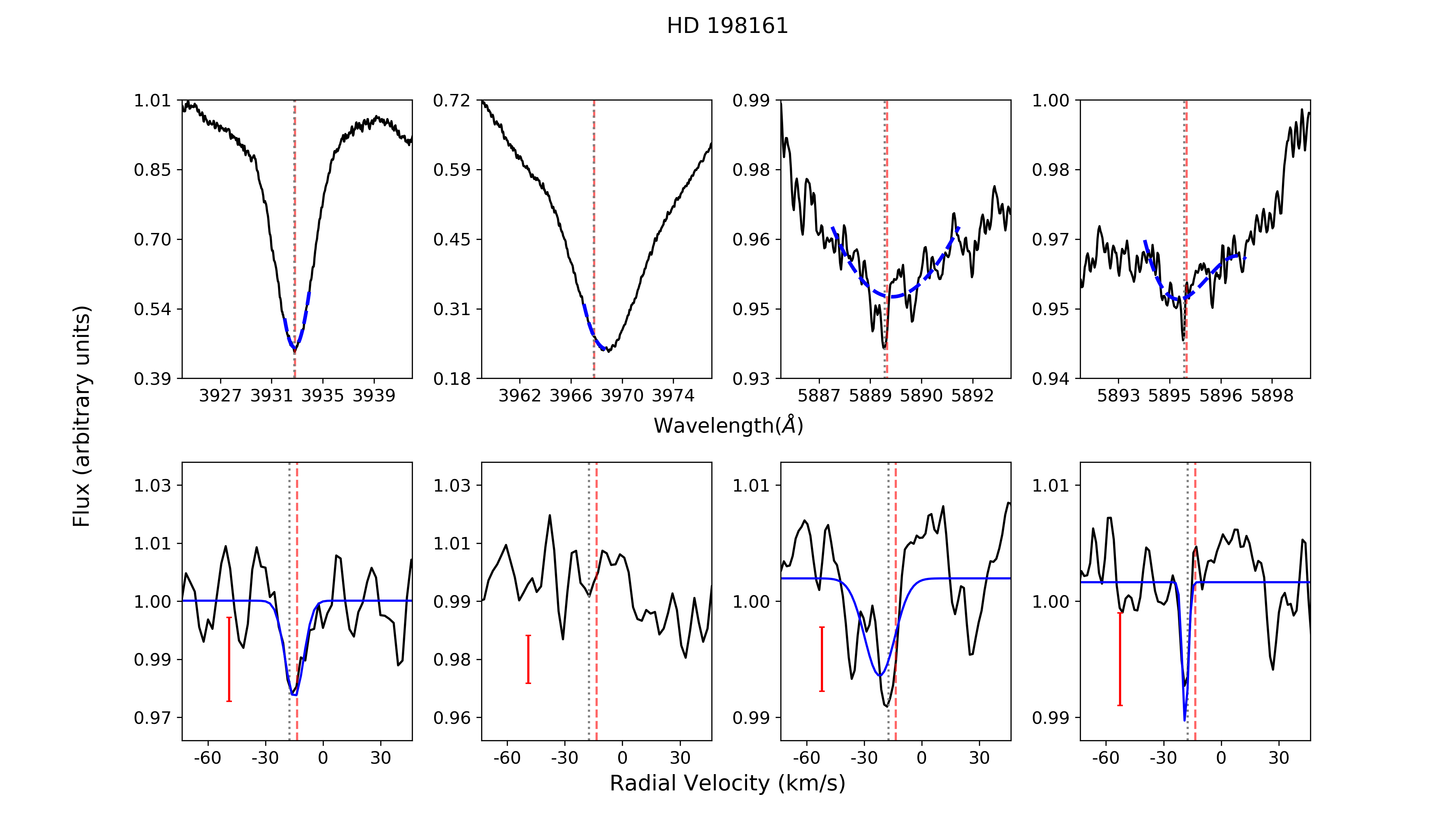} 
\caption{Stars showing narrow non-photospheric absorptions. Top panels: Photospheric Ca {\sc ii} H \& K and Na {\sc i} D lines with fitted modeling dashed blue line, x-axis shows the wavelength. Bottom panels: Residuals once the spectrum is divided by the photosphere, x-axis in velocity. Blue lines mark the fits to the non-photospheric absorptions. Vertical red dashed and grey dotted lines represent the stellar radial velocity and the ISM velocities respectively.  Red error bars show three sigma value measured in the continuum adjacent to the photospheric line. }
\end{figure}

\renewcommand{\thefigure}{\arabic{figure} (Cont.)}
\addtocounter{figure}{-1}

\begin{figure}
\centering 
\includegraphics[width=1.\textwidth]{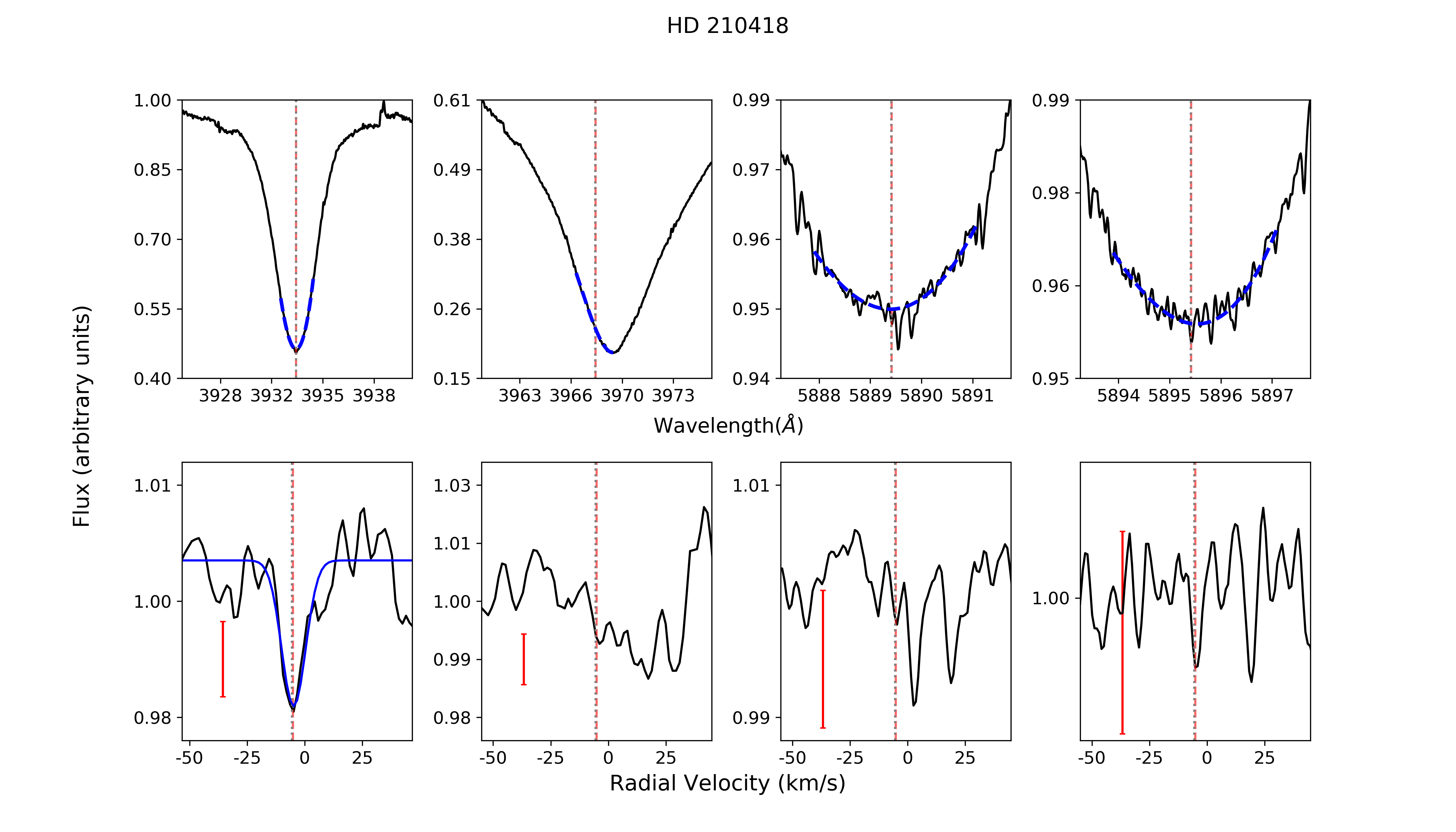}  
\includegraphics[width=1.\textwidth]{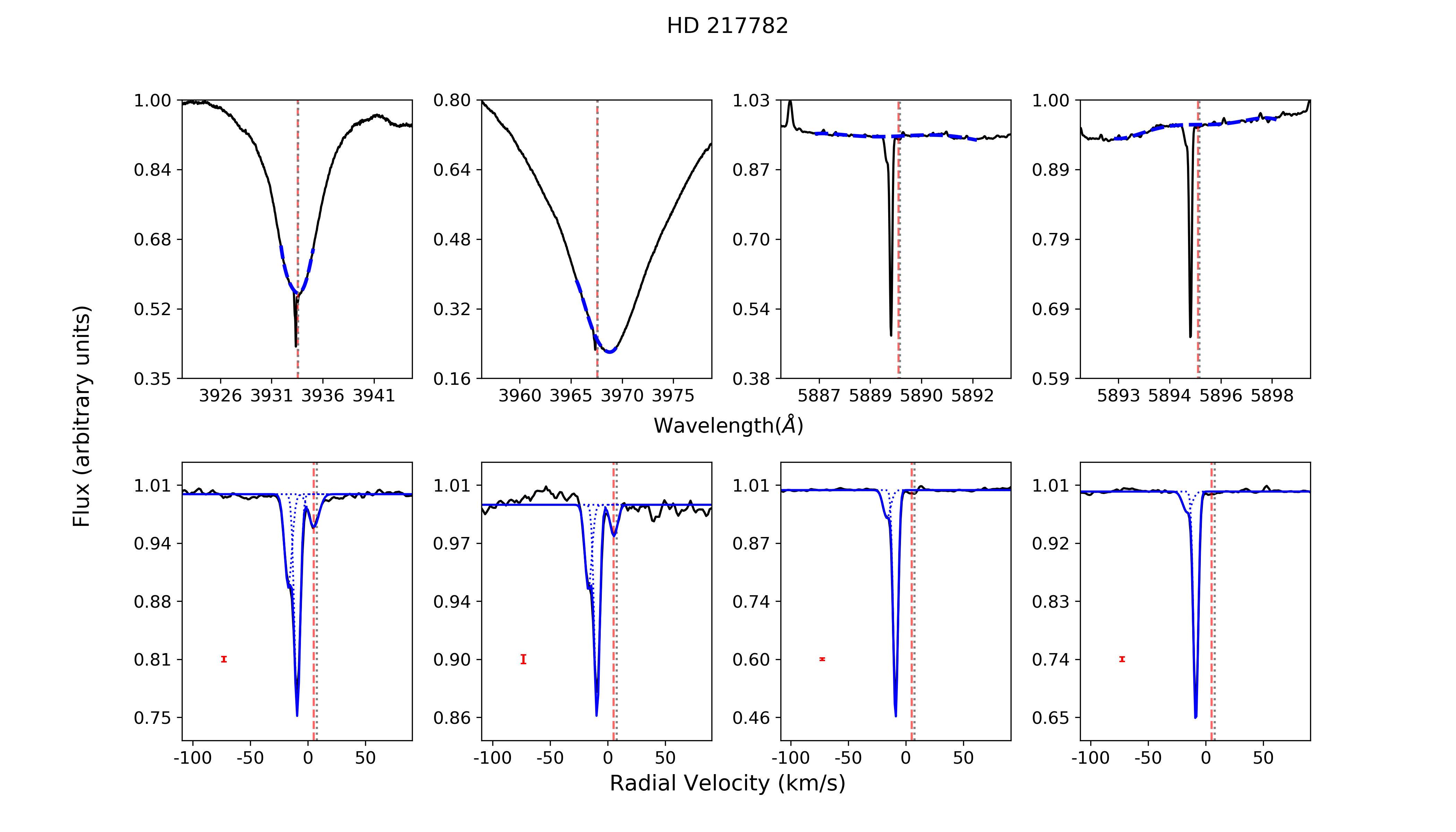}  
\caption{Stars showing narrow non-photospheric absorptions. Top panels: Photospheric Ca {\sc ii} H \& K and Na {\sc i} D lines with fitted modeling dashed blue line, x-axis shows the wavelength. Bottom panels: Residuals once the spectrum is divided by the photosphere, x-axis in velocity. Blue lines mark the fits to the non-photospheric absorptions. Vertical red dashed and grey dotted lines represent the stellar radial velocity and the ISM velocities respectively.  Red error bars show three sigma value measured in the continuum adjacent to the photospheric line. }
\end{figure}

\renewcommand{\thefigure}{\arabic{figure} (Cont.)}
\addtocounter{figure}{-1}

\begin{figure}
\centering
\includegraphics[width=1.\textwidth]{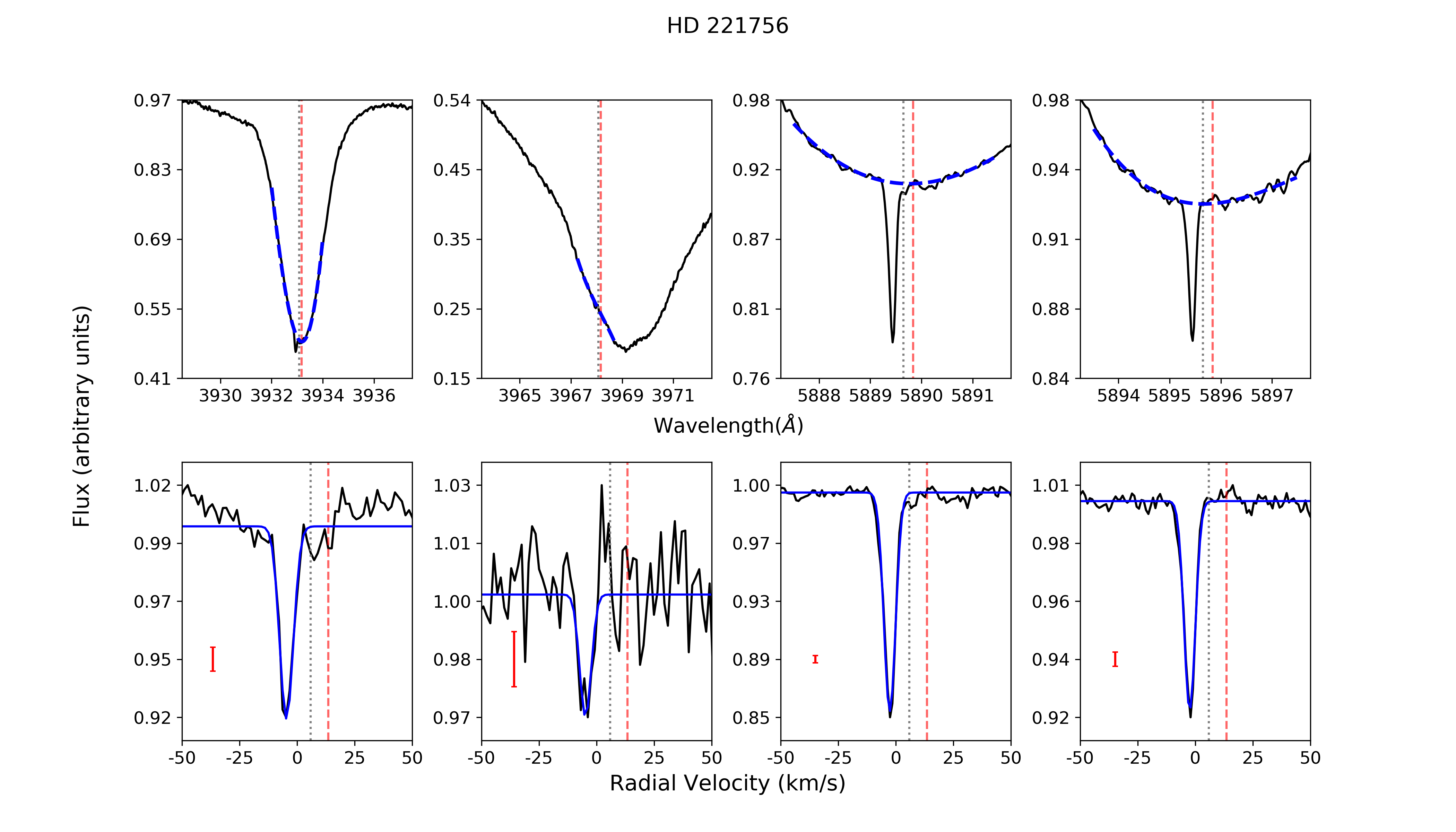} 
\includegraphics[width=1.\textwidth]{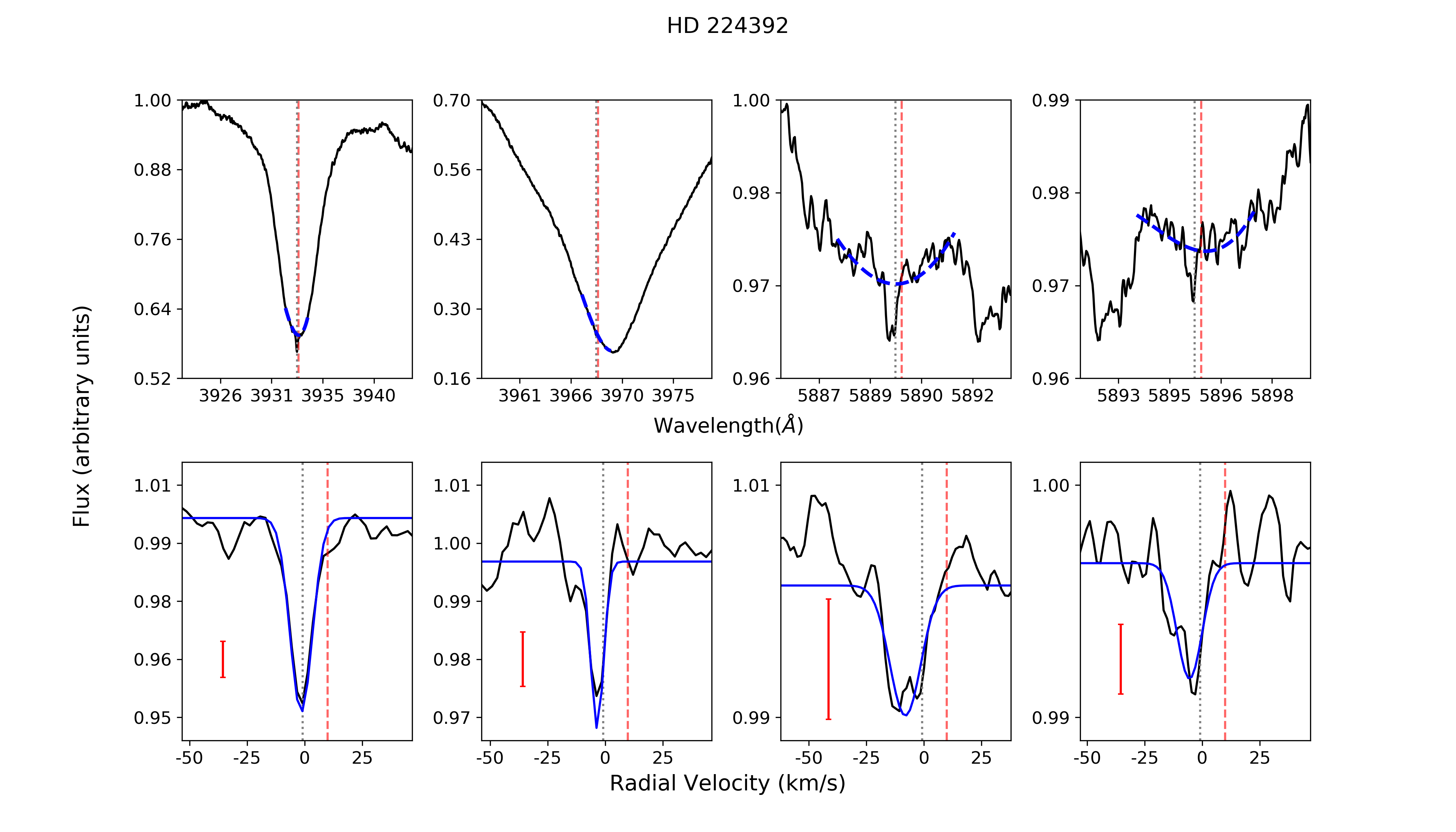} 
\caption{Stars showing narrow non-photospheric absorptions. Top panels: Photospheric Ca {\sc ii} H \& K and Na {\sc i} D lines with fitted modeling dashed blue line, x-axis shows the wavelength. Bottom panels: Residuals once the spectrum is divided by the photosphere, x-axis in velocity. Blue lines mark the fits to the non-photospheric absorptions. Vertical red dashed and grey dotted lines represent the stellar radial velocity and the ISM velocities respectively.  Red error bars show three sigma value measured in the continuum adjacent to the photospheric line. }
\label{fig:na_absorptions}
\end{figure}

\newpage

\onecolumn
\section{Tables}

\begin{landscape}
\begin{tiny}
\begin{center}
\begin{longtable}{llcclrlrrrrrr}
\caption{Sample of observed stars. Columns are self-explanatory. Numbers within parenthesis in columns 10 to 13 denote the corresponding references. Column 13 specifies the primary selection criteria as described in the text: 1. Previously detected exocomets; 2. Debris Discs; 3. Debris discs with cold gas; 4. Near-Infrared excesses; 5. Young discs; 6. Shell Stars; 7. $\lambda$ Boo stars.} \\
\hline
\hline

HD     &  Other Name     &    RA(2000.0) &    Dec(2000.0) &   Sp. Type & Distance & V   & B-V  & V$_{\rm rad}$ & Age  &  L$_{\rm IR}$ / L${_*}$  & Assoc. & Sel.Crit   \\
       &                 &       &        &            &    (pc)  &(mag)&(mag) & (km/s)    & (Myr)&                      &        &            \\\hline 
\endfirsthead
\multicolumn{12}{c}%
{{ \tablename\ \thetable{} -- continued from previous page}} \\
\hline
\hline
HD     &  Other Name     &    RA(2000.0) &    Dec(2000.0) &   Sp. Type & Distance & V   & B-V  & v$_{\rm rad}$ & Age  &  L$_{\rm IR}$ / L${_\star}$  & Assoc. &  Sel.Crit   \\
       &                 &       &        &            &    (pc)  &(mag)&(mag) & (km/s)    & (Myr)&                      &        &             \\\hline 

\endhead
\hline
\multicolumn{12}{r}{{Continued on next page}} \\
\endfoot

\endlastfoot

105    &  HIP 490        &   $00:05:52.54$ & $-41:45:11.0$ & G0V     & 38.85  &  7.53 & 0.6    &   1.7    & 45  (9)      &   2.4$\cdot$ 10$^{-4}$ (1)   & Tuc-Hor (4) &  5 (1)   \\
203    &  HR 9           &   $00:06:50.09$ & $-23:06:27.1$ & F3V     & 39.96  &  6.19 & 0.344  &   6.5    & 23  (2)      &   1.6$\cdot$ 10$^{-4}$ (2)   & BPMG (3)    &  5 (2)   \\
256    &  HR 10          &   $00:07:18.27$ & $-17:23:13.2$ & A2IV/V  & 145.18 &  6.23 & 0.1    &   -10.2  & 549  (3)     &   <6.1$\cdot$ 10$^{-6}$ (3)  &             &  1 (3)   \\
1466   &  HIP 1481       &   $00:18:26.12$ & $-63:28:39.0$ & F8V     & 42.97  &  7.46 & 0.54   &   6.4    & 45  (9)      &   6.3$\cdot$ 10$^{-5}$ (1)   & Tuc-Hor (4) &  5 (1)   \\
2262   &  $\kappa$ Phe   &   $00:26:12.20$ & $-43:40:47.4$ & A5IV    & 23.81  &  3.94 & 0.17   &   11.3   & 200  (1)     &   7.4$\cdot$ 10$^{-6}$ (4)   & Castor (1)  &  4 (4)   \\
2884   &  $\beta$ 1 Tuc  &   $00:31:32.67$ & $-62:57:29.6$ & B9V     & 41.41  &  4.33 & -0.045 &   14.0   & 45  (9)      &   <6.7$\cdot$ 10$^{-6}$ (1)  & Tuc-Hor (4) &  5 (1)   \\
2885   &  $\beta$ 2 Tuc  &   $00:31:33.48$ & $-62:57:56.0$ & A2V     & 51.04  &  4.51 & 0.135  &   9.8    & 45  (9)      &   <1.8$\cdot$ 10$^{-5}$ (1)  & Tuc-Hor (4) &  5 (1)   \\
3003   &  $\beta$ 3 Tuc  &   $00:32:43.91$ & $-63:01:53.4$ & A0V     & 45.90  &  5.09 & 0.04   &   7.7    & 45  (9)      &   1.1$\cdot$ 10$^{-4}$ (1)   & Tuc-Hor (4) &  5 (1)   \\
\end{longtable}
\end{center}
\end{tiny}
\end{landscape}

\newpage

\begin{tiny}
\begin{center}
\begin{longtable}{l|c|c|c|r|r|c|r|c}
\caption{Stellar Parameters estimated  according to Sect. 3.2.}\\
\hline
\hline

   HD    &    RA(2000.0)         &    Dec(2000.0)        & Sp. Type   & \multicolumn{1}{c|}{$v_{\rm rad}$} &  \multicolumn{1}{c|}{$T_{\rm eff}$}  & $\log g$  & \multicolumn{1}{c|}{[M/H]} & $v\sin i$  \\
         &         &         &            & \multicolumn{1}{c|}{(km/s)}     &   \multicolumn{1}{c|}{(K)}       &   [cgs]   &                            & (km/s)          \\\hline
\endfirsthead
\multicolumn{9}{c}%
{{ \tablename\ \thetable{} -- continued from previous page}} \\
\hline\hline
HD    &    RA(2000.0)         &    Dec(2000.0)        & Sp. Type   & \multicolumn{1}{c|}{$v_{\rm rad}$} &  \multicolumn{1}{c|}{$T_{\rm eff}$}  & $\log g$  & \multicolumn{1}{c|}{[M/H]} & $v\sin i$  \\
      &         &        &            & \multicolumn{1}{c|}{(km/s)}     &    \multicolumn{1}{c|}{(K)}      &   [cgs]   &                            & (km/s)          \\\hline
\endhead
\hline
\multicolumn{9}{r}{{Continued on next page}} \\
\endfoot

\endlastfoot
  105    & $00:05:52.54$ & $-41:45:11.0$ & G0V     &$ +1.6\pm0.4$        &  6000$\pm$60   &  4.38$\pm$0.14 & $-0.03$  & 14.6   \\
  203    & $00:06:50.09$ & $-23:06:27.1$ & F3V     &$ +7.1\pm1.7$        &  6850$\pm$200  &  4.25$\pm$0.30 & $ 0.00$  & 170    \\
  256    & $00:07:18.27$ & $-17:23:13.2$ & A2IV/V  &$ -2.2\pm0.3$        &  8690$\pm$50   &  3.48$\pm$0.20 & $ 0.00$  & 270     \\
  1466   & $00:18:26.12$ & $-63:28:39.0$ & F8V     &$ +6.4\pm0.4$        &  6270$\pm$60   &  4.45$\pm$0.15 & $-0.01$  & 18     \\
  2262   & $00:26:12.20$ & $-43:40:47.4$ & A5IV    &$+11.3\pm3.7$        &  8110$\pm$30   &  4.01$\pm$0.13 & $ 0.00$  & 200    \\
  2884   & $00:31:32.67$ & $-62:57:29.6$ & B9V     &$ +8.3\pm1.6$        & 11620$\pm$30   &  4.45$\pm$0.10 & $-0.05$  & 135    \\
  2885   & $00:31:33.48$ & $-62:57:56.0$ & A2V     &$+13.5\pm3.9$        &  9630$\pm$30   &  4.27$\pm$0.10 & $+0.20$  & 40     \\
  3003   & $00:32:43.91$ & $-63:01:53.4$ & A0V     &$ +8.2\pm2.3$        &  9490$\pm$50   &  4.35$\pm$0.14 & $ 0.00$  & 95     \\
\end{longtable}
\end{center}
\end{tiny}

\newpage

\begin{small}
\begin{landscape}
\begin{longtable}{lcrccccccccc}

\caption{Plausible origin of the observed narrow non-photospheric absorptions  for each star. Equivalent widths (EW), radial velocities (RV), full width half maximum (FWHM), and column densities (N) are given for each absorption as estimated from the median spectra. Uncertainty estimates of these parameters are detailed in Sect. 4.1. as follows: 10\% for EW; 2.5 and 1.5 km/s for RV and FWHM in the Ca and Na regime respectively; and propagated errors of those values for N (approx. $\sim$ 10\%). Radial velocities of the stars (v$_{\rm rad}$) and the radial velocity of the interstellar clouds (v$_{\rm ISM}$), if present, are also given. The convention for the plausible origin is the following. CS: circumstellar; ISM: interstellar medium; CS+ISM: both types origins are plausible. A question mark indicates that the origin is ambiguous. Absorptions marked with an asterisk (*) in the EW are triangular shell-like absorptions.} \\

\hline 
\hline

Star         &   V$_{rad}$   &      V$_{ISM}$  & \multicolumn{4}{c}{Ca II K/H}                  & \multicolumn{4}{c}{Na I D2/D1}   & Origin                     \\ 
\hline
             &               &                 &    EW     &    RV   &  FWHM   & N            &    EW     &    RV   &  FWHM   & N             &            \\ 
             &   (km/s)      &    (km/s)       &   ($m\AA$)&  (km/s) &  (km/s) & (cm$^{-2}$)  & ($m\AA$)  &  (km/s) &  (km/s) & (cm$^{-2}$)  &           \\
\hline
\hline
\endfirsthead

\multicolumn{12}{c}%
{{ \tablename\ \thetable{} -- continued from previous page}} \\
Star         &   V$_{rad}$   &      V$_{ISM}$  & \multicolumn{4}{c}{Ca II K/H}                  & \multicolumn{4}{c}{Na I D2/D1}   & Origin     \\ 
\hline
             &               &                 &    EW     &    RV   &  FWHM   & N            &    EW     &    RV   &  FWHM   & N          &             \\ 
             &   (km/s)      &    (km/s)       &   ($m\AA$)&  (km/s) &  (km/s) & (cm$^{-2}$)  & ($m\AA$)  &  (km/s) &  (km/s) & (cm$^{-2}$)  &             \\
\hline
\hline
\endhead

\multicolumn{12}{r}{{Continued on next page}} \\
\endfoot

\endlastfoot

HD 256       & -2.2$\pm$0.3 & 3.16  (LIC)     &   190.9*/180.3*  &  -1.8/-0.82   &   10.67/11.05  & 4.88$\cdot$ 10$^{12}$ &  1.2/0.4     &   3.2/3.3   &  4.17/1.80 & 5.91$\cdot$ 10$^{9}$  &   CS     \\
             &              & -2.73 (Mic)     &           &         &          &                      &          &         &         &                   &          \\
             &              & 1.01  (Cet)     &           &         &          &                      &          &         &         &                   &          \\
                                                                                                                                                         
HD 2884      &  8.3$\pm$1.6 & 13.85 (Dor)     &  2.1/1.66      &  0.5/4.3    &  12.13/7.1   & 4.28$\cdot$ 10$^{10}$    &          &         &         &                   &   ? \\
             &              & 2.46  (Vel)     &           &         &          &                      &          &         &         &                   &          \\
                                                                                                                                    
HD 2885      &  13.5$\pm$3.9& 13.84 (Dor)     &           &         &          &                      &   56.8/52.3   &   -2.0/-2.5  &   17.68/16.1 & 1.85$\cdot$ 10$^{12}$ &   ? \\
             &              & 2.46  (Vel)     &           &         &          &                      &          &         &         &                   &          \\ 
                                                                                                                                     
HD 3003      &  8.2$\pm$2.3 & 13.84 (Dor)     &  4.5/<1.5 &  0.6/-    &  24.55/-   & 4.82$\cdot$ 10$^{10}$&   4.8/2.9    &   -3.7/-3.0  &   6.84/6.57  & 4.61$\cdot$ 10$^{10}$ &   ? \\
\end{longtable}
\end{landscape}
\end{small}

\newpage

\section{Observing Log}

\begin{center}
\begin{longtable}{cllllll}
\caption{Instruments and observing campaigns. In brackets below the star's name SNR in the Ca {\sc ii} K continuum measured in the median for all the available spectra. HEROS observations were taken on service mode during several months as complementary observations. *On March 3$^{rd}$ and 21$^{st}$ during service mode only one spectra was obtained each night.}\\

\endfirsthead
\caption{Observing log} \\
\endhead

\hline
\hline

 Star        & Date (UT)       & Instrument      & Date (UT)       & Instrument  &  Date (UT)       & Instrument        \\
             &                 &                 &                 &             &                  &                   \\
 \hline                            
 \endhead
             &                 &                 &                 &             &                  &                   \\
 HD 105      &  20151022T0403  & FEROS           &  20151023T0354  & FEROS       & 20151024T0421    & FEROS             \\
 (108)       &                 &                 &                 &             &                  &                   \\
 \hline
 HD 203      &  20150803T0314  & HEROS           &  20150803T0400  & HEROS       &  20150804T0427  & HEROS             \\
 (212)       &  20150805T0311  & HEROS           &  20150806T0230  & HEROS       &  20150919T2344  & HEROS             \\
             &  20150920T2324  & HEROS           &  20150923T2307  & HEROS       &  20150924T2300  & HEROS             \\
             &  20151022T0401  & FEROS           &  20151023T0338  & FEROS       &  20151024T0428  & FEROS             \\       
 \hline      
 HR 10       &  20150904T0007  & HERMES          &  20150904T0020  & HERMES      &  20150904T0051  & HERMES            \\
\end{longtable}
\end{center}

\twocolumn

\end{appendix}

\end{document}